\documentclass[a4paper,11pt]{article}
\pdfoutput=1
\usepackage{jheppub}
\usepackage{amsmath,amssymb,bm,graphicx}
\usepackage[framemethod=TikZ]{mdframed}

\allowdisplaybreaks 
\def\tr{\mathop{\rm tr}\nolimits}

\newcommand{\df}{\mathrm{d}}
\newcommand{\nn}{\nonumber}
\newcommand{\img}{{\rm i}}

\newcommand{\cC}{\mathcal{C}}
\newcommand{\cE}{\mathcal{E}}
\newcommand{\cN}{\mathcal{N}}

\newcommand{\si}{\sigma}

\newcommand\as{\alpha_s}

\newcommand{\nf}{n_f}
\newcommand{\nc}{N_c}
\newcommand{\cf}{C_F}
\newcommand{\ca}{C_A}
\newcommand{\tf}{T_F}
\newcommand{\dabcd}{d^{abcd}}

\newcommand{\gammacusp}{\Gamma_\text{cusp}^q}

\newcommand{\EEC}{{\text{EEC}}}

\def\Fig#1{fig.~{\ref{#1}}}
\DeclareRobustCommand{\Sec}[1]{sec.~\ref{sec:#1}}
\DeclareRobustCommand{\sec}[1]{sec.~\ref{sec:#1}}
\DeclareRobustCommand{\Eq}[1]{eq.~(\ref{eq:#1})}
\DeclareRobustCommand{\eq}[1]{eq.~(\ref{eq:#1})}

\newcommand{\fd}[2]{\parbox{#1}{\includegraphics[width=#1]{#2}}}

\title{From DGLAP to Sudakov:\\ Precision Predictions for Energy-Energy Correlators}
\author[1,2,3]{Max Jaarsma,}
\author[3]{Yibei Li,}
\author[4]{Ian Moult,}
\author[1,2]{Wouter Waalewijn,}
\author[5]{Hua Xing Zhu}
\affiliation[1]{Nikhef, Theory Group,
	Science Park 105, 1098 XG, Amsterdam, The Netherlands}
\affiliation[2]{Institute for Theoretical Physics Amsterdam and Delta Institute for 
 Theoretical Physics, University of Amsterdam, Science Park 904, 1098 XH Amsterdam, The Netherlands}
\affiliation[3]{PRISMA$^+$ Cluster of Excellence \& Mainz Institute for Theoretical Physics, \\
Johannes Gutenberg University, Staudingerweg 7, 55099 Mainz, Germany}
\affiliation[4]{Department of Physics, Yale University, New Haven, CT 06511, USA\vspace{0.5ex}}
\affiliation[5]{School of Physics, Peking University, Beijing 100871, China}

\emailAdd{m.jaarsma@uva.nl}
\emailAdd{yibei.li@uni-mainz.de}
\emailAdd{ian.moult@yale.edu}
\emailAdd{w.j.waalewijn@uva.nl}
\emailAdd{zhuhx@pku.edu.cn}

\abstract{Correlations in the distribution of energy produced in collider experiments provide a snapshot of the microscopic dynamics of QCD, and its evolution from asymptotically free quarks and gluons, to confined hadrons. 
There has recently been considerable progress in the interpretation and precision calculation of these correlations, using a specific class of observables called energy correlators (EECs). 
These observables are most cleanly studied in $e^+e^-$ collisions, where they can be measured over their full angular range. 
Of particular interest are kinematic limits of the correlator, both collinear, and back-to-back, where the correlator exhibits scaling behaviors governed by specific operators in QCD. 
Resolving these scalings requires measurements with exceptional angular resolution, which can be achieved by performing measurements on tracks (charged particles).
In this paper we perform the first calculation of the track-based EEC over its entire kinematic range, achieving a record precision of of NNLL (collinear) + NNLO (fixed order) + NNNNLL (back-to-back) for the track-based EEC, and additionally incorporate the leading non-perturbative corrections and their resummation, including the Collins-Soper kernel computed using lattice QCD.
We describe the breadth of physics probed by this observable, and highlight the impact of different components of our factorization theorem on the final distribution.
Combined with recent measurements of the track-based EEC with archival LEP data, our calculation initiates the precision study of track-based observables at LEP, which will lead to new insights into the dynamics of QCD, and the precision extraction of its underlying parameters.}

\begin{document}
\begin{flushright}
{\small
MITP-24-091\\
}
\end{flushright}

\maketitle

\section{Introduction}

The richness and complexity of Quantum Chromodynamics (QCD) arises from its vastly different behavior as a function of scale: at high energies it is an asymptotically free theory of quarks and gluons, while at low energies it is a gapped theory with quarks and gluons confined to hadrons. The complete flow from high to low energies, as well as the confinement transition connecting the two regions, can be studied in collider experiments, where it can be related to the patterns in the energy flux at different angular scales. Detailed measurements of patterns of energy flux in collider experiments provide one of our best means of improving our understanding of QCD, as well as for performing precision measurements of its parameters. These in turn affect our global understanding of the Standard Model, and our ability to search for new physics.

The cleanest type of collisions, both theoretically and experimentally, are $e^+e^-$ collisions, where the final state is produced by the action of the electroweak current, $J^\mu$, on the QCD vacuum. The asymptotic measurement of energy flux at a specific point on the celestial sphere, characterized by a unit vector $\hat n$, can be expressed in terms of the stress tensor $T_{\mu\nu}$ of the theory as a so called ``detector operator", first introduced via its action on asymptotic states 50 years ago by Sterman \cite{Sterman:1975xv}. In the generic case involving massive radiation, the detector operator for energy flow can be written as~\cite{Sveshnikov:1995vi,Tkachov:1995kk,Korchemsky:1999kt,Bauer:2008dt}
\begin{align}\label{eq:ANEC_op}
\mathcal{E}(\hat n) = \lim_{r\to \infty}  \int\limits_0^\infty \df t\, r^2 n_i T_{0i}(t,r \hat n)=   \fd{3cm}{figures/ANE_operator}\,.
\end{align}
Here we have also illustrated the definition of the energy flow operator in a Penrose diagram. Radiation, illustrated as the squiggly line originating from the origin, is detected by the integrated energy detector, shown in dashed blue prior to taking the $r \to \infty$ limit, and in solid blue after taking the limit so that the integration lies on future null infinity. 
This operator is also referred to as the average null energy operator (ANE). It plays an important role in formal quantum field theory and gravity, due to the fact that it satisfies a positivity condition, the so called average null energy condition (ANEC), namely that $\langle \psi | {\cal E}(n) | \psi \rangle \geq 0$ in any state \cite{Faulkner:2016mzt,Hartman:2016lgu}.

Electron-positron ($e^+e^-$) collider experiments enable the direct experimental measurement of correlation functions of these energy flow operators
\begin{align}
\int \df ^4x\, e^{\img q\cdot x}\, \langle0| J(x) \cE(\hat n_1) \cE(\hat n_2)  \cdots \cE(\hat n_k) J(0)|0 \rangle \equiv \langle \cE(\hat n_1) \cE(\hat n_2)  \cdots \cE(\hat n_k) \rangle \,,
\end{align}
where $q^\mu=(Q,0,0,0)$ is  the momentum of the $e^+e^-$ collision. These observables are referred to as energy correlators \cite{Basham:1979gh,Basham:1978zq,Basham:1978bw,Basham:1977iq}. In a rotationally invariant state, they are functions of the angles, $z_{ij}=\frac{1}{2}(1-\hat n_i \cdot \hat n_j)$ between the detector operators.

In this paper we will focus on the correlation function of two energy flow operators, the so-called ``energy-energy correlator" (EEC)~\cite{Basham:1979gh,Basham:1978zq,Basham:1978bw,Basham:1977iq}
\begin{align}\label{eq:EEC_op_def}
\hspace{-0.35cm}\text{EEC}&=
 \int \df^4x\, \frac{e^{\img q\cdot x}}{\sigma Q^2} L_{\mu \nu} \langle0| J^\mu(x) \cE(\hat{n}_1) \cE(\hat{n}_2) J^\nu(0)|0 \rangle  =\fd{3cm}{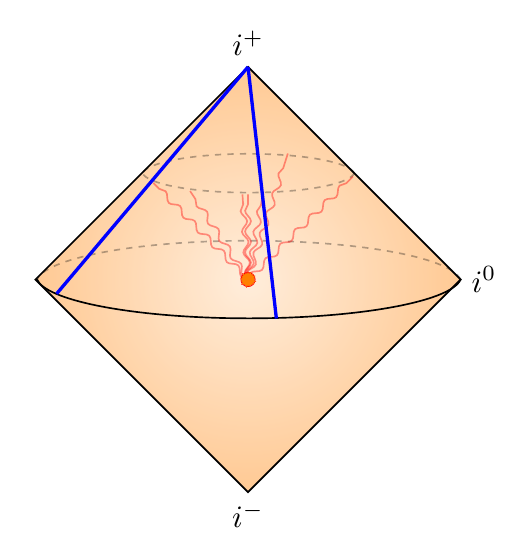}
\,,
\end{align}
which is a function of a single angle $z =\frac{1}{2}(1-\hat{n}_1 \cdot \hat{n}_2)$.

These correlation functions can be measured experimentally in colliders using the action of the energy flow operators on asymptotic states
\begin{align}
\mathcal{E}(\hat{n}) |X\rangle= \sum_i k_i^0 \delta(\Omega_{\vec{n}} -\Omega_{\vec{k}_i})|X\rangle\,,
\end{align}
which allow us to express the energy correlator as
\begin{align}
  \label{eq:EEC_def}
  \text{EEC}(z)= \sum_{i,j}\int \df \sigma\ \frac{E_i E_j}{\sigma Q^2}\, \delta\Bigl(z - \frac{1 - \cos\theta_{ij}}{2}\Bigr) =\fd{5cm}{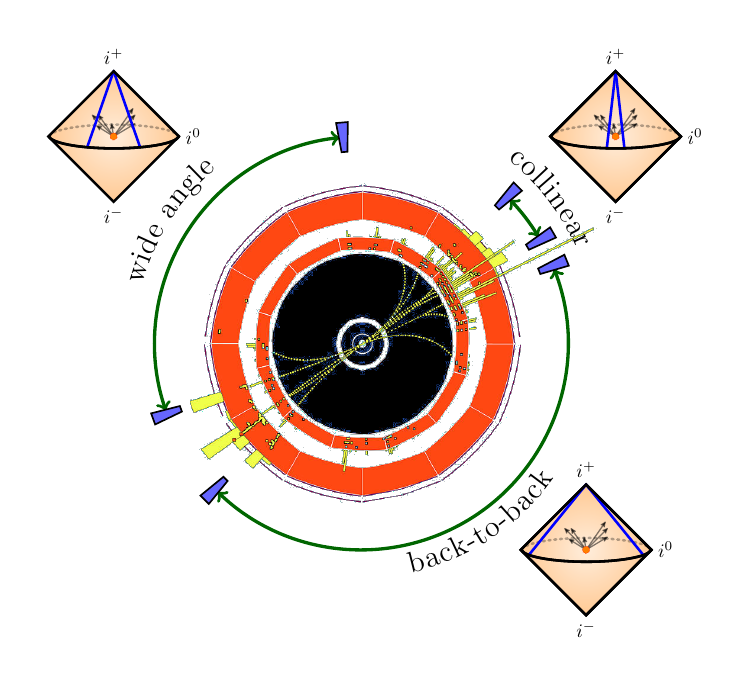} \,.
\end{align}
 Here $E_i$ is the energy of particle $i$, and $\theta_{ij}$ is the angle between a pair of particles $(i,j)$. We have also illustrated the definition of the correlator from a ``collider-centric" perspective, illustrating a pair of correlated particles that contribute to the correlation function at a particular angle.

These correlation functions, being simultaneously measureable in experiment, and directly expressed in terms of operators of the underlying theory, are a prime target for the interaction of theory an experiment. They were extensively measured starting with the PLUTO experiment \cite{PLUTO:1985yzc,PLUTO:1979vfu}, and have since been measured at CELLO \cite{CELLO:1982rca}, JADE \cite{JADE:1984taa}, MAC \cite{Fernandez:1984db}, MARKII \cite{Wood:1987uf}, TASSO \cite{TASSO:1987mcs}, AMY \cite{AMY:1988yrv}, TOPAZ \cite{TOPAZ:1989yod}, ALEPH \cite{ALEPH:1990vew}, L3 \cite{L3:1991qlf,L3:1992btq}, DELPHI \cite{DELPHI:1990sof}, OPAL \cite{OPAL:1990reb,OPAL:1991uui}, SLD \cite{SLD:1994idb}, and attempts to understand these measurements theoretically were key to the development of many techniques in perturbative QFT, as we will review in \sec{history}.

More recently, energy correlators have been extensively developed, and experimentally studied beyond electron-positron colliders. For a detailed review, see \cite{Moult:2025nhu}. This program was initiated in \cite{Dixon:2019uzg,Chen:2020vvp,Komiske:2022enw}, where energy correlators in the collinear limit were identified as phenomenologically powerful jet observables. Energy correlator observables have now been measured in $e^+e^-$ \cite{Bossi:2024qeu,Bossi:2025xsi}, $e p$ \cite{HERA},  proton-proton (p-p) \cite{CMS:2024mlf,ALICE:2024dfl,ALICE:2025igw,talk_Hwang,Tamis:2023guc,STAR:2025jut,talk_Shen}, proton-lead (p-Pb) \cite{talk_Anjali,talk_Anjali2}, and lead-lead (Pb-Pb) \cite{talk_Ananya,CMS-PAS-HIN-23-004,CMS:2025jam,CMS:2025ydi} collisions, and used to study a wide variety of phenomena across particle and nuclear physics. Examples include the strong coupling constant \cite{Chen:2023zlx}, the top quark mass \cite{Holguin:2022epo,Holguin:2023bjf,Holguin:2024tkz,Xiao:2024rol}, and the physics of the quark-gluon plasma \cite{Andres:2022ovj,Yang:2023dwc,Bossi:2024qho}, cold nuclear matter \cite{Devereaux:2023vjz,Barata:2024wsu,Fu:2024pic}, saturation \cite{Liu:2022wop,Liu:2023aqb}, charmonium \cite{Chen:2024nfl,talk_Shen}, and proton structure \cite{Cao:2023oef,Liu:2024kqt}.

Much like for scattering amplitudes, energy correlators exhibit interesting dependencies on the kinematics of the detectors, described by the variable $z$. These detectors can be moved all the way from the collinear limit, $z\to 0$, where they are on top of each other, to the back-to-back limit $z\to 1$, where they are anti-podally separated on the celestial sphere. These confiugrations are illustrated as
\begin{equation}
\fd{3cm}{figures/Penrose_collinear} \quad
   \fd{3cm}{figures/Penrose_central} \quad
   \fd{3cm}{figures/Penrose_b2b}
\end{equation}
We will discuss in detail the physics of these different limits in \sec{overview}. The leading behavior of the energy correlator observables in the collinear and back-to-back limits is controlled by specific operators in QCD, and is universal. At a practical level, this enables extremely precise calculations of the observables in these limits. From a theoretical perspective, it allows one to map out the renormalization group flows of specific QCD operators and directly see the scaling associated with their anomalous dimensions. There has been tremendous recent progress in understanding the behavior the energy correlators in these kinematic limits in a variety of different theories (for a review see \cite{Moult:2025nhu}), making the prospect of studying them in data particularly exciting.

However, unfortunately, it is at this stage that we encounter experimental realities. In a collision at energy $Q$, the scale probed by the correlators is $\mu^2=Q^2 z$ in the collinear limit, and $\mu^2=Q^2 (1-z)$ in the back-to-back limit. At the energy scale of the $Z$-boson mass, $Q=m_Z$, to map out the flow all the way from the collinear to the back-to-back limit,  requires extraordinary angular resolution, well beyond what was achieved in previous measurements.

The renewed study of detector operators has offered an exciting possibility to overcome this longstanding issue. Instead of study energy flow operators, we can study the energy flow on charged particles (tracks), by making the simple substitution
\begin{align}\label{eq:replace_tracks}
\cE(\hat n) \to  \cE_{\text{tr}}(\hat n)\,,
\end{align}
where the subscript ``tr" indicates that this detector operator only measures the energy flux on charged hadrons. Experimentally, this enables the use of high angular resolution tracking detectors, which can achieve the required angular precision to study kinematic limits of the energy correlators. However, correlation functions of the $\cE_{\text{tr}}$ operator are not infrared and collinear (IRC) safe, since they depend on the spectrum of hadrons in QCD, and therefore they cannot be computed solely in perturbation theory. Since our main motivation was to have an observable with a clean connection to theory, this is problematic.

In the last decade, driven primarily by the development of the soft-collinear effective theory (SCET) \cite{Bauer:2000ew,Bauer:2000yr,Bauer:2001ct,Bauer:2001yt,Rothstein:2016bsq,Beneke:2002ph}, there has been tremendous progress in our understanding of non-IRC safe observables. Combining effective field theory and renormalization group techniques, this enables the rigorous separation of perturbative and non-perturbative contributions to observables, as well as the identification of universal non-perturbative matrix elements. A key development in this area was the introduction of the track function formalism \cite{Chang:2013iba,Chang:2013rca}, which enables the systematic calculation of track-based observables. In the context of energy correlator observables, this formalism has been extensively developed in \cite{Jaarsma:2023ell,Chen:2022muj,Chen:2022pdu,Jaarsma:2022kdd,Li:2021zcf} enabling multi-loop calculations and high order resummation in kinematic limits. A key outcome of these studies is that the important properties of the energy correlator observables in kinematic limits, persist under the replacement in eq.~\eqref{eq:replace_tracks}. Correlation functions of $\cE_{\text{tr}}$ therefore provide a genuine meeting between theory and experiment.

Motivated by these developments, there has been a re-analysis of archival LEP data, from both the ALEPH and DELPHI experiments, which achieved a measurement of the energy correlator on tracks with extraordinary angular resolution  \cite{talk_ALEPH,Bossi:2025xsi,Bossi:2024qeu,Zhang:2025nlf}. These measurements motivate an equally precise theoretical calculation of the energy correlator, combining all known theoretical ingredients, and achieving state of the art theoretical precision in all kinematic limits.

In this paper we perform a record precision calculation of the energy correlator on tracks, providing a complete description in all kinematic limits. Our result combines numerous perturbative ingredients to achieve next-to-next-leading order (NNLO) fixed order, combined with next-to-next-to-leading logarithmic (NNLL) resummation in the collinear limit, with next-to-next-to-next-to-next-to-leading logarithmic (NNNNLL) resummation in the back-to-back limit. We denote this as $\text{NNLO}+\text{NNLL}_\text{col}+\text{NNNNLL}_\text{b2b}$. Additionally, we incorporate leading non-perturbative corrections, using both universal parameters extracted from event shape measurements, as well as inputs from lattice QCD. These leading non-perturbative corrections are dressed with LL resummation in the collinear limit, and NLL resummation in the back-to-back limit. This is the most precise calculation ever performed of the energy correlator observable, and we are able to achieve it both for the standard energy correlator, as well as for the energy correlator measured on tracks.   Our final result is shown in \Fig{fig:physics}, compared with data from ALEPH \cite{Electron-PositronAlliance:2025fhk}. This figure was first presented in \cite{Electron-PositronAlliance:2025fhk}. 
The spectacular agreement between theory and data provides a beautiful illustration of the phenomenological impact developments in multi-loop calculations in perturbative QFT, and effective field theory based factorization. The goal of this paper is to describe in detail the theoretical calculation appearing in this result, and its theoretical uncertainties. We also emphasize its dependence on different inputs, as well as highlight directions for future improvement. The calculations in this paper, combined with the re-analysis of archival LEP data \cite{talk_ALEPH,Bossi:2025xsi,Bossi:2024qeu} opens the door to a rich program.

Observables characterizing jets and energy flux have now been studied in QCD for 50 years \cite{Hanson:1975fe}, and there exist innumerable measurements of event shape observables, and their applications to precision studies of QCD. We would therefore like to emphasize why we believe the study of energy correlators using archival data provides a genuine advance.

There are two primary goals in the study of QCD: on the one hand, we want to identify observables that are under extremely good theoretical control, one might call them ``standard candles", which enable us to perform precision extractions of QCD parameters, such as the strong coupling constant, $\alpha_s$. On the other hand, QCD exhibits many phenomena, such as flux tubes and confinement, that we do not understand from first principles. In these cases, measurements of observables which cleanly isolate the underlying physics are particularly valuable, and provide an exciting opportunity to study these phenomena in the laboratory. While studies of QCD event shapes often focus on precision measurements, the fact that these measurements enable the study of such remarkable phenomena as flux tube formation and breaking in relativistic gauge theories, should be more emphasized. Much in analogy with the case of condensed matter physics, where newly observed experimental phenomena drive theory development, we are optimistic that new measurements of phenomena in QCD can motivate exciting theory developments. A unique aspect of the energy correlator observable in this direction, which we will highlight throughout this paper, is that due to its formulation as a correlation function, the \emph{exact same observable} that is measured experimentally in QCD, can also be computed non-perturbatively in related theories, such as $\mathcal{N}=4$ SYM \cite{N4_bootstrap}, effectively providing ``data" in these theories. This allows us to compare the physics of the two theories, which we believe will provide significant insight into phenomena, such as flux tubes, and their differences in confining and conformal gauge theories. In this direction we believe that the complete measurement of the energy correlator performed in \cite{Electron-PositronAlliance:2025fhk}, and reproduced in \Fig{fig:physics} is transformative. The exceptional control over the theoretical interpretation of the energy correlator in different kinematic regions, enabled by its definition as a correlation function, allows us to map out a variety of interesting phenomena in QCD, illustrated by small schematic figures. We will discuss the physics of these different regions in detail in \Sec{overview}.

In the direction of precision physics, one of the primary goals  is to achieve a precision measurement of the strong coupling constant, $\alpha_s$. This can be achieved in numerous ways, including from measurements of jet cross sections at the LHC, precision event shapes in $e^+e^-$, and lattice QCD, see \cite{Huston:2023ofk,dEnterria:2022hzv} for reviews. There is currently a significant discrepancy between extractions from lattice QCD, which gives $\alpha_s=0.1184\pm 0.0008$ \cite{FlavourLatticeAveragingGroupFLAG:2021npn}, and those from precision $e^+e^-$ event shapes computed using field theoretic treatments of non-perturbative power corrections. Such studies have been performed for a number of different event shape observables, including thrust \cite{Abbate:2010xh}, C-parameter \cite{Hoang:2015hka,Hoang:2014wka}, and heavy jet mass \cite{Benitez:2025vsp}, all of which give low values of $\alpha_s$.\footnote{We note that there are similarly large discrepancies in $\alpha_s$ fits from PDFs, with DIS preferring low values of $\alpha_s$, and dijets at the LHC preferring high values of $\alpha_s$.} For example, the most recent determination from the thrust observable gives $\alpha_s=0.1136 \pm 0.0012$ \cite{Benitez:2025vsp}. A number of possible issues with these extractions have been raised, including the treatment of 2-jet and 3-jet power corrections \cite{Nason:2023asn,Nason:2025qbx}, possible issues with the data which was unfolded using old Monte Carlos, or differences in resummation in momentum/position space \cite{Aglietti:2025ezs,Aglietti:2025jdj}. Detailed studies addressing these concerns were performed for the thrust observable in \cite{Benitez:2025vsp,Benitez-Rathgeb:2024ylc}, showing stability under modifications of the fit range and structure of non-perturbative corrections. A recent measurement of the thrust observable using archival data was performed in \cite{Electron-PositronAlliance:2025hze}. Regardless of the ultimate resolution of these current discrepancies, we believe that this indicates a gap in our understanding of QCD, and is important to resolve.

In this respect, we believe that the EEC provides an excellent opportunity to resolve these issues. First, the measurement is on tracks, and is therefore completely independent of previous measurements of event shape observables. Therefore, if one is concerned with the unfolding of the event shape data, it provides an independent check. Second, on the theoretical side, it probes quite different physics than thrust or C-parameter. In the back-to-back limit it is a transverse momentum ($q_T$) type observable with a different structure of non-perturbative power corrections, and the ability to simultaneously fit the collinear and back-to-back limits should be highly constraining. We study the dependence of our EEC prediction on the value of $\alpha_s$, and the leading non-perturbative parameter, $\Omega$. In the back-to-back limit, these parameters have a significant degeneracy, as is familiar for Sudakov observables. However, the different dependence on these parameters in the collinear limit breaks this degeneracy. We are optimistic that this will enable a new and completely independent approach for precision extractions of $\alpha_s$.

An outline of this paper is as follows: In \Sec{overview}, we provide an overview of the physics probed by the energy correlator in different kinematic regions. In \Sec{calc}, we provide a summary of the theoretical framework used in our calculation, highlighting the distinct factorization theorems, and the necessary perturbative and non-perturbative ingredients. This is followed by detailed discussions of the calculations at fixed order, \Sec{fo}, in the collinear limit, \Sec{col}, and in the back-to-back limit, \Sec{b2b}. In \Sec{results}, we present our complete numerical results, and study in detail our sources of uncertainty. 
We discuss numerous avenues for future improvement of our calculation in \Sec{improve}, and conclude in \Sec{conclusion}.

{\bf{Note:}} The complete prediction for the EEC on tracks derived in this paper was first presented in ref.~\cite{talk_ALEPH,Bossi:2024qeu}, where it was compared to a re-analysis of archival LEP data measured with high angular resolution \cite{Bossi:2025xsi}. A more detailed comparison of the data and theory was presented in \cite{Bossi:2025nux}. We thank the authors of \cite{Bossi:2025xsi} for their extensive collaboration.

\section{The Physics of the Energy-Energy Correlator}\label{sec:overview}

In this section we provide a high-level overview of the physics governing the EEC, for which the final result of our calculation is shown in \Fig{fig:physics}. While many observables studied in jet physics are quite involved, the energy correlator is a simple observable with a clean mapping to the underlying physics of QCD. The goal of this section is to provide those who are not experts in QCD an understanding of the physics that enters in \Fig{fig:physics}, without the technical details of the calculation.

In this paper we will focus on the energy correlator measured in $e^+e^-$ collisions, or equivalently, in a state created by a local operator. In the case of $e^+e^-$ collisions in the real world, this local operator is the electromagnetic/electroweak current $J^\mu$.  The total cross section is expressed as the two-point function 
\begin{align}
\si &=\int \df^4x\, e^{\img q\cdot x} L_{\mu \nu} \langle0| J^\mu(x) J^\nu(0)|0 \rangle\,,
\end{align}
where $L_{\mu \nu}$ is the leptonic tensor describing the incoming leptons. Perturbative results for the total cross section, $\sigma$, are collected in App. \ref{sec:total_xsec}.

We define the two-point energy correlator as 
\begin{align}\label{eq:EEC_def_intro}
\text{EEC}(\hat n_1, \hat n_2) &=
\frac{1}{\si Q^2} \int \df^4x\, e^{\img q\cdot x}L_{\mu \nu} \langle0| J^\mu(x) \cE(\hat n_1) \cE(\hat n_2) J^\nu(0)|0 \rangle 
\,.
\end{align}
We will further azimuthally average, defining
\begin{align}
\text{EEC}(z)=
\int \df^2 \hat n_1 \df^2 \hat n_2\, \delta[z -  \tfrac12 (1- \hat n_1\cdot \hat n_2)]\,
\text{EEC}(\hat n_1, \hat n_2)\,,
\end{align}
which reduces the two-point correlator to a function of a single variable
\begin{equation} \label{eq:z_def}
z\equiv \tfrac12 (1- \hat n_1\cdot \hat n_2) = \tfrac12 (1- \cos \chi)
\,,\end{equation}
where $\chi$ is the angle between the two detectors.

As defined in eq.~\eqref{eq:EEC_def_intro}, the EEC is an interesting observable in a generic QFT. Indeed, it is very closely related to the four point function of local operators, $\langle 0| J T T J | 0 \rangle$. This provides a sharp connection between asymptotic collider observables, and the spectrum of operators and their OPE coefficients in the underlying theory.   This connection was advocated, and used to greatly extend the understanding of collider observables in \cite{Hofman:2008ar,Belitsky:2013ofa,Belitsky:2013bja,Belitsky:2014zha,Korchemsky:2015ssa}. Because of these interesting theoretical properties, the energy correlator has been computed in a variety of different theories, and in different states, some of which we will review shortly. For an extensive review, we refer the reader to \cite{Moult:2025nhu}.

In the particular case of QCD, which is a gapped theory of hadrons in the infrared, it is possible to study correlations of detector operators which incorporate properties of the hadrons. In this case, the EEC has a natural generalization where one correlates only energy flow on tracks (charged particles)
\begin{align}
\text{EEC}(\hat n_1, \hat n_2) &=
\frac{1}{\si Q^2} \int \df^4x\, e^{\img q\cdot x}L_{\mu \nu} \langle0| J^\mu(x) \cE_{\text{tr}}(\hat n_1) \cE_\text{tr}(\hat n_2) J^\nu(0)|0 \rangle 
\,. 
\end{align}
This track-based energy correlator is not infrared and collinear safe. However, it exhibits many experimental advantages, since it can be measured with high angular resolution. In this paper, we will show that this track-based energy correlator can be computed to high precision (in fact matching that of the standard energy correlator for the case of QCD). Moreover, we will show that the restriction to tracks does not modify the physics features of the standard energy correlator, namely particular scaling laws in the collinear and back to back region. The track-based energy correlator therefore provides an ideal bridge between theory and real world experimental considerations. Due to this similarity, in this section we focus on developing the physics of the standard energy correlator. We will discuss in detail how tracks are incorporated in our calculations, and their effect on the energy correlator distribution in forthcoming sections.

\subsection{Sum Rules}\label{sec:sum}

An important property of the energy correlator is that it obeys sum rules following from energy and momentum conservation \cite{Korchemsky:2019nzm,Dixon:2019uzg}, or equivalently Ward identities \cite{Kologlu:2019mfz}. In a CFT there are two independent sum rules 
\begin{align}
\int \df z\, (1-z)\, \text{EEC}(z)=\frac{1}{2} \,, \qquad
\int \df z\, z\, \text{EEC}(z)=\frac{1}{2} \,,
\end{align}
which also hold for perturbative QCD, providing important constraints on calculations. In the presence of a mass scale, only the sum rule 
\begin{align}
\int \df z\, \text{EEC}(z)&=1
\end{align}
remains, which also holds non-perturbatively in QCD. While the EEC has linear non-perturbative power corrections, the total cross section (i.e.~the EEC integrated over $z$) has quartic power corrections. This suggests that the sum rule could provide significant constraints on the structure of non-perturbative corrections to the EEC distribution. These sum rules rely crucially on properly incorporating so called contact terms, namely terms proportional to $\delta(z)$ and $\delta(1-z)$ highlighting why it is important to measure them in experiment.

In the case that the EEC is measured only on charged particles, the sum rules also involve properties of the total energy in tracks. In particular, the track-based energy correlator exhibits the following non-perturbative sum rule 
\begin{align}\label{eq:sum_tracks}
\int \df z\, \text{EEC}(z)&=\frac{ \biggl\langle  \Bigl(\sum\limits_{i\in \text{ch}} E_i \Bigr)^2  \biggr\rangle}{Q^2}\,,
\end{align}
where the right-hand side involves the average of the squared  energy fraction of charged particles. This can be computed perturbatively using the track function formalism.

\subsection{The Energy-Energy Correlator in Different Theories}\label{sec:overview_physics}

The primary goal of formulating collider physics in terms of energy correlators, is that these observables can be studied in simplified theories, which exhibit many of the phenomenon of QCD, but in controllable settings. For example, recent progress from the conformal bootstrap is enabling the calculation of four-point functions involving stress tensors in the 3d-Ising model \cite{Chang:2024whx}, and planar \cite{Caron-Huot:2024tzr,Caron-Huot:2022sdy} and even non-planar \cite{Chester:2021aun,Chester:2020dja} $\mathcal{N}=4$ SYM theory. These advances open the door to the non-perturbative calculation of the energy correlator in these theories, with spectacular two-sided bounds recently achieved  for the case of planar $\mathcal{N}=4$ SYM theory \cite{N4_bootstrap}. Combined with precision measurements of the energy correlator in QCD, this will open up the opportunity for a comparison of the energy correlators amongst different theories, allowing precision studies of the impacts of confinement, and other phenomena in QCD. Because of this, we would like to understand the physics contributing to different kinematic regimes of the energy correlator in the sharpest manner possible. In the case of QCD, some of these regimes are not under control, and therefore by studying these kinematic regimes in related theories, we can develop a clearer picture of the phenomena that measurements of the energy correlator in QCD can improve. Therefore, in this section, we discuss in detail the physics of the energy correlator in different theories, building towards the case of real world QCD.

\begin{figure}
    \centering
    \includegraphics[width=0.45\textwidth]{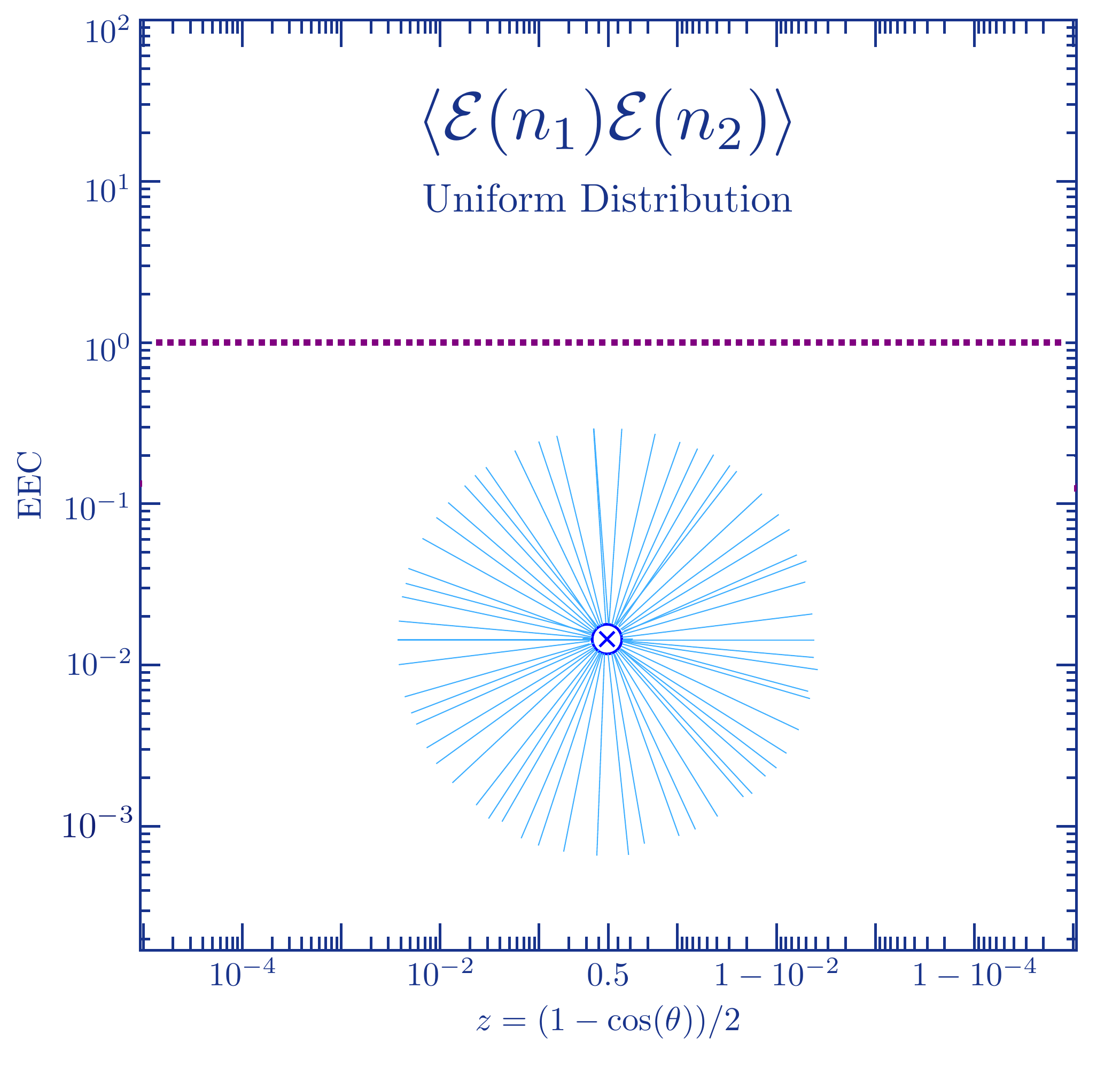}
    \includegraphics[width=0.46\textwidth]{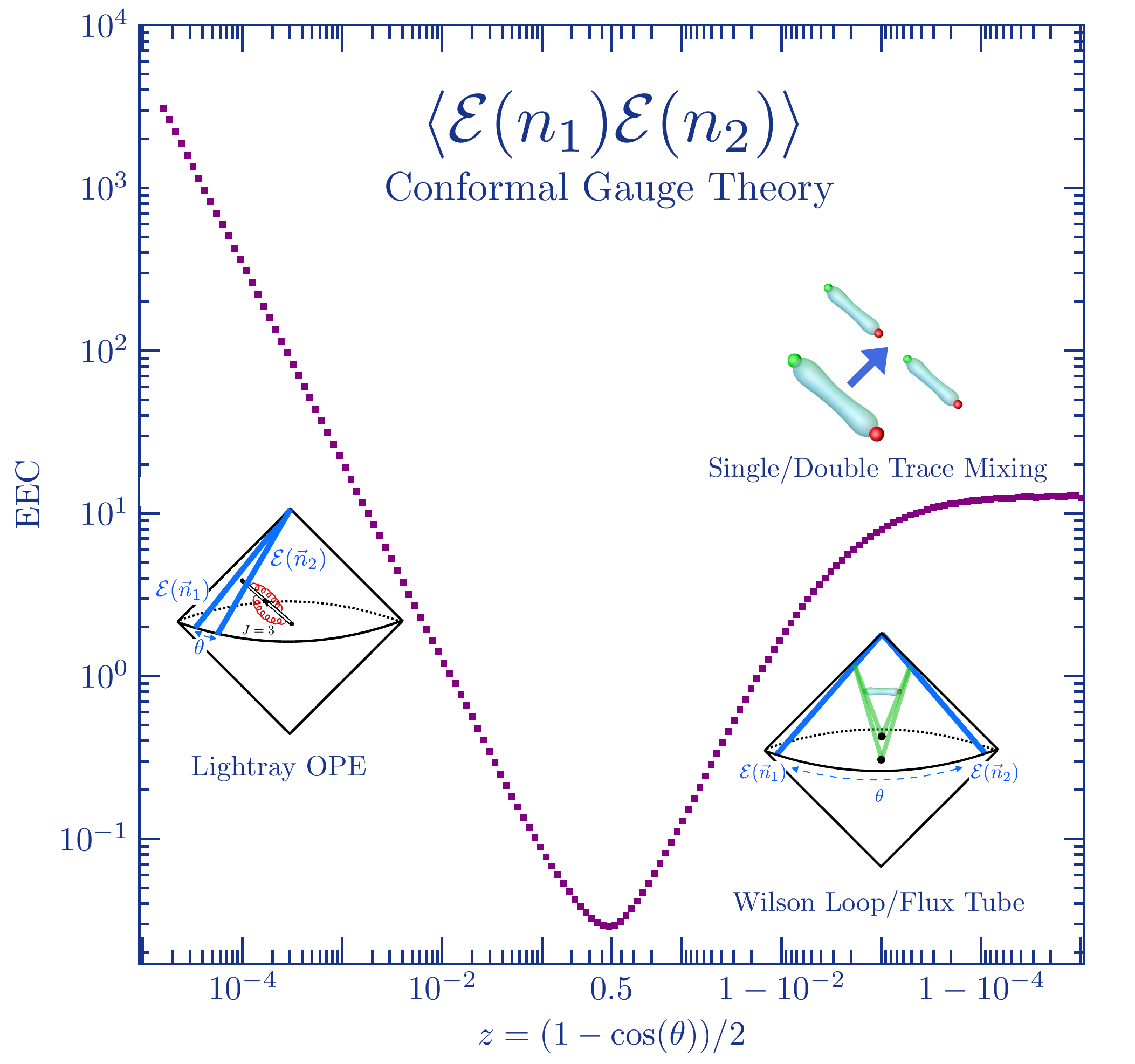}
    \caption{The EEC as ``measured" in toy theories. On the left, we show the distribution arising from a large charge operator in a free scalar theory. On the right, we show the distribution for a conformal gauge theory in $d=4$.}
    \label{fig:toy_physics}
\end{figure}

\noindent {\bf{Free Massless Scalar:}} We begin by considering the simplest case of a free massless scalar in $d=4$. While this may seem quite removed from QCD,  in the infrared, where the measurement of the energy correlators takes place, QCD is a theory of free hadrons. Since the free massless scalar exhibits no dynamics, we can study the behavior of the energy correlator in different states. In particular, we can consider the two-point energy correlator in the state produced by the operator $J=\phi^k$, with $k\to \infty$, mimicking the production of a high multiplicity of particles.  In QCD this high multiplicity of particles (hadrons) is produced through interactions, instead of directly injected by the operator, but nevertheless, we will see that the two theories behave similarly in the deep infrared.

The result for the two-point energy correlator in a $\phi^k$ state takes a particularly simple form at large $k$ \cite{Chicherin:2023gxt}
\begin{align}\label{eq:free_theory}
\langle \cE(\hat n_1) \cE(\hat n_2) \rangle_{\phi^k}=1 +\frac{1}{k}\left( \frac{3}{2}\delta(z) +(9z-6)  \right) +\mathcal{O}(1/k^2)\,.
\end{align}
This distribution is illustrated as toy ``data" in \Fig{fig:toy_physics}.
This highlights two important features. First, the distribution is nearly uniform at large $k$, as expected from the nature of the state. Second, in a states with particle excitations, we generically expect contact terms, illustrated here by the $\delta(z)$ term, which is $1/k$ suppressed. One can also check that the distribution in Eq. \ref{eq:free_theory} satisfies the sum rules to each order in $k$.

Energy correlators were studied more systematically in states produced by heavy half-BPS operators in $\mathcal{N}=4$ SYM in \cite{Chicherin:2023gxt}, and in large charge states in \cite{Firat:2023lbp,Cuomo:2025pjp}.

\medskip

\noindent {\bf{Strongly Coupled Conformal Gauge Theory:}} It is useful to contrast this with the behavior of the energy correlator in a strongly coupled conformal gauge theory. Remarkably, it is possible to calculate the EEC at strong coupling in $\mathcal{N}=4$ SYM using the AdS/CFT correspondence \cite{Gubser:1998bc,Witten:1998qj,Maldacena:1997re}. At strong coupling, one finds that the energy correlator computed in a state produced by a stress tensor is  \cite{Hofman:2008ar}
\begin{align}\label{eq:strong_coupling}
\langle \cE(\hat n_1) \cE(\hat n_2) \rangle_{\mathcal{N}=4}= 1+\frac{1}{\lambda}(1-6z(1-z)) +\mathcal{O}(\lambda^{-3/2})  \,,
\end{align}
where $\lambda$ is the 't Hooft coupling. The leading quantum gravity corrections to this result were computed in \cite{Chen:2024iuv}. We see that this exhibits a similar behavior to \Eq{free_theory}, namely a uniform distribution with small corrections. However, in this case there is no $\delta(z)$ contact term, since the state has no particle excitations.

\medskip

\noindent {\bf{Weakly Coupled Conformal Gauge Theory:}} The energy correlator exhibits much more interesting behavior in the case of a weakly coupled conformal gauge theory. Here we have specifically chosen a gauge theory as the presence of a conserved gauge flux will play a crucial role in the form of the energy correlator distribution, and is also present in the case of QCD. Furthermore, we will restrict ourselves to $d=4$. A prototypical example of such a theory is $\mathcal{N}=4$ SYM, for which their exists a tremendous amount of theoretical data. 

A schematic plot of the energy correlator distribution in a weakly coupled four-dimensional conformal gauge theory is shown in \Fig{fig:toy_physics}. It has a non-trivial shape onto which the physics of the theory is clearly imprinted. This shape is not so different from the case of QCD, and therefore it is worth understanding it in detail.

In the bulk of the distribution, $z\sim 1/2$, no particular state of the theory dominates. The correlator in this regime can either be bootstrapped \cite{Caron-Huot:2024tzr,Caron-Huot:2022sdy}, or computed perturbatively. It has been computed analytically to NLO \cite{Belitsky:2013ofa} and NNLO \cite{Henn:2019gkr}. As bootstrap results for the energy correlator distribution itself become available \cite{N4_bootstrap}, this provides an interesting regime for the comparison of perturbative calculations and the numerical bootstrap.

As we move away from $z\sim 1/2$ to either smaller or larger $z$, the energy correlator becomes dominated by a single state, allowing us to understand its all order form. This greatly simplifies the interpretation of the underlying physics, as well as the calculation.

We first consider the small angle limit of the energy correlator. In a conformal field theory, the small angle limit of the energy correlator is governed by the light-ray OPE \cite{Hofman:2008ar,Kologlu:2019mfz,Chang:2020qpj}, which is an expansion in the twist, $\tau$, of contributing light-ray operators. It takes the schematic form 
\begin{align}
\mathcal{E}(\hat n_1)\mathcal{E}(\hat n_2) &\sim \sum_i C_i \, (n_1\cdot n_2)^{\frac{\tau_i-4}{2}} \mathbb{O}_{i}^{[J=3]} (\hat n_2) + \text{transverse derivatives}\,,
\label{eq:lightray_OPE}
\end{align}
where $\mathbb{O}_{i}^{[J=3]}$ are light-ray operators with spin $J=3$, and $\tau_i$ is their twist. The transverse derivatives encode contributions from descendant operators, and will not play a role in our current discussion. Since this OPE is being done at the level of the detectors, it is convenient to think of it as projecting onto the measurement of specific states, sometimes referred to as light-ray densities, which are the duals of the light-ray detector operators. More formally, we can write the measurement of the energy correlator as $\text{Tr}[\rho \mathcal{E}(n_1) \mathcal{E}(n_2)]$, where $\rho=|\Psi\rangle \langle \Psi |$ is the state produced by the local operator insertion. Specific terms in the OPE therefore project the state $\rho$ onto states of definite quantum number. We can illustrate this schematically as
\begin{align}
\fd{3cm}{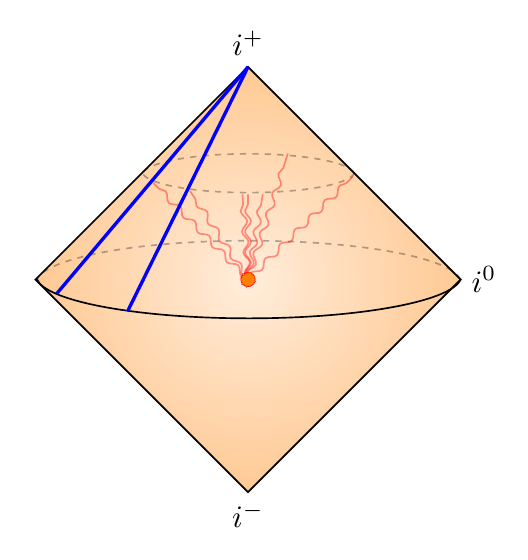}\sim \sum_i \fd{3cm}{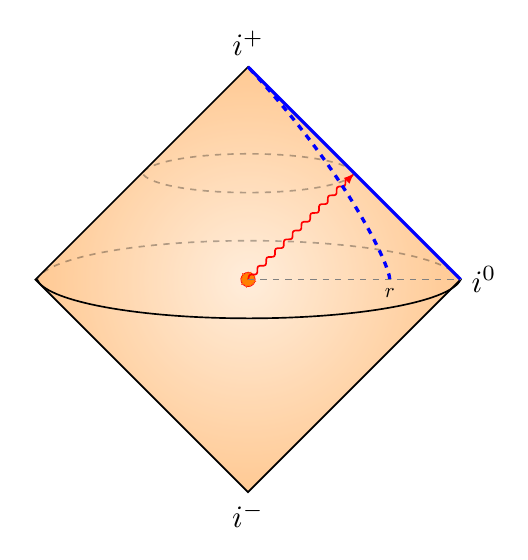}\,.
\end{align}
In the figure on the left, we have a generic state (illustrated by the red radiation), being measured by the product of operators $\mathcal{E}(\hat n_1)\mathcal{E}(\hat n_2)$, which does not have definite quantum numbers. In the figure on the right, we have expanded this into light-ray operators of definite quantum numbers, which project the measured state onto light-ray density states of definite quantum numbers (illustrated by the single squiggly red line). For twist-2 light-ray states at weak coupling, these states are close to single particle states. The scaling behavior of the energy correlator in this limit is therefore a direct probe of the spectrum of light-ray operators of the theory.

In a four dimensional gauge theory at sufficiently weak coupling, the lowest twist operators have twist approximately 2, and the EEC behaves in the collinear limit, as
\begin{align}
\text{EEC}_{z\to 0}\sim \frac{1}{z^{1-\gamma(3)}}\,,
\end{align}
where $\gamma(3)$ is the twist-2 spin-3 anomalous dimension (or more precisely its eigevalues if there are multiple such operators).  Since we are assuming weak coupling, the anomalous dimension $\gamma(3)\ll 1$, so we have a steeply peaked distribution as $z\to 0$. Since $\gamma(3) \geq 0$, this distribution is integrable, but diverges as $z\to 0$.  This is a physical effect, and simply says that there are correlations to arbitrary small scales.

While the behavior in the collinear limit is generic in any CFT, the behavior in the back-to-back limit of the energy correlator depends strongly on the specific CFT, and is particularly interesting in the case of a conformal gauge theory. In the case of a gauge theory, as we move the detectors apart, we become sensitive to the presence of a conserved gauge flux connecting the energetic excitations emitted from the form factor that are charged under the gauge group. This connection can be formalized by noting that the back-to-back limit of the energy correlator is sensitive to the high spin states (the light-cone limit of the four-point correlator \cite{Korchemsky:2019nzm,Chen:2023wah}). In the large-spin limit, the anomalous dimensions of twist-2 operators in gauge theories exhibit a famous $\ln J$ growth \cite{Korchemsky:1988si,Korchemsky:1992xv,Gubser:2002tv,Belitsky:2006en}
\begin{align}\label{eq:large_spin_def}
\Delta-J=2+ \Gamma_{\text{cusp}}(\lambda) (\ln J+\gamma_E) + B_\delta(\lambda) +\mathcal{O}(1/J)\,,
\end{align}
where $\Gamma_{\text{cusp}}(\lambda)$ is the cusp anomalous dimension  \cite{Polyakov:1980ca,Korchemsky:1987wg}.
In planar $\mathcal{N}=4$ SYM, both the cusp anomalous dimension \cite{Eden:2006rx,Beisert:2006ez}, and  $B_\delta$ \cite{Freyhult:2007pz,Freyhult:2009my,Fioravanti:2009xt} can be computed exactly using integrability.

Using the dominance of large-spin operators in the back-to-back limit of the energy correlator, one can write down an all orders expression for its behavior
\begin{align}\label{eq:b2b_factor}
\text{EEC}_{z\to 1}\sim\frac{H(\lambda)}{8 (1-z)} \int\limits_0^\infty \df b\, b\, J_0(b) \exp \biggl[ -\frac{1}{2}\Gamma_{\text{cusp}}(\lambda)\ln^2 \Bigl( \frac{e^{2\gamma_E}\, b^2}{4(1-z)} \Bigr)+2B_\delta(\lambda) \ln \Bigl( \frac{e^{2\gamma_E}\, b^2}{4(1-z)} \Bigr)     \biggr]\,,
\end{align}
where $J_0(b)$ is a Bessel function. This form can be argued from general grounds in any conformal theory with a conserved flux \cite{Alday:2007mf}, it can be derived \cite{Korchemsky:2019nzm} using the duality between correlators and Wilson loops \cite{Alday:2010zy}, or it can be derived using effective field theory techniques \cite{Moult:2018jzp}.

For understanding the physics of this result, it is useful to think of it in terms of the dominant degrees of freedom, just like was done for light-ray states in the collinear limit. In this case, since we probe the large spin limit of the gauge theory, the appropriate state has a size of $\sim \ln(J)$, namely a flux tube state. While the identification with a flux tube state of size $\ln(J)$ is not completely transparent at weak coupling, in \cite{Alday:2007mf} it was shown that the expectation value of the four-point correlator in the light-like limit is dominated by a classical saddle point describing a flux tube state \cite{Alday:2007mf}, whose action gives eq.~\eqref{eq:b2b_factor}.

We can illustrate this in the following manner
\begin{align}
\fd{3cm}{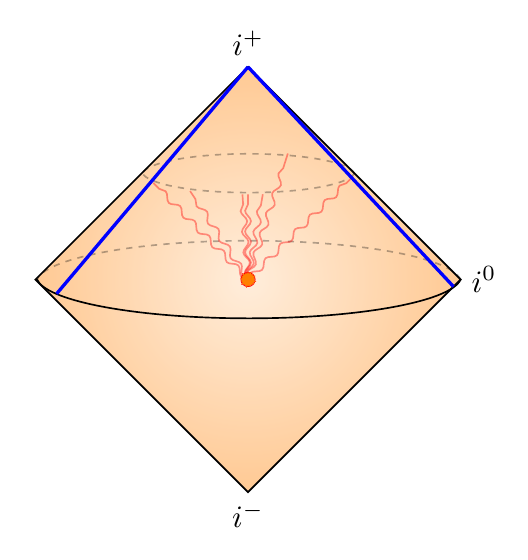}= \fd{3cm}{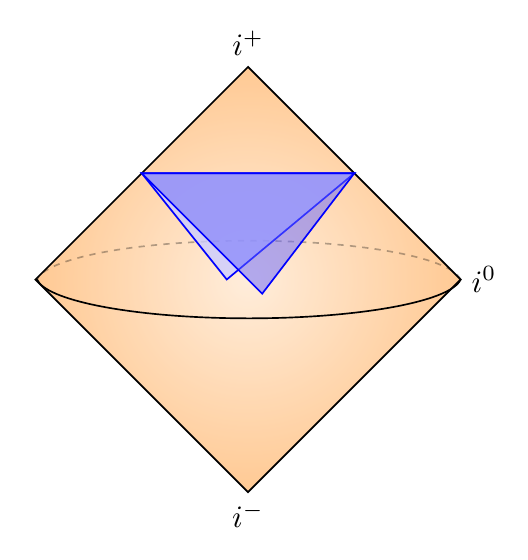}+\cdots
\end{align}
which illustrates two energetic particles framing Wilson lines (defects) in the QCD vacuum with a non-local flux between them, here depicted as the blue triangles. The presence of these Wilson lines can be arrived at from many perspectives, including a direct analysis of the correlator \cite{Alday:2010zy,Chen:2025ffl}, or from the effective field theory factorization \cite{Moult:2018jzp}.

An interesting feature of the result in eq.~\eqref{eq:b2b_factor}, originally observed in \cite{Parisi:1979se} (see also refs.~\cite{Collins:1984kg,Becher:2010tm}), is that it asymptotes to a constant as $z\to 1$ even at weak coupling. This behavior is shown in \Fig{fig:toy_physics}. This is in strong distinction to the behavior of the EEC in the collinear limit. This arises due to the $\ln(J)$ growth of the anomalous dimensions, which strongly modifies the state from its free theory value, or more intuitively, from the presence of the conformal flux tube. This distinction between the behavior in the collinear and back-to-back limits is important for understanding the analogous behavior in QCD,  and the imprints of confinement in these different limits.

Due to the strong resummation, which converts the $1/(1-z)$ leading scaling behavior to a constant, a complete understanding of the plateau region in the back-to-back limit also requires an understanding of power suppressed terms which scale like $(1-z)^0$. In the plateau region, these contribute with an equal scaling to the naive leading terms. These arise from double trace contributions. Improving the understanding of these contributions will particularly interesting for sharpening our understanding of this kinematic limit. At finite $N$, we have a mixing between single trace and double trace operators at large spin \cite{Polchinski:2002jw,Korchemsky:2015cyx}, which can be interpreted as breaking of the conformal flux tube. Forthcoming numerical studies of the energy correlator at finite $N$ can therefore provide insight into string breaking effects, and their comparison with data can lead to an understanding of the differences between string breaking for conformal vs. confining strings. 

\medskip

\noindent {\bf{Asymptotically Free, Confining Gauge Theory:}} Armed with these examples, we now have a much better intuition for the behavior of the EEC, and the physics controlling its kinematic regions. We are therefore ready to move to the more complicated case of an asymptotically free, confining gauge theory, exemplified by real world QCD. While the correlator cannot be understood non-perturbatively in this theory, by combining perturbation theory, with the physical understanding of the states contributing to the correlator in gauge theories, we are able to gain a relatively sharp understanding of the physics controlling the energy correlator in QCD. We believe that sharpening this picture will be important for using recent LEP measurements for improving our understanding of non-perturbative phenomenon in QCD.

\begin{figure}
    \centering
    \includegraphics[width=0.95\textwidth]{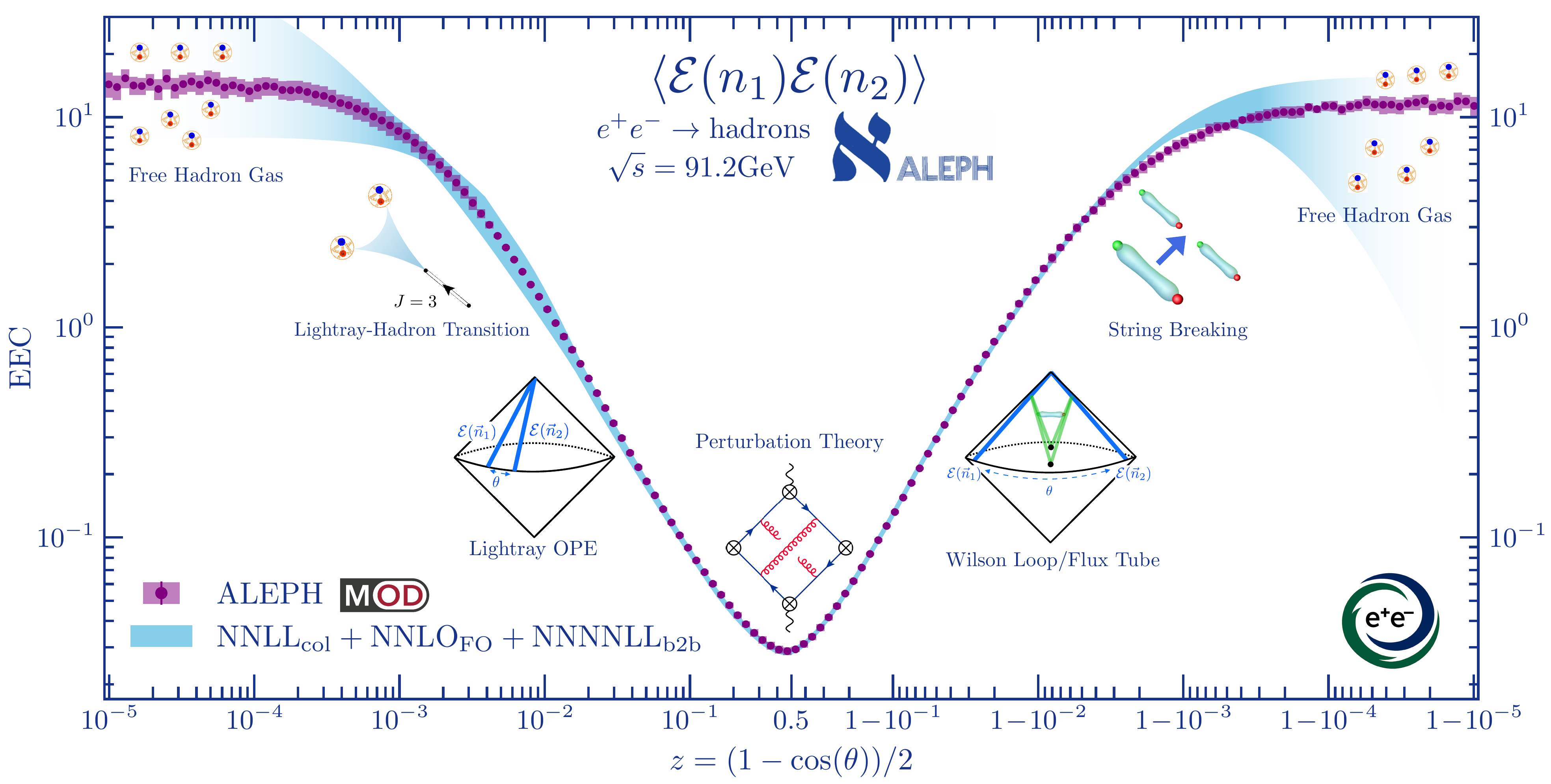}
    \caption{The EEC in $e^+e^-$ collisions: In purple we show data from the ALEPH experiment, and in light-blue we show the calculation developed in this paper. The inset figures show the underlying physics dominating each kinematic region of the energy correlator, and are discussed in the text. This figure originally appeared in \cite{Electron-PositronAlliance:2025fhk}.}
    \label{fig:physics}
\end{figure}

The complete result for the energy correlator in QCD is shown in \Fig{fig:physics}. As compared with \Fig{fig:toy_physics}, in this case, the purple dots are real data from the ALEPH experiment \cite{Electron-PositronAlliance:2025fhk}. The light blue curve represents the theoretical prediction using the calculations of this paper. The small figures illustrate the physics in the different kinematic regions of the energy correlator. We will discuss the underlying physical picture of these different regions here, and the precise calculation will be described in the rest of the paper.

The energy correlator observable is particularly interesting in a scale dependent theory such as QCD, since the measurement probes the theory at a different scale, depending on the kinematics of the detectors.  In QCD, the energy correlator depends on the following physical scales
\begin{itemize}
\item $\mu_Q^2=Q^2$: scale of the momentum injected into the current $J(x)$, given by the total momentum $q^\mu$  of the annihilating $e^+e^-$ pair.
\item $\mu_z^2=Q^2z$ and $\mu_{(1-z)}^2=Q^2(1-z)$: scales associated with the angular measurement $z$, i.e.~the detector configuration, which become distinct from $\mu_Q$ for $z\to0$ and $z\to1$.
\item $\mu_{\Lambda}^2=\Lambda_{\text{QCD}}^2$: the intrinsic scale associated with confinement in QCD.
\end{itemize}
Throughout this paper we will focus on the case $Q\gg \Lambda_{\text{QCD}}$, where we produce high multiplicity final states with interesting patterns of energy flux. Measurements of the energy correlator therefore probe the theory as a function of scale all the way from partons in the UV to hadrons in the IR. This can occur in two distinct ways, namely as a flow to the collinear limit, and a flow to the back-to-back limit. Much like in the case of a conformal theory, these probe different physics, and in the case of QCD, different aspects of the confining transition.

\noindent {\bf{Bulk Distribution}}: We start our discussion in the bulk of the distribution $z\sim 1/2$. In this regime, we have a two scale problem set by $\mu\sim Q$ and $\mu \sim \Lambda_{\text{QCD}}$. Performing an OPE, we are able to separate them into a purely perturbative contribution and a leading non-perturbative contribution.
 \begin{equation}
\hspace{-8.6cm}\text{Perturbative Partons:}\qquad \fd{2cm}{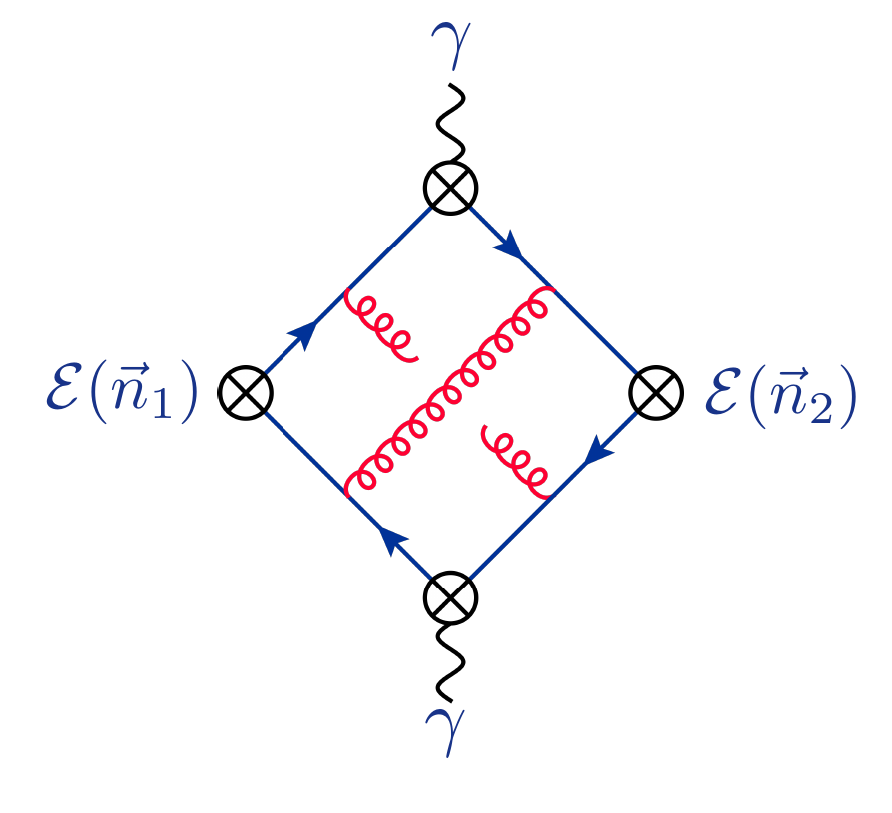}
\end{equation}
The leading perturbative contribution can be obtained from the calculation of the energy correlator on partonic Fock states. We perform our calculation to NNLO$_{\text{FO}}$, which includes the parton states $|q\bar q\rangle$, $|q\bar q g\rangle$, $|q\bar q g g\rangle$, $|q\bar q q\bar q\rangle$, $|q\bar q q\bar qg\rangle$, $|q\bar q g g g\rangle$.

 \begin{equation}
\hspace{-7.3cm}\text{Non-Perturbative Correction:}\qquad \fd{2cm}{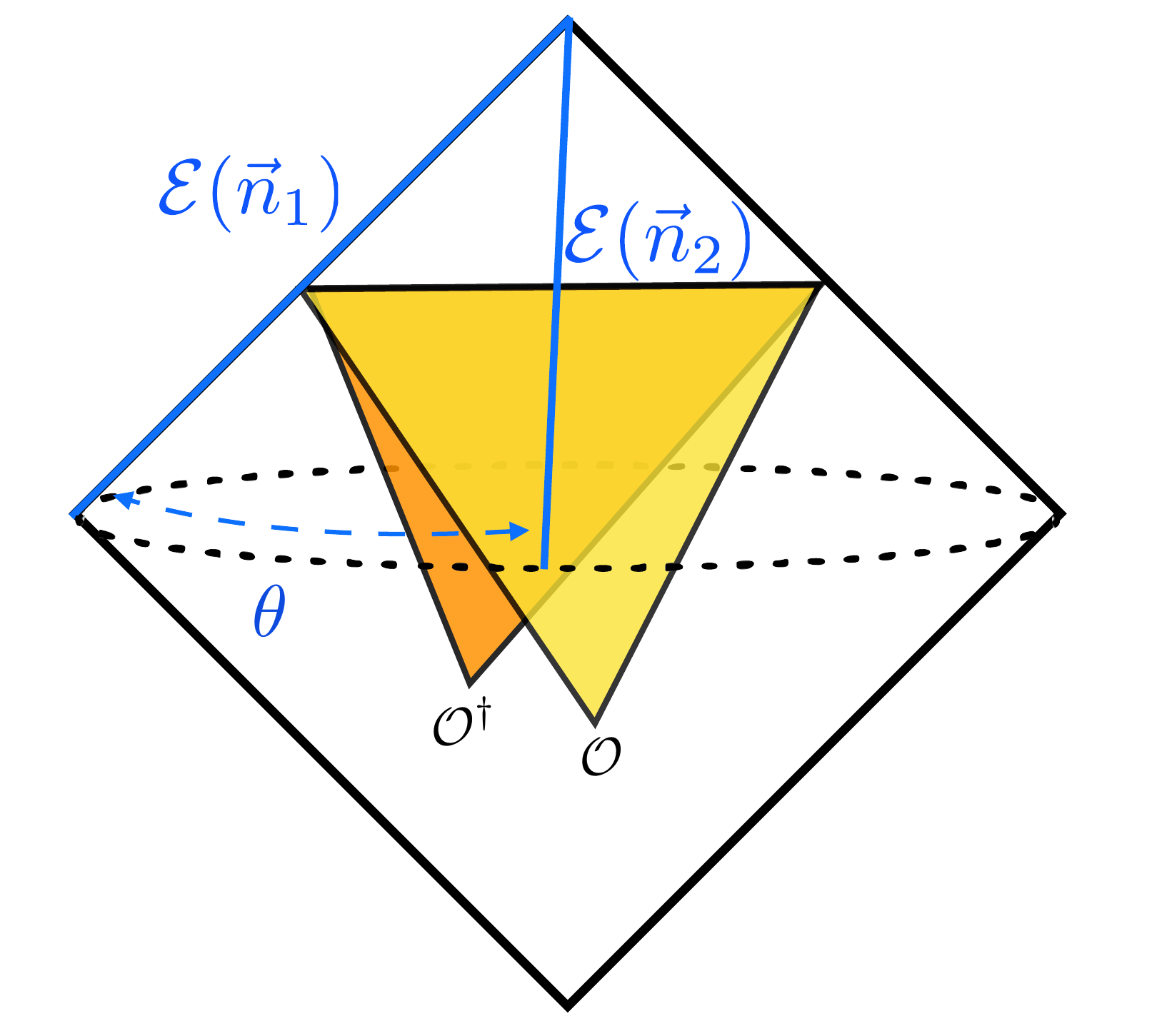}
\end{equation}
We also incorporate the leading non-perturbative correction. As shown in the seminal work of Korchemsky and Sterman \cite{Belitsky:2001ij,Korchemsky:1999kt,Korchemsky:1997sy,Korchemsky:1994is}, this takes the form of the expectation value of a single energy flow operator in a Wilson line state
\begin{align}
\Omega_{1q}=\frac{1}{N_c} \langle 0 | \tr \bar Y_{\bar n}^\dagger Y_n^\dagger \cE_T(0) Y_n \bar Y_{\bar n} |  0 \rangle\,.
\end{align}
Here we use the subscript ``$q$", since the Wilson lines are in the fundamental representation. For precise definitions see \Sec{nppc}. The value of this constant cannot currently be computed from first principles.

We can now move away from the bulk region in two different limits, namely the $z\to 0$ or $z\to 1$ limits. We discuss each of these in turn, highlighting the differences as compared to the case of a conformal gauge theory discussed earlier.
\medskip

\noindent {\bf{Collinear Limit:}}

 \begin{equation}
\hspace{-5.8cm}\text{Lightray OPE}:\qquad \fd{2cm}{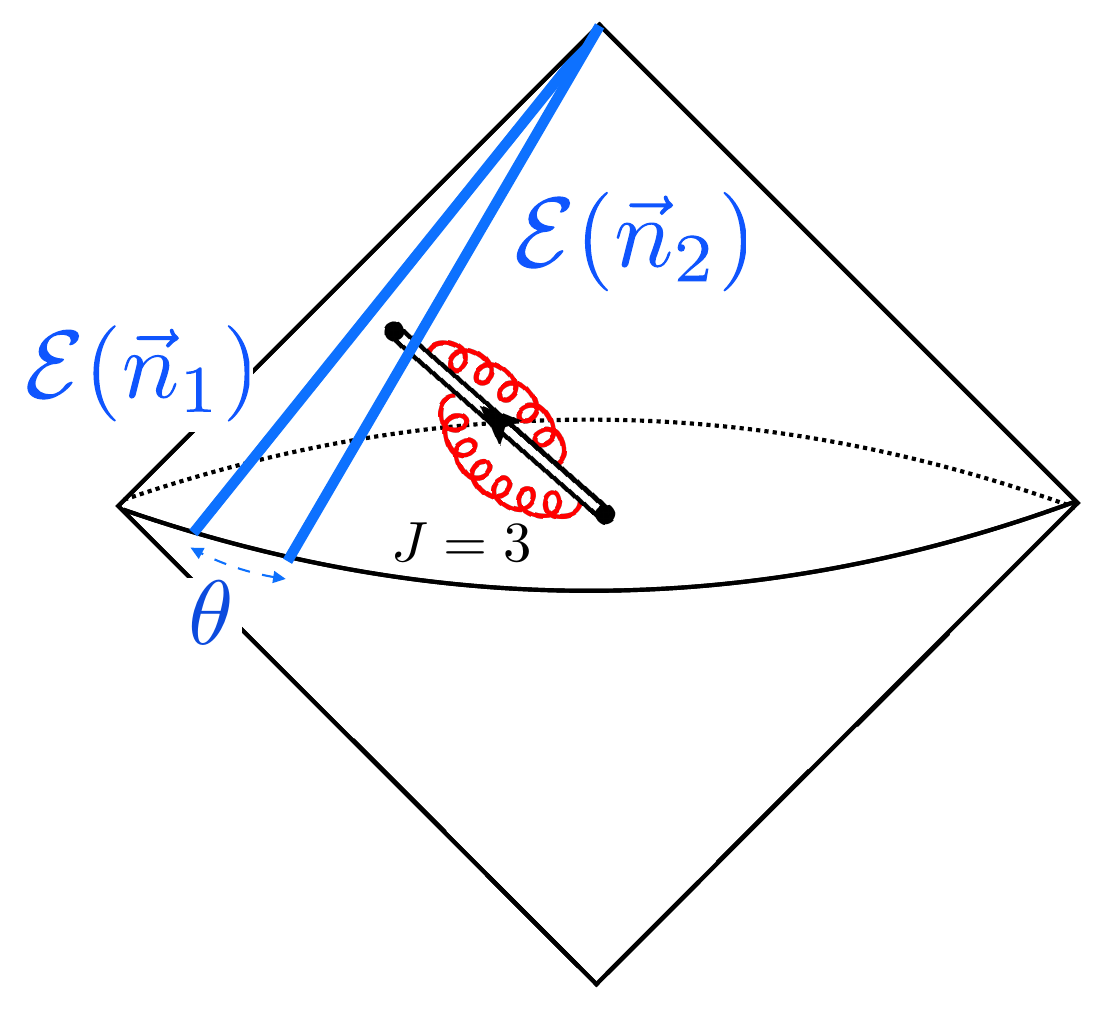}+  \frac{\Omega}{Q}\fd{2cm}{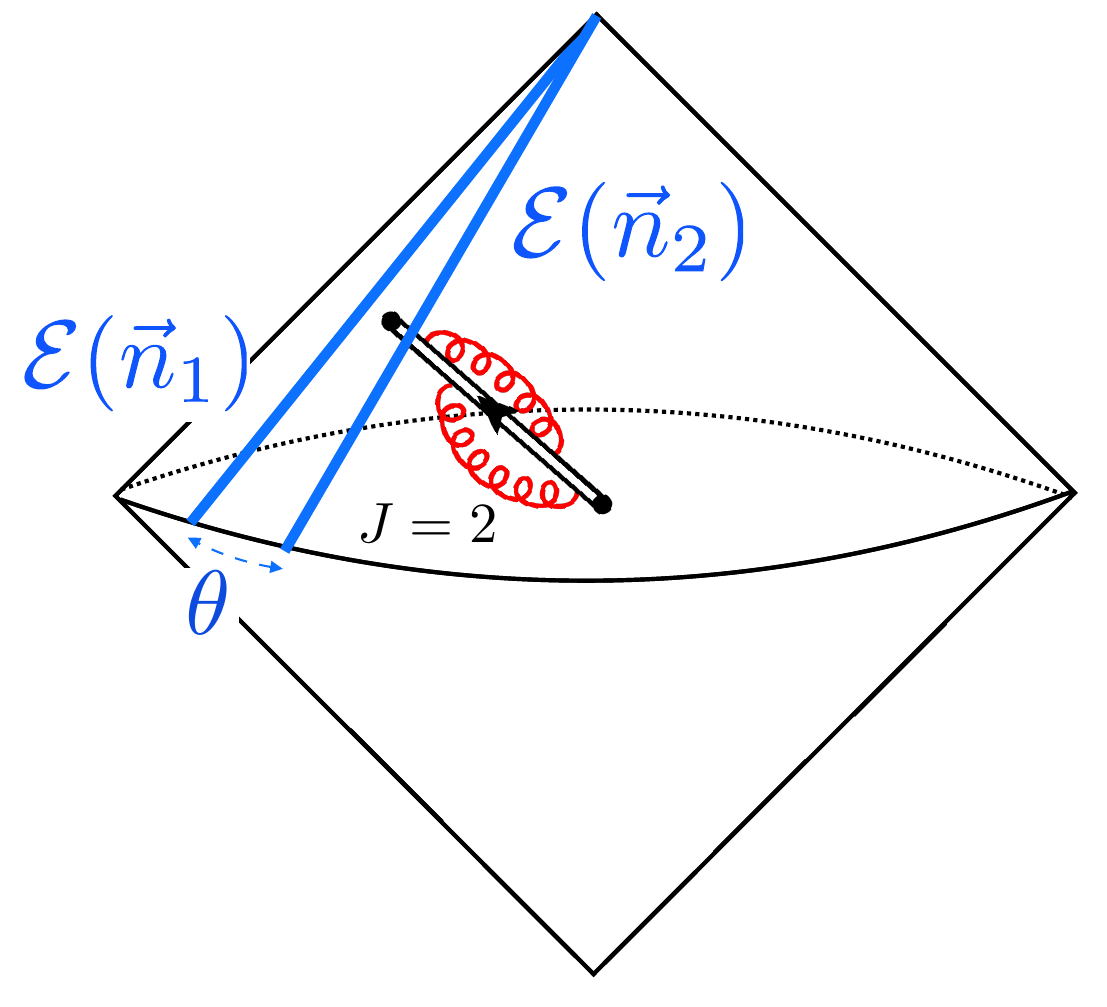}+\cdots
\end{equation}
In the region $\mu_Q \gg \mu_z \gg \mu_{\Lambda}$, the scale $\mu_{\Lambda}$ can be expanded. We have a two-scale problem determined by $\mu_Q \gg \mu_z$ of interacting quarks and gluons, which are nearly conformal, up to corrections due to the running of the coupling described by the $\beta$ function. The leading scaling behavior in this region is determined by the leading operator appearing in the lightray OPE~\cite{Hofman:2008ar}, which gives rise to a power-law behavior with an exponent determined by the twist-2 spin-3 operators (up to $\beta$ function corrections). The scaling behavior in this regime allows us to directly probe the partonic light-ray states.

As $\mu_z $ approaches $\mu_{\Lambda}$ we must incorporate  power corrections in the $\mu_{\Lambda}/\mu_z$ expansion. These give rise to an enhanced scaling, governed by the twist-2 spin-2 operators. In the case of QCD, these power corrections have a quantum scaling, giving rise to a mixing between two non-perturbative parameters $\Omega_{1g}$ and $\Omega_{1q}$ \cite{Chen:2024nyc,Lee:2024esz}.

\begin{equation}
\hspace{-8cm}\text{Confinement Transition}:\qquad \fd{2cm}{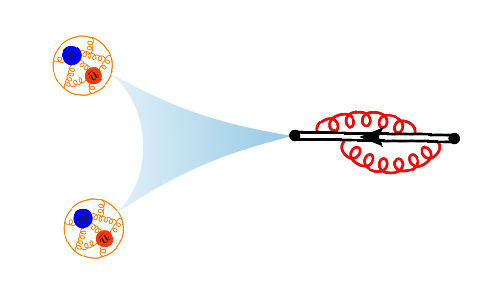}
\end{equation}
As $\mu_z \sim \mu_{\Lambda}$, we have a reorganization of the degrees of freedom from quarks and gluons, to hadrons, namely confinement. However, we can precisely interpret this confinement as the overlap of a twist-2 spin-3 light-ray state with a two-hadron state \cite{Chang:2025kgq}, or in the language of QCD factorization, a di-hadron fragmentation functions \cite{Lee:2025okn,Herrmann:2025fqy,Kang:2025zto}.

 \begin{equation}
\hspace{-9.7cm}\text{Free Hadrons}:\qquad \fd{2cm}{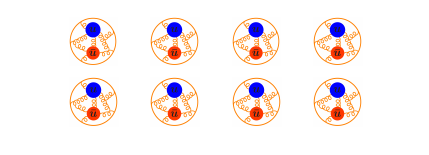}
\end{equation}
When $\mu_z$ goes below $\mu_{\Lambda}$, we enter a gapped theory of non-interacting hadrons, and  $\mu_z$ is no longer a meaningful scale. The EEC in \Fig{fig:physics} becomes constant, with a contact term $\propto \delta(z)$ reflecting the particle-like nature of QCD.

\medskip

\noindent {\bf{Back-to-Back Limit}}

\begin{equation}
\hspace{-4.5cm}\text{Wilson Loop OPE/ Perturbative Flux Tube}:\qquad \fd{2cm}{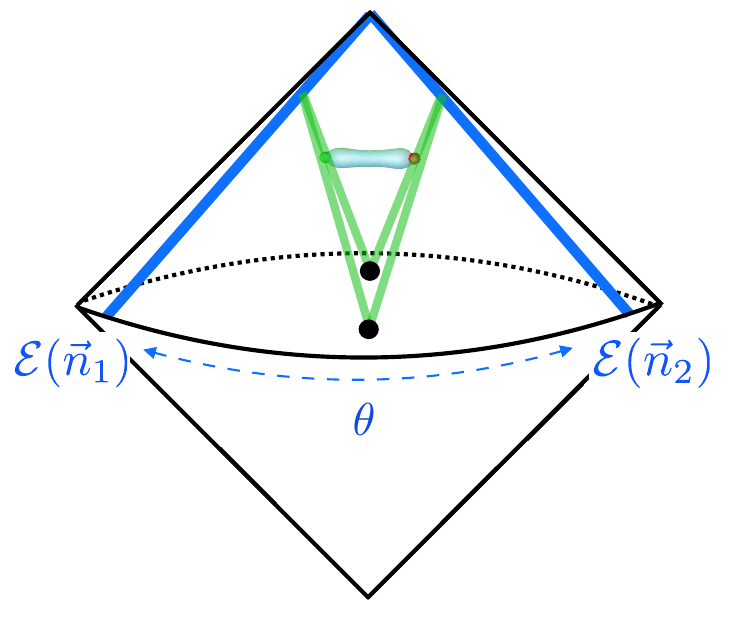}
\end{equation}
In the region $\mu_Q \gg \mu_{(1-z)} \gg \mu_{\Lambda}$ the scale $\mu_{\Lambda}$ can again be expanded, leading to a two-scale problem determined by $\mu_Q \gg \mu_{(1-z)}$. In this case, the leading behavior is determined by high-spin operators, just like in the case of a CFT, discussed above.

As emphasized in the case of a conformal gauge theory, the presence of a conserved gauge flux is able to modify the scaling behavior by an integer amount. This is crucial for interpreting the confinement transition in the back-to-back limit. In QCD, assuming sufficiently high energies so that the turnover to the Sudakov region is in a perturbative regime of the coupling, this turnover is set by perturbative physics.  Therefore, for sufficiently high energies $Q$, this transition, and the height of the plateau, can be computed robustly in perturbation theory.  For the case of LEP, this is at $(1-z)\sim 10^{-3}$, so it is at a $\mu \sim Q \sqrt{10^{-3}}$, which is on the border of the perturbative regime. This allows us to smoothly transition to the flat behavior of free hadrons in perturbative theory.

 \begin{equation}
\hspace{-5.5cm}\text{Leading Non-Perturbative Corrections}:\qquad \fd{2cm}{figures/soft_constant_OPE.pdf}
\end{equation}
As $\mu_{(1-z)}$ approaches $\mu_{\Lambda}$ we must incorporate the leading non-perturbative corrections in $\mu_{\Lambda}/\mu_{(1-z)}$. In the back-to-back limit, in addition to the non-perturbative contributions from $\Omega_{1q}$, we also have non-perturbative contributions to the anomalous dimensions, referred to as the Collins-Soper kernel. Unlike the constant $\Omega_{1q}$, these can be computed on the lattice \cite{Avkhadiev:2024mgd,Avkhadiev:2023poz,Shanahan:2021tst,Shanahan:2019zcq,Shanahan:2020zxr}.

\begin{equation}
\hspace{-8.5cm}\text{Flux Tube Breaking}:\qquad \fd{2cm}{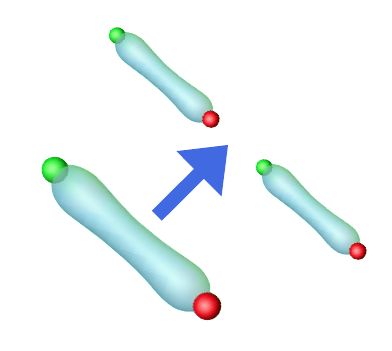}
\end{equation}
As $\mu_{(1-z)} \sim \mu_{\Lambda}$, we lose control of our perturbative calculation. However, given an understanding of the physics on either side of this regime, namely a perturbative flux tube on one side, and a free hadron gas on the other, we are able to interpret this regime in terms of the breaking of the QCD flux tube. It is well known that the QCD flux tube can break, which can be quantitatively studied on the lattice \cite{Bulava:2019iut,Bali:2005fu,Philipsen:1998de}. This is the analog of the single-trace/double-trace mixing for the conformal flux tube. It will be particularly interesting to sharpen the interpretation of this regime in terms of the breaking of confining flux tubes, and to understand better their imprint in data.

 \begin{equation}
\hspace{-9.7cm}\text{Free Hadrons}:\qquad \fd{2cm}{figures/free_hadrons_solo.pdf}
\end{equation}
When $\mu_{(1-z)}$ goes below $\mu_{\Lambda}$, we enter a gapped theory of non-interacting hadrons, and  $\mu_{(1-z)}$ is no longer a meaningful scale. The correlator becomes uniform, $\text{EEC}(z)\sim \text{const}$.

\medskip

In the remainder of this paper, we will describe how we quantitatively compute the energy correlator in each of these distinct regimes. We will also highlight which regimes are under precision theoretical control, and which regimes we hope that precision measurements of the energy correlators can lead to new insights into non-perturbative phenonemenon in QCD. 

\medskip

\subsection{A Brief History of Calculations of the Energy-Energy Correlator}\label{sec:history}

The first (to our knowledge) complete calculation of the energy correlator was presented in \cite{PLUTO:1981gcc}, and compared with the first measurement of the energy correlator. It combined the LO fixed order calculation of the energy correlator in the bulk of the distribution \cite{Basham:1978zq,Basham:1979gh,Ellis:1978ty,Basham:1977iq,Basham:1978bw}, with LL resummation in the collinear limit using the jet calculus \cite{Konishi:1978ax,Konishi:1979cb,Konishi:1978yx}, and LL resummation in the back-to-back limit using the formalism of Parisi and Petronzio \cite{Parisi:1979xd,Parisi:1979se} (see also \cite{Dokshitzer:1978yd,Parisi:1979se,Parisi:1979xd,Ellis:1980my,Ellis:1981sj,Kodaira:1981nh} for other papers developing the resummation of Sudakov double logarithms). In our notation was at the order $\text{LO}+\text{LL}_\text{col}+\text{NLL}_\text{b2b}$. To our knowledge, this was the first matched and resummed calculation of an $e^+e^-$ observable in QCD.

The back-to-back limit of the energy correlators was extensively studied in the early 1980s, along with the development of transverse momentum dependent (TMD) factorization theorems \cite{Chao:1982wb,Soper:1982wc,Kodaira:1982az,Collins:1981va,Collins:1985xx,Collins:1985kw,Collins:1981zc,Kodaira:1982az}.  Comparisons to data in the back-to-back limit in the early 1980s \cite{Collins:1981zc,Kodaira:1982az} already identified the need for, and structure of power corrections in the EEC. These were formalized much later by Korchemsky, Sterman and collaborators~\cite{Belitsky:2001ij,Korchemsky:1999kt,Korchemsky:1997sy,Korchemsky:1994is}.

More recently, there have been a number of calculations of the energy correlators in $e^+e^-$ colliders. Many of these have focused on resummation in the back-to-back limit, due to the ability to use the well developed machinery of TMD resummation. Examples include refs.~\cite{Tulipant:2017ybb,Kardos:2018kth,deFlorian:2004mp,Aglietti:2024zhg,Aglietti:2024xwv,Ebert:2020sfi,Kang:2024dja}, where the state of the art has now reached N$^4$LL resummation \cite{Duhr:2022yyp}. A systematic factorization theorem for the back-to-back limit was first formulated in \cite{Moult:2018jzp}.  There has also been progress in fixed-order calculations, which have been performed at NLO analytically \cite{Dixon:2018qgp,Luo:2019nig}, and NNLO numerically \cite{DelDuca:2016csb}.

The  resummation in the collinear limit was resurrected and systematized in ref.~\cite{Dixon:2019uzg}. This progress set the stage for a precision calculation of the entire EEC spectrum using the full combination of modern perturbative calculations and effective field theory techniques.

Our calculation is the first modern calculation of the energy correlator which achieves a simultaneous description of all kinematic regions, with high order resummation achieved using factorization theorems.  However, it is important to emphasize that this builds on 50 years of development of techniques in perturbative and non-perturbative QCD, without which our analysis would not be possible. For a more detailed review of the history of the energy correlators, and the development of the techniques for their analysis, we refer the reader to \cite{Moult:2025nhu}.

\section{Overview of Theory Framework}\label{sec:calc}

A precise description of the energy correlators on tracks is non-trivial since it simultaneously involves the accurate description of the underlying perturbative process, high order resummation in kinematic limits, the incorporation of the leading non-perturbative power corrections, and the inclusion of the non-perturbative corrections due to performing the measurement on tracks. This requires the accurate description of physics over several orders of magnitude. To achieve this, we build on tremendous progress in our understanding of factorization using SCET \cite{Bauer:2000ew,Bauer:2000yr,Bauer:2001ct,Bauer:2001yt,Rothstein:2016bsq,Beneke:2002ph}.  SCET has enabled operator-based factorization theorems, providing definitions  of perturbative and non-perturbative contributions in terms of matrix elements of operators, and allowing resummation to be performed at high orders using the renormalization group.

For our calculation of the energy correlator, we proceed by breaking up the energy correlator into distinct kinematic regions, performing high precision calculations for these specific regions, and then combining the different kinematic regions. In particular, for our calculation, we have the following regions
\begin{itemize}
\item Bulk Region ($z\sim 1/2$): In this regime we can use fixed order perturbation theory, combined with leading non-perturbative corrections. We denote these as $\EEC_\text{FO}(z)$ and  $\EEC^{\Omega}(z)$, respectively
\item Collinear Region ($z\to 0$): In the collinear limit, we must resum logarithms of $z$ to all orders, which is achieved using a factorization theorem. We denote the perturbative and non-perturbative contributions in this limit as  $\EEC^\text{fact.}_{z\to0}(z)$ and $\EEC_{z\to0}^{\Omega\text{,res.}}(z)$. Additionally, in the deep collinear limit, we have a confinement transition to a plateau region, which is described by $\EEC^\text{\text{plateau}}_{z\to0}(z)$.
\item Back-to-Back Region ($z\to 1$): In the back-to-back limit, we must resum logarithms of $1-z$, as well as incorporate modified non-perturbative corrections, in particular the Collins-Soper kernel. We denote the perturbative and non-perturbative contributions in this limit by $\EEC^\text{fact.}_{z\to1}(z)$ and $\EEC_{z\to1}^{\Omega\text{,res.}}(z)$. Additionally, in the deep back-to-back limit, we have a confinement transition to a plateau region, which is described by $\EEC^\text{\text{plateau}}_{z\to1}(z)$.
\end{itemize}
In addition to this separation into regions, we must also incorporate, throughout the entire EEC distribution, the fact that the distribution is computed on tracks. This is achieved by combining the theoretical description in each region highlighted above with the track function formalism \cite{Chang:2013rca,Chang:2013iba}.

The goal of this section is to provide a high level overview of the different components of our calculation. This is designed to be self contained, highlighting in particular our incorporation of tracks in \Sec{tracks}, our treatment of non-perturbative corrections in \Sec{nppc}, our factorization theorems for the $z\to 0$ and $z\to 1$ limits in \Sec{fact_sum}, and our matching of different kinematic limits in \Sec{match}. We hope that for the reader interested only in an overview, these sections are sufficient. For those interested in the technical aspects of our calculation,  in the following sections we will discuss our calculations in each of the three regimes, detailing the perturbative calculation, resummation, non-perturbative corrections and the merging of the different calculations. In particular, the bulk region, $z\sim 1/2$ is discussed in \Sec{fo}, the collinear region, $z\to 0$, is discussed in \Sec{col}, and the back-to-back limit, $z\to 1$, is discussed in \Sec{b2b}.

\subsection{Track Functions}\label{sec:tracks}

One of the primary theoretical advances that enables the results of this paper is the ability to systematically compute observables on (charged particle) tracks. This is achieved using the track function formalism \cite{Chang:2013rca,Chang:2013iba}, which enables a rigorous factorization into perturbative and non-perturbative physics. It was extended beyond the leading order and systematically understood in refs.~\cite{Jaarsma:2023ell,Chen:2022pdu,Chen:2022muj,Jaarsma:2022kdd,Li:2021zcf}. This has been combined with the realization~\cite{Chen:2020vvp} that only moments are needed for energy correlators, specifically, the $N$-point correlator only requires the integer moments from 1 to $N$.
Here we briefly review the necessary aspects of the track function formalism. Their appearance in the factorization theorem for the energy correlators on tracks will be discussed in the next sections.

The quark and gluon track functions, $T_q(x)$ and $T_g(x)$, respectively, describe the total energy fraction $x$ of charged hadrons resulting from the fragmentation of a quark or gluon. They are defined as~\cite{Chang:2013iba,Chang:2013rca} 
\begin{align} \label{T_def}
T_q(x)&=\!\int\! \df y^+ \df ^{d-2} y_\perp e^{\img k^- y^+/2} \sum_X \delta \biggl( x\!-\!\frac{P_C^-}{k^-}\biggr)  \frac{1}{2N_c}
\text{tr} \biggl[  \frac{\gamma^-}{2} \langle 0| \psi(y^+,0, y_\perp)|X \rangle \langle X|\bar \psi(0) | 0 \rangle \biggr]\,,
 \nn \\
T_g(x)&=\!\int\! \df y^+ \df^{d-2} y_\perp e^{\img k^- y^+/2} \sum_X \delta \biggl( x\!-\!\frac{P_C^-}{k^-}\biggr) \frac{-1}{(d\!-\!2)(N_c^2\!-\!1)k^-}
\nn \\ & \quad \times
 \langle 0|G^a_{- \lambda}(y^+,0,y_\perp)|X\rangle \langle X|G^{\lambda,a}_- (0)|0\rangle, 
\end{align} 
where $P_C$  denotes the momentum of the charged particles in $|X\rangle$, $N_c = 3$ is the number of colors, $\psi$ is the quark field and $G$ the gluon field strength tensor. We use light-cone gauge for simplicity to avoid writing additional Wilson lines. The track functions for quark and antiquark are equal due to charge conjugation symmetry. The use of track functions allows us to rigorously separate perturbative and non-perturbative physics in the calculation of event shape observables measured on tracks.

For the specific case of energy correlators we can perform a matching at the level of the detector operators, for example
\begin{align}
\cE_{\text{tr}}(\vec n_1)=T_{\bar q}(1) \cE_{\bar q} (\vec n_1)+  T_q(1) \cE_q (\vec n_1)+  T_g(1) \cE_g (\vec n_1) \,.
\end{align}
In this case, only moments of the track function appear \cite{Chen:2020vvp}, which we define as 
\begin{align} \label{eq:T_mom}
T_i(n,\mu)=\int\limits_0^1 \df x\, x^n\, T_i(x,\mu)\,.
\end{align}
The zeroth moment satisfies 
\begin{align}\label{eq:T0}
T_i(0,\mu)=1\,,
\end{align}
due to probability conservation,
but the higher moments are non-perturbative parameters of QCD. This simplification allows us to perform high order perturbative calculations for track-based energy correlator observables. We will describe in detail how the incorporation of tracks modifies our calculations in different kinematic regions of the energy correlators.

Another recent advance is that the track functions moments have been measured by the ATLAS collaboration~\cite{ATLAS:2024jrp}. The values of track function moments used in our analysis are given in \Sec{results}. They satisfy non-linear RG equations, which have recently been extended to NLO~\cite{Chen:2022pdu,Chen:2022muj,Jaarsma:2022kdd,Li:2021zcf}. For the first two moments of the track functions, as required for the analysis of the two-point energy correlator, these equations are known to NNLO \cite{Chen:2022pdu,Chen:2022muj,Jaarsma:2022kdd,Li:2021zcf}. This allows us to evolve the track function moments between energy scales, enabling measurements of these parameters in one experiment to be used elsewhere.

We also wish to emphasize that given a complete calculation on tracks, it is trivial to obtain the calculation on all hadrons. This is achieved by performing the replacement
\begin{align}
T_i(x,\mu) \to \delta(1-x)\,\text{, that is, any moment }\,T_i(n,\mu)\to 1.
\end{align}
Our calculations therefore also provide predictions for the standard energy correlator to the same order.

\subsection{Non-Perturbative Power Corrections}\label{sec:nppc}

A precise description of energy flux observables in QCD requires an incorporation of non-perturbative corrections due to hadronization. Factorization theorems allow the leading non-perturbative power corrections to be expressed as universal matrix elements, which can either be extracted from data by measuring multiple observables, or computed using non-perturbative methods such as lattice QCD. There has been tremendous progress in our understanding of non-perturbative power corrections to event shape observables, enabling their definition as field theoretic matrix elements \cite{Lee:2006fn,Hoang:2007vb,Abbate:2010xh,Hoang:2014wka,Benitez:2024nav,Benitez-Rathgeb:2024ylc,Hoang:2025uaa}, and their extraction from precision measurements \cite{Hoang:2015hka,Abbate:2012jh,Abbate:2010xh,Hoang:2014wka,Benitez:2025vsp,Benitez:2024nav,Benitez-Rathgeb:2024ylc}.

In our description of the energy correlator, we incorporate non-perturbative corrections in four different regions of the energy correlator:
\begin{itemize}
\item Bulk Region ($z\sim 1/2$): We incorporate the leading (linear) non-perturbative power corrections described by $\EEC^{\Omega}(z)$.
\item Collinear Resummation Region ($z\to 0$):  We incorporate the leading non-perturbative correction to the collinear factorization theorem $\EEC_{z\to0}^{\Omega\text{,res.}}(z)$. These are dressed with LL resummation. 
\item Back-to-Back Region: We incorporate the linear non-perturbative power correction dressed by NLL resummation, as well as the logarithmically enhanced Collins-Soper kernel. We denote this combination by $\EEC_{z\to1}^{\Omega\text{,res.}}(z)$.
\item Collinear and Back-to-Back Plateaus: We discuss the transition to the non-perturbative collinear and back-to-back plateaus, described by $\EEC^\text{\text{plateau}}_{z\to0}(z)$, $\EEC^\text{\text{plateau}}_{z\to1}(z)$.
\end{itemize}
We briefly discuss each of these.

A remarkable feature of the energy correlator is that the \emph{functional form} of the leading power correction in the bulk of the distribution can be predicted using symmetries \cite{Belitsky:2001ij,Korchemsky:1999kt,Korchemsky:1997sy,Korchemsky:1994is,Chen:2024nyc}. It takes the form
\begin{align} 
\EEC^{\Omega}(z) = 
\frac{1}{2} \frac{\sigma_0}{\sigma} \frac{\Omega_{1q}}{Q [z(1-z)]^{3/2}}\,.
\end{align}
Here $ \Omega_{1q}$ is a universal non-perturbative parameter
\begin{align}
\Omega_{1q}=\frac{1}{N_c} \langle 0 | \tr \bar Y_{\bar n}^\dagger Y_n^\dagger \cE_T(0) Y_n \bar Y_{\bar n} |  0 \rangle\,,
\end{align}
where $q$ refers to the fundamental representation of the outgoing quark and anti-quark. It is defined in terms of the transverse energy flow operator
\begin{align}\label{eq:E_T}
\mathcal{E}_T (\eta) =\frac{1}{\cosh^3 \eta}\int \limits_0^{2\pi} \df \phi~ \mathcal{E}(\hat n)\,,
\end{align}
and $Y$ denote soft Wilson lines in the fundamental representation
\begin{align}
Y_n(z)= \text{P}\, \text{exp} \Biggl[\img g \int\limits_0^\infty \df s\, n \cdot A(n s+z)  \Biggr]\,.
\end{align}
This non-perturbative parameter appears in a number of dijet event shape observables, such as thrust and C-parameter.  It was  extracted from high precision thrust fits in ref.~\cite{Abbate:2010xh}, and recently updated in \cite{Benitez-Rathgeb:2024ylc}. When converted to the non-perturbative correction for the EEC \cite{Schindler:2023cww}, it takes the value\footnote{There is an anti-correlation between $\alpha_s$ and $\Omega_{1q}$ (see e.g.~fig. 1 of \cite{Benitez-Rathgeb:2024ylc}). Therefore, strictly speaking, if we apply the $\Omega_{1q}$ value from the fits in \cite{Abbate:2010xh,Benitez-Rathgeb:2024ylc}, we should also use their value of $\alpha_s(m_Z)$, namely $\alpha_s(m_Z) = 0.114$.}
\begin{align}\label{eq:omega1_R0}
\Omega_{1q}(R_0)=0.895 \pm 0.054\, \text{GeV}\,.
\end{align}
This allows us to \emph{predict} its contribution to the EEC distribution in the bulk region.

The leading non-perturbative corrections to the energy correlator in the collinear limit were recently studied in \cite{Lee:2024esz,Chen:2024nyc}. To leading logarithmic order,
\begin{align}\label{eq:col_resum_omega_LL}
\EEC^{\Omega,\text{LL}}_{z\to 0}(z) =\frac{\sigma_0}{\sigma}\frac{\df}{\df z}\left\{-\frac{1}{2Q\sqrt{z}}(\Omega_{1q},\Omega_{1g}) \!\cdot\! V \left[\left(\frac{\alpha_s(\sqrt{z} Q)}{\alpha_s(\mu)} \right)^{-\frac{\vec{\gamma}_T^{(0)}\!(2)}{\beta_0}}\right] \! \cdot\! V^{-1}\!  \cdot\! \begin{pmatrix}
    2
\\
    0
  \end{pmatrix}\right\} \,,
\end{align}
where $V$ is the matrix that diagonalizes the twist-$2$ spin-$2$ anomalous dimension matrix $\gamma_T^{(0)}(2)=\{\{\gamma^{(0)}_{qq}(2),\gamma^{(0)}_{qg}(2)\},\{\gamma^{(0)}_{gq}(2),\gamma^{(0)}_{gg}(2)\}\}$, i.e., $V^{-1}\!\cdot\!\gamma_T^{(0)}(2)\!\cdot\! V=\vec{\gamma}_T^{(0)}(2)$, with the superscript $(0)$ indicating that it's at leading order. In addition to the parameter, $\Omega_{1q}$, eq.~\eqref{eq:col_resum_omega_LL} additionally involves
\begin{align}
\Omega_{1g}=\frac{1}{N_c^2-1} \langle 0 | \tr \bar {\mathcal{Y}}^\dagger_{\bar n} {\mathcal{Y}}_n^\dagger \cE_T(0) \mathcal{Y}_n \bar{ \mathcal{Y}}_{\bar n} |  0 \rangle\,,
\end{align}
where $\mathcal{Y}$ denote adjoint Wilson lines.
Non-perturbative corrections in the collinear limit are important to describe the data, since they exhibit an enhanced scaling, $\sim1/z^{3/2}$, compared to $\sim1/z$ from the perturbative contributions. Additionally, we will find that the mixing between $\Omega_{1g}$ and $\Omega_{1q}$ is numerically important at the level of precision required to compare with the LEP re-analyses. While  $\Omega_{1g}$ is currently not known, the sensitivity of our EEC prediction to this parameter is quite interesting, and it should be extractable from precision measurements of the EEC. For the predictions of this paper, we will generally use the naive assumption of Casimir scaling, $\Omega_{1g}\sim C_A/C_F\,\Omega_{1q}$, however, we will study sensitivity to this parameter in our final results.

In the back-to-back limit there are multiple non-perturbative effects. First, there is a leading linear non-perturbative correction to the energy correlators \cite{Dokshitzer:1999sh}. In this paper we study this leading correction within the context of our factorization theorem, and show that to NLL order, it arises from a linear shift to the jet function in conjugate, $b$ space,
\begin{align}
J_q(b_\perp)&\to J_q(b_\perp)+ J_{q,\rm NP}(b_\perp) = J_q(b_\perp) -b_\perp \Omega_{1q}\,,
\end{align}
described by the same universal non-perturbative parameter $\Omega_{1q}$. At $\mathcal{O}(\alpha_s)$, this receives corrections from $\Omega_{1g}$. The appearance of $\Omega_{1g}$ in the collinear limit, but not the back-to-back limit, at the order we work arises from the different forms of resummation (single vs.\ double logarithmic) in the two limits. At higher logarithmic orders, they will both appear in both limits. At lowest order in perturbation theory, the non-perturbative corrections in the back-to-back limit give rise to a scaling of $1/(1-z)^{3/2}$, and our factorization theorem allows us to dress this with Sudakov logarithms to NLL.

Additionally, the energy correlator has logarithmically enhanced quadratic power corrections in the back-to-back limit. These arise from non-perturbative corrections to the anomalous dimensions describing the scaling in the back-to-back limit, in particular, a non-perturbative contribution to the rapidity anomalous dimension  $\gamma_\nu^{q,\text{NP}}(b_\perp)$. Compared to $\Omega$, this non-perturbative correction is a non-trivial \emph{function} of $b_\perp$. Remarkably, this non-perturbative function can now be computed from first principles using lattice QCD \cite{Avkhadiev:2024mgd,Avkhadiev:2023poz,Shanahan:2021tst,Shanahan:2019zcq,Shanahan:2020zxr}, and the large momentum effective theory \cite{Ji:2020ect,Izubuchi:2018srq,Ji:2014gla,Ji:2013dva}. In this paper we will use the lattice extraction of the Collins-Soper kernel from \cite{Avkhadiev:2024mgd,Avkhadiev:2023poz,Shanahan:2021tst,Shanahan:2019zcq,Shanahan:2020zxr} in our prediction, highlighting an exciting interaction with the lattice. 

Therefore, in summary, combining extractions of $\Omega$ from previous experiments with recent lattice data, we are able to provide a complete description of all leading non-perturbative corrections to the energy correlator, with no-independent parameters.

\subsection{Factorization Theorems and Resummation}\label{sec:fact_sum}

A precise description of the energy correlator in the kinematic limits $z\to 0$ and $z\to 1$ requires the resummation of all orders logarithmic correction, as well as the inclusion of non-perturbative corrections. We achieve this through the use of factorization theorems, which cleanly separate the dynamics at different scales, and enable the resummation of logarithmic corrections using the renormalization group. We use two distinct factorization theorems, one for the back-to-back limit, which provides a description of $\EEC^\text{fact.}_{z\to1}(z)$, and one of the collinear limit, which provides a description of $\EEC^\text{fact.}_{z\to0}(z)$.

The leading power dynamics in the back-to-back limit of the energy correlator, $\EEC^\text{fact.}_{z\to1} (z)$, is described by a factorization theorem for the energy correlator derived in SCET in ref.~\cite{Moult:2018jzp}, building on the seminal works of \cite{Collins:1981uk,Collins:1981va}. In this paper we extend it to include tracks.  It takes the form of a transverse momentum dependent (TMD) factorization theorem
\begin{mdframed}[linewidth=1.5pt, roundcorner=10pt]
\begin{align}\label{eq:b2bfactorization}
    \EEC_{z\to1}^{\text{fact.}}(z)
    &=
    \frac{\sigma_0 Q^2}{4\sigma}
    \int\df b_\perp\, b_\perp J_0\bigl(\sqrt{1-z}\,b_\perp Q\bigr)\,
    \nonumber\\
    &\qquad\times
    H(Q,\mu)\,S(b_\perp,\mu,\nu)
    \sum_q J_q(b_\perp,Q,\mu,\nu)\,J_{\bar{q}}(b_\perp,Q,\mu,\nu)\,.
\end{align}
\end{mdframed}\vspace{0.3cm}
This factorization theorem is expressed in terms of a hard function $H$, TMD jet functions $J_{q,\bar q}$ and a soft function $S$. The use of tracks enters only in the jet functions, since the soft radiation is not directly measured but only contributes through its recoil. Resummation is achieved by evaluating each of these ingredients at their natural $\mu$ ($\nu$) scales and using the (rapidity) renormalization group to evolve them to a common scale.

In this paper we compute the resummed result in the back-to-back limit at N$^4$LL. Since the logarithmic counting is distinct in the collinear and back-to-back limits of the EEC, we will use the notation N$^4$LL$_\text{b2b}$ to indicate the logarithmic counting in the back-to-back limit.\footnote{Note that we use a different counting for the logarithmic accuracy in the collinear and back-to-back limit due to the fact that the collinear limit is single logarithmic, while the back-to-back limit is double logarithmic. Our conventions for resummation are discussed in detail in \Sec{params}.} Resummation at N$^4$LL$_\text{b2b}$ uses the following state-of-the art perturbative ingredients
\begin{itemize}
\item Four \cite{Moch:2018wjh,Moch:2017uml,Davies:2016jie,Henn:2019swt} and approximate five \cite{Herzog:2018kwj} loop cusp anomalous dimension.
\item Five loop beta function \cite{Baikov:2016tgj,Herzog:2017ohr},
\item Four-loop rapidity anomalous dimension \cite{Duhr:2022yyp,Moult:2022xzt},
\item Three-loop jet function on tracks (new in this paper using \cite{Luo:2019hmp,Luo:2019bmw,Ebert:2020qef}), 
\item Three-loop TMD soft function \cite{Li:2016ctv}.
\end{itemize}
This matches the state of the art calculation achieved for the back-to-back limit of the energy correlator computed on all hadrons \cite{Duhr:2022yyp}, and extends it to a track-based calculation.

The leading power dynamics in the collinear limit of the EEC,
$\EEC^\text{fact.}_{z\to0}(z)$, is described by a factorization theorem for the collinear limit of the energy correlator derived in ref.~\cite{Dixon:2019uzg}. It is a collinear factorization theorem similar to those for timelike fragmentation, and takes the form
\begin{mdframed}[linewidth=1.5pt, roundcorner=10pt]
\begin{align}
\EEC_{z\to0}^{\text{fact.}}(z)=\frac{\sigma_0}{\sigma}\frac{\df}{\df z}\int_0^1\! \df x\, x^2 \vec J\Bigl(\ln \frac{z x^2 Q^2}{\mu^2},\mu\Bigr) \cdot \vec H\Bigl(x, \ln\frac{Q^2}{\mu^2},\mu\Bigr) 
\,.
\end{align}
\end{mdframed}\vspace{0.3cm}
The hard function $H$ and jet function $J$ differ from those in \eq{b2bfactorization} and are vectors in flavor space.
In this paper we perform the calculation at NNLL. To distinguish this resummation from the resummation in the back-to-back limit, we denote it as $\text{NNLL}_\text{col}$. Resummation at $\text{NNLL}_\text{col}$ uses the following perturbative inputs 
\begin{itemize}
\item Two-loop inclusive hard function \cite{Mitov:2006ic,Mitov:2006wy},
\item Three-loop timelike DGLAP anomalous dimensions  \cite{Mitov:2006ic,Mitov:2006wy,Chen:2020uvt},
\item Two-loop collinear EEC jet function on tracks (new in this paper),
\item Three-loop renormalization group equations for first two moments of the track functions \cite{Jaarsma:2022kdd}.
\end{itemize}
Combined, these two factorization theorems allow us to achieve high order perturbative accuracy in both the $z\to 0$ and $z\to 1$ limits of the EEC.

\subsection{Matching}\label{sec:match}

A theoretical description which simultaneously describes all kinematic limits of the energy correlator is non-trivial due to the necessity of performing distinct resummations in different kinematic limits. Additionally, distinct non-perturbative corrections must be incorporated in different regions of the distribution.

To achieve this, we must smoothly combine the descriptions the asymptotic expansions of the energy correlators provided by the factorization theorems, with the standard perturbative expansion. We can write the total expression for the energy correlator as the sum
\begin{align}
\EEC(z)&=\EEC_\text{bulk}(z)+ \EEC_{z\to0}(z)+  \EEC_{z\to1}(z) +  \EEC^\text{\text{plateau}}_{z\to0}(z)+\EEC^\text{\text{plateau}}_{z\to1}(z) \,.
\end{align}
When combining these different terms, we must avoid double counting. This is achieved by subtracting overlapping contributions obtained the factorized descriptions to fixed order in perturbation theory. This enables us to write the complete expression for the EEC as
\begin{align}\label{eq:eecdecomposition}
    \EEC(z)&=
    \Bigl[\EEC_\text{FO}(z)-\EEC_\text{FO}^\text{0-sing}(z)-\EEC_\text{FO}^\text{1-sing}(z)\Bigr]
    \\
    &
    +\Bigl[\EEC^{\Omega}(z) -\EEC_\text{bulk}^{\Omega,\text{0-sing}}(z) -\EEC_\text{bulk}^{\Omega,\text{1-sing}}(z) \Bigr]
    \nn\\
    &
    +\EEC^\text{fact.}_{z\to0}(z)
    + \EEC_{z\to0}^{\Omega,\text{res.}}(z) 
    \nonumber\\
    &
    +\EEC^\text{fact.}_{z\to1}(z)  +\EEC_{z\to1}^{\Omega,\text{res.}}(z) 
    \nonumber\\
    &
    + \EEC^\text{\text{plateau}}_{z\to0}(z)+ \EEC^\text{\text{plateau}}_{z\to1}(z)\,,
    \nonumber
\end{align}
where we have grouped terms in distinct lines to purposely highlight the structure. When performing the transitions between different kinematic regions, to ensure that no discontinuities arise between the resummed and fixed-order region, the resummation is smoothly turned off as a function of $z$ as the collinear and back-to-back regions merge into the bulk region. This is achieved through the use of profile scales \cite{Ligeti:2008ac,Abbate:2010xh}. We will discuss this in detail for both the collinear and back-to-back regions in subsequent sections.

\section{Bulk Region: $z\sim 1/2$}\label{sec:fo}

In this section we discuss in detail the calculation in the ``bulk region" of the energy correlator, namely when there are no scale hierarchies introduced by  $z$. In terms of our master formula, in eq.~\eqref{eq:eecdecomposition}, this corresponds to the terms 
\begin{align}
\EEC_{\text{bulk}}=\EEC_\text{FO}(z)    +\EEC^{\Omega}(z)\,.
\end{align}
As described in \Sec{overview}, this region probes primarily the UV of the theory, where we can reliably perform perturbative calculations for the matching onto track functions, with (other) non-perturbative effects suppressed by $\Lambda_{\text{QCD}}/Q$. In \Sec{bulktracks} we describe this fixed-order calculation, presenting our analytic expressions in terms of track function moments in \Sec{FOresult}. In \Sec{bulkcorrection} we discuss the leading non-perturbative correction, extending previous discussions in the literature to incorporate the effect of tracks. We present and discuss numerical results for the bulk region in \Sec{numerical_bulk}, focusing on the combination of all regions in \Sec{results}.

\subsection{Perturbative Calculations on Tracks}\label{sec:bulktracks}

State-of-the-art perturbative calculations for event shapes in $e^+e^-$ have achieved NNLO accuracy~\cite{Gehrmann-DeRidder:2007vsv,Gehrmann-DeRidder:2007nzq,Gehrmann-DeRidder:2007foh}, see in particular ref.~\cite{DelDuca:2016ily} for an NNLO calculation of the EEC. However these calculations are performed numerically, and require the observable to be infrared- and collinear safe. Since the EEC on tracks is not collinear safe, numerical techniques are not currently available.

The key advantage of the energy correlator is that there is a simple factorization theorem relating the observable on tracks to perturbative matching coefficients, which can be analytically calculated using modern integration techniques. We write the energy flow operator which detects only charged particles as $\cE_R$. This operator admits an OPE onto energy flow operators involving quarks and gluons, 
\begin{align}
\cE_R(\vec n_1)=T_{\bar q}(1) \cE_{\bar q} (\vec n_1)+  T_q(1) \cE_q (\vec n_1)+  T_g(1) \cE_g (\vec n_1) \,,
\end{align}
in terms of track function moments.
At the level of the two-point correlator, the relation reads 
\begin{align}\label{eq:match_track}
\langle \cE_R(n_1) \cE_R(n_2) \rangle &=\sum_{a_1,a_2}   T_{a_1}(1) T_{a_2}(1)  \langle \cE_{a_1} (\vec n_1) \cE_{a_2} (\vec n_2) \rangle \\
&+\left(\sum\limits_{a_1,a_2} T_{a_1}(1) T_{a_2}(1) \langle \cE_{a_1,a_2}^{(1,1)}(\vec n_1)\rangle+\sum\limits_a T_a(2) \langle \cE_a^{(2)}(\vec n_1)\rangle \right) \delta(\vec n_1-\vec n_2)\,. \nn
\end{align}
Here  $\langle \cE_{a_1} (\vec n_1) \cE_{a_2} (\vec n_2) \rangle$, $\langle \cE_{a_1,a_2}^{(1,1)}(\vec n_1)\rangle$, and $\langle \cE_a^{(2)}(\vec n_1)\rangle$ are perturbatively calculable matching coefficients. The terms proportional to $\delta(\vec n_1-\vec n_2)$ are contact terms, and do not contribute to the bulk of the distribution.  The formalism for performing this matching was described in detail in \cite{Li:2021zcf}.

The simple analytic structure of the energy correlator has enabled its analytic calculation at NLO in both $\cN=4$ SYM~\cite{Belitsky:2013ofa,Henn:2019gkr} and QCD~\cite{Dixon:2018qgp,Luo:2019nig}.  In ref.~\cite{Li:2021zcf} we extended this to a calculation of the matching coefficients in \eq{match_track}. We will use this analytic result in this paper, presenting the expressions here for the first time. The ability to obtain a finite result when measured on tracks illustrates the ability to factorize collinear divergences associated with track-based measurements. These fixed-order calculations exhibits unphysical behavior in the collinear ($z\to 0$) and back-to-back ($z\to 1$) regions that will be addressed  by resummation in sections~\ref{sec:col} and~\ref{sec:b2b}.

\subsection{Fixed-Order Results}\label{sec:FOresult}
In this section, we present the full fixed-order results of track EEC in $e^+e^-$ up to two loops\footnote{These results are available in a Mathematica notebook attached with the
submission of this article.}: 
\begin{align}
    \EEC_{\rm FO}(z) =\frac{\sigma_0}{\sigma}\left(\frac{\df\Sigma^{(0)}}{\df z}+a_s\frac{\df\Sigma^{(1)}}{\df z}+a_s^2\frac{\df\Sigma^{(2)}}{\df z}+\mathcal{O}(a_s^3)\right)\,,
\end{align}
where $\df\Sigma/\df z$ is the track EEC normalized to the Born-level cross section $\sigma_0$, with the superscripts denoting orders in $a_s \equiv \alpha_s/(4\pi)$. 
The leading order track EEC is given by 
\begin{align}
    \frac{\df\Sigma^{(0)}}{\df z}=\frac{1}{4}\bigl(T_q(2) + T_{\bar{q}}(2)\bigr)\,\delta(z)
    +\frac{1}{2}T_q(1)T_{\bar q}(1)\,\delta(1-z)\,,
\end{align}
which integrates to 1 on replacing $T_i(n)=1$ due to this normalization. 

In sections~\ref{sec:LO} and \ref{sec:NLO}, we will split $\df\Sigma^{(L)}/\df z$ ($L=1,2$) into three pieces, 
denoted by $\df\Sigma^{(L)}_\text{col}/\df z$, $\df\Sigma^{(L)}_\text{b2b}/\df z$ and $\df\Sigma^{(L)}_\text{bulk}/\df z$ respectively: 
the collinear ($z\to 0$) part with $\delta(z)$ and plus distributions in $z$, the back-to-back ($z\to 1$) part with $\delta(1-z)$ and plus distributions in $1-z$, 
and the bulk contribution which is non-singular. We emphasize that although our fixed-order calculations include these delta function contributions at 0 and 1, resummation is required to obtain reliable results in the $z\to0,1$ limit. Nevertheless, this provides crucial perturbative data for performing the  resummation.

\subsubsection{LO}\label{sec:LO}
The full one-loop result reads
\begin{align}
    \frac{\df\Sigma^{(1)}}{\df z}=\frac{\df\Sigma^{(1)}_\text{col}}{\df z}+\frac{\df\Sigma^{(1)}_\text{b2b}}{\df z}+\frac{\df\Sigma^{(1)}_\text{bulk}}{\df z}\,,
\end{align}
where
\begin{align}\label{eq:trackEEC_1loop_col}
    \frac{\df\Sigma^{(1)}_\text{col}}{\df z}
    &=
    T_g(2)\,C_F  \delta (z) \left(-\frac{7 L}{6}-\frac{71}{48}\right) 
    +T_q(2)\,C_F  \delta (z)\left(\frac{25 L}{6}+\frac{131}{16}\right)\\
    &\quad
    +T_g(1) T_q(1)\,C_F \left[\left(-3 L-\frac{37}{6}\right) \delta (z)+\frac{3}{2}\Bigl[\frac{1}{z}\Bigr]_+\right]
    \,,\nn
\end{align}
\begin{align}\label{eq:trackEEC_1loop_b2b}
    \frac{\df\Sigma^{(1)}_\text{b2b}}{\df z}
    &=
    T_g(1) T_q(1)\, C_F \left[\left(-\frac{16 L}{3}-\frac{52}{9}\right) \delta (1-z)
    +\frac{8}{3}\Bigl[\frac{1}{1-z}\Bigr]_+\right]\\
    &\quad
   +T_q(1)T_q(1)\, C_F \left[\left(\frac{16 L}{3}-\frac{\pi
   ^2}{3}+\frac{16}{9}\right) \delta (1-z)
   -\frac{17}{3}\Bigl[\frac{1}{1-z}\Bigr]_+ 
   -2\Bigl[\frac{\ln(1-z)}{1-z}\Bigr]_+\right]\,,\nn
\end{align}
and 
\begin{align}\label{eq:trackEEC_1loop_bulk}
    &\frac{\df\Sigma^{(1)}_\text{bulk}}{\df z}\,\\
    &=
    T_g(1) T_q(1)\,C_F  \biggl(\frac{7 z^3+16 z^2-90 z+156}{6 z^4}+\frac{2 \left(4 z^2-14
   z+13\right) \ln (1-z)}{z^5}\biggr)\,\nn\\
   &\quad
   +T_q(1)T_{{q}}(1) C_F \biggl(-\frac{17 z^3+17 z^2+18 z+24}{3z^4}-\frac{2 \left(z^4+z^3+z^2+z+4\right) \ln (1-z)}{z^5}\biggr)\nn\,,
\end{align}
with $L\equiv \ln(\mu/Q)$ and the subscript ``$+$'' indicates that this is a plus distribution. 
In these expressions we assumed $T_q = T_{\bar q}$, which holds for track functions. However, these expressions can also be used when the measurements are performed on other subsets of final state particles, e.g.~positively charged particles only.
In this case, the quark vs.~anti-quark information can straightforwardly be recovered by replacing 
$T_q(1)T_{{q}}(1)$ with $T_q(1)T_{\bar{q}}(1)$, $T_g(1) T_q(1)$ with $[T_g(1) T_q(1)+T_g(1) T_{\bar{q}}(1)]/2$, and $T_q(2)$ with $[T_q(2)+T_{\bar{q}}(2)]/2$. 
Setting all the moments in eqs.~\eqref{eq:trackEEC_1loop_col}-\eqref{eq:trackEEC_1loop_bulk} to one, yields the all-particle expression at one loop which has no $\mu$ dependence except  from $a_s(\mu)$. 

\subsubsection{NLO}\label{sec:NLO}
Due to the length of the full expression at two-loop order, 
we first separate the scale-dependent logarithmic part from the constant part, 
and then
divide each part into the aforementioned collinear, back-to-back and bulk contributions. 
The full two-loop result thus reads
\begin{align}
    \frac{\df\Sigma^{(2)}}{\df z}
    & =\frac{\df\Sigma^{(2,0)}_\text{col}}{\df z}+\frac{\df\Sigma^{(2,0)}_\text{b2b}}{\df z}+\frac{\df\Sigma^{(2,0)}_\text{bulk}}{\df z}
    +L\biggl[\frac{\df\Sigma^{(2,1)}_\text{col}}{\df z}+\frac{\df\Sigma^{(2,1)}_\text{b2b}}{\df z}+\frac{\df\Sigma^{(2,1)}_\text{bulk}}{\df z}\biggr]\\
    &\quad
    +L^2\biggl[\frac{\df\Sigma^{(2,2)}_\text{col}}{\df z}+\frac{\df\Sigma^{(2,2)}_\text{b2b}}{\df z}+\frac{\df\Sigma^{(2,2)}_\text{bulk}}{\df z}\biggr]
    \, \nn
\end{align}
with again $L\equiv\ln(\mu/Q)$. 
The ingredients are presented below.
\medskip

\noindent {\bf Constant terms: }

\noindent We split each $\df\Sigma^{(2,0)}_{\text{reg}}/\df z$ with $\text{``reg''}=\text{``col'', ``b2b'', ``bulk''}$ into nine terms according to the track function combinations they involve:
\begin{align}
    \frac{\df\Sigma^{(2,0)}_{\text{reg}}}{\df z}
    &=
    T_g(2)\,\frac{\df\Sigma^{(2,0)}_{\text{reg}}}{\df z}\bigg|_{T_g(2)}
    +T_g(1)T_g(1)\,\frac{\df\Sigma^{(2,0)}_{\text{reg}}}{\df z}\bigg|_{T_g(2)}
    \\
    &\quad
    +\Big(T_q(2)+T_{\bar{q}}(2)\Big)\,\frac{\df\Sigma^{(2,0)}_{\text{reg}}}{\df z}\bigg|_{T_q(2)}
    +\sum_{Q\neq q}\Big(T_Q(2)+T_{\bar{Q}}(2)\Big)\,\frac{\df\Sigma^{(2,0)}_{\text{reg}}}{\df z}\bigg|_{T_Q(2)}
    \nn\\
    &\quad
    +\Big(T_g(1)T_q(1)+T_g(1) T_{\bar{q}}(1)\Big)
    \,\frac{\df\Sigma^{(2,0)}_{\text{reg}}}{\df z}\bigg|_{T_g(1)T_q(1)}
    +T_q(1)T_{\bar{q}}(1)\,\frac{\df\Sigma^{(2,0)}_{\text{reg}}}{\df z}\bigg|_{T_q(1)T_{\bar q}(1)}
    \nn\\
    &\quad
    +\Big(T_q(1)T_q(1)+T_{\bar{q}}(1)T_{\bar{q}}(1)\Big)\,\frac{\df\Sigma^{(2,0)}_{\text{reg}}}{\df z}\bigg|_{T_q(1)T_q(1)\!}
    +\sum_{Q\neq q}T_Q(1) T_{\bar{Q}}(1)\,\frac{\df\Sigma^{(2,0)}_{\text{reg}}}{\df z}\bigg|_{T_Q(1)T_{\bar Q}(1)\!}
    \nn\\
    &\quad
    +\sum_{Q\neq q}\Big(T_q(1) T_Q(1)+T_q(1) T_{\bar{Q}}(1)+T_{\bar{q}}(1)T_Q(1)+T_{\bar{q}}(1) T_{\bar{Q}}(1)\Big)\,\frac{\df\Sigma^{(2,0)}_{\text{reg}}}{\df z}\bigg|_{T_q(1)T_Q(1)}
    \,.\nn
\end{align}
In this case we do not assume charge conjugation symmetry, treating $T_q$ and $T_{\bar q}$ as independent.
The perturbative ingredients in the collinear limit are 
\begin{align}\label{eq:treec2L_col0_tg2}
    &\frac{\df\Sigma^{(2,0)}_\text{col}}{\df z}\bigg|_{T_g(2)}
    \\
    &
    =
    \delta (z) \bigg[\left(-\frac{19}{3}\zeta_3-\frac{29802739}{1296000}+\frac{47 \pi^2}{270}\right) C_A C_F
   +\left(\frac{31}{3}\zeta_3-\frac{674045}{20736}+\frac{523 \pi^2}{432}\right) C_F^2\bigg]
    \,,\nn
    \\
    &\frac{\df\Sigma^{(2,0)}_\text{col}}{\df z}\bigg|_{T_g(1)T_g(1)}
    \nn\\
    &=
    \delta (z) \left[\left(\frac{91}{6} \zeta_3-\frac{11059849}{1296000}+\frac{65 \pi
   ^2}{72}\right) C_A C_F+\left(\frac{46613}{864}-\frac{4 \pi ^2}{3}\right)
   C_F^2\right]
   \nn\\
   &\quad
   +\left[\frac{1}{z}\right]_+\bigg[\left(\frac{52681}{10800}-\frac{7 \pi ^2}{18}\right) C_A
   C_F-\frac{659}{36}C_F^2\bigg]
   +\left[\frac{\ln z}{z}\right]_+ \left(4 C_F^2-\frac{49}{30}C_AC_F\right)
    \,,\nn
    \\
    &\frac{\df\Sigma^{(2,0)}_\text{col}}{\df z}\bigg|_{T_q(2)}
    \nn\\
    &=
    \delta (z) \bigg[
    \left(-\frac{293}{6}\zeta_3 +\frac{2386397}{20736}-\frac{83 \pi
   ^2}{24}+\frac{4 \pi ^4}{45}\right) C_A C_F
   +\left(\frac{3479299}{1296000}-\frac{73 \pi ^2}{1080}\right) C_F T_F
   \nn\\
   &\quad 
   +\left(4 \zeta_3-\frac{116287}{5184}+\frac{25 \pi ^2}{108}\right) C_F n_fT_F
   +\left(\frac{127}{3}\zeta_3-\frac{1105289}{41472}+\frac{1751\pi ^2}{864}-\frac{8 \pi ^4}{45}\right)C_F^2
   \bigg]
   \,,\nn
   \\
    &\frac{\df\Sigma^{(2,0)}_\text{col}}{\df z}\bigg|_{T_Q(2)}
    =
    \delta (z)\left(\frac{3479299}{1296000}-\frac{73 \pi^2}{1080}\right) C_F T_F 
   \,,\nn
   \\
    &\frac{\df\Sigma^{(2,0)}_\text{col}}{\df z}\bigg|_{T_g(1)T_q(1)}
    \nn\\
    &
    =
    \delta (z) \bigg[\left(\frac{61}{6} \zeta_3-\frac{72811}{1728}-\frac{5 \pi^2}{18}\right) C_A C_F
   +\left(15 \zeta_3-\frac{74551}{768}+\frac{137 \pi^2}{36}\right) C_F^2
   \bigg]
   \nn\\
   &\quad
   +\left[\frac{1}{z}\right]_+\left(\frac{1043}{72} C_A C_F+\frac{617}{72}C_F^2\right)
   +\left[\frac{\ln z}{z}\right]_+ \left(\frac{9}{8}C_F^2-\frac{11}{4} C_A C_F\right)
    \,,\nn
    \\
    &\frac{\df\Sigma^{(2,0)}_\text{col}}{\df z}\bigg|_{T_q(1)T_{\bar q}(1)}
    \nn\\
    &
    =
    \delta (z) \bigg[
    \left(\frac{1069 \zeta_3}{6}-\frac{261119}{1152}+\frac{1675 \pi^2}{216}-\frac{59 \pi^4}{90}\right) C_A C_F
   +\left(\frac{396557}{24000}+\frac{\pi^2}{9}\right) C_F T_F
   \nn\\
   &\quad
   +\left(-\frac{1069 \zeta_3}{3}+\frac{261119}{576}-\frac{1675 \pi
   ^2}{108}+\frac{59 \pi ^4}{45}\right) C_F^2
   \bigg]
   +\frac{23 }{30}C_F T_F\left[\frac{\ln(z)}{z}\right]_+
   \nn\\
   &\quad
   +\left[\frac{1}{z}\right]_+
   \bigg[\left(8 \zeta_3 +\frac{14057}{432}-\frac{77 \pi ^2}{18}\right) C_A C_F-\frac{4801}{900}C_FT_F+\left(-16 \zeta_3 -\frac{14057}{216}+\frac{77 \pi ^2}{9}\right)C_F^2\bigg]
   \,,\nn
   \\
    &\frac{\df\Sigma^{(2,0)}_\text{col}}{\df z}\bigg|_{T_q(1)T_{q}(1)}
    \nn\\
    &
    =
    \delta (z)\left[
    C_F \left(C_A-2 C_F\right) \left(-\frac{221}{6} \zeta_3 +\frac{485129}{10368}-\frac{37 \pi^2}{27}+\frac{7 \pi^4}{60}\right)
    +C_FT_F\left(\frac{1537}{192}+\frac{\pi^2}{18}\right)
   \right]
   \nn\\
   &\quad
   +\left[\frac{1}{z}\right]_+\left[C_F\left(C_A-2 C_F\right) \left(-2 \zeta_3-\frac{3023}{432}+\frac{17 \pi^2}{18}\right) -\frac{67}{24}C_F T_F\right]
   +\frac{1}{2}C_F T_F \left[\frac{\ln z}{z}\right]_+
   \,,\nn
   \\
    &\frac{\df\Sigma^{(2,0)}_\text{col}}{\df z}\bigg|_{T_q(1)T_{Q}(1)}
    =
    C_F T_F \left[\left(\frac{1537}{192}+\frac{\pi ^2}{18}\right) \delta (z)
    -\frac{67}{24}\left[\frac{1}{z}\right]_++\frac{1}{2}\left[\frac{\ln z}{z}\right]_+\right]
   \,,\nn
   \\
    &\frac{\df\Sigma^{(2,0)}_\text{col}}{\df z}\bigg|_{T_Q(1)T_{\bar Q}(1)}
    =
    C_F T_F \left(
    \frac{12307}{24000}\delta (z)+\frac{56}{225}\left[\frac{1}{z}\right]_+
    -\frac{7}{30}\left[\frac{\ln z}{z}\right]_+
   \right)
    \,.\nn
\end{align}
The perturbative ingredients in the back-to-back limit are
\begin{align}\label{eq:treec2L_b2b0_tg2}
    &\frac{\df\Sigma^{(2,0)}_\text{b2b}}{\df z}\bigg|_{T_g(2)}
    =\frac{\df\Sigma^{(2,0)}_\text{b2b}}{\df z}\bigg|_{T_q(2)}
    =\frac{\df\Sigma^{(2,0)}_\text{b2b}}{\df z}\bigg|_{T_Q(2)}
    =0\,,
    \\
    &\frac{\df\Sigma^{(2,0)}_\text{b2b}}{\df z}\bigg|_{T_g(1)T_g(1)}
    =
    \frac{1352}{81} \,C_F^2\, \delta (1-z)
    +C_F^2 \left(\frac{64}{9}\left[\frac{\ln(1-z)}{1-z}\right]_+-\frac{416}{27}\left[\frac{1}{1-z}\right]_+\right)
   \,,\nn
   \\
    &\frac{\df\Sigma^{(2,0)}_\text{b2b}}{\df z}\bigg|_{T_g(1)T_q(1)}
    \nn\\
    &
    =
    \delta (1-z) \bigg[
    \left(16 \zeta_3 -\frac{10427}{324}-\frac{8 \pi ^2}{9}\right)C_AC_F
   +\left(\frac{16}{3} \zeta_3 -\frac{12061}{324}+\frac{104 \pi ^2}{27}\right)C_F^2
   \bigg]
   \nn\\
   &\quad
   +C_F^2 \bigg\{
   \left(\frac{1088}{27}-\frac{8 \pi ^2}{3}\right)\left[\frac{1}{1-z}\right]_+
   -\frac{136}{9}\left[\frac{\ln (1-z)}{1-z}\right]_+
   -8\left[\frac{\ln^2 (1-z)}{1-z}\right]_+
   \bigg\}
   \nn\\
   &\quad
   +C_A C_F \left(\frac{158}{9}\left[\frac{1}{1-z}\right]_+ 
   -\frac{44}{9}\left[\frac{\ln (1-z)}{1-z}\right]_+\right)
    \,,\nn
    \\
    &\frac{\df\Sigma^{(2,0)}_\text{b2b}}{\df z}\bigg|_{T_q(1)T_{\bar q}(1)}
    \nn\\
    &
    =
    \delta (1-z) \bigg[
    \left(\frac{\zeta_3}{3}+\frac{1717}{24}-\frac{139 \pi^2}{27}+\frac{23 \pi^4}{90}\right) C_A C_F
    +\left(\frac{8 \zeta_3}{3}-\frac{704}{81}-\frac{16 \pi^2}{27}\right) C_F n_f T_F
    \nn\\
    &\quad
    +\left(\frac{874}{81}+\frac{4 \pi^2}{9}\right) C_F T_F
    +\left(-34 \zeta_3 -\frac{9013}{324}+\frac{565 \pi^2}{54}-\frac{7 \pi^4}{45}\right) C_F^2
   \bigg]
   \nn\\
   &\quad
   +C_F^2 \bigg\{
   \left(\frac{50 \pi^2}{9}-\frac{91}{6}\right)\left[\frac{1}{1-z}\right]_+
   +\left(\frac{4 \pi^2}{3}+\frac{514}{9}\right)\left[\frac{\ln(1-z)}{1-z}\right]_+
   \nn\\
   &\quad
   +34\left[\frac{\ln^2(1-z)}{1-z}\right]_+
   +4\left[\frac{\ln^3(1-z)}{1-z}\right]_+
   \bigg\}
   \nn\\
   &\quad
   +C_A C_F \bigg\{
   \left({8 \zeta_3}+\frac{50 \pi ^2}{9}-\frac{1792}{27}\right) \left[\frac{1}{1-z}\right]_+
   +\left(\frac{2 \pi ^2}{3}+\frac{53}{9}\right)\left[\frac{\ln(1-z)}{1-z}\right]_+
   \nn\\
   &\quad
   +\frac{22}{3}\left[\frac{\ln^2(1-z)}{1-z}\right]_+
   \bigg\}
   +C_F n_f T_F \bigg\{
   \left(-\frac{4 \pi ^2}{3}+\frac{166}{9}\right)\left[\frac{1}{1-z}\right]_+
   -\frac{28}{9}\left[\frac{\ln(1-z)}{1-z}\right]_+
   \nn\\
   &\quad
   -\frac{8}{3}\left[\frac{\ln^2(1-z)}{1-z}\right]_+
   \bigg\}
   +C_F T_F \left(\frac{16}{9}\left[\frac{\ln(1-z)}{1-z}\right]_+
   -\frac{56}{9}\left[\frac{1}{1-z}\right]_+\right)
    \,,\nn
    \\
    &\frac{\df\Sigma^{(2,0)}_\text{b2b}}{\df z}\bigg|_{T_q(1)T_{q}(1)}
    \nn\\
    &
    =
    \delta (1-z) \left[C_F \left(C_A-2 C_F\right)\left(\frac{85 \zeta_3}{6}-\frac{32155}{1296}+\frac{5 \pi^2}{2}-\frac{31 \pi^4}{180}\right)
   +C_F T_F\left(\frac{437}{81}+\frac{2 \pi^2}{9}\right)\right]
   \nn\\
   &\quad 
   +C_F\left(C_A-2 C_F\right)\left(2\zeta_3+\frac{743}{108}-\frac{17 \pi^2}{18}\right)
   \left[\frac{1}{1-z}\right]_+
   \nn\\
   &\quad
   +C_F T_F \left(\frac{8}{9}\left[\frac{\ln(1-z)}{1-z}\right]_+
   -\frac{28}{9}\left[\frac{1}{1-z}\right]_+\right)
    \,,\nn
    \\
    &\frac{\df\Sigma^{(2,0)}_\text{b2b}}{\df z}\bigg|_{T_q(1)T_{Q}(1)}
    =
    C_F T_F \left[\left(\frac{437}{81}+\frac{2 \pi ^2}{9}\right) \delta (1-z)
    -\frac{28}{9}\left[\frac{1}{1-z}\right]_+
    +\frac{8}{9}\left[\frac{\ln(1-z)}{1-z}\right]_+\right]
    \,,\nn
    \\
    &\frac{\df\Sigma^{(2,0)}_\text{b2b}}{\df z}\bigg|_{T_Q(1)T_{\bar Q}(1)}
    =
    0\,.\nn
\end{align}
The perturbative ingredients for the bulk contribution are
\begin{align}\label{eq:treec2L_bulk0_tg2}
    &\frac{\df\Sigma^{(2,0)}_\text{bulk}}{\df z}\bigg|_{T_g(2)}
    =\frac{\df\Sigma^{(2,0)}_\text{bulk}}{\df z}\bigg|_{T_q(2)}
    =\frac{\df\Sigma^{(2,0)}_\text{bulk}}{\df z}\bigg|_{T_Q(2)}
    =0\,,
    \\
    &\frac{\df\Sigma^{(2,0)}_\text{bulk}}{\df z}\bigg|_{T_g(1)T_g(1)}
    \nn\\
    &
    =
    C_A C_F \bigg[
    \frac{2 (1-z)^2}{z^5}g_5^{(3)}-\frac{2 (1-z)^2}{3 z^5}g_2^{(3)}
    +\frac{71 z^2-216 z+124}{15z^5}g_2^{(2)} 
    -\frac{203 z^2-795 z+480}{60 z^{9/2}}g_3^{(2)}
    \nn\\
    &\quad
    +\frac{12 z^5-25 z^4+40 z^3-30 z^2-140 z+113}{30 z^5}g_1^{(2)} 
    +\frac{35 z^4-221 z^2+476z-237}{15 z^5}\zeta_2
    \nn\\
    &\quad
    -\frac{(1-z) \left(24z^3-14 z^2-1057 z+1569\right)}{30 z^5}g_1^{(1)}
    +\frac{49 z^3+623 z-488}{30 z^4}g_2^{(1)} 
   \nn\\
   &\quad
   -\frac{44041 z^3+11520 z^2-317520 z+381240}{10800 z^4}
   \bigg]
   +C_F^2
   \bigg[
   \frac{4 (1-z)^2}{3 z^5}g_2^{(3)}
   -\frac{4 (1-z)^2}{z^5}g_5^{(3)}
   \nn\\
   &\quad 
   -\frac{2 \left(5 z^6-9 z^5+5 z^4+20 z^3-110 z^2+245 z-231\right)}{15z^5}g_1^{(2)}
   -\frac{4 \left(21 z^2-91 z+40\right)}{15 z^5}g_2^{(2)}
   \nn\\
   &\quad
   +\frac{129 z^2-400 z+240}{15 z^{9/2}}g_3^{(2)}
   +\frac{4 \left(z^2+94z-191\right)}{15 z^5}\zeta _2
   \nn\\
   &\quad
   -\frac{2 \left(30z^5-199 z^4-147 z^3+19 z^2-2846 z+3628\right)}{45 z^5}g_1^{(1)}
   -\frac{4 \left(15 z^3+151 z-100\right)}{15 z^4}g_2^{(1)} 
   \nn\\
   &\quad
   -\frac{720 z^4-2321 z^3+8508 z^2-70668 z+140496}{540 z^4}
   \bigg]
    \,,\nn
    \\
    &\frac{\df\Sigma^{(2,0)}_\text{bulk}}{\df z}\bigg|_{T_g(1)T_q(1)}
    \nn\\
    &
    =
    C_A C_F \bigg[
    -8 \left(5 z^2-5 z+1\right)^2 g_1^{(3)} 
    +\frac{2 z^2-10z+5}{3 z^5}g_2^{(3)}
   \nn\\
   &\quad
   +\frac{8 z^4+4 z^3+94 z^2-172 z+51}{6 z^5}g_1^{(2)}
   -\frac{1641 z^2-4566 z+2626}{30 z^5}\zeta_2
   \nn\\
   &\quad
   -\frac{12000 z^8-18000 z^7+8800 z^6-1400 z^5+80 z^4+40
   z^3-701 z^2+2846 z-2116}{30 z^5}g_2^{(2)}
   \nn\\
   &\quad
   -\bigg(400 z^3-700 z^2+\frac{23}{9 z^2}+\frac{5063}{180 z^3}-\frac{145}{3 z^4}-\frac{137}{20 z^5}+\frac{3640 z}{9}+\frac{82}{9 z}-\frac{755}{9}
   \bigg)g_1^{(1)}
   \nn\\
   &\quad
   -\frac{1}{1\!-\!z}
   \bigg(400 z^4-900 z^3+\frac{6340 z^2}{9}+\frac{2657}{45 z^2}-\frac{1952}{15 z^3}+\frac{1058}{15 z^4}-225 z-\frac{37}{12 z}+\frac{991}{36}
   \bigg)g_2^{(1)}
   \nn\\
   &\quad
   -\frac{21600 z^6-21600 z^5+6540 z^4-663 z^3-4048 z^2+13635 z-28278}{216 z^4}
   \bigg]
   \nn\\
   &\quad
   +C_F^2 \bigg[
   -8 \left(35 z^4-70 z^3+46 z^2-11 z+1\right)g_1^{(3)}
   +\frac{4 \left(z^2-2 z+4\right)}{3 z^5}g_2^{(3)}
   \nn\\
   &\quad
   +\frac{4 z^6-2 z^5-16 z^4-20 z^3-88 z^2+106 z-141}{3 z^5}g_1^{(2)}
   \nn\\
   &\quad
   -\bigg(560 z^3-840 z^2+\frac{8}{3 z^2}+\frac{167}{15 z^3}-\frac{517}{15 z^4}+\frac{997}{30
   z^5}+\frac{1088}{3}z+\frac{16}{3 z}-\frac{124}{3}\bigg)
   g_2^{(2)} 
   \nn\\
   &\quad
   -\frac{16}{1-z}g_4^{(2)}
   +\frac{480 z^4+480 z^3+2094 z^2-3154 z+3817}{30 z^5}\zeta_2
   \nn\\
   &\quad
   -\bigg(560 z^3-980 z^2+\frac{484}{45 z^2}-\frac{9143}{180 z^3}+\frac{3689}{15 z^4}-\frac{41453}{180 z^5}+\frac{4664 z}{9}+\frac{958}{45 z}-\frac{757}{9}\bigg)g_1^{(1)}
   \nn\\
   &\quad
   +\bigg(560 z^3-700 z^2+\frac{224}{45 z^2}-\frac{357}{20 z^3}+\frac{997}{30 z^4}+\frac{2144}{9}z+\frac{1819}{360 z}-\frac{151}{9}\bigg)g_2^{(1)} 
   \nn\\
   &\quad
   -\frac{151200 z^6-151200 z^5+29940 z^4-34985 z^3-48188 z^2+102207 z-278454}{1080 z^4}
   \bigg]
    \,,\nn
    \\
    &\frac{\df\Sigma^{(2,0)}_\text{bulk}}{\df z}\bigg|_{T_q(1)T_{\bar q}(1)}
    \nn\\
    &
    =
    C_F^2
    \bigg[
    -\frac{4 \left(1320 z^6-3960 z^5+4548 z^4-2505 z^3+687 z^2-82 z+8\right)}{3(1-z) z}g_1^{(3)}
    -\frac{4}{(1-z) z^5}g_7^{(3)}
    \nn\\
   &\quad
   -\frac{2 \left(6 z^7-12 z^6+29 z^5-4 z^4-18 z+21\right)}{9 (1-z)z^5}g_2^{(3)}
   +\frac{2 (z-4) (z+4) }{3 (1-z)}g_3^{(3)}
   -\frac{8 (19 z-2)}{(1-z) z}\zeta _3
   \nn\\
   &\quad
   -\frac{4 (z+1) \left(3 z^2-9 z-2\right)}{3 (1-z) z}\left(2g_4^{(3)}+g_5^{(3)}\right)
   -\frac{2 \left(2 z^5-1\right)}{(1-z)z^5}g_6^{(3)}
   -\frac{4 \left(3 z^2-20 z+1\right)}{3 (1-z) z}g_8^{(3)}
    \nn\\
    &\quad
   -\frac{86 z^6+214 z^5-735 z^4-500 z^3-360 z^2-386 z-459}{15 z^5}g_1^{(2)}
   +\frac{68}{1-z}g_4^{(2)}
   \nn\\
   &\quad
   +4 \left(880 z^3-1320 z^2+\frac{2056
   z}{3}-\frac{386}{3}+\frac{23}{z}+\frac{5}{z^2}+\frac{27}{5 z^3}+\frac{14}{5
   z^4}+\frac{5}{z^5}\right) g_2^{(2)}
   \nn\\
   &\quad
   -\frac{240 z^3-400 z^2+128 z+7}{15 z^{5/2}}g_3^{(2)}
   +\frac{2 \left(180 z^5-1425 z^4-650 z^3-522 z^2-470 z-609\right)}{15 z^5}\zeta _2
   \nn\\
   &\quad
   +\left(3520 z^3-6160 z^2+\frac{33472 z}{9}-\frac{40036}{45}+\frac{6569}{45z}
   +\frac{3139}{45 z^2}+\frac{1123}{15 z^3}+\frac{1283}{45 z^4}-\frac{538}{15
   z^5}\right) g_1^{(1)}
   \nn\\
   &\quad
   +\frac{2}{1-z}\left(1760 z^4-3960 z^3+\frac{28616 z^2}{9}-1094 z+\frac{7013}{45}-\frac{83}{9 z}-\frac{24}{5 z^2}+\frac{7}{5
   z^3}-\frac{10}{z^4}\right)g_2^{(1)}
   \nn\\
   &\quad
   +880 z^2-880 z
   +\frac{52321}{1080
   z}+\frac{389}{18 z^2}-\frac{799}{18 z^3}-\frac{856}{15 z^4}+\frac{12914}{45}
   \bigg]
   \nn\\
   &\quad
   +C_A C_F \bigg[
   \frac{2 \left(1320 z^6-3960 z^5+4548 z^4-2505 z^3+687 z^2-101 z+8\right)}{3 (1-z)z} g_1^{(3)}
   \nn\\
   &\quad
   +\frac{\left(6 z^7-12 z^6+10 z^5-4 z^4+18 z-27\right)}{9 (1-z) z^5}g_2^{(3)}
   -\frac{2}{3 (1-z)}g_3^{(3)}
   +\frac{2 \left(3z^2-z+1\right)}{3 (1-z) z}g_8^{(3)}
   \nn\\
   &\quad
   +\frac{2 \left(3 z^3-6 z^2+8 z-2\right)}{3 (1-z) z}\left(2 g_4^{(3)}+g_5^{(3)}\right)
   -\frac{g_6^{(3)}-2 g_7^{(3)}}{(1-z) z^5}
   -\frac{8 (1+z)}{(1-z)z}\zeta_3
   \nn\\
   &\quad
   \!+\!\frac{36 z^6+134 z^5-115z^4+40 z^3+250 z^2+284 z+301}{30 z^5}g_1^{(2)}
   \!+\!\frac{240 z^3\!-\!400 z^2\!+\!128 z\!+\!7}{30 z^{5/2}}g_3^{(2)}
   \nn\\
   &\quad 
   -2 \left(880 z^3-1320 z^2+\frac{2056
   z}{3}-\frac{386}{3}+\frac{23}{z}+\frac{5}{z^2}+\frac{27}{5 z^3}+\frac{14}{5
   z^4}+\frac{5}{z^5}\right) g_2^{(2)}
   +\frac{44}{3 (1-z)}g_4^{(2)}
   \nn\\
   &\quad
   -\left(1760 z^3-3080 z^2+\frac{16736 z}{9}-\frac{19778}{45}+\frac{2117}{45 z}+\frac{214}{15 z^2}+\frac{1102}{45 z^3}-\frac{871}{45 z^4}+\frac{4943}{45 z^5}\right) g_1^{(1)}
   \nn\\
   &\quad
   -\frac{1}{1-z}\left(1760 z^4-3960 z^3+\frac{28616}{9}z^2-1094 z+\frac{7013}{45}-\frac{83}{9 z}-\frac{24}{5 z^2}+\frac{7}{5 z^3}-\frac{10}{z^4}\right)g_2^{(1)}
   \nn\\
   &\quad
   -\frac{120 z^5-805 z^4-110 z^3+88 z^2+200 z+151}{15 z^5}\zeta_2 
   \nn\\
   &\quad
   -440 z^2+440 z
   -\frac{79567}{720 z}-\frac{955}{9 z^2}-\frac{299}{4 z^3}-\frac{6746}{45
   z^4}-\frac{6787}{45}
   \bigg]
   \nn\\
   &\quad
   + C_F n_f T_F \bigg[
   \frac{4 \left(z^6+z^5-z^4-z^3-z^2-z-4\right)}{3 z^5}g_1^{(2)}
   \!-\!\frac{16g_4^{(2)}}{3(1\!-\!z)}
   \!+\!\frac{8\left(z^4\!+\!z^3\!+\!z^2\!+\!z\!+\!4\right)}{3 z^5}\zeta _2
   \nn\\
   &\quad
   +\frac{4\left(6 z^5+2 z^4+4 z^3+7 z^2-5 z+78\right)}{9 z^5}g_1^{(1)}
   +\frac{4 \left(18 z^4+156 z^3+167 z^2+174 z+306\right)}{27 z^4}
   \bigg]
   \nn\\
   &\quad
   + C_F T_F \bigg[
   \frac{4}{3}\left(240 z^4-480 z^3+338 z^2-98 z+7\right) g_1^{(3)}
   -\frac{4}{9}\left(2 z^2-2 z+1\right) g_2^{(3)}
   \nn\\
   &\quad
   -\frac{2}{3} \left(4 z^2-6 z+3\right) g_3^{(3)}
   -\frac{4}{3} \left(2 z^2-2 z+1\right) \left(2 g_4^{(3)}+g_5^{(3)}-2 g_8^{(3)}\right)
   \nn\\
   &\quad
   -\frac{4356 z^7-5306 z^6+1449 z^5-560 z^4-350 z^3+3444 z^2-2919 z+586}{105 z^5}g_1^{(2)}
   \nn\\
   &\quad
   +\frac{2 \left(33600 z^8-50400 z^7+25480 z^6-4340 z^5-973 z^2+2583 z-2189\right)}{105 z^5}g_2^{(2)}
   \nn\\
   &\quad
   +\frac{\left(6720 z^5-4480 z^4+2464 z^3+736 z^2-119 z+35\right)}{210 z^{7/2}}g_3^{(2)}
   \nn\\
   &\quad
   +\left(640 z^3-1120 z^2+\frac{182744 z}{315}-\frac{22252}{315}-\frac{1894}{315 z}+\frac{3833}{315 z^2}+\frac{3152}{105 z^3}-\frac{465}{7 z^4}+\frac{359}{45 z^5}\right) g_1^{(1)}
   \nn\\
   &\quad
   +\frac{1}{1-z}\left(640 z^4-1440 z^3+\frac{10000 z^2}{9}-\frac{1012 z}{3}+\frac{189}{10}+\frac{2003}{630 z}+\frac{1219}{35 z^2}-\frac{488}{7 z^3}+\frac{4378}{105 z^4}\right)g_2^{(1)}
   \nn\\
   &\quad
   -\frac{2 \left(1680 z^6+560 z^4+350 z^3-4417 z^2+5502 z-2775\right) }{105 z^5}\zeta _2
   \nn\\
   &\quad
   +160 z^2-\frac{7384 z}{35}+\frac{76721}{18900 z}+\frac{2012}{189 z^2}-\frac{20653}{630 z^3}-\frac{20239}{315 z^4}+\frac{30854}{315}\bigg]
    \,,\nn
    \\
    &\frac{\df\Sigma^{(2,0)}_\text{bulk}}{\df z}\bigg|_{T_q(1)T_{q}(1)}
    \nn\\
    &
    =
    C_F T_F \bigg[
    \frac{2}{3} \left(360 z^4-720 z^3+454 z^2-94 z+5\right)g_1^{(3)}
    \nn\\
   &\quad
   +\frac{2}{9} \left(2 z^2-2 z+1\right) \left(g_2^{(3)}+6g_4^{(3)}+3 g_5^{(3)}-6g_8^{(3)}\right)
   +\frac{1}{3} \left(4 z^2-6 z+3\right)g_3^{(3)}
    \nn\\
    &\quad
    +\left(\frac{726 z^2}{35}-\frac{379 z}{15}+\frac{69}{10}-\frac{8}{3 z}-\frac{5}{3 z^2}+\frac{286}{15 z^3}-\frac{697}{30 z^4}+\frac{401}{35 z^5}\right) g_1^{(2)}
    \nn\\
   &\quad
   +\left(480 z^3-720 z^2+280 z-20+\frac{67}{15 z^3}-\frac{49}{5 z^4}+\frac{176}{21 z^5}\right) g_2^{(2)}
   \nn\\
   &\quad
   -\frac{3360 z^4\!-\!2240 z^3\!+\! 1232 z^2\!+\! 368 z\!-\!7}{210 z^{5/2}}g_3^{(2)}
   \!+\!
   \left(16 z\!+\!\frac{16}{3 z}\!+\!\frac{10}{3z^2}\!-\!\frac{213}{5 z^3}\!+\!\frac{844}{15 z^4}\!-\!\frac{3286}{105 z^5}\right) \zeta_2
    \nn\\
    &\quad
   +\left(480 z^3-840 z^2+\frac{47756 z}{105}-\frac{2956}{35}+\frac{1927}{315 z}-\frac{317}{70
   z^2}-\frac{12319}{630 z^3}+\frac{8525}{126 z^4}-\frac{124}{3 z^5}\right)g_1^{(1)}
   \nn\\
   &\quad
   +\frac{1}{1\!-\!z}\left(480 z^4\!-\!1080 z^3+\frac{2368 z^2}{3}\!-\! \frac{604}{3}z+\frac{1609}{90}\!-\!\frac{709}{315 z}-\frac{4979}{630 z^2}+\frac{1469}{105 z^3}-\frac{176}{21 z^4}\right)g_2^{(1)}
   \nn\\
   &\quad
   +120 z^2-\frac{3308 z}{35}-\frac{464}{105}-\frac{4223}{1512
   z}-\frac{2084}{189 z^2}+\frac{19172}{315 z^3}-\frac{1662}{35 z^4}
   \bigg]
   \nn\\
   &\quad
   +\left(C_A-2 C_F\right) C_F \bigg[
   \frac{6 z^7\!-\!6 z^6+6 z^5\!-\!2 z^4+9}{18 (1-z) z^5}g_2^{(3)}
   \!-\!\frac{\left(3 z^3-3 z^2+3 z-1\right)}{3 (1-z) z}\!\left(g_1^{(3)}\!-\!2 g_4^{(3)}\right)
   \nn\\
   &\quad
   +\frac{3 z^7-3 z^6+3 z^5-z^4-3}{3 (1-z) z^5}g_5^{(3)}
   +\frac{g_6^{(3)}-2 g_7^{(3)}}{2 (1-z) z^5}
   -\frac{3 z^2-3 z+1}{3 (1-z) z}g_8^{(3)}
   +\frac{2 (1-2 z)}{(1-z)z}\zeta_3
   \nn\\
   &\quad
   -\frac{(1-z) \left(73 z^5-144 z^4+36 z^3+111 z^2-194 z+148\right)}{30 z^5}g_1^{(2)}
   \nn\\
   &\quad
   +\frac{30 z^5+45 z^4+55 z^3+10 z^2+142 z-93}{15 z^5}g_2^{(2)}
   \nn\\
   &\quad
   +\frac{4 (1\!+\!z) \left(15 z^4\!+\!5 z^3\!+\!18 z^2\!+\!5 z\!+\!15\right)}{15 z^{9/2}}g_3^{(2)}
   \!-\!\frac{60 z^5\!+\!350 z^4\!+\!240 z^3\!-\!205 z^2\!+\!544 z\!-\!241}{15 z^5}\zeta_2
   \nn\\
   &\quad
   +\frac{438 z^5-543 z^4+710 z^3+1408 z^2-1628 z+1800}{90 z^5}g_1^{(1)}
   \nn\\
   &\quad
   -\frac{180 z^5+105 z^4+74 z^3-547 z^2+801 z-918}{90 (1-z) z^4}g_2^{(1)}
   \nn\\
   &\quad
   +\frac{10512 z^4+32171 z^3+36080 z^2-33456 z+29088}{2160 z^4}
   \bigg]
    \,,\nn
    \\
    &\frac{\df\Sigma^{(2,0)}_\text{bulk}}{\df z}\bigg|_{T_q(1)T_{Q}(1)}
    \nn\\
    &
    =
    C_F T_F \bigg[
    4 \left(50 z^4-100 z^3+66 z^2-16 z+1\right)g_1^{(3)}
   +\frac{4 z^2-14 z+13}{3 z^5}g_1^{(2)} 
   \nn\\
   &\quad
   +\frac{6000 z^8\!-\!9000 z^7\!+\!3920 z^6\!-\!460 z^5\!-\!33 z^2\!+\!108 z\!-\!88}{15 z^5}g_2^{(2)}
   -\frac{7 z^2-32 z+42}{15 z^5}\zeta_2
   \nn\\
   &\quad
   +\left(400 z^3-700 z^2+\frac{7}{9 z^2}-\frac{161}{90 z^3}+\frac{143}{9 z^4}-\frac{802}{45 z^5}+\frac{3352 z}{9}+\frac{14}{9 z}-\frac{539}{9}\right)g_1^{(1)}  
   \nn\\
   &\quad
   +\frac{1}{1\!-\!z}\left(400 z^4\!-\!900 z^3\!+\!\frac{6052 z^2}{9}\!+\!\frac{217}{45 z^2}\!-\!\frac{152}{15 z^3}\!+\!\frac{88}{15 z^4}\!-\!185 z\!-\!\frac{7}{18 z}\!+\!\frac{247}{18}\right)g_2^{(1)}
   \nn\\
   &\quad 
   +\frac{21600 z^6-21600 z^5+4812 z^4-69 z^3-616 z^2+4596 z-8256}{216 z^4}
   \bigg]
    \,,\nn
    \\
    &\frac{\df\Sigma^{(2,0)}_\text{bulk}}{\df z}\bigg|_{T_Q(1)T_{\bar Q}(1)}
    \nn\\
    &
    =
    C_F T_F \bigg[
    \frac{2 \left(6 z^2-6 z+11\right)}{15 z^5}\left(\zeta_2-g_2^{(2)}\right)
    -\frac{3 z-1}{6 z^{7/2}}g_3^{(2)}
    +\frac{(1-z) (29 z-51)}{15 z^5}g_1^{(1)}
    \nn\\
    &\quad
    +\frac{7 z^3+8 z+44}{30 z^4}g_2^{(1)}
    -\frac{112 z^3-1725 z+2850}{450 z^4}
   \bigg]
    \,,\nn
\end{align}
with the $g_m^{(n)}$ denoting weight-$n$ functions 
\begin{align}
    g^{(1)}_1
    &=
    \ln (1-z)\,,
    \quad 
    g_2^{(1)}
    =
    \ln z\,,
    \\
    g_1^{(2)}
    &= 2 \big(\zeta_2+\text{Li}_2(z)\big)+\ln^2(1-z)\,,
    \quad
    g_2^{(2)}
    = \text{Li}_2(1-z)-\text{Li}_2(z)\,,
    \nn\\
    g_3^{(2)}
    &=
   -2\, \text{Li}_2\left(-\sqrt{z}\right)+2\, \text{Li}_2\left(\sqrt{z}\right)+\ln\left(\frac{1-\sqrt{z}}{1+\sqrt{z}}\right) \ln z\,,
   \quad
   g_4^{(2)}
   = \text{Li}_2(z)-\zeta_2
   \,,
   \nn\\
   g_1^{(3)}
   &= 
   -6 \left[\text{Li}_3\left(-\frac{z}{1-z}\right)-\zeta _3\right]
   +\ln\left(\frac{1-z}{z}\right) \left[2 \big(\zeta _2+\text{Li}_2(z)\big)+\ln^2(1-z)\right]
   \,,
   \nn\\
   g_2^{(3)}
   &= 
   -12\left[\text{Li}_3(z)+\text{Li}_3\left(-\frac{z}{1-z}\right)\right]
   +6 \,\text{Li}_2(z) \ln(1-z)+\ln^3(1-z)
   \,,
   \nn\\
   g_3^{(3)}
   &=
   6 \ln (1-z) \big(\text{Li}_2(z)-\zeta_2\big)
   -12\,\text{Li}_3(z)+\ln^3(1-z)\,,
   \nn\\
   g_4^{(3)}
   &= \text{Li}_3\left(-\frac{z}{1-z}\right)+8 \zeta _3-3\zeta_2 \ln z\,,
   \nn\\
   g_5^{(3)}
   &=
   -8\, \text{Li}_3\left(-\frac{\sqrt{z}}{1-\sqrt{z}}\right)
   -8\,\text{Li}_3\left(\frac{\sqrt{z}}{1+\sqrt{z}}\right)
   +2\, \text{Li}_3\left(-\frac{z}{1-z}\right)+4 \zeta _2 \ln(1-z)
   \nn\\
   &\quad
   +\ln\left(\frac{1-z}{z}\right) \ln^2\left(\frac{1-\sqrt{z}}{1+\sqrt{z}}\right)\,,
   \nn\\
   g_6^{(3)}
   &=
   \ln^3(1-z)-15 \zeta _2 \ln (1-z)\,,
   \nn\\
   g_7^{(3)}
   &= 
   \ln (1-z) \left(\text{Li}_2(z)+\ln (1-z) \ln z-\frac{15}{2}\zeta_2\right)
   \,,
   \nn\\
   g_8^{(3)}
   &=
   29\, \text{Li}_3(1-z)
   -76\, \text{Li}_3\left(1-\sqrt{z}\right)
   -42\,\text{Li}_3\left(\frac{1}{1+\sqrt{z}}\right)
   +\frac{82}{9}\left[\text{Li}_3\left(-\sqrt{z}\right)+\text{Li}_3\left(\sqrt{z}\right)\right]
   \nn\\
   &\quad
   -\frac{5}{18}\text{Li}_3(z)
   +34\bigg[
   \text{Li}_3\left(\frac{1-\sqrt{z}}{1+\sqrt{z}}\right)
   +\text{Li}_3\left(\frac{1-\sqrt{z}}{2}\right)
   +\text{Li}_3\left(\frac{1+\sqrt{z}}{2}\right)
   +\text{Li}_3\left(\frac{\sqrt{z}}{1+\sqrt{z}}\right)
   \nn\\
   &\quad
   +\text{Li}_3\left(-\sqrt{z}\right)
   \bigg]
   -2 \big(\text{Li}_2(z)+\ln^2(1-z)\big)\ln z
   -23 \big(\text{Li}_2(z)+\ln(1-z) \ln z\big)\ln(1-z)
   \nn\\
   &\quad
   -29\,\text{Li}_2(1-z) \ln(1-z)
   -\frac{64}{3} \ln^3\left(1+\sqrt{z}\right)
   +\frac{17}{2}\big(2 \ln (1-z)+\ln z\big) \ln^2\left(1+\sqrt{z}\right)
   \nn\\
   &\quad
   -\frac{17}{2} \zeta_3
   -50 \zeta_2 \ln\left(1+\sqrt{z}\right)
   +17 \zeta_2\ln (1-z)
   +4 \zeta_2 \ln z
   +34 \ln 2\, \ln^2\left(1+\sqrt{z}\right)
   \nn\\
   &\quad
   +17 \ln 2\, \big(\ln 2-2 \ln\left(1+\sqrt{z}\right)\big) \ln(1-z)
   +34 \zeta_2 \ln 2
   -\frac{34}{3} (\ln 2)^3
   \,.\nn
\end{align}

\par\vspace{1em}
\noindent {\bf $\ln(\mu/Q)$-terms: }
\begin{align}\label{eq:trackEEC_twoloop_col_l1}
    &\frac{\df\Sigma^{(2,1)}_\text{col}}{\df z}
    \\
    &=
    T_g(2)\, \delta (z)
   \left[\left(-\frac{35689}{1350}-\frac{7 \pi ^2}{9}\right) C_A C_F+\left(\frac{14 \pi
   ^2}{9}-\frac{5615}{216}\right) C_F^2\right]
   \nn\\
   &\quad
   +T_g(1)T_g(1) \bigg\{
   \delta (z)
   \left[\left(\frac{43331}{5400}+\frac{7 \pi ^2}{9}\right) C_A C_F+\frac{659
   C_F^2}{18}\right]
   -8 C_F^2\left[\frac{1}{z}\right]_+
   \bigg\}
   \nn\\
   &\quad
    +\Big(T_q(2)+T_{\bar{q}}(2)\Big)\,\delta (z) 
    \bigg[
    \left(4 \zeta_3+\frac{1409}{24}-\frac{43 \pi ^2}{18}\right) C_A C_F
    +\left(-8 \zeta_3+\frac{791}{48}+2 \pi ^2\right)C_F^2
   \nn\\
   &\quad
   -\frac{797}{54} C_F n_f T_F+\frac{5923}{2700} C_F T_F
   \bigg]
   \nn\\
   &\quad
   +\Big(T_g(1) T_q(1)+T_g(1) T_{\bar{q}}(1)\Big)
   \bigg\{\delta (z)
   \left[\left(2 \pi ^2-\frac{471}{8}\right) C_F^2-\frac{1043}{36} C_A C_F\right]
   \nn\\
   &\quad
   +\left[\frac{1}{z}\right]_+\left(\frac{11}{2}C_A C_F+4 C_F^2\right) 
   \bigg\}
   \nn\\
   &\quad
   +T_q(1) T_{\bar{q}}(1) \bigg\{\delta (z)
   \bigg[\left(-16 \zeta_3-\frac{14057}{216}+\frac{77 \pi ^2}{9}\right) C_F(C_A-2C_F)
   +\frac{313}{25}C_F T_F
   \bigg]
   \nn\\
   &\quad
   -2 C_F T_F\left[\frac{1}{z}\right]_+\bigg\}
   \nn\\
   &\quad
   +\Big(T_q(1)T_q(1)+T_{\bar{q}}(1)T_{\bar{q}}(1)\Big)
   \bigg\{\delta (z) \bigg[\left(4 \zeta_3+\frac{3023}{216}-\frac{17 \pi ^2}{9}\right)
   C_F(C_A-2C_F)
   \nn\\
   &\quad
   +\frac{67}{12}C_F T_F\bigg]
   -C_F T_F\left[\frac{1}{z}\right]_+\bigg\}
   \nn\\
   &\quad
   +\sum_{Q\neq q}C_FT_F
   \bigg[
   \Big(T_q(1) T_Q(1)+T_q(1) T_{\bar{Q}}(1)+T_{\bar{q}}(1)T_Q(1)+T_{\bar{q}}(1) T_{\bar{Q}}(1)\Big)\bigg(\frac{67}{12}\, \delta(z)
   \nn\\
   &\quad
   -\left[\frac{1}{z}\right]_+\bigg)
   +\frac{203}{150} \,T_Q(1) T_{\bar{Q}}(1)\, \delta (z)
   +\frac{5923}{2700} \Big(T_Q(2)+T_{\bar{Q}}(2)\Big) \delta (z)\bigg]
    \,,\nn
\end{align}

\begin{align}\label{eq:trackEEC_twoloop_b2b_l1}
    &\frac{\df\Sigma^{(2,1)}_\text{b2b}}{\df z}
    \\
    &=
    T_g(1)T_g(1) \left(\frac{832}{27} C_F^2\, \delta (1-z)-\frac{128}{9}C_F^2\left[\frac{1}{1-z}\right]_+\right)
    \nn\\
   &\quad
   +\Big(T_g(1)T_q(1)+T_g(1) T_{\bar{q}}(1)\Big) 
   \bigg\{\delta (1-z) \bigg[\left(\frac{16 \pi^2}{3}-\frac{1240}{27}\right) C_F^2-\frac{316}{9}C_A C_F\bigg]
   \nn\\
   &\quad
   +\frac{88}{9} C_A C_F\left[\frac{1}{1-z}\right]_+ 
   +C_F^2 \left(\frac{112}{3}\left[\frac{1}{1-z}\right]_+ 
   +\frac{32}{3}\left[\frac{\ln(1-z)}{1-z}\right]_+\right)\bigg\}
   \nn\\
   &\quad
   +
    T_q(1) T_{\bar{q}}(1) \bigg\{
    \delta (1-z) \bigg[\left(8 \zeta_3 +\frac{1847}{27}-\frac{56 \pi ^2}{9}\right) C_A C_F+\left(\frac{8 \pi^2}{9}-\frac{128}{9}\right) C_F n_f T_F
    \nn\\
    &\quad 
    +\frac{112}{9}C_F T_F+\left(-16 \zeta_3+6-\frac{28 \pi ^2}{9}\right) C_F^2\bigg]
   +C_A C_F \left(-\frac{374}{9}\left[\frac{1}{1-z}\right]_+ 
   -\frac{44}{3}\left[\frac{ \ln (1-z)}{1-z}\right]_+\right)
   \nn\\
   &\quad
   +C_F^2\left(-\frac{544}{9}\left[\frac{1}{1-z}\right]_+ 
   -\frac{64}{3}\left[\frac{\ln(1-z)}{1-z}\right]_+\right)
   -\frac{32}{9} C_F T_F\left[\frac{1}{1-z}\right]_+ 
   \nn\\
   &\quad
   +C_F n_f T_F \left(\frac{136}{9}\left[\frac{1}{1-z}\right]_+ 
   +\frac{16}{3}\left[\frac{\ln (1-z)}{1-z}\right]_+\right)
   \bigg\}
   \nn\\
   &\quad
   +\Big(T_q(1)T_q(1)+T_{\bar{q}}(1)T_{\bar{q}}(1)\Big) 
   \bigg\{
   \delta (1-z)\left[\left(\frac{17 \pi ^2}{9}-4 \zeta_3-\frac{743}{54}\right) C_F \left(C_A-2C_F\right)+\frac{56}{9} C_F T_F\right]
   \nn\\
   &\quad
   -\frac{16}{9}C_F T_F\left[\frac{1}{1-z}\right]_+
   \bigg\}
   +\sum_{Q\neq q} 
   C_F T_F 
   \Big(T_q(1) T_Q(1)+T_q(1)T_{\bar{Q}}(1)+T_{\bar{q}}(1)T_Q(1)+T_{\bar{q}}(1)T_{\bar{Q}}(1)\Big)
   \nn\\
   &\quad\quad
   \times \left(\frac{56}{9} \delta (1-z)-\frac{16}{9}\left[\frac{1}{1-z}\right]_+\right)
   \,,\nn
\end{align}
and 
\begin{align}\label{eq:trackEEC_twoloop_bulk_l1}
    \frac{\df\Sigma^{(2,1)}_\text{bulk}}{\df z}
    &=
    \biggl[
    \Bigl(T_g(1) T_q(1)+T_g(1)T_{\bar{q}}(1)\Bigr)\left(\frac{11}{3} C_A C_F+\frac{8}{3} C_F^2\right) 
    \\
    &\quad
    -\frac{16}{3} T_g(1)T_g(1)C_F^2 
   -\frac{4}{3} T_q(1)T_{\bar{q}}(1)C_F T_F 
   -\frac{2}{3} \Bigl(T_q(1)T_q(1)+T_{\bar{q}}(1)T_{\bar{q}}(1)\Bigr)C_F T_F 
   \nn\\
   &\quad
   -\frac{2}{3} \sum_{Q\neq q}\Bigl(T_q(1) T_Q(1)+T_q(1)T_{\bar{Q}}(1)+ T_{\bar{q}}(1)T_Q(1)+T_{\bar{q}}(1)T_{\bar{Q}}(1)\Bigr) C_F T_F
   \biggr]f_1
   \nn\\
   &\quad
   +\biggl[
   T_q(1)T_{\bar{q}}(1) \left(\frac{22}{3} C_A C_F-\frac{8}{3} C_F n_f T_F+\frac{32}{3}C_F^2\right)
   \nn\\
   &\quad
   -\frac{16}{3} \Bigl(T_g(1) T_q(1)+T_g(1)T_{\bar{q}}(1)\Bigr)C_F^2 
   \biggr]f_2
    \,,\nn
\end{align}
where the functional basis associated with the one-loop result reads
\begin{align}
    f_1&=
    \frac{7 z^3+16 z^2-90 z+156}{6 z^4}+\frac{2 \left(4
   z^2-14 z+13\right) \ln (1-z)}{z^5}
    \,,\\
    f_2&=
    -\frac{17 z^3+17 z^2+18
   z+24}{3 z^4}-\frac{2 \left(z^4+z^3+z^2+z+4\right) \ln (1-z)}{z^5}
   \,.\nn
\end{align}

\par\vspace{1em}
\noindent {\bf $\ln^2(\mu/Q)$-terms: }
\begin{align}\label{eq:trackEEC_twoloop_col_l2}
    &\frac{\df\Sigma^{(2,2)}_\text{col}}{\df z}
    \\
    &=\delta (z) \Biggl\{
    T_g(2) \left(-\frac{679}{90}C_A C_F-\frac{175}{36} C_F^2\right)
   +T_g(1)T_g(1) \left(\frac{49}{15} C_A C_F +8 C_F^2\right)
   \nn\\
   &\quad
    +\Bigl(T_q(2)+T_{\bar{q}}(2)\Bigr)\left(\frac{275}{36}C_A C_F-\frac{25}{9}
   C_F n_f T_F+\frac{49}{90} C_F T_F +\frac{625}{72}C_F^2\right)
   \nn\\
   &\quad
   +\Bigl(T_g(1) T_q(1)+T_g(1)T_{\bar{q}}(1)\Bigr) \left(-\frac{11}{2}C_A C_F-\frac{41}{4}C_F^2\right)
   \nn\\
   &\quad
   +\frac{37}{15} C_F T_F T_q(1)T_{\bar{q}}(1) 
   +C_F T_F 
   \Bigl(T_q(1)T_q(1)+T_{\bar{q}}(1)T_{\bar{q}}(1)\Bigr)
   \nn\\
   &\quad
   +\sum_{Q\neq q} \biggl[
   \Bigl(T_q(1) T_Q(1)+T_q(1)
   T_{\bar{Q}}(1)+T_{\bar{q}}(1) T_Q(1) +T_{\bar{q}}(1)T_{\bar{Q}}(1)\Bigr)C_F T_F
   \nn\\
   &\quad
   +\frac{7}{15} T_Q(1) T_{\bar{Q}}(1)C_F T_F
   +\frac{49}{90}\Bigl(T_Q(2)+T_{\bar{Q}}(2)\Bigr) C_F T_F
   \biggr]
   \Biggr\}
    \,,\nn
\end{align}
\begin{align}\label{eq:trackEEC_twoloop_b2b_l2}
    \frac{\df\Sigma^{(2,2)}_\text{b2b}}{\df z}
    &=
    \delta (1-z) \Biggl\{
    \frac{128}{9}T_g(1)T_g(1)\,C_F^2 +\frac{16}{9}\Bigl(T_q(1)T_q(1) + T_{\bar{q}}(1)T_{\bar{q}}(1)\Bigr)C_F T_F
    \\
    &\quad
    +
    T_q(1) T_{\bar{q}}(1) \left(\frac{176}{9}C_A C_F -\frac{64}{9} C_Fn_f T_F
    +\frac{32}{9} C_F T_F+\frac{256}{9} C_F^2\right)
    \nonumber \\
    &\quad
    +\Bigl(T_g(1) T_q(1)+T_g(1)T_{\bar{q}}(1)\Bigr)\left(-\frac{88}{9} C_AC_F-\frac{64}{3} C_F^2\right)
    \nn\\
    &\quad
    +\sum_{Q\neq q} \Bigl(
    T_q(1) T_Q(1)+T_q(1)
   T_{\bar{Q}}(1)+T_{\bar{q}}(1)T_Q(1)+T_{\bar{q}}(1)T_{\bar{Q}}(1)
   \Bigr)\frac{16}{9} C_F T_F 
   \Biggr\}
    \,,\nn
\end{align}
\begin{align}\label{eq:trackEEC_twoloop_bulk_l2}
    \frac{\df\Sigma^{(2,2)}_\text{bulk}}{\df z}
    =0\,.
\end{align}
As expected from the scale invariance of physical observables, the $[\ln(\mu/Q)]^2$-terms only appear with $\delta(z)$ and $\delta(1-z)$, and vanish in the all-particle case (i.e.~setting all track function moments to one). 

\subsubsection{NNLO}\label{sec:NNLO}

To match the precision of theoretical calculations for current extractions of $\alpha_s$ from thrust \cite{Abbate:2010xh} or the C-parameter \cite{Hoang:2014wka,Hoang:2015hka}, it is necessary to also include NNLO $(\alpha_s^3)$ corrections to the fixed order calculation. Such corrections can be computed numerically for generic infrared and collinear safe dijet event shapes in $e^+e^-$, due to the seminal work of \cite{Gehrmann-DeRidder:2007foh,Gehrmann-DeRidder:2007vsv,Weinzierl:2008iv}. However, since the energy correlator measured on tracks is not IRC safe, standard NNLO numerical calculations are not applicable. In the future, it will be important to either develop numerical techniques for the calculation of event shapes on tracks, or to analytically compute the energy correlator at NNLO on tracks. In the direction of numerical calculations, there has been some recent progress in subtractions schemes involving identified particles \cite{Bonino:2024adk,Liu:2023fsq,Gehrmann:2022pzd,Czakon:2024tjr,Czakon:2022pyz,Czakon:2021ohs}. The energy correlator was computed analytically at NNLO in $\mathcal{N}=4$ SYM \cite{Henn:2019gkr}, and its calculation in QCD should be within reach using modern integration techniques.

However, we can make an extremely good approximation, and use the NNLO calculation on all hadrons rescaled by products of the track functions. This is possible due to the fact that the first moments of the track functions for quarks and gluons are extremely numerically close. In particular, our input values for the first moments of the track functions are
\begin{align*}
    T_g(1,100\,\text{GeV})&=0.617936\,,\quad 
    T_u(1,100\,\text{GeV})=0.604025\,, \quad
    T_d(1,100\,\text{GeV})=0.624608\,,
    \\
    T_c(1,100\,\text{GeV})&=0.627222\,,\quad
    T_s(1,100\,\text{GeV})=0.622424\,, \quad
    T_b(1,100\,\text{GeV})=0.622983\,.
\end{align*}
In a complete calculation of the track-based energy correlator at NNLO, different correlations are weighted by different products of the track functions, for example $T_g(1) T_g(1)$ for a gluon-gluon correlation and $T_q(1) T_g(1)$ for a quark-quark correlation. Due to their numerical similarity, we estimate that reweighting by an overall factor introduces an $\sim 5\%$ error on the NNLO coefficient. This is smaller than our numerical uncertainty on this quantity, and is therefore a reasonable approximation.

The energy correlator was computed using the ColorfulNNLO subtraction scheme \cite{DelDuca:2016ily,DelDuca:2016csb,Tulipant:2017ybb}, and we use these results in the bulk of the distribution.

\subsection{Non-Perturbative Power Corrections}\label{sec:bulkcorrection}

In the bulk of the distribution ($z\sim 1/2$), the leading non-perturbative power corection is linear, $\Lambda_{\text{QCD}}/Q$. We first review the case of the energy correlator measured on all hadrons, where this non-perturbative power correction is well studied, before extending this to a track-based measurement. 

A remarkable feature of the energy correlator is that the  functional form of the leading power correction is fixed, as originally shown in the seminal work of refs.~\cite{Korchemsky:1999kt,Korchemsky:1997sy,Korchemsky:1995zm,Korchemsky:1994is,Belitsky:2001ij}. The result is 
\begin{align} 
 \EEC^{\Omega}(z) = 
\frac{1}{2} \frac{\sigma_0}{\sigma} \frac{\bar \Omega_{1q}}{Q (z(1-z))^{3/2}}\,.
\end{align}
The size of the non-perturbative correction is controlled by the parameter $\bar{\Omega}_{1q}$ (the bar is due to the scheme, discussed below), which can be given a field theory definition as \cite{Korchemsky:1999kt,Korchemsky:1997sy,Korchemsky:1995zm,Korchemsky:1994is,Belitsky:2001ij,Lee:2006fn} 
\begin{align}\label{eq:fundamental_wilson}
\Omega_{1q}=\frac{1}{N_c} \langle 0 | \tr \overline{Y}_{\bar n}^\dagger Y_n^\dagger \cE_T(0) Y_n \overline{Y}_{\bar n} |  0 \rangle\,.
\end{align}
with $Y_{n,\bar n}$ Wilson lines in the fundamental representation along outgoing (anti-)quark, and $\cE_T$ the transverse (to $n$ and $\bar n$) energy-flow operator.

This universal non-perturbative parameter of QCD also appears in e.g.~the calculation of the thrust and C-parameter event shapes, and has been extracted in ref.~\cite{Abbate:2010xh}. Note that there is a slight non-universality from hadron mass effects \cite{Salam:2001bd,Mateu:2012nk}, but it is understood how to account for this. As defined, the energy flow operator acting on hadron states computes the total energy $E=\sqrt{\vec p^2 +m^2}$, including the hadron mass.

There has been significant discussion in the literature regarding applicability of the leading non-perturbative power correction for thrust, and its extension into the three-jet region \cite{Nason:2023asn,Luisoni:2020efy,Caola:2021kzt,Caola:2021kzt,Benitez-Rathgeb:2024ylc}. An interesting feature of the energy correlator is that it doesn't have a genuine three-jet region in the same way. It is for this reason that the full functional dependence of the leading power correction is understood. We believe that this makes it promising, or at least complementary, for extractions of $\alpha_s$.

It is well known that depending on the renormalization scheme used, the division between the perturbative and non-perturbative physics can introduce a renormalon ambiguity. This can be improved by using specific renormalon-free schemes, for example the R-scheme \cite{Hoang:2007vb,Hoang:2008fs,Hoang:2009yr,Hoang:2017suc,Bachu:2020nqn}. This has been discussed for the energy correlator in refs.~\cite{Schindler:2023cww,Lee:2024esz}. While it is not the primary focus of this paper, we discuss it briefly in an attempt to make this paper a self-contained explanation of the physics of the energy correlator. Furthermore, the most accurate fits for this non-perturbative parameter are performed in the R-scheme~\cite{Abbate:2010xh}.

We can move to a renormalon free scheme by performing a renormalon subtraction 
\begin{align}
\Omega_{1\kappa}(R)=\bar \Omega_{1\kappa}-R \sum\limits_{n=1}^\infty d_{\kappa n}(\mu/R)\, a_s^n(\mu)\,.
\end{align}
The coefficients $d_{\kappa n}$ are entirely determined by the properties of the parameter $\Omega_{1\kappa}$,
and are known to two loops for $\kappa=q$~\cite{Schindler:2023cww} 
\begin{align}
d_{q1}(\mu/R)&=d_{10}=-8.357\,, \\
d_{q2}(\mu/R)&=d_{20}+2 \beta_0 d_{10}\ln \frac{\mu}{R}=-72.443-16.713\,\beta_0 \ln \frac{\mu}{R}\,.
\end{align}
One can use the renormalization group evolution to evaluate the perturbative coefficients at the scale $Q$, and the non-perturbative parameter at a low scale, often chose to be $R_0=2$ GeV:
\begin{align}
\Omega_{\kappa1}(R_1)= \Omega_{\kappa 1}(R_0) + K_\kappa (R_1,R_0)
= \Omega_{\kappa 1}(R_0)- \sum\limits_{n=0}^\infty \gamma_{\kappa n}^{\Omega_1,R} \int\limits_{R_0}^{R_1} \df R \biggl[ \frac{\alpha_s(R)}{4\pi}  \biggr]^{n+1}
\,,\end{align}
where
\begin{align}
\gamma_{q0}^{\Omega_1,R}&=d_{10}=-8.357\,,\\
\gamma_{q1}^{\Omega_1,R}&=d_{20}=-2\beta_0 d_{10}=55.693\,.
\end{align}
This allows us to write the leading non-perturbative power correction in the R-scheme as 
\begin{align}
\frac{\sigma}{\sigma_0} \EEC^{\Omega}_R(z) = 
\sum_{n=1}^\infty \frac{R_1}{2Q} \frac{d_{qn}(\mu/R_1)}{[z(1-z)]^{3/2}} \biggl[ \frac{\alpha_s(\mu)}{4\pi}  \biggr]^n
+\frac{1}{2}\frac{K_q(R_1,R_0)}{Q (z(1-z))^{3/2}}+\frac{1}{2}\frac{ \Omega_{1q}(R_0)}{Q (z(1-z))^{3/2}}\,.
\end{align}

From fits to thrust \cite{Abbate:2010xh} finds
\begin{align}
\Omega_{1q}(R_0=2\,\text{GeV})=0.739 \pm 0.045\, \text{GeV}\,.
\end{align}
This can be converted to the EEC incorporating hadron mass effects \cite{Mateu:2012nk}, which was estimated in ref.~\cite{Schindler:2023cww} to be
\begin{align}
\Omega_{1q}(R_0)=0.895 \pm 0.054\, \text{GeV}\,.
\end{align}
Ref.~\cite{Schindler:2023cww} also provided an estimate in the $\overline{\text{MS}}$ scheme, which for the EEC gives 
\begin{align}
\bar \Omega^{\text{EEC}}_{1q}&=0.305 ~\text{GeV}\,. 
\end{align}
We will use this $\overline{\text{MS}}$ value in our calculations, leaving the incorporation of renormalon subtractions for the future.

We now extend this to tracks. In this case, the leading non-perturbative correction is still determined by $ \Omega_{1q}$, and is modified only by the product of track functions
\begin{align}
    \EEC_\text{bulk}^{\Omega}(z)
    &=
    \frac{1}{2}\frac{\sigma_0}{\sigma}\frac{\Omega_{1q}(\mu_\Omega)}{Q}\frac{1}{[z(1-z)]^{\frac{3}{2}}} 
    \frac{1}{2n_f}\sum_{q}T_q(1,Q)\,T_g(1,\mu_\Omega)\,.
\end{align}
The reason this combination of track functions appears, is that the leading non-perturbative effects arise due to the emission of a soft gluon from a quark. Here $\mu_\Omega$ denotes a low scale, to indicate that the gluon track function should be evaluated at the scale associated with $\Omega$. However, since the first moment of the gluon track function, $T_g(1,\mu_\Omega)$, evolves slowly, we can evaluate it at $\mu_\Omega=Q$ in practice. 

The physically interesting feature of the power correction is that it is even more singular as $z\to0,1$ than the perturbative contribution. Therefore, to get an accurate description of the energy correlator in the resummation region we must also understand how the non-perturbative power correction behaves in the collinear and back-to-back region, which will be discussed in the next sections.

\subsection{Numerical Results and Discussion}\label{sec:numerical_bulk}

\begin{figure}
\centering
\includegraphics[width=0.65\linewidth]{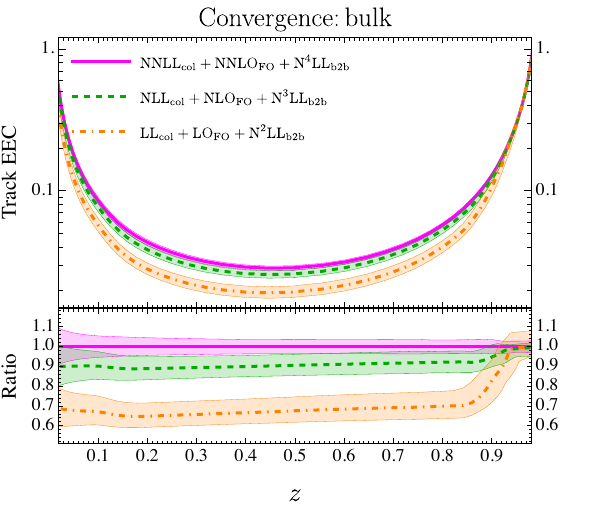}
\caption{The perturbative convergence of the track-EEC at $Q=m_Z$ in the bulk region of the distribution, which is dominated by fixed order perturbation theory.}
\label{fig:convergence_bulk}
\end{figure}

In \Fig{fig:convergence_bulk} we show numerical results for the EEC, highlighting the bulk region of the distribution. While these results are obtained using our complete calculation, they are dominated in this region by fixed order perturbation theory. We show results at three different perturbative orders, with the highest being $\text{NNLO}+\text{NNLL}_\text{col}+\text{NNNNLL}_\text{b2b}$.

In this perturbative region, we see large perturbative corrections at each order in perturbation theory, although the gap between NLO and NNLO is much smaller. Our final prediction achieves quite good theoretical uncertainty in the bulk region, at the level of several percent. 

It has been observed that improved perturbative convergence in the bulk of the distribution can be achieved through the incorporation of renormalon subtractions \cite{Schindler:2023cww}, as discussed in the previous section. We intend to revisit their inclusion in future work.

\section{Collinear Factorization Region: $z\to 0$}\label{sec:col}

In the collinear limit, large logarithms of $z$ spoil the perturbative convergence and thus require resummation. To enable the resummation, we proceed by splitting off the singular terms of the EEC as $z\to0$ and construct a factorization formula, enabling us to resum these most singular terms to all orders in the coupling. In terms of our master formula, eq.~\eqref{eq:eecdecomposition}, this component of the factorization is described by the terms
\begin{align}
    \EEC(z)_{z\to 0}&=
    \EEC^\text{fact.}_{z\to0}(z)
    + \EEC_{z\to0}^{\Omega,\text{res.}}(z) 
    + \EEC^\text{\text{plateau}}_{z\to0}(z)\,.
    \nonumber
\end{align}
In this section we present the details of our calculation of each of these components, and their matching to the non-singular fixed order distribution: In \Sec{colfactorization} we describe the factorization theorem for the EEC in the collinear limit, including tracks. In \Sec{colnpregion} we discuss the non-perturbative power corrections in the collinear limit. We discuss the profiles and matching in \Sec{colprofile}, and the collinear plateau  in \Sec{colPlateau}. We summarize our procedure for estimating the uncertainty in \Sec{colUncertainty}. In \Sec{numerical_coll} we present and discuss numerical results for the collinear region.

\subsection{Factorization and Resummation}\label{sec:colfactorization}

The factorization theorem for the collinear limit of the energy correlator was developed in ref.~\cite{Dixon:2019uzg}, and was extended to the case of tracks in  ref.~\cite{Jaarsma:2023ell}. (Early approaches to the resummation in the collinear limit were pioneered using the jet calculus \cite{Konishi:1979cb}.) Here we briefly review the structure of the factorization theorem, and present the NNLO constants for the jet function computed on tracks. These are primary new ingredients presented in this paper allowing us to achieve NNLL accuracy (of these single logarithms) in the collinear limit.

The factorization at leading power (order $z^{-1}$) is given by 
\begin{align}\label{eq:factorEEC_col}
\EEC^{\text{fact.}}_{z\to0}(z)=\frac{\sigma_0}{\sigma}\frac{\df}{\df z}\int_0^1\! \df x\, x^2 \vec J\Bigl(\ln \frac{z x^2 Q^2}{\mu^2},\mu\Bigr) \cdot \vec H\Bigl(x, \ln\frac{Q^2}{\mu^2},\mu\Bigr) 
\,,\end{align}
where $Q$ is the c.o.m.~energy, and the hard function $\vec{H}\equiv \{H_q, H_g\}^{\text{t}}$ is the standard coefficient function for the semi-inclusive hadron fragmentation~\cite{Dixon:2019uzg,Mitov:2006ic}. 
The jet function $\vec{J}\equiv \{J_q, J_g\}$ accounts for the measurement of the correlations in energy flow, which in our case includes tracking information through the track function formalism\footnote{This is the jet for the cumulative distribution in $z$, which is why there is a derivative of $z$ in \eq{factorEEC_col}.}. 
There is an implicit dependence on $\mu$ through $\alpha_s(\mu)$, as indicated by the final argument $\mu$ in $\vec H$ and $\vec J$.
Both $\vec{H}$ and $\vec{J}$ are vectors in flavor space. While the hard function does not depend on the (anti-)quark flavor when quark mass effects are ignored, the jet function in principle does through its matching onto track functions. 
The perturbative power corrections to the factorization in \eq{factorEEC_col} are higher twist. 

We will achieve resummation using the renormalization group evolution, which for the hard and jet functions is of DGLAP type, 
\begin{align}
    &\frac{\df}{\df\ln\mu^2}\vec{H}\Bigl(x,\ln\frac{Q^2}{\mu^2},\mu\Bigr)
    =
    -\int_0^1\df y\,\df y^\prime\,\widehat{P}\bigl(y,\mu\bigr)\cdot \vec{H}\Bigl(y^\prime,\ln\frac{Q^2}{\mu^2},\mu\Bigr)
    \delta(x-yy^\prime)\,,\\
    &\frac{\df}{\df\ln\mu^2}\vec{J}\Bigl(\ln\frac{zQ^2}{\mu^2},\mu\Bigr)
    =
    \int_0^1\df y\,y^2\vec{J}\Bigl(\ln\frac{zy^2Q^2}{\mu^2},\mu\Bigr)
    \cdot\widehat{P}\bigl(y,\mu\bigr)\,.
    \label{eq:J_evo}
\end{align}
Here $\widehat{P}$ denotes the singlet timelike splitting kernel matrix, 
\begin{align}\label{eq:splittingkernel}
    \widehat{P}=
    \begin{pmatrix}
    P_{qq} &\quad P_{qg}\\
    P_{gq} &\quad P_{gg}
    \end{pmatrix}
    ,
\end{align}
with $P_{qg}$ involving the splitting processes $\sum_q [(q\leftarrow g)+(\bar{q}\leftarrow g)]$
and $P_{qq}$ involving $(q\leftarrow q)+(\bar{q}\leftarrow q)+\sum_{Q\neq q} [(Q\leftarrow q)+(\bar{Q}\leftarrow q)]$.

As mentioned before, the jet function accounts for the fact that the measurement is performed on tracks. The all-order jet function can be written as~\cite{Jaarsma:2023ell}
\begin{align}\label{eq:ansatz_jet_track}
J_i \Bigl(\ln \frac{z Q^2}{\mu^2},a_s(\mu), {\bf T}_2(\mu) \Bigr)
=
\sum\limits_{L=0}^\infty a_s^L \biggl[\, \sum\limits_{m=0}^L  {\bf j}_{i,m}^{\,[2],(L)} \cdot {\bf T}_2 ~ \ln^m \Bigl( \frac{z Q^2}{\mu^2}  \Bigr)   \biggl]\,.
\end{align}
Here $i=q,g$ is the flavor index, $a_s(\mu)\equiv\alpha_s(\mu)/(4\pi)$, and the superscript ``2" on ${\bf j}_{i,m}^{\,[2],(L)}$ and subscript ``$2$'' on ${\bf T}_2$ indicates that this is for the two-point energy correlator (EEC). Both the perturbative coefficient ${\bf j}_{i,m}^{\,[2],(L)}$ and the non-perturbative track function ${\bf T}_2$ are vectors in the track function moment space of weight two, 
\begin{align}
    {\bf T}_2\equiv\{T_g(2),T_{q}(2),T_{g}(1)T_{g}(1),T_{g}(1)T_{q}(1),T_{q}(1)T_{q}(1)\}
\,.\end{align}

The hard function has its natural scale $\mu_H^{\text{nat.}}= Q$, while the jet function has $\mu_J^{\text{nat.}}=\sqrt{z}Q$ for which all logarithms in eq.~\eqref{eq:ansatz_jet_track} drop out. 
The gap between the hard scale $\sim Q$ and the jet scale $\sim \sqrt{z}Q$ leads to 
 large logarithms in the cumulative and differential EEC:
\begin{align}
&\text{\bf cumulative: }\text{LL: }\alpha_s^L \ln^L z\,,\quad 
  \text{NLL: }\alpha_s^L\ln^{L-1}z\,,\quad
  \text{NNLL: }\alpha_s^L\ln^{L-2}z\,;
\nn\\
&\text{\bf differential: }
\text{LL: }\alpha_s^L\left[\frac{\ln^{L-1}z}{z}\right]_+\,,\quad 
  \text{NLL: }\alpha_s^L\left[\frac{\ln^{L-2}z}{z}\right]_+\,,\quad
  \text{NNLL: }\alpha_s^L\left[\frac{\ln^{L-3}z}{z}\right]_+\,,\nn
\end{align} 
with LL abbreviating leading logarithms, NLL next-to-leading logarithms, etc.
These logarithmically enhanced contributions should be resummed to all orders in $\alpha_s$ when $\ln z$ becomes large.
We achieve this by evolving the jet function from its natural scale $\mu_J\sim\sqrt{z}Q$ to the hard scale $\mu_H\sim Q$ through the RGE ~\cite{Dixon:2019uzg,Chen:2020vvp,Jaarsma:2023ell}. Combining this with the  hard function, yields the resummed EEC in the collinear limit.

The resummation accuracy achieved depends on the order of the renormalization group equations (RGE) used. At LL we need the one-loop (timelike) DGLAP anomalous dimensions for the jet function, with the one-loop evolution kernels of track functions (in order to simultaneously evolve the track function from the jet scale to the hard scale), as well as the tree-level fixed-order hard and jet functions. At NLL we need the two-loop DGLAP anomalous dimensions and track function kernels and one-loop hard and jet function, etc.

\medskip
\noindent\textbf{Track function evolution} \\
The evolution of $T_g(2,\mu)$ and $T_{q_i}(2,\mu)$ appearing in our EEC case reads~\cite{Li:2021zcf}
\begin{align}\label{eq:RGE4Tg2Tq2}
    \frac{\df}{\df\ln\mu^2}T_g(2)
    &=-\gamma_{gg}(3)T_g(2)
    -\sum_{i=1}^{n_f}\gamma_{qg}(3)\left[T_{q_i}(2)+T_{\bar{q}_i}(2)\right]\\
&\quad +\text{other terms involving the first moments},\nn\\
\frac{\df}{\df\ln\mu^2}T_{q_i}(2)
&=-\gamma_{gq}(3)T_g(2)-\gamma_{qq}(3)T_{q_i}(2)-\gamma_{\bar{q}q}(3)T_{\bar{q}_i}(2)
-\sum_{j\neq i}\bigl[\gamma_{Qq}(3)T_{q_j}(2)\nn\\
&\quad +\gamma_{\bar{Q}q}(3)T_{\bar{q}_j}(2)\bigr]
+\text{other terms involving the first moments},\nn
\end{align}
and the evolution of $T_g(1,\mu)T_g(1,\mu), T_g(1,\mu)T_{q_i}(1,\mu), T_{q_i}(1,\mu)T_{q_i}(1,\mu)$ and $T_{q_i}(1,\mu)T_{q_j}(1,\mu)$ $(i\neq j)$ in ${\bf T}_2(\mu)$
can be derived from the RGEs of the first moment: 
\begin{align}\label{eq:RGE4Tg1Tq1}
    \frac{\df}{\df\ln\mu^2}T_g(1)
    &=-\gamma_{gg}(2)T_g(1)
    -\sum_{i=1}^{n_f}\gamma_{qg}(2)\left[T_{q_i}(1)+T_{\bar{q}_i}(1)\right]\,,\\
    \frac{\df}{\df\ln\mu^2}T_{q_i}(1)
    &=-\gamma_{gq}(2)T_g(1)-\gamma_{qq}(2)T_{q_i}(1)-\gamma_{\bar{q}q}(2)T_{\bar{q}_i}(1)
    \nn\\
    &\quad -\sum_{j\neq i}
    \bigl[\gamma_{Qq}(2)T_{q_j}(1)+\gamma_{\bar{Q}q}(2) T_{\bar{q}_j}(1)\bigr]
    \,,\nn 
\end{align}
where $\gamma_{ji}(n)$ denotes the moment of the timelike DGLAP function $P_{ji}(z)$~\cite{Almasy:2011eq,Chen:2020uvt},\footnote{Note the different convention for the moment than used for the track functions in \eq{T_mom}.}
\begin{align}
    \gamma_{ji}(n)=-\int_0^1\df z\,z^{n-1}P_{ji}(z)\,,
\end{align}
and $q_i,q_j$ with $i\neq j$ are used to specify distinct quark flavors.
For convenience, we list all the $\gamma_{ji}(n)$'s up to three-loop order in App.~\ref{sec:ingredients}. 
Given the above eqs.~\eqref{eq:RGE4Tg2Tq2} and \eqref{eq:RGE4Tg1Tq1}, clearly, the evolution of the first moments (and their combinations) is fully controlled by the DGLAP kernels, 
while that of the second moments involves the non-linear terms of $T(2)\to T(1)T(1)$, 
which can be considered corrections to the linear, DGLAP part. 

Shift invariance~\cite{Li:2021zcf,Jaarsma:2022kdd} allows the RGEs of the track function moments to be expressed more compactly, avoiding redundancy among the evolution kernels.
In terms of the shift-invariant objects, the evolution of the first and second moments of the gluon and quark track functions can be rewritten as 
\begin{align}
\frac{\df}{\df\ln\mu^2}\Delta &=\left[-\gamma_{qq}(2)-\gamma_{gg}(2)\right]\Delta\ ,\\
\frac{\df}{\df\ln\mu^2}\begin{bmatrix}
    \sigma_g(2)\\
    \sigma_q(2)
\end{bmatrix}
&=\begin{bmatrix}
    -\gamma_{gg}(3) &\quad -\gamma_{qg}(3)\\
    -\gamma_{gq}(3) &\quad -\gamma_{qq}(3)
\end{bmatrix}
\begin{bmatrix}
    \sigma_g(2)\\
    \sigma_q(2)
\end{bmatrix}
+\begin{bmatrix}
    \gamma_{\Delta^2}^g\\
    \gamma_{\Delta^2}^q
\end{bmatrix}
\Delta^2\,,
\end{align}
where $\Delta\equiv T_q(1)-T_g(1), \sigma_i(2)\equiv T_i(2)-T_i(1)^2$. (We remind the reader that for simplicity we assume all (anti-)quark track functions are equal, which is numerically a good approximation in our case.) The $\gamma$'s denote moments of the singlet splitting functions~\cite{Dixon:2019uzg}, and the kernels $\gamma_{\Delta^2}^g,\gamma_{\Delta^2}^q$ have been calculated up to $\mathcal{O}(\alpha_s^2)$ in refs.~\cite{Li:2021zcf,Jaarsma:2022kdd}. 
Thus, the only missing ingredient to the NNLL resummation of the track EEC in the collinear limit is the three-loop $\gamma_{\Delta^2}^g,\gamma_{\Delta^2}^q$.  
Fortunately, the small difference between $T_g(1)$ and $T_q(1)$ implies that the $\sigma(2)$ evolution is dominated by the DGLAP kernels, telling us that the missing piece can be safely discarded; see ref.~\cite{Jaarsma:2022kdd} for detailed discussions\footnote{In our resummed results we vary the values of the three-loop $\gamma_{\Delta^2}^g,\gamma_{\Delta^2}^q$ from $\sim -10^5$ to $\sim 10^5$ to account for the uncertainty from not known these ingredients., We find that the energy correlator hardly changes, with the maximum relative difference of $\sim 0.1\% $ at $z=10^{-4}$ (far below the perturbative region) at $Q=91.2$~GeV. For larger $z$, the difference is even smaller: at $z=10^{-2}$, it reduces to order of $0.01\% $. If the size of the three-loop $\gamma_{\Delta^2}^g,\gamma_{\Delta^2}^q$ were beyond $10^6$, the effect of these $\Delta^2$ terms would be noticeable in numerical results and plots; however, $10^6$ is much larger than the DGLAP anomalous dimensions.}. In this way, we are able to extend the track EEC to NNLL accuracy using the known three-loop DGLAP anomalous dimensions. 

\medskip
\noindent \textbf{Recurrence relation for resummation} \\
The way we organize the resummation up to a given accuracy is to calculate the perturbative coefficients ${\bf j}^{[2],(L)}_{i,m}$ ($1\leq m\leq L$) iteratively, order by order. 
The resummed jet function to N${}^l$LL evolved to the scale $\mu$ has the following form 
\begin{align}\label{eq:resum_jet_track}
J_i^{\text{N}^l\text{LL}} \biggl(\ln \frac{\mu_J^2}{\mu^2},a_s(\mu), {\bf T}_2(\mu)\biggr)
=
\sum\limits_{L=0}^\infty a_s^L(\mu) \biggl[\, \sum\limits_{m=L-l}^L  {\bf j}_{i,m}^{\,[2],(L)} \cdot {\bf T}_2 ~ \ln^m \Bigl( \frac{\mu_J^2}{\mu^2}  \Bigr)   \biggl]\,,
\end{align}
where $l=0,1,2,...$, and $\mu_J$ denotes the initial scale for the evolution (resummation) which is set to the intrinsic scale of the jet function, $\mu_J \sim \sqrt{z}Q$.
To get the recurrence relation for ${\bf j}^{[2],(L)}_{i,m}$ ($0\leq m\leq L$), we insert this with the RGEs of track function moments in \eqref{eq:RGE4Tg2Tq2}-\eqref{eq:RGE4Tg1Tq1} into the evolution equation of the jet function in \eq{J_evo}. 
Then, equating the coefficients on both sides for the terms with the same powers in $a_s$ and $\ln(\mu_J^2/\mu^2)$, 
we obtain the recurrence relation for the perturbative coefficients, 
\begin{align}\label{eq:recur_all}
    (m+1){\bf j}^{[2],(L)}_{i,m+1}\cdot {\bf T}_2
    &=
    -\sum_{k=m}^{L-1}\sum_{n=m}^k\binom{n}{m}\sum_r{\bf j}^{[2],(k)}_{r,n}\cdot {\bf T}_2
    \int_0^1\df y\,y^2\bigl(\ln y^2\bigr)^{n-m}P_{ri}^{(L-1-k)}(y) \nn \\
    &\quad
    -\sum_{n=m}^{L-1}n\cdot\beta_{L-1-n}\,{\bf j}^{[2],(n)}_{i,m}\cdot {\bf T}_2
    +\sum_{n=m}^{L-1}{\bf j}_{i,m}^{[2],(n)}\cdot\widehat{R}_2^{(L-n)}\cdot {\bf T}_2
    \,,
\end{align}
where $L\geq 1, 0\leq m\leq L-1$, and $\beta_i$ the  coefficients of the beta function, 
\begin{align}
    \frac{\df\alpha_s}{\df\ln\mu}\equiv \beta(\alpha_s)=-2\alpha_s\sum_{n=0}^\infty\beta_n\left(\frac{\alpha_s}{4\pi}\right)^{n+1}\,.
\end{align}
The superscript on timelike splitting function $P$ denotes its order in $a_s$, 
\begin{align}
    P_{ji}(y)=\sum_{L=0}^\infty a_s^{L+1}P^{(L)}_{ji}(y)\,,
\end{align}
and $\widehat{R}_2$ denotes the evolution kernel (matrix) of the (vector of) track function moments,
\begin{align}
    \frac{\df}{\df\ln\mu^2}{\bf T}_2=\widehat{R}_2{\bf T}_2\,,\quad
    \text{with }\widehat{R}_2=\sum_{L=1}^\infty a_s^L\widehat{R}_2^{(L)}\,.
\end{align}
For the LL resummation, we set $m=L-1$ ($L\geq 1$) in eq.~\eqref{eq:recur_all}; for NLL accuracy, we need $m=L-2$ ($L\geq 2$), and at NNLL also the terms with $m=L-3$ ($L\geq 3$) need to be included. 
Note that the recurrence relation at N${}^l$LL with $m=L-1-l$ requires the knowledge of ${\bf j}^{[2],(n)}_{i,n^\prime}$ at lower logarithmic accuracy, i.e., the known ${\bf j}^{[2],(n)}_{i,n^\prime}$ with $n-l<n^\prime\leq n$. 
In practice, our resummed result is a truncated solution of eq.~\eqref{eq:resum_jet_track} up to $\mathcal{O}(\alpha_s^{25})$. 

Although presented in this paper for the two-point case, as indicated by the superscript and subscript ``2'' of the perturbative coefficients ${\bf j}$ and track function moments ${\bf T}$, respectively, eq.~\eqref{eq:recur_all} can be straightforwardly extended to any (projected) $N$-point energy correlator ($N=2,3,4,...$) with ``2'' replaced by $N$. 
The recursive equations at LL and NLL for $N$-point correlators are explicitly given in sec.~4.1 of ref.~\cite{Jaarsma:2023ell}. 
Another interesting feature of eq.~\eqref{eq:recur_all} is that if one sets ${\bf T}_2$ (or ${\bf T}_N$ for the $N$-point case) to $\{1,1,...,1\}^{\text{t}}$ the contribution from the track function evolution is naturally dropped, and that the equation reduces to the all-particle case. This follows because the all-particle case corresponds to track functions $T_i(x)=\delta(1-x)$.

\subsubsection{NNLO Jet Function Constants}

To solve eq.~\eqref{eq:recur_all} requires the boundary (initial) condition. 
We rewrite eq.~\eqref{eq:ansatz_jet_track} as
\begin{align}\label{eq:jet_track_form2}
    J_i \Bigl(\ln \frac{z Q^2}{\mu^2},a_s(\mu), {\bf T}_2(\mu) \Bigr)
    =
    \sum\limits_{L=0}^\infty a_s^L 
    \biggl[\, \sum\limits_{m=0}^L  J_{i,m}^{[2],(L)} ~ \ln^m \Bigl( \frac{z Q^2}{\mu^2}  \Bigr) \biggl]\,.
\end{align}
At LL, we need the leading order (order-$a_s^0$) jet functions as the initial condition, 
\begin{align}
    J_{q,0}^{[2],(0)}=T_q(2)\frac{1}{4},\quad J_{g,0}^{[2],(0)}=T_g(2)\frac{1}{4}\,,
\end{align}
where the $1/4$ arises because in the EEC definition we divide by $Q^2$ rather than the squared jet energy.
At NLL, we calculate the next-to-leading order (NLO) jet function constants from the $1\to 2$ splitting functions, following ref.~\cite{Chen:2020vvp}. We can also extract the NLO constants from the order $a_s$ EEC on tracks, shown in sec.~\ref{sec:fo}, by using the factorization formula in \eq{factorEEC_col} and the known hard functions. The results from the two different methods agree, resulting in:
\begin{align}
    J_{q,0}^{[2],(1)}&=T_g(1)T_q(1)C_F\left(-\frac{37}{12}\right)\,,\\
    J_{g,0}^{[2],(1)}&=T_g(1) T_g(1) C_A\left(-\frac{449}{150}\right)+\sum_q T_q(1)T_{\bar{q}}(1) T_F\left(-\frac{7}{25}\right).
\end{align}

The NNLL requires the next-to-next-to-leading order (NNLO) constants. We extract them from the two-loop result shown in sec.~\ref{sec:fo}, given the factorization formula eq.~\eqref{eq:factorEEC_col} and the known hard functions up to two loops (see appendix A of ref.~\cite{Chen:2023zlx}). Here we present the quark jet function constant at NNLO,  
\begin{align}\label{eq:jetconstNNLO_q}
J_{q,0}^{[2],(2)}
&=T_q(1)T_q(1) \left\{C_AC_F \left(-\frac{221\zeta_3}{6}+\frac{485129}{10368}-\frac{37\pi^2}{27}+\frac{7 \pi ^4}{60}\right)+C_F^2\left(\frac{221\zeta_3}{3}-\frac{485129}{5184}\right.\right.\nonumber\\
&\quad\left.\left.+\frac{74\pi^2}{27}-\frac{7\pi ^4}{30}\right)+C_FT_F\left(\frac{1537}{192}+\frac{\pi^2}{18}\right)\right\}\nonumber\\
&\quad+T_q(1)T_{\bar{q}}(1)\left\{C_AC_F \left(\frac{1069 \zeta_3 }{12}-\frac{261119}{2304}+\frac{1675 \pi^2}{432}-\frac{59\pi^4}{180}\right)+C_F^2\left(-\frac{1069 \zeta_3 }{6}\right.\right.\nonumber\\
&\quad\left.\left.+\frac{261119}{1152}-\frac{1675 \pi ^2}{216}+\frac{59\pi^4}{90}\right)+C_FT_F\left(\frac{319597}{48000}+\frac{17\pi^2}{180}\right)\right\}\nonumber\\
&\quad+T_g(1)T_g(1) \left\{C_AC_F\left(\frac{91\zeta_3 }{12}-\frac{48074329}{2592000}+\frac{521\pi^2}{720}\right)+C_F^2\left(\frac{46613}{1728}-\frac{2 \pi ^2}{3}\right)\right\}\nonumber\\
&\quad+T_g(1)T_q(1) \left\{C_AC_F\left(\frac{61 \zeta_3 }{6}-\frac{72811}{1728}-\frac{5 \pi ^2}{18}\right)+C_F^2 \left(-15 \zeta_3 +\frac{6649}{768}+\frac{91\pi^2}{72}\right)\right\}\nonumber\\
&\quad+\sum_{\substack{Q\\ (Q\neq q)}}T_{q}(1)T_{Q}(1)C_FT_F\left(\frac{1537}{192}+\frac{\pi ^2}{18}\right)
+\sum_{\substack{Q\\ (Q\neq q)}}T_{q}(1)T_{\bar{Q}}(1) C_FT_F\left(\frac{1537}{192}+\frac{\pi ^2}{18}\right)\nonumber\\
&\quad+\sum_{\substack{Q\\ (Q\neq q)}}T_{Q}(1) T_{\bar{Q}}(1)\ C_FT_F\left(\frac{7 \pi^2}{180}-\frac{21551}{16000}\right)\, ,
\end{align}
as well as the gluon jet function constant, 
\begin{align}\label{eq:jetconstNNLO_g}
    J_{g,0}^{[2],(2)}
&=T_g(1)T_g(1) \left\{C_A^2\left(-\frac{527 \zeta_3}{10}+\frac{133639871}{3240000}-\frac{2159 \pi ^2}{1800}+\frac{19 \pi^4}{90}\right)+C_An_fT_F\frac{139}{270}\right\}\nn\\
&\quad+\sum_qT_g(1)T_q(1) \left\{C_AT_F\left(\frac{4\zeta_3}{3}+\frac{1585969}{54000}-\frac{9 \pi^2}{10}\right)+C_FT_F\left(\frac{35707}{8000}+\frac{\pi ^2}{18}\right)\right\}\nn\\
&\quad+\sum_q T_q(1)T_q(1) \left\{C_AT_F\left(\frac{46 \zeta_3}{15}-\frac{28714837}{3240000}+\frac{143 \pi ^2}{300}\right)+C_FT_F\left(2 \zeta_3-\frac{623189}{54000}\right.
\right.\nonumber\\
&\quad\left.\left.+\frac{8 \pi^2}{45}\right)+n_fT_F^2\left(\frac{2344}{1125}-\frac{4 \pi ^2}{45}\right)\right\}
\,,
\end{align}
where for simplicity we assume $T_q=T_{\bar{q}}$. The contributions from $T_q$ and $T_{\bar q}$ can be separated straightforwardly by replacing $T_g(1)T_q(1)$ in eq.~\eqref{eq:jetconstNNLO_g} with $[T_g(1)T_q(1)+T_g(1)T_{\bar q}(1)]/2$ and $T_q(1)T_q(1)$ with $T_q(1)T_{\bar{q}}(1)$, because the two-loop gluon jet function only involves the splitting processes $g\to gg, g\to q\bar{q}, g\to ggg$ and $g\to gq\bar{q}$. 

\subsubsection{NNLO Jet Function Logarithmic Terms}
When considering scale variations in the initial scale $\mu_J\sim\sqrt{z}Q$ of the resummed jet function, 
we should also include the logarithmic terms induced by these variations, in addition to the constants above.
Concretely this means that for eqs.~\eqref{eq:resum_jet_track} and \eqref{eq:recur_all} at N${}^l$LL, the boundary condition ${\bf j}^{[2],(l)}_{i,0}$ should be replaced by the full jet function of that order, namely
\begin{align} \label{eq:J_var}
    {\bf j}^{[2],(l)}_{i,0}\to 
    \sum_{m=0}^l{\bf j}^{[2],(l)}_{i,m}
    \ln^m\Big(\frac{zQ^2}{\mu_J^2}\Big)\,,
\end{align}
which is then the new initial value for the constant in the recurrence relation.
These logarithmic terms 
can be calculated directly from the splitting functions, 
or extracted from the fixed-order results in sec.~\ref{sec:fo} with the $\mu$-dependent hard functions. 
They can also be quickly derived through the recurrence relation in eq.~\eqref{eq:recur_all} with the jet function constants presented above. 
At NLL, the coefficients we need for \eq{J_var} are
\begin{align}
    J^{[2],(1)}_{q,1}&=\frac{3}{4} C_F\, T_g(1) T_q(1)\,,\\
    J^{[2],(1)}_{g,1}&=\frac{7}{10} C_A\, T_g(1)T_g(1)+\frac{1}{10} n_f T_F\, T_q(1)T_q(1)\nn\,.
\end{align}
At NNLL we need
\begin{align}
    J^{[2],(2)}_{q,2}
    &=
    T_g(1)T_g(1)\left(C_F^2-\frac{49}{120}C_A C_F\right)
    +T_g(1) T_q(1) \left(\frac{9}{16}C_F^2-\frac{11}{8}C_A C_F\right)
    \\
    &\quad
    +\frac{53}{120} C_F n_f T_F\,T_q(1)T_q(1)
    \,,\nn\\
    J^{[2],(2)}_{q,1}
    &=
    T_q(1)T_q(1) \biggl[\left(2 \zeta_3+\frac{8011}{864}-\frac{43 \pi^2}{36}\right) C_A C_F
    +\left(-4 \zeta_3 -\frac{8011}{432}+\frac{43 \pi ^2}{18}\right) C_F^2
    \nn\\
    &\quad
    -\frac{18587 }{3600}C_F n_f  T_F
    \biggr]
   +T_g(1) T_q(1)
   \left(\frac{1043}{72}C_A C_F -\frac{1069 }{288}C_F^2\right)
   \nn\\
   &\quad
   +T_g(1)T_g(1)
   \biggl[\left(\frac{97411}{21600}-\frac{7 \pi ^2}{36}\right) C_A
   C_F-\frac{659}{72}C_F^2\biggr]
    \,,\nn\\
    J^{[2],(2)}_{g,2}
    &=
    -\frac{91}{300} C_A^2\,T_g(1)T_g(1)
    +T_q(1)T_q(1)
   \biggl[n_f T_F \left(-\frac{13 }{300}C_A-\frac{4}{15}C_F\right)+\frac{2}{15} n_f^2T_F^2\biggr]
   \nn\\
   &\quad
   +T_g(1) T_q(1) \left(\frac{14}{15}C_A-\frac{1}{12}C_F\right)n_f T_F
    \,,\nn\\
    J^{[2],(2)}_{g,1}
    &=
    T_g(1)T_g(1) \biggl[\left(-2 \zeta_3 -\frac{44887}{13500}+\frac{97 \pi ^2}{90}\right) C_A^2-\frac{1}{9} C_A n_f T_F\biggr]
   \nn\\
   &\quad 
   +
   T_q(1)T_q(1)
   \biggl\{n_f T_F
   \left[\left(\frac{42703}{27000}-\frac{7 \pi ^2}{45}\right) C_A+\frac{718
   }{225}C_F\right]
   -\frac{68}{75} n_f^2 T_F^2\biggr\}
   \nn\\
   &\quad
   +
T_g(1) T_q(1) \left(-\frac{1297}{150}C_A-\frac{1181}{1800}C_F\right)n_f T_F
    \,,\nn
\end{align}
where we assume all quark and anti-quark flavors have the same track function $T_q$.

\subsection{Non-Perturbative Power Corrections}\label{sec:colnpregion}

Non-perturbative power corrections in the collinear limit of  generic $N$-point projected correlators (defined in ref.~\cite{Chen:2020vvp}) have recently been studied: Ref.~\cite{Chen:2024nyc} from the perspective of the light-ray OPE, and ref.~\cite{Lee:2024esz} in the context of factorization. The results from both papers are consistent, but since our analysis is also based on factorization, we will build on the latter.

We begin by reviewing the case where the measurement is made on all hadrons. We will then discuss the extension to tracks. The leading non-perturbative power correction in the collinear limit appears in the jet function, and for the two-point correlator takes the form 
\begin{align}\label{eq:LNP_jet_col_all}
J_i\Bigl(\ln \frac{z x^2 Q^2}{\mu^2},\mu,\Lambda_{\text{QCD}}\Bigr) = \hat J_i \Bigl(\ln \frac{z x^2 Q^2}{\mu^2},\mu   \Bigr) - \frac{\bar \Omega_{1i}}{2\sqrt{z} x Q} \mathcal{J}_i \Bigl(\ln \frac{z x^2 Q^2}{\mu^2},\mu   \Bigr)
\,.\end{align}
The first term, $\hat J$ corresponds to the perturbative contribution, while the second term, $\mathcal{J}$ encodes the leading non-perturbative correction. The non-perturbative parameter $\Omega_{1q}$ is as defined in eq.~\eqref{eq:fundamental_wilson}, while $\Omega_{1g}$ is defined analogously with adjoint Wilson lines
\begin{align}
\Omega_{1g}=\frac{1}{N_c} \langle 0 | \tr \bar {\mathcal{Y}}_{\bar n} {\mathcal{Y}}_n^\dagger \cE_T(0) \mathcal{Y}_n \bar{ \mathcal{Y}}_{\bar n} |  0 \rangle\,.
\end{align}
As compared to $\Omega_{1q}$ which has been extracted from $e^+e^-$ event shapes, we are not aware of precision extractions of $\Omega_{1g}$. It is often approximated via a so called ``Casimir scaling", namely $\Omega_{1g}\sim C_A/C_F \Omega_{1q}$, however, this should not be taken seriously. As with the non-perturbative power corrections in the bulk, there is also a renormalon ambiguity in the collinear limit, which can again be treated using the $R$-scheme~\cite{Lee:2024esz}.

In the bulk of the distribution, the EEC is only sensitive to $\Omega_{1q}$ at leading order. Sensitivity to $\Omega_{1g}$ enters at $\mathcal{O}(\alpha_s \Lambda_{\text{QCD}})$. In the collinear limit of the EEC, we get sensitivity to $\Omega_{1g}$ at leading logarithmic order, due to collinear resummation. The parameter $\Omega_{1g}$ will also enter in the back-to-back limit, but beyond the NLL order considered in this paper. Additionally, there it will be suppressed by Sudakov double logarithms. Precision measurements of the collinear limit of the EEC therefore provide an interesting opportunity to study $\Omega_{1g}$ in data.

In this paper, we work to LL accuracy for the non-perturbative power correction. It would be interesting to compute perturbative corrections to $\mathcal{J}$, but they are currently not known. At LL, we take $H=\{2\delta(1-x),0\}^{\text{t}}$, and we evolve the jet function with the one-loop DGLAP. For the case of the two-point energy correlator, the leading non-perturbative corrections to the jet function evolve with the twist-2 spin-2 anomalous dimensions. Explicitly, we have
\begin{align}
\EEC^{\Omega,\text{LL}}_{z\to 0}(z) =\frac{\sigma_0}{\sigma}\frac{\df}{\df z}\!\left\{\!-\frac{1}{2Q\sqrt{z}}(\bar{\Omega}_{1q},\bar{\Omega}_{1g}) \!\cdot\! V \left[\left(\frac{\alpha_s(\sqrt{z} Q)}{\alpha_s(\mu)} \right)^{-\frac{\vec{\gamma}_T^{(0)}\!(2)}{\beta_0}}\right] \! \cdot\! V^{-1}\!  \cdot\! \begin{pmatrix}
    2
\\
    0
  \end{pmatrix}\!\right\} \,,
\end{align}
where $V$ is the matrix that diagonalizes $\gamma_T^{(0)}(2)$,
and $\vec{\gamma}_T^{(0)}(2)$ is the diagonal vector of the diagonalized matrix. Due to conservation of energy, this evolution cancels if $\bar{\Omega}_{1q}=\bar{\Omega}_{1g}$, but exhibits a mixing in the case where they are not equal. We will see that due to the single logarithmic evolution in the collinear limit, the mixing between $\Omega_{1q}$ and $\Omega_{1g}$ is important at the level of accuracy to which we work.

For this paper, we should extend this factorization to incorporate the measurement on tracks. This was first considered in \cite{Jaarsma:2023ell}, which we extend to incorporate the renormalization group evolution of the non-perturbative contributions.  For the two-point energy correlator jet function on tracks, we have
\begin{align}
J_{i}\Bigl(\ln \frac{z x^2 Q^2}{\mu^2},a_s(\mu), {\bf T}_2(\mu),\Lambda_{\text{QCD}}\Bigr) = &\hat J_{i} \Bigl(\ln \frac{z x^2 Q^2}{\mu^2},a_s(\mu), {\bf T}_2(\mu)\Bigr)\\
&- \frac{ \Omega_{1i}(\mu_\Omega) T_g(1,\mu_\Omega)}{2\sqrt{z} x Q} \mathcal{J}_{i} \Bigl(\ln \frac{z x^2 Q^2}{\mu^2},a_s(\mu),{\bf T}_1(\mu)\Bigr)
\,,\nn
\end{align}
where we have put the explicit argument ${\bf T}$ to indicate that this is the track-based jet function, and ${\bf T}_1(\mu)$ indicates that $\mathcal{J}_i$ involves only the first moments of track functions. In this expression, the first moment of the gluon track function, $T_g(1,\mu_\Omega)$, should be evaluated at the scale, $\mu_\Omega$, associated with the non-perturbative parameter $\Omega_i$. However, due to the extraordinarily slow running of the first moments of the track functions, the choice of evaluation scale has a numerically negligible impact on the final result. At lowest order, the perturbative matching coefficient 
\begin{align}
\mathcal{J}_{i} \Bigl(\ln \frac{z x^2 Q^2}{\mu^2},a_s(\mu),{\bf T}_1(\mu)\Bigr)= T_{i}(1,\mu)+\mathcal{O}(a_s).
\end{align}
It would be interesting to compute higher perturbative corrections to this relation.

\subsection{Scale Setting and Matching}\label{sec:colprofile}

A complete description, as given in eq.~\eqref{eq:eecdecomposition}, requires the matching between the resummed result and the non-singular fixed order calculation.
A smooth transition from the resummation region into the fixed-order region is ensured by using a profile function for the jet scale: 
\begin{align}\label{eq:muJ_profile}
    \mu_J&\rightarrow\tilde\mu_J(z)=
    \mu_{\text{FO}}\,f_{\text{run}}(\sqrt{z},\sqrt{z})
\,,\end{align}
where we use tildes to denote profile scales. Here $f_{\text{run}}$ is given by~\cite{Lustermans:2019plv,Cal:2023mib}
\begin{align}\label{eq:profilefrun}
    f_{\text{run}}(x,y)=1+g_{\text{run}}(x)(y-1)\,,
\end{align}
in terms of $g_\text{run}(x)$, a smooth function that changes from $1$ in the resummation region to $0$ in the fixed-order region. Both in the collinear and back-to-back region, we use the following quadratic interpolation for $g_\text{run}$,
\begin{align} \label{eq:profile1}
g_\text{run}\bigl(x,\{x_1,x_2,x_3\}\bigr) &= 
\begin{cases}
1 & 0 \leq x \leq x_1 \,, \\
1- \frac{(x-x_1)^2}{(x_2-x_1)(x_3-x_1)} & x_1 < x \leq x_2\,, \\
\frac{(x-x_3)^2}{(x_3-x_1)(x_3-x_2)} & x_2 < x \leq x_3\,, \\
0 & x_3 \leq x \leq 1\,.
\end{cases}
\end{align}
It is characterized by three parameters $x_i$ that provide the boundaries of the transition region, with $x_2=(x_1+x_3)/2$ our default choice.
In this paper, we use $x_1=0.3,x_3=1/\sqrt{5}$ for the collinear region in the c.o.m energy $Q=91.2$~GeV case\footnote{From ref.~\cite{Jaarsma:2023ell} 
we know that for $Q=91.2$~GeV the border between the transition (to confinement) and the perturbative region is around $z=0.02$. 
We choose $x_1=0.3$ corresponding to $z=0.09$, which is sufficiently above $z=0.02$ but still clearly in the resummation region.}. 
With the profile function, the initial resummation scale runs between $\mu_{\text{FO}}\sqrt{z}$ at $z<x_1^2$ and $\mu_{\text{FO}}$ at $z>x_3^2$. 
And thus in the fixed-order region ($z>x_3^2$), this reproduces the pure fixed-order EEC:  
\begin{align}
    \EEC_{z\to0}^\text{res}( \mu_{\text{res}}=\mu_{\text{FO}}, \mu_{\text{FO}} )
    +\left[\EEC_\text{FO}( \mu_{\text{FO}} )
    -\EEC_\text{FO}^\text{0-sing}( \mu_{\text{FO}} )\right]
    &=\EEC_\text{FO}( \mu_{\text{FO}} )\,.\nn
\end{align}

While we resum the large logarithms at the cumulant level, taking the deriviative with respect to $z$ only after resummation in \eq{factorEEC_col}, we implement the profile scale into the differential EEC. More specifically, in the process of resummation of the cumulant, we use a symbol $L_\rho$ to trace the initial jet scale $\mu_J$, where $L_\rho\equiv\ln[\mu_J/(Q\sqrt{z})]$, which means $\ln\mu_J=L_\rho + \ln(Q\sqrt{z})$; taking the derivative with respect to the explicit $z$, we obtain the differential EEC, and then implement
\begin{align}
    L_\rho\to\ln\frac{\tilde\mu_J(z)}{Q\sqrt{z}}=
    \ln\frac{\mu_{\text{FO}}\,\big[1+(\sqrt{z}-1)g_{\text{run}}(\sqrt{z})\big]}{Q\sqrt{z}}
\end{align}
to incorporate the profile scale\footnote{For scale variations, we further replace $\tilde\mu_J$ with $\tilde\mu_J^{\text{vary}}$; see section~\ref{sec:colUncertainty}.}. 
As a consequence, the collinear LL resummed result matches onto the one-loop fixed order, the NLL matches onto the two-loop, etc,
\begin{align}
    \EEC_{z\to0}^\text{N$^k$LL}( \mu_{\text{res}}=\mu_{\text{FO}},\mu_{\text{FO}} )=\EEC_\text{N$^k$LO}^\text{0-sing}( \mu_{\text{FO}} )
    \,.\nn
\end{align} 
If one incorporated the explicit profile function $f_{\text{run}}(\sqrt{z},\sqrt{z})$ into the cumulant first and then the derivative acted on all $z$'s, this derivative acting on the profile function would introduce artifacts, ruining the structure of the energy correlator. One therefore needs to be more careful about how to choose profile functions (see e.g.~\cite{Bertolini:2017eui}).

\begin{figure}
    \centering
    \includegraphics[width=0.6\textwidth]{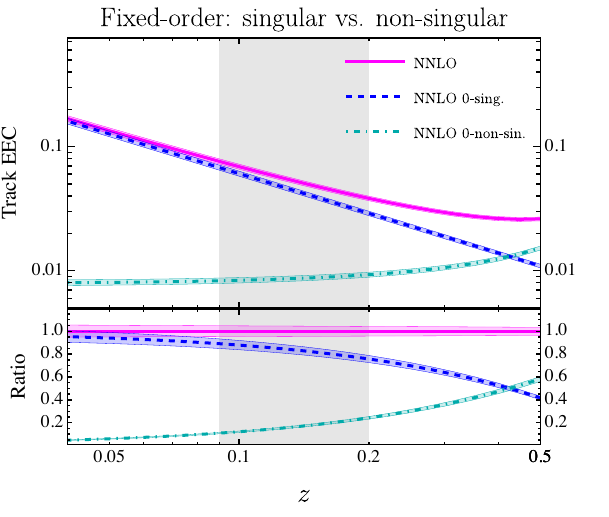}
    \caption{The NNLO fixed order result, along with the singular and non-singular contributions in the collinear limit. The shaded region indicates where we choose our profiles.}
    \label{fig:colTransition}
\end{figure}

In \Fig{fig:colTransition} we show a comparison of the full NNLO result, along with the leading power singular expansion in the collinear limit, and the non-singular. We observe that the power expansion is well behaved. The shaded region shows the region in which we perform we perform the matching using the profiles described in this section.

\subsection{Uncertainties}\label{sec:colUncertainty}

In this section we summarize the uncertainties in our calculation in the collinear factorization regime, which we use to make the error bands in our predictions. Here we are only attempting to estimate uncertainties within the scope of our calculation, which is performed in QCD with massless quarks. We do not consider additional physical uncertainties, such as due to quark masses, or QED effects. These are important, and will be included in future work. Furthermore, we do not discuss the uncertainties associated with the transition to the collinear plateau, which are the focus of \Sec{colPlateau}.

The first class of uncertainties are standard for resummed calculations.  They arise from:
\begin{itemize}
    \item Fixed-order uncertainties,
    \item Resummation uncertainties,
    \item Matching uncertainties,
    \item Non-perturbative power corrections.
\end{itemize}

The fixed-order and resummation uncertainties are each obtained by taking the envelope of different scale variations. To implement the scale variations, we replace the hard and jet scales by 
\begin{align}
\mu_H \rightarrow\mu_H^\text{vary}&=\mu_\text{FO}^\text{vary}=2^{v_\text{FO}}Q\,,\\
\tilde \mu_J \rightarrow\tilde{\mu}_J^\text{vary}&
=
f_\text{vary}^{v_J}\left(\sqrt{z},\{x_1,x_2,x_3\}\right)
\left[1+(\sqrt{z}-1)g_{\text{run}}(\sqrt{z})\right]\mu_{\text{FO}}^{\text{vary}}
\,.\nn
\end{align}
The fixed-order uncertainty is obtained by setting $v_\text{FO}=0, \pm1$ and taking the envelope of the resulting distribution. Similarly, the resummation uncertainties are obtained by taking $v_J=0,\pm 1$ and taking the envelope. We use the function $f_\text{vary}$ given by 
\begin{align} 
f_\text{vary}\bigl(x,\{x_1,x_2,x_3\}\bigr) &= \begin{cases}
2\left(1-\frac{x^2}{x_3^2}\right) & 0 \leq x < \frac{x_3}{2} \,, \\
1 + 2\left(1-\frac{x}{x_3}\right)^2 & \frac{x_3}{2} \leq x < x_3
\,, \\
1 & x_3 \leq x \,,
\end{cases}
\end{align}
which smoothly transitions between $1$ in the fixed-order region $x \geq x_3$, where there should be no resummation uncertainty, and $\sim 2$ in the resummation region $x \leq x_3/2$. The parameters $x_1$, $x_2$ and $x_3$ are chosen to match the profile scales given in \Eq{profile1}.

The matching uncertainty is evaluated by 
varying $x_2$ around the central value $(x_1+x_3)/2$ with fixed $x_1$ and $x_3$, following the approach discussed in sec.~3.2 of ref.~\cite{Cal:2023mib}. 
In particular, we vary $x_2$ by
\begin{align}
    x_2\in \left\{\frac{4x_1+x_3}{5},\frac{2x_1+x_3}{3},\frac{x_1+2x_3}{3},\frac{x_1+4x_3}{5}\right\}\,,
\end{align}
and the list of $x_1$ and $x_3$ values for different c.o.m.~energies $Q$ is presented below: 
\begin{align}
\centering
\begin{tabular}{|c||c|c|}
\hline
$Q$ (GeV) & 
$x_1$ &
$x_3$ 
\\ \hline\hline
91.2 &
0.3 & 
$1/\sqrt{5}$ 
\\ \hline
200.0 & 
0.3 & 
$\sqrt{3/10}$ 
\\ \hline
35.0 &
$1/\sqrt{5}$ &
$\sqrt{3/10}$
\\ \hline
10.0 &
$1/\sqrt{5}$ &
$\sqrt{3/10}$
\\ \hline 
\end{tabular}
\end{align}
Here we choose $x_1$ to be sufficiently higher than the transition (to confinement) region\footnote{This cannot be satisfied for the $Q=10$~GeV case. At $Q=91.2$~GeV, the border between the transition region and the perturbative region is $z\approx 0.02$, corresponding to the energy scale $Q\sqrt{z}\approx 12.9$~GeV. In this sense, we estimate the lower boundary of the perturbative region $z=(12.9/200)^2\approx0.004$ at $Q=200$~GeV and $z=(12.9/35)^2\approx 0.136$ at $Q=35$~GeV while at $Q=10$~GeV the boundary $z=(12.9/10)^2>1$. }, 
and $x_3$ to be relatively high (but still reasonable as the cutoff turning off the resummation) to keep the resummation region wide enough for implementation of perturbative resummation techniques while transitioning to the fixed order region efficiently.

\begin{figure}
    \centering
    \includegraphics[width=0.65\textwidth]{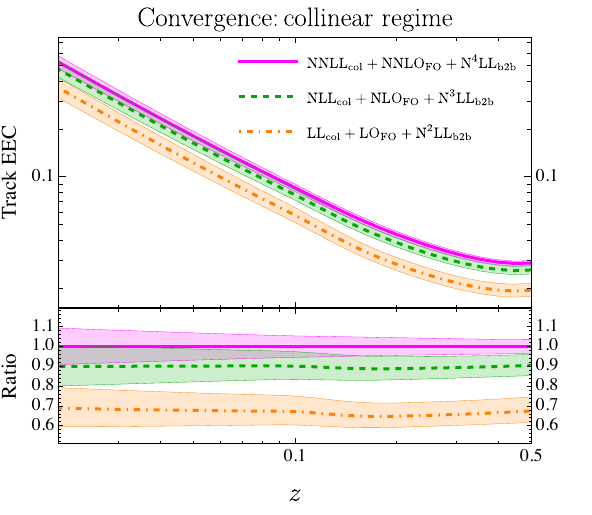}
    \caption{Convergence in the perturbative collinear limit of the energy correlator.}
    \label{fig:zoomB_collinear}
\end{figure}

The theoretical uncertainty coming from the non-perturbative power corrections in the collinear region is achieved by varying both the non-perturbative parameters, $\Omega_{1q}$ and $\Omega_{1g}$, as well as the scales (including profile parameters) of the resummed contribution
\begin{align}
\EEC^{\Omega,\text{LL}}_{z\to 0}(z) =&\frac{\sigma_0}{\sigma}\frac{\df}{\df z}\!\Biggl\{\!-\frac{1}{2Q\sqrt{z}}\Big({\Omega}_{1q}(\mu_\Omega)T_g(1,\mu_\Omega)T_q(1,\mu_J),{\Omega}_{1g}(\mu_\Omega)T_g(1,\mu_\Omega)T_g(1,{\mu}_J)\Big) \nn\\
& \cdot V \cdot\left(\frac{\alpha_s({\mu}_J)}{\alpha_s(\mu_{\text{FO}})} \right)^{-\frac{\vec{\gamma}_T^{(0)}\!(2)}{\beta_0}} \! \cdot\! V^{-1}\!  \cdot\! \begin{pmatrix}
    2
\\
    0
  \end{pmatrix}\!\Biggr\} \,,
\end{align}
The value of $\Omega_{1g}$ is weakly constrained. We take as a central value the result predicted by Casimir scaling, and vary the result in the range $0.7\cdot C_A/C_F \bar{\Omega}_{1q}\leq \Omega_{1g} \leq 1.3\cdot C_A/C_F \bar{\Omega}_{1q}$, which corresponds to
\begin{align}
\Omega_{1g}=0.686^{+0.206}_{-0.206} ~\text{GeV}\,.
\end{align}
We emphasize that this does not take into account the presence of higher-order non-perturbative power corrections, however, due to the fairly large uncertainty on $\Omega_{1g}$, we feel that this is reasonable at this stage.

\begin{figure}
\centering
\includegraphics[width=0.4\textwidth]{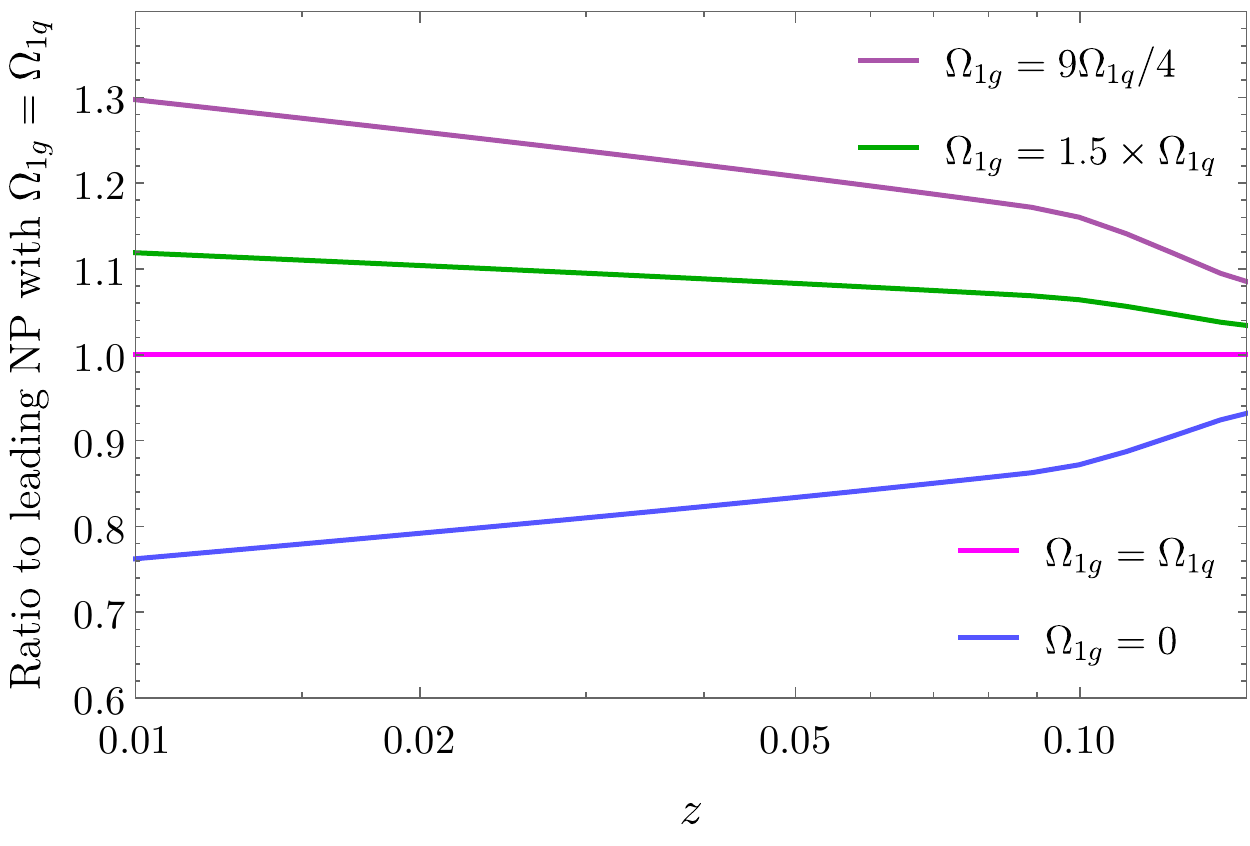}\qquad
\includegraphics[width=0.4\textwidth]{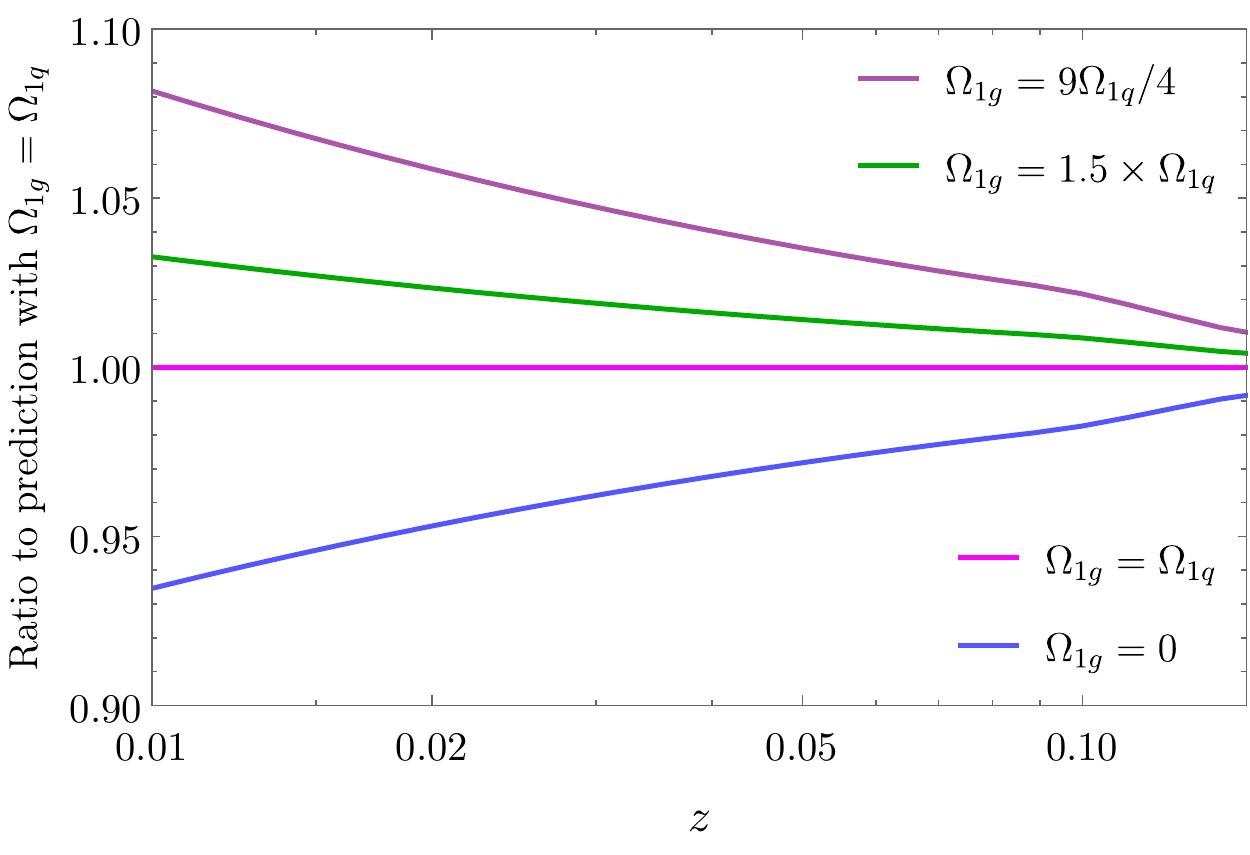}
\caption{The anomalous dimensions of the quark and gluon jet functions in the collinear limit lead to a contribution from $\Omega_{1g}$ already at leading logarithmic accuracy in the collinear limit. On the left, we show the ratio to the case of $\Omega_{1q}=\Omega_{1g}$ for the linear NP power correction alone, $\EEC_{z\to0}^{\Omega,\text{res.}}(z)$. On the right we show the impact on the full prediction. A sizeable impact is observed. Both plots use a common $z$-range of 0.01-0.15.}
\label{fig:Omega_g_ratio}
\end{figure}

\begin{figure}
\centering
\includegraphics[width=0.6\textwidth]{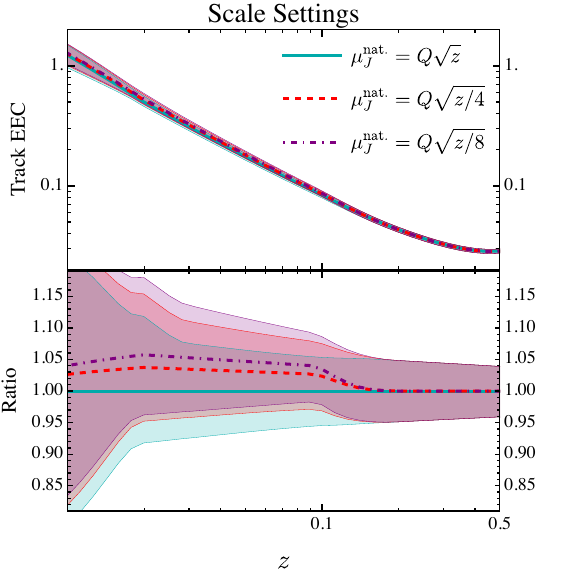}
\caption{A comparison of the energy correlator distributions with different choices for the factorization scale in the factorization theorem for the collinear limit. Lower scales choices are motivated due to the inclusive nature of the collinear limit. To study the variation in perturbative scale choices, we have kept fixed $\Omega_{1g}=\Omega_{1q}$ which eliminates scale variation from resummation for the linear power correction. The dependence on this scale choice motivates higher order calculations in this limit. }
\label{fig:scalesetting}
\end{figure}

\subsection{Numerical Results and Discussion}\label{sec:numerical_coll}

In \Fig{fig:zoomB_collinear} we show numerical results for the energy correlator in the perturbative regime of the collinear limit. While these results are obtained from our complete prediction, this region of the energy correlator is dominated by collinear resummation. Results are shown at three different levels of perturbative accuracy, with the highest being $\text{NNLO}+\text{NNLL}_\text{col}+\text{NNNNLL}_\text{b2b}$. We see large corrections in this region, although the results at $\text{NNL}_\text{col}$ and $\text{NNLL}_\text{col}$ begin to overlap. Due to the single logarithmic nature of this regime, these corrections are largely a total offset, and are being inherited from the large fixed order corrections in the bulk of the distribution. As with the fixed order predictions in the bulk, these could be improved through the incorporation of renormalon subtractions.

An interesting feature of the collinear limit of the energy correlator is that, since it is a single logarithmic observable, collinear evolution mixes $\Omega_{1q}$ and $\Omega_{1g}$ already at leading logarithmic order. In the back-to-back limit, $\Omega_{1g}$ is expected to contribute at NLL$'$, consistent with sum rules. The non-perturbative parameter $\Omega_{1g}$ is poorly constrained from data. It has been fit from parton shower simulations in \cite{Stewart:2014nna,Mo:2017gzp}, however, we are not aware of direct extractions from data. In \Fig{fig:Omega_g_ratio}, we show the impact of variations in the value of $\Omega_{1g}$ on our predictions. In the left panel, we show the impact on the effective non-perturbative contribution in the collinear limit, here expressed as a ratio to the case $\Omega_{1q}=\Omega_{1g}$. When $\Omega_{1q}=\Omega_{1g}$ the evolution of the non-perturbative contribution cancels. However, for $\Omega_{1q}\neq\Omega_{1g}$ it evolves due to a mixing between quark and gluon jet functions. We see from \Fig{fig:Omega_g_ratio} that this introduces a sizeable impact on the size of the non-perturbative correction, which is $z$ dependent. In the right panel we show the impact on the total energy correlator distribution (including perturbative contributions). Again this is shown as a ratio to the case $\Omega_{1q}=\Omega_{1g}$. We see that this leads to an important $z$-dependent effect at the order of $\sim 5\%$ in the perturbative regime. Due to the incredible precision of the recent LEP data in this regime, at the level of $\sim 1\%$, this effect is crucial to incorporate. 

We view it as an interesting feature of the energy correlator that the collinear limit is so sensitive to difference of $\Omega_{1q}$ and $\Omega_{1g}$. As compared to previous studies of non-perturbative effects, it gives a much more refined view. Since there is a long history of distinct techniques for the treatment of non-perturbative corrections (e.g. dispersive approaches \cite{Dokshitzer:1995qm,Dokshitzer:1995zt,Dokshitzer:1997ew} or operator based approaches
\cite{Belitsky:2001ij,Korchemsky:1999kt,Korchemsky:1997sy,Korchemsky:1994is,Lee:2006fn}), it would be interesting to understand if the re-analysis of LEP data can distinguish between them. We hope that this motivates
attempts to further constrain the magnitude of $\Omega_{1g}$, either using LEP data, or LHC data.

Compared to the resummation in the back-to-back limit, the convergence of higher order resummation in the collinear limit of the EEC is much less well understood. We believe that it is worth exploring in more detail, now that we have achieved this level of precision.  Here we highlight one particular aspect that we think would be interesting to explore in more detail. Recall the factorization theorem for the energy correlator in the collinear limit, which takes the form
\begin{align}\label{eq:fact_formula_discuss}
\EEC^{\text{fact.}}_{z\to0}(z)=\frac{\sigma_0}{\sigma}\frac{\df}{\df z}\int_0^1\! \df x\, x^2 \vec J\Bigl(\ln \frac{z x^2 Q^2}{\mu^2},\mu\Bigr) \cdot \vec H\Bigl(x, \ln\frac{Q^2}{\mu^2},\mu\Bigr) 
\,.\end{align}
If we were to try and solve the RG evolution for the jet function exactly, taking as natural scale $\mu_J=\sqrt{z}x Q$, we would have the following schematic solution for the jet function (taking for simplicity the case of a single flavor)
\begin{align}\label{eq:ansatz_YM}
&J\left(\frac{\mu^2}{z x^2 Q^2}, \alpha_s(\mu) \right) =\ C_J (\alpha_s(\sqrt{z} x Q)) \exp \Biggl[
 - \! \!\!\int\limits_{\alpha_s(\sqrt{z}x Q)}^{\alpha_s(\mu)}   \!\!\!\df\bar \alpha_s \frac{\gamma^{\text{YM}}_J(\bar \alpha_s)}{\beta(\bar \alpha_s)} \Biggr] \,.
\end{align}
However, here we see that  we hit the Landau pole for every value of $z$, when we integrate over $x$.  

In our analysis, we therefore choose the canonical scale for the jet function to be $\mu_J\sim Q\sqrt{z}$ as its initial scale. This treats the evolution from $\mu_J\sim Q\sqrt{z}x$ to $\mu_J\sim Q\sqrt{z}$ as a fixed order expansion. However, this is slightly concerning, since the hard function exhibits small-$x$ logarithms, behaving as $\ln^n(x)/x$, which becomes increasingly singular as $x\to0$ at higher loop orders.  This suggests that one may want to use an effective lower scale for the jet function. The intuition behind this is that the parton sourcing the jet function has an average energy much less than $Q/2$. In \Fig{fig:scalesetting} we show a comparison of the energy correlator distributions in the collinear limit with different scale settings. Choosing a lower scale enhances the correlator in this region, as expected.

Similar sensitivity was discussed in \cite{Dixon:2019uzg}, where it was shown by studying the behavior of the energy correlator at the Banks-Zaks fixed point that the large perturbative corrections arise primarily from $\beta$ function contributions.  It would be interesting to explore if this could be stabilized by performing the small-$x$ resummation for the hard function. This has been explored for the case of fragmentation in 
\cite{Albino:2011si,Vogt:2011jv,Albino:2011cm,Kom:2012hd,Anderle:2016czy}. This might help to stabilize the perturbative series in the collinear limit, and we believe deserves further investigation.

\subsection{Collinear Plateau and Contact Term}\label{sec:colPlateau}

In this section, we briefly discuss our treatment of the non-perturbative collinear plateau, and contact term. As discussed in \Sec{overview_physics} there are two primary distinctions between the collinear limit in QCD, as compared to a conformal gauge theory. First, there is a non-perturbative transition to a plateau, associated with the scaling of a free-hadron gas. Second, there is a contact term, $\langle E^2 \rangle \delta(z)$ associated with the presence of particle states, which is absent in the case of a conformal field theory. In QCD, the distribution in this region cannot be reliably computed in perturbation theory, but rather gives insight into the confining transition. The transition is described by the matrix element between a twist-2 light-ray operator and a di-hadron state \cite{Chang:2025kgq}, or in the language of QCD factorization, a di-hadron fragmentation function \cite{Lee:2025okn,Herrmann:2025fqy,Kang:2025zto}. Unfortunately, such matrix elements are not known, and introduce a non-perturbative function into the description of the energy correlator. See \cite{Lee:2025okn,Herrmann:2025fqy,Kang:2025zto,Liu:2024lxy} for different parameterizations. 

Instead of introducing an unknown  non-perturbative model function, we take a different approach. Much in analogy with the conformal bootstrap, we believe that we should shift from parameterizations of non-perturbative shape functions to placing model independent bounds on observables in regimes where they are not understood. While these bounds may be weak initially, we believe that they can be strengthened in the future. This is illustrated in our predictions, where we place a central curve only in the perturbative regime, and only bounds in the collinear and back-to-back limits. Our optimism in applying this approach for the case of the energy correlator arises from the fact that it can be achieved for the energy correlator in $\mathcal{N}=4$ SYM \cite{N4_bootstrap}. Basic positivity properties should also enable constraints in the case of QCD.

\begin{figure}
    \centering
    \includegraphics[width=0.65\textwidth]{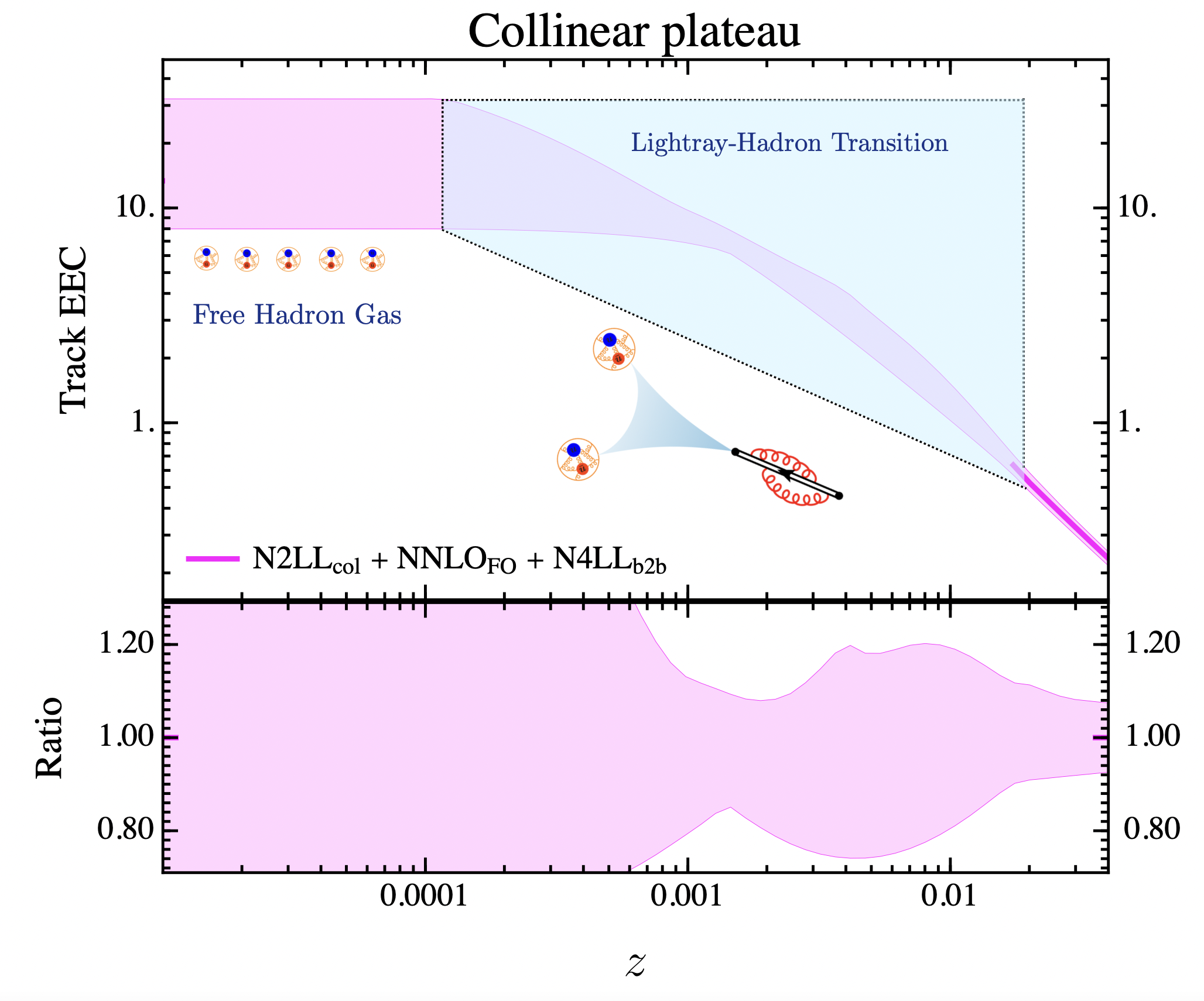}
    \caption{The interpolation between the plateau and the non-perturbative scaling regime. We parameterize the transition with a family of monotonic curves. These are then constrained by the knowledge of the cumulant distribution in the NP scaling regime. In this region we do not plot a central value, but rather an allowed region.}
    \label{fig:zoom_collinear_plateau}
\end{figure}

Here we take an extremely elementary approach, applying this philosophy. Our approach is illustrate in \Fig{fig:zoom_collinear_plateau}. For sufficiently small, $z$, the true EEC distribution behaves as $\langle E^2\rangle/Q^2\,\delta(z)+h$, where $h$ is the height of the plateau. We will assume that $\langle E^2\rangle/Q^2$ is a non-perturbative number that can be extracted from experiment. Indeed, it has been measured at LEP, and takes the value $0.0527^{+0.0027}_{-0.0027}$  for $Q=91.2$ GeV. At large values of $z$ we have our prediction for the perturbative region. Our goal is to consider the most general interpolating function between these two regimes, and impose consistency with the perturbative calculation of the cumulant distribution
\begin{align} \label{eq:cum_condition}
\Sigma(z)= \int\limits_0^z\df z'\, \text{EEC}(z')\,.
\end{align}
While the differential EEC can't be calculated perturbatively for small values of $z$, the cumulant \emph{can} be calculated as long as a sufficiently large region is integrated over. To make progress, we can impose some physical constraints on the interpolating functions, expecting that in time these will be made more rigorous.  First, we assume that the interpolation for the  EEC($z$) should be monotonic. Second, we assume that it is bounded below by the linear interpolation between the two points.  This can easily be relaxed, but it corresponds to having only a peak instead of also a dip, when plotted as $\text{EEC}(\theta)$. This property has been seen in experiment~\cite{CMS:2024mlf,ALICE:2024dfl}, and we believe it to be physical.  We also note that these properties hold both in calculations of the EEC using holographic models of confinement \cite{Csaki:2024joe,Csaki:2024zig}, as well as in calculations with an explicit mass \cite{Craft:2022kdo}. However, it would certainly be interesting to prove them from first principles. Positivity properties for amplitudes were recently considered in ref.~\cite{Henn:2024qwe}, and it would be interesting to explore these for energy correlators. Under these assumptions, for fixed values of $h$, the interpolation corresponds to a monotonic curve contained within the shaded blue region in \Fig{fig:zoom_collinear_plateau}.  

We can now consider families of curves filling this region, and eliminate inconsistent curves by demanding consistency with the cumulant condition in \Eq{cum_condition}.
In this paper we are conservative about the uncertainty in this region. We parameterize the transition as
\begin{align}
  \text{EEC}(z)=
  \begin{cases}
    \langle E_i^2\rangle/Q^2\delta(z)+h\,, 
    & z<z_{\Lambda_{\text{QCD}}}\,,
    \\
    \frac{c_1}{(c_2+z)^b}
    \,,
    & z_{\Lambda_{\text{QCD}}}\leq z<z^*\,,
    \\
    {
      \text{EEC}_{\text{pert}+\Omega_1}(z)\,,}
      & { z^* \leq z \,,}
  \end{cases}
\end{align}
where $z_{\Lambda_{\text{QCD}}}$ and $z^*$ set the lower and upper bounds, respectively, for the transition,
and $\text{EEC}_{\text{pert}+\Omega_1}(z)$ is our full result consisting of the fixed-order, perturbatively resummed and leading NP contributions. 
These curves fill the shaded blue region, parameterizing all transitions and plateau heights. 

In our simplified initial implementation of this idea, to reduce the number of variables, we set $z_{\Lambda_{\text{QCD}}}\sim\Lambda_{\text{QCD}}^2/Q^2=(1~\text{GeV})^2/Q^2$, 
which is reasonable given that $z_{\Lambda_{\text{QCD}}}$ is the upper bound of the free-hadron region. To make the transition function join with the plateau, we have the plateau height $h=c_1/(c_2+z_{\Lambda_{\text{QCD}}})^b$. 
In the following, we write $c_1$ in terms of $c_2$ and $h$, i.e., $c_1=h(c_2+z_{\Lambda_{\text{QCD}}})^b$. We then have two unknown variables to fix and two constraints:
\begin{itemize}
  \item continuity: 
  \begin{align}
    \frac{c_1}{(c_2+z^*)^b}={ \text{EEC}_{\text{pert}+\Omega_1}(z)\Big|_{z=z^*}}\,;
  \end{align}
  \item the cumulant sum rule: 
  \begin{align}\label{eq:cumulant_sum_rule}
    &\langle E_i^2\rangle/Q^2+h\,z_{\Lambda_{\text{QCD}}}
    +\int_{z_{\Lambda_{\text{QCD}}}}^{z^*}\!\df z\, \frac{c_1}{(c_2+z)^b}
    +{\int_{z^*}^{1/2}\df z\, \text{EEC}_{\text{pert}+\Omega_1}(z) }
     \nn\\
    &\quad =
\int_0^{1/2}\df z\,\text{EEC}_{\text{FO}+\Omega_1}(z)\,.
  \end{align}
\end{itemize}
Here $\text{EEC}_{\text{FO}+\Omega_1}(z)$ denotes the fixed order calculation of the EEC, including the leading non-perturbative power correction.

An estimate of the uncertainty associated with this treatment of the plateau is given by varying
\begin{enumerate}
  \item $z^*$, for which the range is determined by considering the ratio of the leading nonperturbative correction to the perturbation result, as well as $\Lambda_{\rm QCD}/(\sqrt{z}Q)$.
  \item $b\in\{1,3/2,5/2\}$, for $b$ significantly larger than this range, we find inconsistency with the cumulant sum rule, given reasonable values of $z^*$. 
  \item $z_{\Lambda_{\text{QCD}}}\in[0,(1\text{~GeV})^2/Q^2]$.
  \item  $\langle E_i^2\rangle/Q^2$ up and down by a factor of $5\%$ corresponding to the experimental uncertainty. 
  \item as well as variations from the uncertainty of {$\text{EEC}_{\text{pert}+\Omega_1}(z)$}.
\end{enumerate}
These variations give rise to the shaded region in \Fig{fig:zoom_collinear_plateau}. 

We emphasize that this is an extremely crude proof of principle, but we believe that it should be possible to constrain the transition region from first principles, which would in turn enable first principles constraints on di-hadron fragmentation functions.

\section{Back-to-Back Region: $z\to 1$}\label{sec:b2b}

To achieve a precision description of the back-to-back limit, we must resum logarithms of $1-z$. In terms of our master formula, this corresponds to the calculation of the component
\begin{align}
    \EEC_{z\to1}(z)
    &=\EEC^\text{fact.}_{z\to1}(z)  +\EEC_{z\to1}^{\Omega,\text{res.}}(z) 
    + \EEC^\text{\text{plateau}}_{z\to1}(z)\,.
\end{align}
In this section we present the details of our calculation of each of these components, as well as the matching to the fixed-order result.

This section is structured as follows: In \Sec{b2bfactorization} we extend the factorization theorem of ref.~\cite{Moult:2018jzp} to incorporate tracks. In \Sec{b2bNP} we discuss the non-perturbative  corrections in the back-to-back limit, which include both the linear $\Omega_{1q}$ power correction, as well as the non-perturbative corrections to the Collins-Soper kernel. The plateau in the deeply non-perturbative region is discussed in \Sec{b2bPlateau}.
We discuss the profile scales and matching onto the fixed-order results in \Sec{b2bMatch}, and summarize our procedure for assessing the uncertainty in \Sec{b2bUncertainty}. We present and discuss numerical results for the back-to-back region in \Sec{numerical_b2b}.

\subsection{Factorization and Resummation}\label{sec:b2bfactorization}

The factorization theorem for the leading power contribution to the energy correlator in the back-to-back limit was developed in ref.~\cite{Moult:2018jzp}, using the rapidity renormalization group \cite{Chiu:2011qc,Chiu:2012ir} with the exponential rapidity regulator~\cite{Li:2016axz}. This has been used to achieve N$^4$LL resummation \cite{Ebert:2020sfi,Duhr:2022yyp}. Here we extend this factorization theorem to the case of tracks. While the focus of this paper is on understanding the EEC in $e^+e^-$ collisions, this will also be useful for understanding closely related observables, such as the transverse energy energy correlator (TEEC) at the LHC~\cite{Gao:2019ojf,Gao:2023ivm} on tracks. 

At leading power, only energetic (collinear) particles directly contribute to the measurement of the energy correlators, with the effect of soft radiation limited to recoil (through momentum conservation). This makes deriving the factorization theorem on tracks in the back-to-back limit particularly simple, since only the jet functions need to be modified to account for the measurement on tracks, in contrast to e.g.~the case of track thrust where also the soft function is modified~\cite{Chang:2013iba}.
We can therefore view the restriction to tracks as an IR modification of the jet functions, with an otherwise identical factorization theorem as in ref.~\cite{Moult:2018jzp} 
\begin{align}
    \EEC_{z\to1}^{\text{fact.}}(z)
    &=
    \frac{\sigma_0}{2\sigma}
    \int\df^2\mathbf{k}\,
    \delta\Bigl(1-z-\frac{\mathbf{k}^2}{Q^2}\Bigr)
    \int\frac{\df^2\mathbf{b}}{(2\pi)^2}\,
    e^{-\img\mathbf{k}\cdot\mathbf{b}}\,
    \\
    &\qquad\times
    H(Q,\mu)\,S(\mathbf{b},\mu,\nu)
    \sum_q J_q(\mathbf{b},Q,\mu,\nu)\,J_{\bar{q}}(\mathbf{b},Q,\mu,\nu)\,
    \nonumber
    \\
    &=
    \frac{\sigma_0 Q^2}{4\sigma}
    \int\df b_\perp\, b_\perp J_0\bigl(\sqrt{1-z}\,b_\perp Q\bigr)\,
    \nonumber\\
    &\qquad\times
    H(Q,\mu)\,S(b_\perp,\mu,\nu)
    \sum_q J_q(b_\perp,Q,\mu,\nu)\,J_{\bar{q}}(b_\perp,Q,\mu,\nu)\,.
    \nonumber
\end{align}
In going to the second line we performed the angular integration over $\mathbf{b}$, leaving $b_\perp \equiv |\mathbf{b}|$, and integrated over $\mathbf{k}$.

The jet function encodes the collinear splittings inside a jet. While it is in principle a non-perturbative object, its evolution and matching onto fragmentation functions can be calculated perturbatively. To this end we split the jet function up into a multiplicative non-perturbative piece described by a free parameter $\tau\sim \Lambda_{\rm QCD}^2$, and a perturbative piece that contains the matching onto track function moments,
\begin{align} \label{eq:J_match}
    J_i(\mathbf{b},Q,\mu,\nu)&=e^{-\frac{1}{2}\tau\mathbf{b}^2}\sum_j T_j(1,\mu)\,\tilde{\cC}_{ji}(1,\mathbf{b},Q,\mu,\nu)\,.
\end{align}
Here
\begin{align}
    \tilde{\cC}_{ji}(n,\mathbf{b},Q,\mu,\nu)
    &=\int \df x\,x^n\,\cC_{ji}(x,\mathbf{b},Q,\mu,\nu)\,,
\end{align}
where $\cC_{ji}$ are the coefficients for matching transverse-momentum dependent (TMD) fragmentation functions onto collinear fragmentation functions, which have been calculated up to three-loop order~\cite{Luo:2020epw,Ebert:2020qef}. The non-perturbative parameter $\tau$ has to be fit to data, but in the current prediction we use \cite{Sun:2014dqm}
\begin{align}
    \tau=0.212\,\text{GeV}^2\,.
\end{align}

In the back-to-back limit the resummation of large logarithms of $1-z$ is needed to improve the  convergence. Because the back-to-back limit is governed by Sudakov-like logarithms, two sets of RGEs govern the resummation of these logarithms, namely the usual virtuality RGE and an additional rapidity RGE. Since the track functions only appear in the jet function, this implies that the RG structure of the factorization is unaffected, i.e.~the track functions only appear in the constant terms of the jet function. This is similar to case of the azimuthal angular decorrelation considered in refs.~\cite{Chien:2020hzh,Chien:2022wiq}. However, in \eq{J_match} only the first moment of the track function appears, which is even simpler. The virtuality RGEs read
\begin{align}
    \frac{\df}{\df\ln\mu^2} \ln H(\mu,Q)
    &=
    \gammacusp\bigl[a_s(\mu)\bigr]
        \ln\Bigl(\frac{Q^2}{\mu^2}\Bigr)
    +\gamma_H\bigl[a_s(\mu)\bigr]\,,
    \nn \\
    \frac{\df}{\df\ln\mu^2} \ln J_q(\mathbf{b},Q,\mu,\nu)
    &=
    -\tfrac{1}{2} \gammacusp\bigl[a_s(\mu)\bigr]
        \ln\Bigl(\frac{Q^2}{\nu^2}\Bigr)
    +\gamma_J\bigl[a_s(\mu)\bigr]\,,
    \nn \\
    \frac{\df}{\df\ln\mu^2} \ln S(\mathbf{b},\mu,\nu)
    &=
    -\gammacusp\bigl[a_s(\mu)\bigr]
        \ln\Bigl(\frac{\nu^2}{\mu^2}\Bigr)
    -\gamma_S\bigl[a_s(\mu)\bigr]\,,
\end{align}
with $\gammacusp$ the cusp anomalous dimension~\cite{Polyakov:1980ca,Korchemsky:1987wg}. The perturbative expressions for all anomalous dimensions used in this paper are collected in appendix~\ref{sec:ingredients}. The rapidity RGE reads
\begin{align}\label{eq:rapidity_RG}
    \frac{\df}{\df\ln\nu}\ln J_q(\mathbf{b},Q,\mu,\nu)
    &=-\tfrac{1}{2}\gamma_\nu^q(b_\perp,\mu)
    \,, \nn\\
    \frac{\df}{\df\ln\nu}\ln S(\mathbf{b},\mu,\nu)
    &=\gamma_\nu^q(b_\perp,\mu)\,,
\end{align}
where $\gamma_\nu$ is the Collins-Soper kernel~\cite{Collins:1981uk,Collins:1981va,Collins:1984kg}, expressed in the rapidity renormalization group formalism~\cite{Chiu:2012ir,Chiu:2011qc}. The evolution of the Collins-Soper kernel is governed by the cusp anomalous dimension,
\begin{align}\label{eq:CS_evolve}
    \frac{\df}{\df\ln\mu^2}\gamma_\nu^q(b_\perp,\mu)
    &=-2\gammacusp\bigl[a_s(\mu)\bigr]\,.
\end{align}

To resum the logarithms, we evaluate each ingredient of the factorization formula at their natural scale and evolve them to a common scale
\begin{align}
    &H(Q,\mu)\,J_q(\mathbf{b},Q,\mu,\nu)\,
    J_{\bar{q}}(\mathbf{b},Q,\mu,\nu)\,S(\mathbf{b},\mu,\nu)
    \\
    &=H(Q,\mu_H)\,J_q(\mathbf{b},Q,\mu_J,\nu_J)\,
    J_{\bar{q}}(\mathbf{b},Q,\mu_J,\nu_J)\,S(\mathbf{b},\mu_S,\nu_S)
    \nonumber\\
    &\quad\qquad\times
    U\bigl(Q,b_\perp,\{\mu_H,\mu_J,\mu_S,\mu_0\},\{\nu_J,\nu_S\}\bigr)\,,
    \nonumber
\end{align}
where the evolution kernel is given by 
\begin{align}
    &U\bigl(Q,b_\perp,\{\mu_H,\mu_J,\mu_S,\mu_0\},\{\nu_J,\nu_S\}\bigr)
    \\
    &=\exp\biggl[
    \int_{\mu_H}^{\mu}\df\ln\mu^{\prime2}\,
    \biggl(\gammacusp\bigl[a_s(\mu^\prime)\bigr]
        \ln\Bigl(\frac{Q^2}{\mu^{\prime2}}\Bigr)
    +\gamma_H\bigl[a_s(\mu^\prime)\bigr]\biggr)
    \nonumber\\
    &\qquad
    +\int_{\mu_J}^{\mu}\df\ln\mu^{\prime2}\,
    \biggl(-\gammacusp\bigl[a_s(\mu^\prime)\bigr]
        \ln\Bigl(\frac{Q^2}{\nu_J^2}\Bigr)
    +2\gamma_J\bigl[a_s(\mu^\prime)\bigr]\biggr)
    \nonumber\\
    &\qquad
    +\int_{\mu_S}^{\mu}\df\ln\mu^{\prime2}\,
    \biggl(-\gammacusp\bigl[a_s(\mu^\prime)\bigr]
        \ln\Bigl(\frac{\nu_S^2}{\mu^{\prime2}}\Bigr)
    -\gamma_S\bigl[a_s(\mu^\prime)\bigr]\biggr)
    \nonumber\\
    &\qquad
    +\gamma^q_\nu(\mathbf{b},\mu)\,\ln\Bigl(\frac{\nu_J}{\nu_S}\Bigr)
    +\int_{\mu_0}^{\mu}\df\ln\mu^{\prime2}\,
    \gammacusp\bigl[a_s(\mu^\prime)\bigr]
        \ln\Bigl(\frac{\nu_S^2}{\nu_J^2}\Bigr)
    \biggr]\,.
    \nonumber
\end{align}
Consistency of the factorization, i.e.~that the sum of  anomalous dimensions vanishes, explains why the $\mu$-dependence on the right-hand side of the above equation drops out. While the evolution kernel is formally independent of the chosen path in $(\mu,\nu)$ space, there is a small dependence due to truncation.

The integrals that appear in the above evolution kernel are non-trivial and cannot be computed analytically without some approximation. There are however several methods to compute these integrals either numerically or semi-analytically, with a summary presented in ref.~\cite{Billis:2019evv}. Here we opt for the so-called unexpanded analytic method, which is an approximate but fully analytic way of evaluating the above integrals. It involves rewriting the integrals over $\mu$ as integrals over the coupling using the beta function, and then evaluating these integrals using an iterative solution for the running coupling.

We perform the resummation to N$^4$LL order. The complete set of perturbative ingredients required to achieve this order are provided in App. \ref{sec:b2bingredients}. These include the three-loop jet \cite{Luo:2019hmp,Luo:2019bmw,Ebert:2020qef}, soft \cite{Li:2016ctv} and hard function \cite{Baikov:2009bg,Lee:2010cga,Gehrmann:2010ue} constants, the four \cite{Moch:2018wjh,Moch:2017uml,Davies:2016jie,Henn:2019swt} and approximate five \cite{Herzog:2018kwj} loop cusp anomalous dimension, the five loop beta function \cite{Baikov:2016tgj}, and the four-loop rapidity anomalous dimension \cite{Duhr:2022yyp,Moult:2022xzt}.

\subsection{Non-Perturbative Power Corrections}\label{sec:b2bNP}

The factorization theorem for the energy correlator in the back-to-back limit has a close relation to the factorization theorem for the transverse momentum $p_T$ spectrum at small $p_T$. Non-perturbative corrections to the $p_T$ spectrum scale like $b^2$, i.e.~$\Lambda_{\text{QCD}}^2/Q^2$. However, there are logarithmically enhanced contributions to these power corrections, corresponding to non-perturbative corrections to the Collins-Soper kernel. They can be thought of as non-perturbative corrections to the anomalous dimensions appearing in our factorization theorem. Since the soft function, and anomalous dimensions for the factorization theorem for the energy correlator in the back-to-back limit are identical to those for $p_T$, these non-perturbative corrections also occur in the energy correlator. These are well studied, and there has been significant recent progress computing them on the lattice \cite{Avkhadiev:2024mgd,Avkhadiev:2023poz,Shanahan:2021tst,Shanahan:2019zcq,Shanahan:2020zxr}, using large momentum effective theory \cite{Ji:2020ect,Izubuchi:2018srq,Ji:2014gla,Ji:2013dva}.

Although the soft function in the energy correlator and $p_T$ are identical, the jet functions appearing in the factorization theorem for the energy correlators have a different structure than the beam functions appearing in the factorization theorem for the $p_T$ spectrum. In particular, the energy weighting in the jet function gives rise to a $\Lambda_{\text{QCD}}/Q$ power correction. Power corrections to jet functions are much less explored than for soft functions, making the energy correlator an interesting observable to explore their structure. One other example where they have been studied is the broadening distribution \cite{Becher:2013iya}. We will be able to show that this power correction is determined by the same $\Omega_{1q}$ as in the bulk of the distribution. Since this correction occurs only in the jet function, it cannot be logarithmically enhanced. These two types of power corrections for the energy correlator in the back-to-back limit were identified early on in ref.~\cite{Dokshitzer:1999sh}.

In our prediction, we include both the linear power correction in the jet function, as well as the Collins-Soper kernel. We now discuss each of these in turn.

\subsubsection{Collins-Soper Kernel}\label{sec:b2bCSkernel}

The Collins-Soper kernel, also known as the rapidity anomalous dimension, is a non-perturbative object that describes the rapidity scale dependence of TMD parton distributions and fragmentation functions. In this work, the Collins-Soper kernel $\gamma_\nu^q$ appears in the rapidity RGE for the soft and jet functions, as shown in eq.~\eqref{eq:rapidity_RG}.
The kernel is itself scale dependent, and its evolution is described by the the cusp anomalous dimension, as shown in eq.~\eqref{eq:CS_evolve}.
With its dependence on $b_\perp$, the Collins-Soper kernel evidently becomes non-perturbative for $b_\perp^{-1}\sim\Lambda_\text{QCD}$, regardless of the choice of $\mu$.  Since its scale dependence is known, the Collins-Soper kernel can be uniquely determined provided we have a non-perturbative boundary condition. 

The Collins-Soper kernel is conventionally split up into three pieces: an evolution piece governed by the cusp anomalous dimension, a fixed-order perturbative boundary condition, and a non-perturbative boundary condition, 
\begin{align}\label{eq:CS_boundary}
    \gamma_\nu^q(b_\perp,\mu)
    &=
    \gamma_\nu^{q,\text{NP}}(b_\perp)
    +2\gamma^q_r[a_s(\mu_0)]
    -2\int_{\mu_0}^{\mu}\df\ln\mu^{\prime2}\,
    \gammacusp\bigl[a_s(\mu^\prime)\bigr]
\end{align}
The fixed-order boundary conditions and the cusp anomalous dimension are available at 4-loop order \cite{Henn:2019swt,Duhr:2022yyp,Moult:2022xzt}, sufficient for resummation at N$^4$LL. We provide them in App. \ref{sec:ingredients}.

The remaining term, $\gamma_\nu^{q,\text{NP}}(b_\perp)$ is inherently non-perturbative. Due to its universality, it can in principle be extracted from data. This has been extensively pursued, and a variety of fits exist in the literature \cite{Landry:2002ix,Scimemi:2019cmh,Bacchetta:2019sam,Isaacson:2023iui,Bacchetta:2022awv,Moos:2023yfa}. These fits are based on theoretically motivated phenomenological models for the functional form of $\gamma_\nu^{q,\text{NP}}(b_\perp)$. For example, renormalon analyses suggest it behaves like $b^2$ \cite{Scimemi:2016ffw}. One possibility is that precision measurements of energy correlators in the back-to-back limit will enable a new way to extract the Collins-Soper kernel. This direction was recently pursued in ref.~\cite{Kang:2024dja,Cuerpo:2025zde}. 
An exciting recent development has been the possibility to directly compute the Collins-Soper kernel from first principles QCD using lattice QCD \cite{Ji:2013dva,Lee:2013mka,Izubuchi:2018srq,Ji:2020ect}. This approach is being actively pursued by a number of groups. For a review, we refer the reader to \cite{Ji:2020ect}. Joint fits, combining the lattice and experimental data have also recently been performed \cite{Avkhadiev:2025wps}. This recent activity suggests our understanding of the Collins-Soper kernel will improve significantly in the coming years.

In this paper, we will use the results of 
\cite{Avkhadiev:2024mgd,Avkhadiev:2023poz,Shanahan:2021tst,Shanahan:2020zxr,Shanahan:2019zcq}, which suggest that for large $b_\perp$ the Collins-Soper kernel grows linearly with $b_\perp$. We will use the following parameterization \cite{Moos:2023yfa}
\begin{align}\label{eq:CS_param}
    \gamma_\nu^{q,\text{NP}}(b_\perp)
    &=
    -4\, \frac{b_\perp^2}{\sqrt{1+b_\perp^2/b_\text{max}^2}}
    \Biggl[c_0 + c_1\ln\biggl(\frac{b_\perp}{\sqrt{b_\text{max}^2+b_\perp^2}}\biggr)\Biggr]\,,
\end{align}
with coefficients fit to the lattice results of \cite{Avkhadiev:2024mgd}\footnote{Note that the results of \cite{Avkhadiev:2024mgd} are computed with $n_f=2+1+1$ dynamical quark flavors. In our perturbative calculations we use $n_f=5$. Strictly speaking, to use the results of \cite{Avkhadiev:2024mgd}, we should integrate out the b-quark, giving rise to a perturbatively calculable, mass dependent correction to the Collins-Soper kernel \cite{Pietrulewicz:2017gxc}. We leave the proper treatment of this to future work.}
\begin{align}
    b_\text{max}&=1.56^{+0.13}_{-0.09}\,\text{GeV}^{-1}\,,\\
    c_0&=0.0369^{+0.0061}_{-0.0065}\,,\\
    c_1&=0.0582^{+0.0064}_{-0.0088}\,.
\end{align}
This form incorporates a quadratic behavior that dies off into a linear growth at large $b_\perp$.
We find it quite exciting that we are able to directly use this lattice input in our factorization theorem for the back-to-back limit of the energy correlator.  We should also emphasize that the lattice results agree well with recent extractions of the Collins-Soper kernel from data, in particular, the fits of \cite{Bacchetta:2022awv,Moos:2023yfa}. For a comparison, see \cite{Avkhadiev:2024mgd}, and for a joint fit, see \cite{Avkhadiev:2025wps}.

We note that the split into a perturbative and non-perturbative component introduces a renormalon ambiguity. For a discussion of renormalons in the Collins-Soper kernel or rapidity anomalous dimensions, see  refs.~\cite{Moos:2023yfa,Scimemi:2016ffw,Gracia:2021nut,Avkhadiev:2024mgd,Liu:2023onm}. While this is certainly interesting to investigate in more detail for lattice extractions of the Collins-Soper kernel, since the Collins-Soper kernel itself already represents a suppressed power correction to the energy correlator, these renormalons are expected to have a minor numerical impact. We therefore simply use the parameterization of \cite{Avkhadiev:2024mgd}, along with the perturbative and non-perturbative components of the Collins-Soper kernel in the $\overline{\text{MS}}$ scheme.

The current state of the art calculations of the Collins-Soper kernel, e.g.~\cite{Avkhadiev:2024mgd} are able to compute the Collins-Soper kernel to around a scale of 1 fm. As lattice calculations improve, and the Collins-Soper kernel is computed to larger values of $b$, we will be able to further constrain the behavior of the energy correlator in the back-to-back limit.  It will be interesting to see the interplay of lattice calculations and phenomenological fits going forward, and we believe that the energy correlator will play a key role in this story.

\subsubsection{Linear Power Corrections}\label{sec:linear}

In addition to the quadratic power corrections, which the EEC shares with $q_T$, the energy correlator also has linear power corrections in the back-to-back limit due to the measurement of the energy flow operator. In the factorization theorem of ref.~\cite{Moult:2018jzp}, these appear in the jet function, since this is where the detector operator is inserted. We now discuss this, showing that the linear power correction in the back-to-back limit is governed by the same $\Omega_{1q}$ parameter as in the fixed-order and collinear region.

 Non-perturbative corrections to the EEC in the back-to-back limit were analyzed in ref.~\cite{Dokshitzer:1999sh} using dispersive methods \cite{Dokshitzer:1995qm}, where it was argued that the linear shift in the back-to-back limit is equal to $\Omega_{1q}$. Here we show how this arises from the SCET factorization theorem~\cite{Moult:2018jzp}, and generalize it to incorporate in addition, the higher order terms including $\Omega_{1g}$.

 To identify the action of the energy flow operator, it is convenient to use the expression for the EEC factorization in momentum space. For this purpose we can return to the starting point of the derivation of the factorization theorem \cite{Moult:2018jzp} for the back-to-back limit, where it was expressed in terms of inclusive di-hadron production in the TMD limit 
 \begin{align}
\EEC_{z \to 1} &= \frac{1}{2} \sum\limits_{ij} \int \df x_i\, \df x_j\,  x_i x_j    \int \df^2 \vec k_\perp\,    \delta \biggl(  1-z -\frac{\vec k_\perp^2}{Q^2}  \biggr) \nn \\
& \quad \times \frac{H(Q,\mu) \sigma_0}{\sigma}    \int \df^2 \vec k_{\perp,i} \int \df^2 \vec k_{\perp,j} \int \df^2 \vec k_{\perp,s}\,  \delta^{(2)} \Biggl(  \vec k_\perp -\biggl(    \frac{\vec k_{\perp,i}}{x_i}+\frac{\vec k_{\perp,j}}{x_j}-\vec k_{\perp,s} \biggr)  \Biggr) \nn \\
&\quad \times F_{q\to i}(\vec k_{\perp,i}, x_i)\,  F_{q\to j}(\vec k_{\perp,j}, x_j)\,  S_\EEC(\vec k_{\perp,s}) \,.
\end{align}
We can now remove the momentum conserving $\delta$-function, but instead of writing the functions in $b_\perp$ space, it is convenient to keep them in $k_\perp$ space.
The cross section can be then written in a factorized form as
\begin{align}
\EEC_{z \to 1} &=\frac{1}{2} \sum\limits_{ij}\frac{H(Q,\mu)\sigma_0}{\sigma} \int\! \df^2 k_\perp \frac{\df^2 b_\perp}{(2\pi)^2}\, \delta \Bigl( 1-z-\frac{k_\perp^2}{Q^2}  \Bigr) e^{-\img b_\perp \cdot k_\perp}
\biggl [ \int\! \df^2 k_{\perp,s}e^{-\img b_\perp\cdot k_{\perp,s}} S(k_{\perp,s})  \biggr]\nn
 \\
& \times \biggl[\int \df x_i x_i  \int \df^2 k_{\perp i} e^{\img b_\perp \cdot \frac{k_\perp}{x_i}}    F_{q\to i}(k_{\perp,i},x_i )  \biggr]
\biggl[\int \df x_j x_j  \int \df^2 k_{\perp j} e^{\img b_\perp \cdot \frac{k_\perp}{x_j}}    F_{q\to j}(k_{\perp,j},x_j )  \biggr]
\,.\end{align}
The terms in the square brackets are identified with the position space soft function, $S(b_\perp)$, and jet functions $J(b_\perp)$. However, for our purposes, we find it convenient to write them as Fourier transforms of the momentum space objects. 

The linear power correction arises from the jet function, 
\begin{align}
J(b_\perp)= \sum_i\int\! \df x_i~ x_i  \int \df^2 k_{\perp i}\, e^{\img  b_\perp \cdot \frac{k_\perp}{x_i}}    F_{q\to i}(k_{\perp,i},x_i )\,.
\end{align}
Writing it in this particular way makes it manifest that it is a one-point energy correlator integrated with a particular angular weight.

To study the leading non-perturbative contribution to the jet function, we can now factorize it into contributions from perturbative and non-perturbative modes
\begin{align}
J(b_\perp)&\to J(b_\perp)+ J_{\rm NP}(b_\perp)\,.
\end{align}
This factorization follows closely those in SCET$_+$ \cite{Bauer:2011uc}. After performing the BPS field redefinition of the non-perturbative modes, we obtain
\begin{align} \label{eq:J_NP}
J_{q_i,\rm NP}(b_\perp)
&=\sum_{j}\int \df x~ x  \int \df^2 k_{\perp}\, e^{\img b_\perp \cdot \frac{k_\perp}{x}} \int \df^2k_{\perp J} \df^2k_{\perp \text{NP}} \delta(k_\perp-k_{\perp J}-k_{\perp\text{NP}}) 
\nn \\ & \qquad \times
\mathcal{J}_{q_i\to q_j} (k_{\perp J}) \langle Y_{\bar n}^\dagger Y_n  \cN(k_{\perp \text{NP}},x)  Y_n^\dagger Y_{\bar n} \rangle\nn\\
&\quad +\int \df x~ x  \int \df^2 k_{\perp}\, e^{\img b_\perp \cdot \frac{k_\perp}{x}} \int \df^2k_{\perp J} \df^2k_{\perp \text{NP}} \delta(k_\perp-k_{\perp J}-k_{\perp\text{NP}}) 
\nn \\ & \qquad \times
\mathcal{J}_{q_i\to g}(k_{\perp J}) \langle {\mathcal{Y}}_{\bar n}^\dagger {\mathcal{Y}}_n  \cN(k_{\perp \text{NP}},x)  {\mathcal{Y}}_n^\dagger {\mathcal{Y}}_{\bar n} \rangle\,,
\end{align}
where $\mathcal{J}_{q_i\to q_j} (k_\perp)$ and $\mathcal{J}_{q_i\to g}(k_\perp)$ are perturbative matching coefficients. In the matrix elements involving Wilson lines, $\cN(k_\perp,x)$ indicates a detector measuring the $k_\perp$ and $x$, namely with $\delta$ functions  $\delta^{(2)}(k_\perp -\hat k_\perp)\, \delta (x-\hat x)$ (We make this explicit, since we will have to be careful with Jacobians when converting these matrix elements to those defining the standard universal non-perturbative power corrections).

In this paper, we will only consider NLL accuracy for the non-perturbative corrections, and thus only need these matching coefficients at LO. These are given by
\begin{align}
\mathcal{J}_{q_i\to q_j}(k_\perp)&=\delta^{(2)}(k_\perp) \delta_{ij}+\mathcal{O}(\alpha_s)\,, \nn \\
\mathcal{J}_{q_i\to g}(k_\perp)&=0+\mathcal{O}(\alpha_s)\,.
\end{align}
It would be extremely interesting to compute them to higher perturbative orders, and to study more systematically the structure of the factorization in \Eq{J_NP}. In particular, we will now show that the matrix elements of Wilson lines can be reduced to $\Omega_{1g}$ and $\Omega_{1q}$. The fact that $\Omega_{1g}$ does not appear in the back-to-back limit to the order that we work, while it does appear in the collinear limit, is a result of the different logarithmic structure in the factorization theorems.  Additionally, we will see that the lack of inclusion of these terms results in one of the leading uncertainties in the back-to-back limit. Therefore, further development of this factorization theorem is clearly motivated. We leave this to future work.

We will now show that the non-perturbative matrix elements reduces to $\Omega_{1q}$. As shown above, working to NLL we have
\begin{align}
J_{q,\rm NP}(b_\perp)=\int \df x~ x  \int \df^2 k_{\perp}\, e^{\img b_\perp \cdot \frac{k_\perp}{x}}  \langle Y_{\bar n}^\dagger Y_n  \cN(k_\perp,x)  Y_n^\dagger Y_{\bar n} \rangle\,.
\end{align}
We would now like to transform this into the standard definition of the non-perturbative parameter $\Omega_{1q}$, by re-expressing the result in terms of the energy flow operator. This can be achieved by rewriting the integrals over $k_\perp$ and $x$ as integrals over energies and angles. We can write
\begin{align}
\frac{\vec k_\perp}{E}=\frac{\hat n_\perp}{\cosh \eta}\,,
\end{align}
where $\hat n_\perp$ is a unit vector parameterized by the azimutahl angle $\varphi$, and pseudorapidity $\eta$ with respect to the $n$.  This allows us to rewrite the leading NP correction as
\begin{align} \label{eq:J_NP_int}
J_{q,\rm NP}(b_\perp)=\frac{1}{Q} \int \df E E \int \df \Bigl( \frac{1}{\cosh \eta}  \Bigr) \df \phi\, e^{\img Q \vec b_\perp \cdot \frac{\hat n_{\perp}}{\cosh \eta}} \frac{1}{\tanh \eta} \langle  Y_{\bar n}^\dagger Y_n  \cN(\theta (\eta),E)  Y_n^\dagger Y_{\bar n} \rangle\,,
\end{align}
where now $\cN(\theta (\eta),E)$ denotes a detector operator defined using delta functions in $\theta = 2 \arctan(e^{-\eta})$ and $E$. By separating the $\eta$ and $E$ integrals, this allows us to recognize this as a matrix element of the energy flow operator at some boost angle $\eta$.  It is now a straightforward exercise to use the boost properties of eikonal matrix elements \cite{Lee:2006fn}, to write it in terms of the universal non-perturbative parameter $\Omega_{1q}$, and an integral over $\eta$. To do so,  we can rewrite the two dimensional integral over $\eta$ and $\phi$ in a more standard way, by going back to a vector 
\begin{align}
\vec t=\frac{\hat n_\perp}{\cosh \eta}\,,
\end{align} 
which leads us to 
\begin{align}
J_{q,\rm NP}(b_\perp)=\frac{\Omega_{1q}}{Q}\int \df^2 t\, e^{\img Q \vec b_\perp \cdot \vec t} \cosh^3 \eta \, \frac{1}{\tanh \eta}\,.
\end{align}
Rewriting this in terms of $t$ at forward rapidity, where $\tanh \eta \sim 1$, $\cosh \eta=1/|t|$, gives
\begin{align}
J_{q,\rm NP}(b_\perp)=\frac{\Omega_{1q}}{Q} \int \df^2 t\, e^{\img Q b_\perp \cdot t}\frac{1}{t^3}\,.
\end{align}
This integral is power divergent, with a divergence from the collinear region. This divergence should be regulated since we have factorized this matrix element from the full jet function. This can be achieved using dimensional regularization
\begin{align}
\frac{1}{Q} \int \df^{2-2\epsilon} t\, \frac{1}{2\pi} \frac{1}{t^3} e^{\img Q b_\perp \cdot t}=-b + \mathcal{O}(\epsilon)\,,
\end{align}
which gives
\begin{align}
J_{q,\rm NP}(b_\perp)=\frac{\Omega_{1q}}{Q} \int \df^2 t\, \frac{1}{2\pi} \frac{1}{t^3}\, e^{\img  Q b_\perp \cdot t}=-\Omega_{1q} b\,.
\end{align}
This same power divergent integral is also obtained within the dispersive approach, leading to the same result \cite{Dokshitzer:1999sh}. This derivation gives the linear power correction in the back-to-back limit, and relates it to the universal non-perturbative power correction $\Omega_{1q}$. The negative sign in the power correction is physical in that it shifts the height of the plateau in the back-to-back region down, whereas in the bulk region the linear power correction shifts the cross section to higher values. This aligns with the intuition the non-perturbative corrections should flatten the distribution, as discussed in \Sec{overview}.

We emphasize that this result is only true to NLL. As emphasized above, at beyond NLL, one must also compute the perturbative matching coefficients, which induces a dependence on $\Omega_{1g}$ as well. This is expected from the sum rule, since $\Omega_{1g}$ also appears in the collinear limit. It will be interesting to study the non-perturbative power corrections in the back-to-back limit in more detail, particularly in light of recent analyses of archival LEP data which allow precision measurements with high angular resolution. Beyond the calculation of the $\alpha_s$ matching coefficients,  it will also be important to move to a proper  renormalon free scheme. We note that while renormalons in the back-to-back limit of SCET$_{\text{II}}$ observables have received less attention than their SCET$_{\text{I}}$ counterparts, the presence of the universal $u=1/2$ renormalon equal to $\Omega_1$ in the jet function is quite interesting. Typically the $R$-evolution anomalous dimensions are extracted from soft functions for thrust \cite{Hoang:2009yr,Hoang:2007vb}. However, since we know the three-loop jet function for the EEC, we can use it to extract the three-loop coefficients for the renormalon subtraction, and do a renormalon subtraction based on the EEC jet function. We leave this for future work.

Adding in the effects of tracks, just as was done in the bulk of the distribution, and incorporating the next-to-leading logarithmic resummation from the factorization theorem, we finally arrive at the expression for the linear power correction in the back-to-back region
\begin{align}
    \EEC_{z\to1}^{\Omega\text{,res.}}(z)&=\frac{\sigma_0 Q^2}{4\sigma}
    \int_0^\infty\df b_\perp\,b_\perp J_0\bigl(\sqrt{1-z}\,b_\perp Q\bigr)\,
    \Bigl(-2\Omega_{1q}(\mu_\Omega) b_\perp e^{-\frac{1}{2}\tau b_\perp^2} T_g (1,\mu_\Omega)\Bigr)\\\nn
    &\quad\times U^\text{NLL}\bigl(Q,b_\perp,\{\mu_H,\mu_J,\mu_S,\mu_0\},\{\nu_J,\nu_S\}\bigr)\\\nn
    &\quad\times H(Q,\mu_H)\,S(b_\perp,\mu_S,\nu_S)\,\frac{1}{2n_f}\sum_q \bigl[J_q(b_\perp,Q,\mu_J,\nu_J)+J_{\bar{q}}(b_\perp,Q,\mu_J,\nu_J)\bigr].
\end{align}
To merge with the description in the bulk region, the complete expression for the non-perturbative corrections in the back-to-back limit is given by
\begin{align}
    \EEC_{z\to1}^{\Omega}(z)
    &=
    \EEC^{\Omega}(z)
    -\EEC_\text{bulk}^{\Omega,\text{1-sing}}(z)
+\EEC_{z\to1}^{\Omega,\text{res.}}(z)\,,
\end{align}
where the singular part of the non-perturbative power correction as $z\to1$ is given by
\begin{align}
    \EEC_\text{bulk}^{\Omega\text{,1-sing}}(z)
    &=
    \frac{1}{2}\frac{\sigma_0}{\sigma}\frac{\Omega_{1q}(\mu_\Omega)}{Q}\frac{1}{(1-z)^{\frac{3}{2}}} 
    \frac{1}{2n_f}\sum_{q}T_q(1,Q)\,T_g(1,\mu_\Omega)\,.
\end{align}

\subsection{Scale Setting and Matching}\label{sec:b2bMatch}

The factorization theorem captures all terms that are of leading power in $1-z$ in the fixed-order result. It enables the resummation of large logarithms of $1-z$ in the back-to-back limit, which we achieve by evaluating the ingredients of the factorization theorem at their natural scales $\mu_i^{\rm nat.}$ and then evolving them to a common scale. However, in the fixed-order region, when $1-z$ is not small, we need to set all scales  equal to the fixed-order scale $\mu_\text{FO}$ such that the factorized result agrees exactly with the fixed-order singular result. We use profile scales to ensure a smooth transition between the resummation and fixed-order region, as discussed below. In this section we restrict ourselves to the central scale choice, discussing the scale variations used to assess uncertainties in \sec{b2bUncertainty}.

In the resummation region, the natural scales that minimize the logarithms of the hard, jet and soft function are
\begin{align}
    \mu_H^\text{nat.}=\nu_J^\text{nat.}&=Q\,,\\
    \mu_J^\text{nat.}=\mu_S^\text{nat.}=\nu_S^\text{nat.}&= b_0/b_\perp\,,
\nn \end{align}
where $b_0=2 e^{-\gamma_E}$. To avoid the Landau pole in evaluating the coupling, we use the $b^*$ prescription 
\begin{align}
b^*(b_\perp)=\frac{b_\perp}{\sqrt{1+b_\perp^2/b^2_\text{max}}}\,,
\end{align}
ensuring that scales do not go below a certain minimal value. With the $b^*$ prescription, the minimal value that the scales can take is
\begin{align}    \mu_S^\text{min}=\mu_J^\text{min}=b_0/b_\text{max}\,.
\end{align}
For all the other scales we do not impose a minimal value. The fixed-order scale is equal to the hard scale
\begin{align}
    \mu_\text{FO}=\mu_H=Q\,.
\end{align}

\begin{figure}
    \centering
    \includegraphics[width=0.6\textwidth]{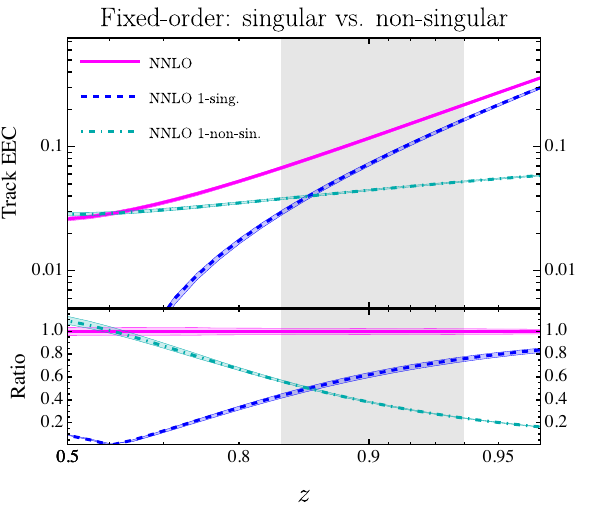}
    \caption{The contributions of the singular and non-singular parts of the EEC in the back-to-back limit for the perturbative contribution. The transition region of $z\in[0.84,0.94]$ is indicated by the gray shaded band}
  \label{fig:b2bTransition}
\end{figure}

To transition from the resummation region to the fixed-order region we smoothly turn off the resummation using profile scales 
\begin{align}
    \mu_X\rightarrow\tilde \mu_X(z)&=
    g_\text{run}\bigl(\sqrt{1-z}\bigr)
    \sqrt{(\mu_X^\text{nat.})^2 + (\mu_X^\text{min})^2}
    +\bigl[1-g_\text{run}\bigl(\sqrt{1-z}\bigr)\bigr]\,\mu_\text{FO}\,,
    \\
    \nu_X\rightarrow\tilde \nu_X(z)&=
    g_\text{run}\bigl(\sqrt{1-z}\bigr)\,\nu_X^\text{nat.}
    +\bigl[1-g_\text{run}\bigl(\sqrt{1-z}\bigr)\bigr]\,\mu_\text{FO}\,,
\end{align}
where $X=J,S$. We choose the activation function to be the same as in the collinear region, namely 
\begin{align} \label{eq:profile}
g_\text{run}\bigl(x,\{x_1,x_2,x_3\}\bigr) &= 
\begin{cases}
1 & 0 \leq x \leq x_1 \,, \\
1- \frac{(x-x_1)^2}{(x_2-x_1)(x_3-x_1)} & x_1 < x \leq x_2\,, \\
\frac{(x-x_3)^2}{(x_3-x_1)(x_3-x_2)} & x_2 < x \leq x_3\,, \\
0 & x_3 \leq x \leq 1\,,
\end{cases}
\end{align}
The transition region is specified through the parameters $\{x_1,x_2,x_3\}$. Specifically, the parameters $x_1$ and $x_3$ describe the boundaries of the transition region while $x_2$ affects the shape or "quickness" of the transition. Given a transition region of $z\in[z_\text{FO},z_\text{res}]$ we have
\begin{align}
    x_1^2=1-z_\text{res}\qquad\text{and}\qquad x_3^2=1-z_\text{FO}\,.
\end{align}
The additional variable $x_2$ is taken to be equal to the average, $x_2=(x_1+x_3)/2$.

To choose the transition parameters we study the relative size of the fixed-order singular and non-singular pieces, shown in \Fig{fig:b2bTransition}. 
We take $z\in[0.84,0.94]$ as the transition region. At $z=0.94$ the singular is about a factor 5 larger than the non-singular, which is definitely in the resummation region. Since the singular and non-singular have the same sign, we can delay turning off the resummation until $z=0.7.$\footnote{When the singular and non-singular have opposite signs, the transition must be completed before they are of similar size, since there will then be large cancellations between the two.} This corresponds to
\begin{align}\label{eq:b2bxs}
    z\in[0.84,0.94]\qquad\text{with}
    \qquad\{x_1,x_2,x_3\}=\{0.244949, 0.322474, 0.4\}\,,
\end{align}
for the central value of our prediction.

For reference, it is interesting to convert this transition region to the case of $q_\perp$ for the $Z$ $p_T$ distribution. Using $q_\perp \simeq Q \sqrt{1-z}$, our transition region translates to $q_\perp\in[22.3, 36.5]$ GeV for $Q=m_Z$.

\subsection{Uncertainties}\label{sec:b2bUncertainty}

In the back-to-back limit we take into account five different sources of uncertainty: missing higher perturbative orders of fixed-order and resummation ingredients, non-perturbative uncertainties arising from the Collins-Soper kernel and the linear power correction, and a matching uncertainty. In this section we provide details about how we estimate the size of each source of uncertainty.

Both the fixed-order and resummation uncertainties are  estimated using scale variations. The profile scales in \sec{b2bMatch} were constructed in terms of the fixed-order scale $\mu_{\rm FO}$, natural scales $\mu_i^{\rm nat.}$ and minimal scales $\mu_i^{\rm min}$. The fixed-order scale variation varies both the natural scales and the fixed-order scale by a factor of 2, keeping the minimal scale fixed. This ensures that in the fixed-order region, the scale in the singular contribution (described by the factorization) and non-singular contribution are simultaneously varied up and down by a factor of 2. 

To assess the resummation uncertainty, the resummation scales are varied a $z$-dependent factor $f_\text{vary}$,
\begin{align} 
f_\text{vary}\bigl(x,\{x_1,x_2,x_3\}\bigr) &= \begin{cases}
2\left(1-\frac{x^2}{x_3^2}\right) & 0 \leq x < \frac{x_3}{2} \,, \\
1 + 2\left(1-\frac{x}{x_3}\right)^2 & \frac{x_3}{2} \leq x < x_3
\,, \\
1 & x_3 \leq x \,,
\end{cases}
\end{align}
which is equal to $2$ in the resummation region, and smoothly goes to 1 as the fixed-order region is approached. The parameters $\{x_1,x_2,x_3\}$ describe the transition to the fixed-order region, and are taken equal to those for the central scale in \Eq{b2bxs}. Explicitly, the variations for the profile scales can be written as 
\begin{align}
    \mu_X\rightarrow\tilde \mu_X^\text{vary}(z)&=
    g_\text{run}\bigl(\sqrt{1-z}\bigr)\sqrt{\bigl(2^{m_\text{FO}} f_\text{vary}^{m_X}(\sqrt{1-z})\mu_{J,S}^\text{nat.}\bigr)^2 +\bigl(\mu_{J,S}^\text{min}\bigr)^2}
    \\
    &\quad
    +\bigl[1-g_\text{run}\bigl(\sqrt{1-z}\bigr)\bigr]\, 2^{m_\text{FO}}\,\mu_\text{FO}\,.
    \nonumber
    \\
    \nu_X\rightarrow\tilde \nu_X^\text{vary}(z)&=
    f(z)\,2^{m_\text{FO}} f_\text{vary}^{n_X}(\sqrt{1-z})\nu_{J,S}^\text{nat.}
    +\bigl[1-g_\text{run}(z)\bigr]\,2^{w_\text{FO}}\,\mu_\text{FO}\,.
\end{align}
In this formula, the profile scales for the central curve are obtained by setting $m_\text{FO}$ and all $m_X$ and $n_X$ to zero. We have also suppressed the arguments $\{x_1,x_2,x_3\}$ in $g_{\rm run}$ for simplicity.  To obtain the fixed-order uncertainties we simply vary $m_\text{FO}=\pm1$. The scale variations for the resummation uncertainty are governed by four parameters, namely $m_J$, $m_S$, $n_J$ and $n_S$. Each of these four parameters can take 3 values, $\pm1$ and $0$. This would give $3^4=81$ distinct scale variations. Following \cite{Cal:2023mib}, the variations that result in logarithms whose argument is varied by more than 2 are discarded. Additionally we discard the variations where $\mu_J$ is varied down and the ratio of $\nu_J$ and $\nu_S$ is varied up. This leaves a total of 36 distinct scale variations. To obtain the total resummation uncertainty we take the envelope of these 36 scale variations. 

Next we consider the uncertainty from matching the resummed result to the fixed-order result. To obtain the theoretical uncertainty related to the matching we vary the parameters $\{x_1,x_2,x_3\}$ that describe the transition region and the shape of the profile functions. We consider two sets of variations: one where the boundaries of the transition region are varied but the overall shape is kept constant and one where the boundaries are kept constant but the shapes of the profile functions are varied. Recall that for the central curve we have chosen the transition region as $z\in[0.84,0.94]$ and set $x_2=(x_1+x_3)/2$, this amounts to
\begin{align}
    z\in[0.84,0.94]\qquad\text{with}
    \qquad\{x_1,x_2,x_3\}=\{0.244949, 0.322474, 0.4\}\,.
\end{align}
First we vary the boundaries of the transition regions, described by $x_1$ and $x_3$, while keeping $x_2$ fixed as $x_2=(x_1+x_3)/2$,
\begin{align}
    &z\in[0.80,0.92]\qquad\text{with}
    \qquad\{x_1,x_2,x_3\}=\{0.282843, 0.365028, 0.447214\}\,,\\
    &z\in[0.88,0.92]\qquad\text{with}
    \qquad\{x_1,x_2,x_3\}=\{0.282843, 0.314626, 0.34641\}\,,\\
    &z\in[0.80,0.96]\qquad\text{with}
    \qquad\{x_1,x_2,x_3\}=\{0.2, 0.323607, 0.447214\}\,,\\
    &z\in[0.88,0.96]\qquad\text{with}
    \qquad\{x_1,x_2,x_3\}=\{0.2, 0.273205, 0.34641\}\,.
\end{align}
Next we vary the shape of the profile functions while keeping the boundaries of the transition region fixed. We do this by varying $x_2$ and keeping $x_1$ and $x_3$ fixed. For the central curve $x_2$ is fixed as the average of $x_1$ and $x_3$. We vary $x_2$ by
\begin{align}
    x_2&=\frac{x_1+2x_3}{3}\qquad\text{with}
    \qquad\{x_1,x_2,x_3\}=\{0.244949, 0.348316, 0.4\}\,,\\
    x_2&=\frac{2x_1+x_3}{3}\qquad\text{with}
    \qquad\{x_1,x_2,x_3\}=\{0.244949, 0.296633, 0.4\}\,,\\
    x_2&=\frac{x_1+3x_3}{4}\qquad\text{with}
    \qquad\{x_1,x_2,x_3\}=\{0.244949, 0.361237, 0.4\}\,,\\
    x_2&=\frac{3x_1+x_3}{4}\qquad\text{with}
    \qquad\{x_1,x_2,x_3\}=\{0.244949, 0.283712, 0.4\}\,.
\end{align}

Finally, we consider the three sources of non-perturbative uncertainties, one coming from the Collins-Soper kernel, one from the model for the non-perturbative part of the TMD jet function, and one from the non-perturbative power correction. For the theory uncertainty coming from the non-perturbative part of the Collins-Soper kernel we vary the model parameters $b_\text{max}$, $c_0$ and $c_1$ within the given fit uncertainties \cite{Moos:2023yfa},
\begin{align}
    b_\text{max}&=1.56^{+0.13}_{-0.09}\,\text{GeV}^{-1}\,,\\
    c_0&=0.0369^{+0.0061}_{-0.0065}\,,\\
    c_1&=0.0582^{+0.0064}_{-0.0088}\,.
\end{align}
We vary each of these parameters between their upper and lower value while keeping the other parameters at their central value. This results in a set of six curves of which we take the envelope. For the TMD model uncertainty, we vary the parameter $\tau$ up and down by a factor of 2. 

In our treatment of the theoretical uncertainties, we treat all uncertainties related to the non-perturbative linear power correction as independent from the rest. To estimate the uncertainty from the power correction, we combine all variations discussed above, combined with varying $\Omega_{1q}$ within its fit uncertainty given in ref.~\cite{Schindler:2023cww}, and take the envelope of all variations. To emphasize, for all above sources of uncertainty we keep all parameters of all $\Omega_{1q}$ terms at their central values. Instead, we combine all the above variations for the $\Omega_{1q}$ terms into one combined uncertainty.

\subsection{Numerical Results and Discussion}\label{sec:numerical_b2b}

\begin{figure}
\centering
\includegraphics[width=0.45\textwidth]{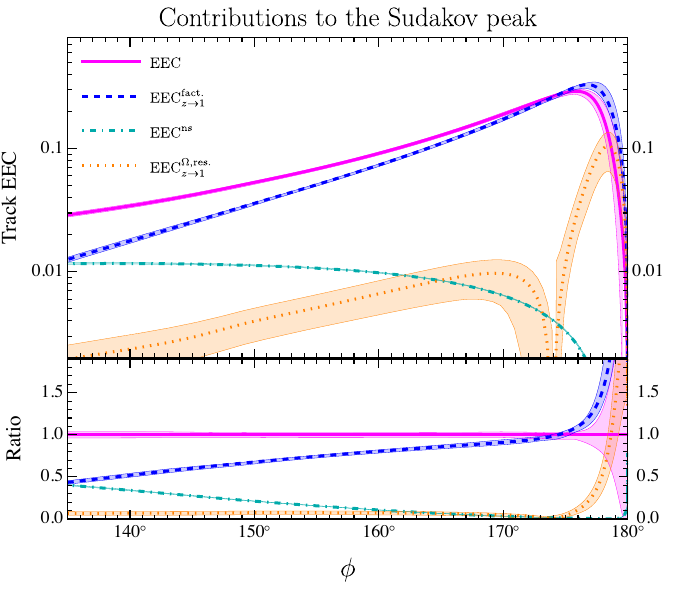}
\includegraphics[width=0.45\textwidth]{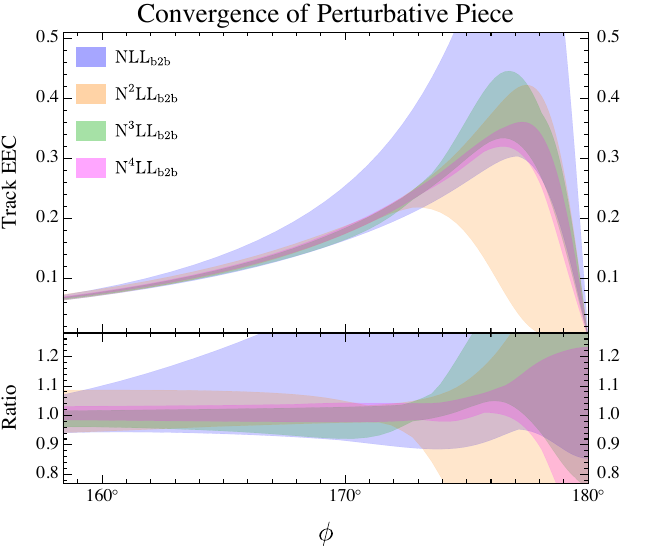}
\caption{In the left panel, we show a comparison of different contributions to the total result in the back-to-back limit. The linear power correction provides a large contribution in the peak region, emphasizing the need to improve our control of this contribution. In the right panel, we show the convergence of the perturbative contribution in the back-to-back limit.}
\label{fig:sudakov_contributions}
\end{figure}

In the left panel of \Fig{fig:sudakov_contributions} we show our highest order prediction, $\text{NNLO}+\text{NNLL}_\text{col}+\text{NNNNLL}_\text{b2b}$, in the back-to-back limit. We have decomposed it into different contributions, to show the result from the factorized singular contributions (blue), the non-singular contributions (green), and the leading linear power correction (orange). The non-singular contributions are small in the peak region. However, the contributions from the leading non-perturbative power correction are large.

In the right panel of \Fig{fig:sudakov_contributions} we show the convergence of our purely perturbative prediction in the back-to-back limit. We observe excellent convergence, highlighting the importance of high order perturbative ingredients in the back-to-back limit.

There is an important lesson that can be drawn from \Fig{fig:sudakov_contributions}, which highlights the need for improvement of our theoretical calculations. The non-perturbative power corrections in the back-to-back limit scale like $1/(1-z)^{3/2}$ dressed by Sudakov logarithms, as compared to the perturbative result, which scales like $1/(1-z)$ dressed by Sudakov logarithms. Due to this sharp growth, the non-perturbative corrections are extremely sensitive to the order at which the Sudakov that dresses them is evaluated. In this paper, we have only incorporated the non-perturbative corrections into our factorization theorem at NLL.  While the convergence of these logarithms for the non-perturbative corrections may be slightly different, the right plot of \Fig{fig:sudakov_contributions} suggests that it is important to achieve at least NNLL for these contributions. This emphasizes that to improve understanding of the peak will require improving our understanding of the non-perturbative power corrections in the back-to-back limit.

\begin{figure}
    \centering
     \includegraphics[width=0.45\textwidth]{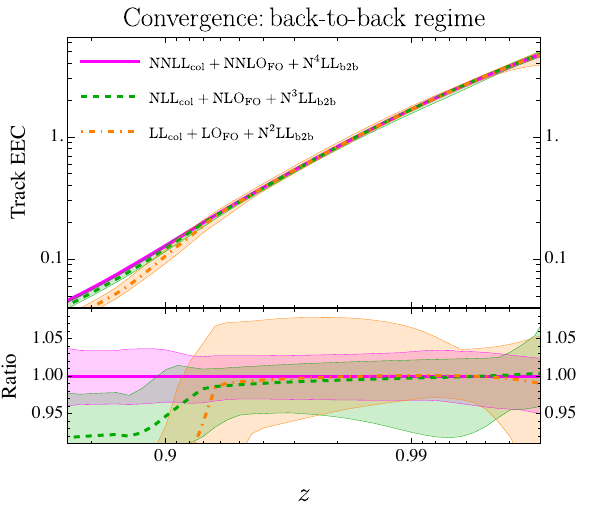}
      \includegraphics[width=0.45\textwidth]{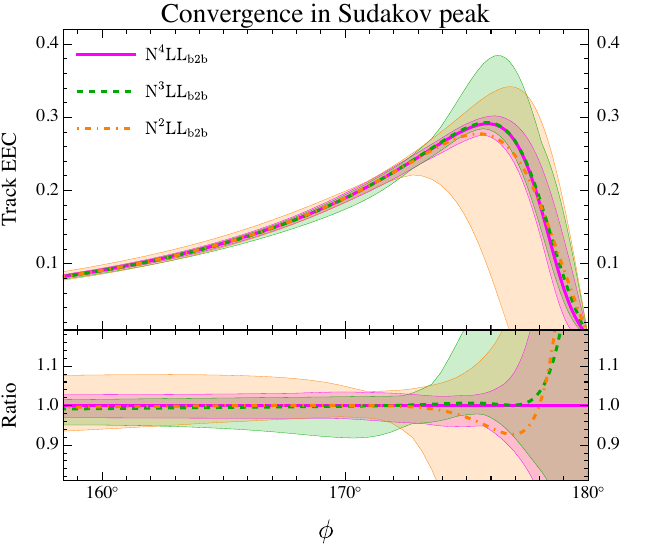}
    \caption{Convergence of our results in the back-to-back limit, shown both as a function of $z$, and $\phi$. In $\phi$ coordinates it exhibits the famous Sudakov peak.}
    \label{fig:zoomB_b2b}
\end{figure}

In \Fig{fig:zoomB_b2b} we show our  predictions in the back-to-back limit at different perturbative orders. While these numerical results are obtained from our complete calculation, in this regime they are dominated by resummation in the back-to-back limit. Results are shown both in the $z$ variable, as well as in the $\phi$ variable. Overall, we observe quite good convergence in the back-to-back limit, particularly compared with the bulk and collinear regions. We observe quite impressive uncertainties in this regime, at the order of a few percent. This highlights the impact of the remarkable amount of high-loop perturbative data required to describe this region. Ref.~\cite{Duhr:2022yyp} also performed a study of the perturbative convergence of the singular contributions in the back-to-back limit at N$^4$LL order. The results in \cite{Duhr:2022yyp} show a smaller uncertainty in the peak region due to the fact that they only study the perturbative results. Our results incorporate non-perturbative contributions, whose resummation is only known to lower orders. It is these effects which dominate the uncertainty in the peak region.

\begin{figure}
\centering
\includegraphics[width=\textwidth]{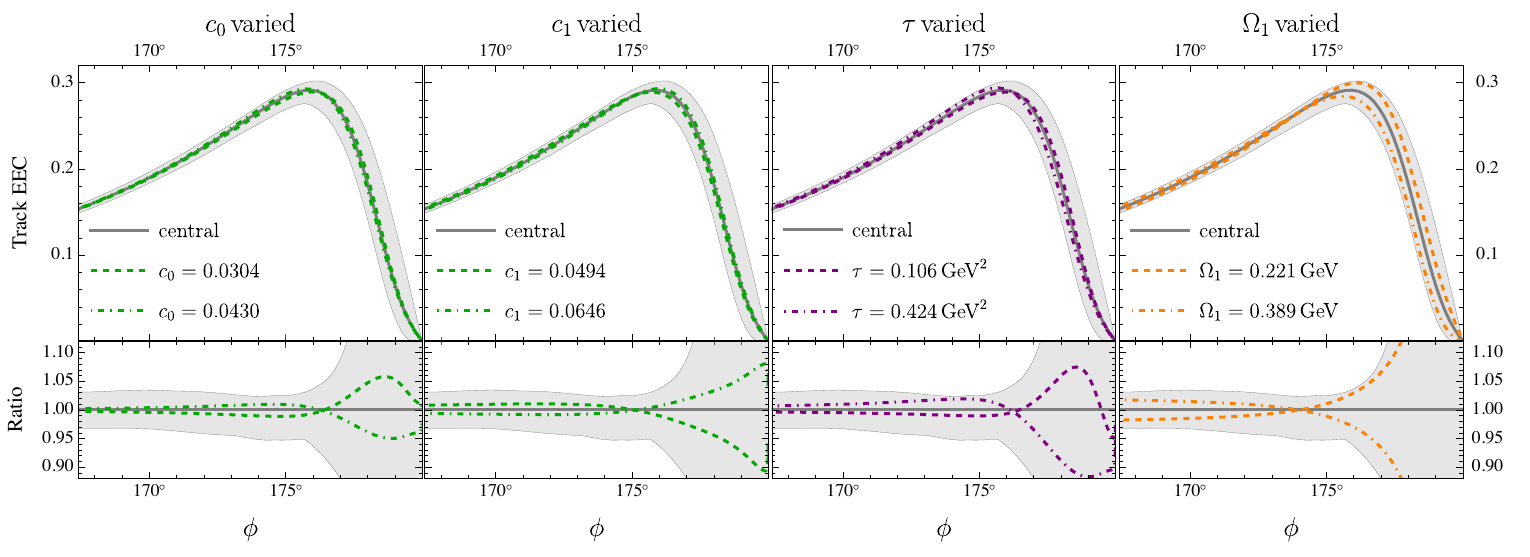}
\caption{An illustration of variations in the different non-perturbative parameters in the Sudakov limit. The parameters $c_0$, $c_1$ and $\tau$ control the parameterization of the Collins-Soper kernel and have minimal effect. On the other hand, we see significant sensitivity to the value of $\Omega_1$ in the peak region.  }
\label{fig:sudakov_variations}
\end{figure}

It is also interesting to study the sensitivity of our result to variations in the non-perturbative parameters. In \Fig{fig:sudakov_variations} we show the dependence of the peak region of the EEC on the variation of three parameters in the Collins-Soper kernel ($c_0$, $c_1$, $\tau$), as well as variations in the linear power correction, $\Omega_1$. We see that the peak is extremely sensitive to variations in the linear power correction. These have an effect at smaller angles as compared to variations in the parameters of the Collins-Soper kernel. The variations associated with parameters of the Collins-Soper kernel arise at very small angles, and are much smaller than our uncertainties, which in that region are dominated by the perturbative resummation dressing the leading non-perturbative parameter. This highlights that we will first have to control the resummation associated with the leading linear non-perturbative power correction, before we can achieve a level of sensitivity to the parameters of the Collins-Soper kernel.

\subsection{Back-to-Back Plateau}\label{sec:b2bPlateau}

\begin{figure}
    \centering
    \includegraphics[width=0.535\textwidth]{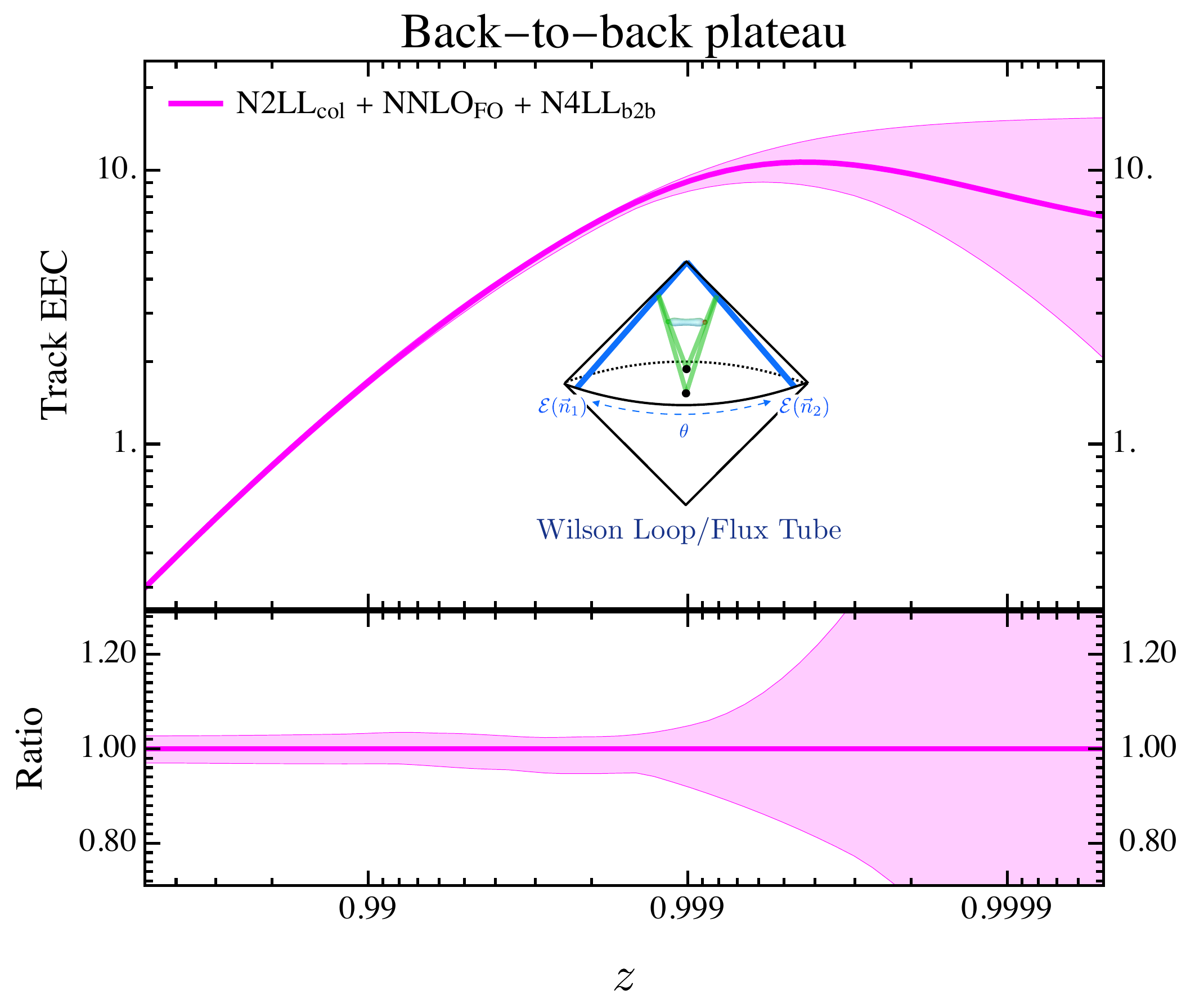}
    \includegraphics[width=0.325\textwidth]{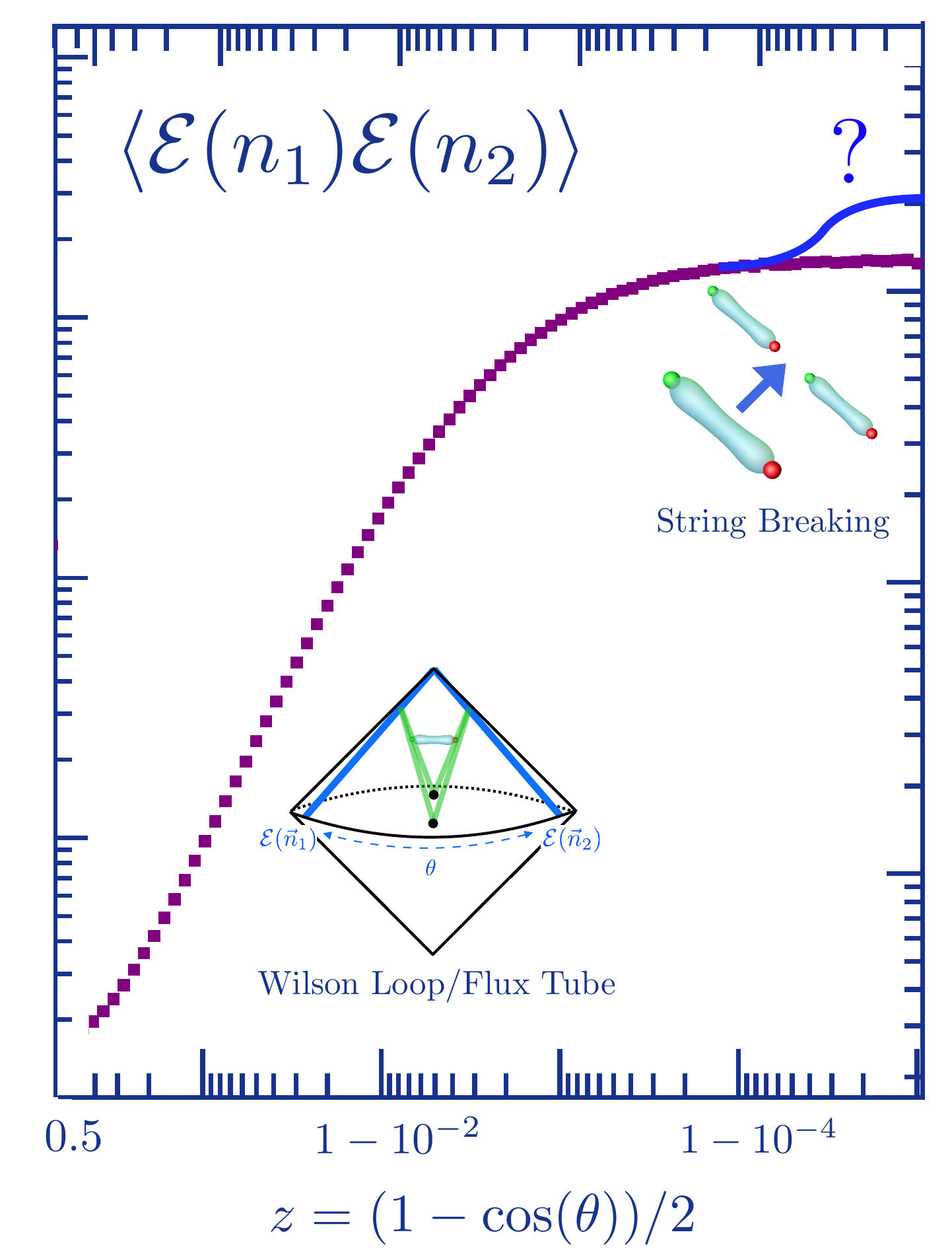}
    \caption{In the back-to-back limit of the EEC, the turnover to a plateau occurs perturbatively. In the left plot we use our pure factorization theorem result, in which the string does not break, pulling too hard, and causing the distribution to decrease. In the right plot we show the data, which has string breaking. It would be interesting to understand if there can be an imprint of this, leading to some non-trivial behavior of the distribution in the deep IR, as illustrated by the blue line with a question mark next to it. }
    \label{fig:zoom_b2b_plateau}
\end{figure}

In this section, we briefly comment on the interpretation of our calculation in the extreme back-to-back limit, $z\to 1$, as well as the physics that might be learned from precision measurements in this region. Here we are interested in the regime below the confinement transition. As such, our goal is not a precision understanding of the physics in this region, but rather an understanding of the physics which is accessed by measurements of the confining transition in this region.

As discussed in \Sec{overview_physics}, the physical picture of the back-to-back limit is of the formation of a flux tube of length $\ln(J)$, where $J\sim 1/(1-z)$. The formation of this flux tube significant modifies the naive perturbative scaling of $1/(1-z)$, converting it into a flat plateau. This point was originally emphasized by Parisi and Petronzio \cite{Parisi:1979se}, namely that double logarithms are strong enough to turn over the distribution in a fixed coupling theory.

In the case of real world QCD, there are two major modifications as compared to the case of a conformal gauge theory. The first is that due to the running coupling, we lose control of our perturbative prediction as $Q^2(1-z)\sim \Lambda_{\text{QCD}}^2$. In the left panel of \Fig{fig:zoom_b2b_plateau} we show our prediction extended all the way to very small values of $(1-z)$, well beyond its regime of validity. We can see that for $Q=m_Z$, our uncertainties begin to rapidly grow right before the transition to the plateau region. One motivation for extending measurements to higher energies is that it will enable the transition to be purely within the perturbative regime. However, even at $Q=m_Z$, we believe that we are able to use our leading power predictions to make a prediction for the height of the plateau in the back-to-back limit. 

A shortcoming of our calculation is that it includes only the leading non-perturbative power correction, $\Omega_1$. Recall that our expansion was in $\Omega_1/Q\sqrt{1-z}$. Once $Q\sqrt{1-z}\sim \Omega_1$, we have to resum the complete set of non-perturbative corrections. As such, we do not expect our result to exhibit a physical behavior when extrapolated into the extreme $(1-z)\to 0$ limit.  Indeed, one can see in \Fig{fig:zoom_b2b_plateau} that our central prediction does not asymptote to a flat behavior as $z\to 1$. This is in contrast to what is expected physically, namely that the energy correlator distribution should become flat as $z\to 1$ corresponding to the behavior of a free hadron gas. Because of this, in our final prediction, shown in \Fig{fig:physics}, we do not plot a central value for the curve in the region $z\to 1$, since this would involve additional modeling. Instead, we simply use our leading power results to estimate upper and lower bounds on the behavior of the correlator in this regime.

There are a number of aspects of the extreme back-to-back limit that would be worth understanding in more detail, and where our treatment could be improved.  First, we have made a crude approximation for the value of the plateau height in the back-to-back limit using our leading power factorization theorem. This makes the assumption that once the plateau is reached, it has no additional features. In \Fig{fig:zoom_b2b_plateau}, we also show the data in this regime, where this feature seems to be borne out. However, it would be interesting to try and understand this in more detail. For example, at higher energies, is it possible to first have a perturbative transition associated with the perturbative flux tube, and then have a second transition (illustrated with the blue question mark) associated with hadronization? This would be particularly interesting for understanding how string breaking in QCD manifests in the behavior of the transition region in the back-to-back limit. In the collinear region confinement imprints itself in a clear way in the behavior of the distribution. This is much less clear for the back-to-back limit. 

Another way of placing bounds on the height of the plateau in the back-to-back limit is through the study of cumulant constraints. The cumulant
\begin{align}
\Sigma_{b2b}(y)=\int\limits_0^y \df z\, \text{EEC}(1-z)\,,
\end{align}
is computable using our leading power factorization, and can bound extreme deviations in the plateau region. It would be interesting to explore it in more detail, as was done in the collinear region \Sec{colPlateau}. 

More generally, we believe that the data in the back-to-back limit of the energy correlator represents an opportunity to improve our understanding of the physics in this limit, and to go beyond the standard paradigm. We believe that this is particularly interesting due to the forthcoming data from the conformal bootstrap \cite{N4_bootstrap}. Combined with the LEP data which will enable the study of this regime for theories with both conformal and confining flux tubes. To maximize the understanding of this region, we believe that it will be important to clarify the relationship between non-perturbative parameters used in the QCD description, and properties of the underlying field theory, or effective string description. For example, our calculation uses the Collins-Soper kernel extracted from the lattice. It would be interesting to understand how string breaking effects are encoded in this object. Can it be understood using the effective field theory of long strings \cite{Aharony:2013ipa,Aharony:2010cx}, or effective string interaction vertices \cite{Komargodski:2024swh}? Could one bootstrap these interactions using sum rules, or extract them from measurements? It would also be interesting to search for other manifestations of the QCD flux tube in LEP data. For example, in \cite{Gallicchio:2010sw,Chien:2011wz} it was shown that patterns in soft hadrons resemble the QCD flux tube. Can these patterns be sharply connected to the flux tube in our calculations? We believe that their remains much to understand about this limit of the correlator.

We would also like to highlight that the study of this region of the energy correlator should be particularly interesting for improving the understanding of hadronization models in parton shower Monte Carlo programs, particularly those that include higher order soft resummation \cite{Dasgupta:2020fwr}.

\section{Numerical Predictions for the Full EEC Spectrum}\label{sec:results}

In this section we present our numerical results for the full spectrum of the energy-energy correlator computed on tracks. This result was compared with ALEPH data in \cite{Bossi:2025nux}, and is shown in \Fig{fig:physics}.

In \Sec{params} we provide a summary of the numerical input parameters.   We present results for the perturbative convergence of the distribution in \sec{convergence}. In \sec{sources} we present a detailed study of different sources of theory uncertainties. In \Sec{compare_as} we study the dependence on $\alpha_s$, highlighting why we believe the EEC provides an ideal observable for precision extractions of the strong coupling constant. In \sec{resultsQs}, we study the $Q$ dependence of our results. In \sec{compare} we present comparisons of the results computed on tracks with those computed on all particles.

\subsection{Summary of Input Parameters}\label{sec:params}

Before presenting our numerical results we summarize all input parameters, and the order counting for our different predictions. For all our predictions we treat the $b$ quark as massless, and neglect QED effects. We hope to improve the treatment of these, as well as other effects discussed in \Sec{improve}, in future work. The parameters used as inputs to our calculation are as follows:

{\bf{Strong Coupling Constant:}} We take the strong coupling constant as \cite{ParticleDataGroup:2024cfk} \footnote{Note that we use the PDG value, instead of the value extracted in joint fits with $\Omega_1$. In future work, we hope that both $\alpha_s$ and $\Omega_1$ can be simultaneously extracted from fits to the EEC.}
\begin{align}
\alpha_s(m_Z)=0.118\,.
\end{align}

{\bf{Linear Power Correction:}} We take the linear power correction extracted from thrust \cite{Abbate:2010xh} and converted to the EEC in \cite{Schindler:2023cww} 
\begin{align*}
    \bar\Omega_{1q}&=0.305\pm0.084\,\text{GeV}\,.
\end{align*}
The value of $\Omega_{1g}$ is weakly constrained. We take as a central value the result predicted by Casimir scaling, and vary the result in the range $0.7\cdot C_A/C_F \bar{\Omega}_{1q}\leq \Omega_{1g} \leq 1.3\cdot C_A/C_F \bar{\Omega}_{1q}$, which corresponds to
\begin{align}
\Omega_{1g}=0.686^{+0.206}_{-0.206} ~\text{GeV}\,.
\end{align}
We believe that this is reasonable, and we hope that this uncertainty can be significantly reduced in the near future.

{\bf{Collins-Soper parametrization:}}
We use the parameterization given in \Eq{CS_param}, with parameters
\begin{align}
    b_\text{max}&=1.56^{+0.13}_{-0.09}\,\text{GeV}^{-1}\,,\\
    c_0&=0.0369^{+0.0061}_{-0.0065}\,,\\
    c_1&=0.0582^{+0.0064}_{-0.0088}\,.
\end{align}

{\bf{Non-perturbative jet function model:}}
For the multiplicative non-perturbative piece in the jet function in the back-to-back factorization formula we use the parameterization given in \Eq{J_match}, with
\begin{align}
    \tau&=0.212^{+0.212}_{-0.106}\,\text{GeV}^{2}\,.
\end{align}

{\bf{Track Functions:}} We take the track function moments extracted from \cite{Chang:2013rca}. Using charge conjugation, we have $T_{\bar{q}}(n,\mu)=T_q(n,\mu)$. The first moments of the track function, which enter the prediction for the EEC (apart from contact terms) are
\begin{align*}
    T_g(1,100\,\text{GeV})&=0.617936\,,\quad 
    T_u(1,100\,\text{GeV})=0.604025\,, \quad
    T_d(1,100\,\text{GeV})=0.624608\,,
    \\
    T_c(1,100\,\text{GeV})&=0.627222\,,\quad
    T_s(1,100\,\text{GeV})=0.622424\,, \quad
    T_b(1,100\,\text{GeV})=0.622983\,.
\end{align*}

{\bf{Perturbative Order Counting:}} In the bulk we use a strict order counting
\begin{align}
    \frac{\df\Sigma}{\df z}= \sum\limits_{L=0} a_s^L \frac{\df\Sigma^{(L)}}{\df z}\,,
\end{align}
where we denote the $a_s^L$ term in this expansion as N$^{L-1}$LO.

In the back-to-back limit, we use a standard Sudakov counting \cite{Almeida:2014uva}. We denote it by  N$^k$LL$_{\text{b2b}}$ to distinguish it from the resummation in the collinear limit. We perturbatively expand the anomalous dimensions and boundary conditions as 
\begin{align}
    \gamma_X(\mu) 
    &=
    \sum_{k=0}^\infty
    \biggl(\frac{\alpha_s(\mu)}{4\pi}\biggr)^{k+1} \gamma_k^X\,,
\end{align}
and
\begin{align}
X(b_\perp,\mu,\nu)&=1+\sum_{n=1}^{\infty}\biggl(\frac{\as(\mu)}{4\pi}\biggr)^n X^{(n)}(b_\perp,\mu,\nu)\,.
\end{align}
The orders required to achieve different resummation accuracies are given in the following table:
\begin{align}
\centering
\begin{tabular}{|c||c|c|c|}
\hline
Order & 
$H,\mathcal{C},S$ &
$\gamma_H,\gamma_J,\gamma_S$ & 
$\gammacusp,\beta$
\\ \hline\hline
NNLL$_{\text{b2b}}$ &
$H^{(1)},S^{(1)},\mathcal{C}^{(1)}$ & 
$\gamma^H_1,\gamma^S_1,\gamma^J_1$ &
$\Gamma_2,\beta_2$
\\
NNNLL$_{\text{b2b}}$ & 
$H^{(2)},S^{(2)},\mathcal{C}^{(2)}$ & 
$\gamma^H_2,\gamma^S_2,\gamma^J_2$ &
$\Gamma_3,\beta_3$ 
\\
NNNNLL$_{\text{b2b}}$ & 
$H^{(3)},S^{(3)},\mathcal{C}^{(3)}$ & 
$\gamma^H_3,\gamma^S_3,\gamma^J_3$ &
$\Gamma_4,\beta_4$ 
\\ \hline 
\end{tabular}
\end{align}
In this paper we achieve N$^4$LL$_{\text{b2b}}$ accuracy. Perturbative expansions of the relevant anomalous dimensions are given in appendix~\ref{sec:ingredients}.

In the collinear limit, we use a standard logarithmic counting used for DGLAP evolution. We denote it as N$^k$LL$_{\text{col}}$ to distinguish it from the resummation in the back-to-back limit. The orders to achieve different resummation accuracies are given in the following table:
\begin{align}
\centering
\begin{tabular}{|c||c|c|c|}
\hline
Order & 
$H,J$ &
$\gamma_H,\gamma_J$ & 
$\beta$
\\ \hline\hline
NLL$_{\text{col}}$ &
$H^{(0)},J^{(0)}$ & 
$\gamma^H_0,\gamma^J_0$ &
$\beta_0$
\\
NLL$_{\text{col}}$ &
$H^{(1)},J^{(1)}$ & 
$\gamma^H_1,\gamma^J_1$ &
$\beta_1$
\\
NNLL$_{\text{col}}$ & 
$H^{(2)},J^{(2)}$ & 
$\gamma^H_2,\gamma^J_2$ &
$\beta_2$ 
\\ \hline 
\end{tabular}
\end{align}
In this paper we achieve N$^2$LL$_{\text{col}}$ accuracy. Perturbative expansions  of the relevant anomalous dimensions are given in appendix~\ref{sec:ingredients}.

Therefore, to summarize, our best prediction in this paper is at
NNNNLL$_{\text{b2b}}$+NNLL$_{\text{col}}$ +NNLO. This is the state-of-the-art for any event shape observable, and we are able to achieve it both for the energy correlator computed on all hadrons, as well as for the energy correlator computed on tracks.

\subsection{Perturbative Convergence}\label{sec:convergence}

\begin{figure}
\centering
\includegraphics[width=0.65\linewidth]{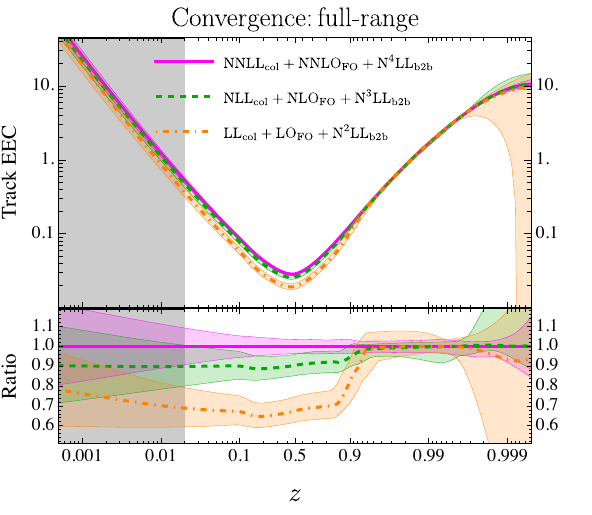}
\caption{The perturbative convergence of the track-EEC at $Q=m_Z$ in a log-log plot. In this plot we have not included the transitions to the plateau regions, which is why we have shaded out the regime in the collinear region. The transition regions, are not described by perturbation theory, and their uncertainties and stability should be considered separately.}
\label{fig:convergence}
\end{figure}

In \Fig{fig:convergence} we show linear-log and log-log plots of the EEC on tracks computed at increasing orders in perturbation theory. Since our goal here is to emphasize the perturbative convergence of our result, we have not incorporated the transitions into the collinear and back-to-back plateau regions in this plot, as these are not described by perturbation theory

We observe large perturbative corrections in the bulk region of the distribution and collinear regions of the distribution, which are highly correlated due to the weak nature of the resummation in the collinear limit. Good convergence is observed in the back-to-back region. It is important to emphasize that we have not implemented a renormalon subtraction. We expect that this would significantly improve the convergence, as has been illustrated at lower orders in 
\cite{Schindler:2023cww}. The perturbative behavior in the collinear limit also motivates pushing to higher perturbative orders.

\subsection{Breakdown of Theoretical Uncertainties}\label{sec:sources}

\begin{figure}
\centering
\includegraphics[width=0.85\textwidth]{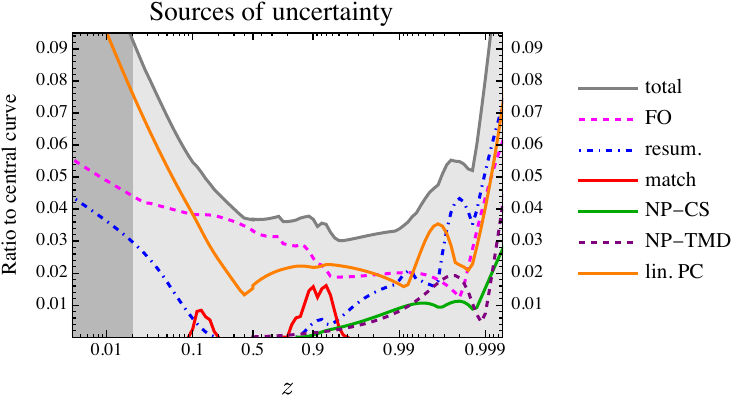}
\caption{Contributions of different sources of uncertainty to our highest precision calculation, $\text{NNLO}_{\text{FO}}+\text{NNLL}_\text{col}+\text{NNNNLL}_\text{b2b}$, at $Q=91.2$ GeV. The total uncertainty is obtained by taking the quadrature sum of the uncertainties for the different sources.}
\label{fig:uncertainties_sources}
\end{figure}

\begin{figure}
    \centering
    \includegraphics[width=\textwidth]{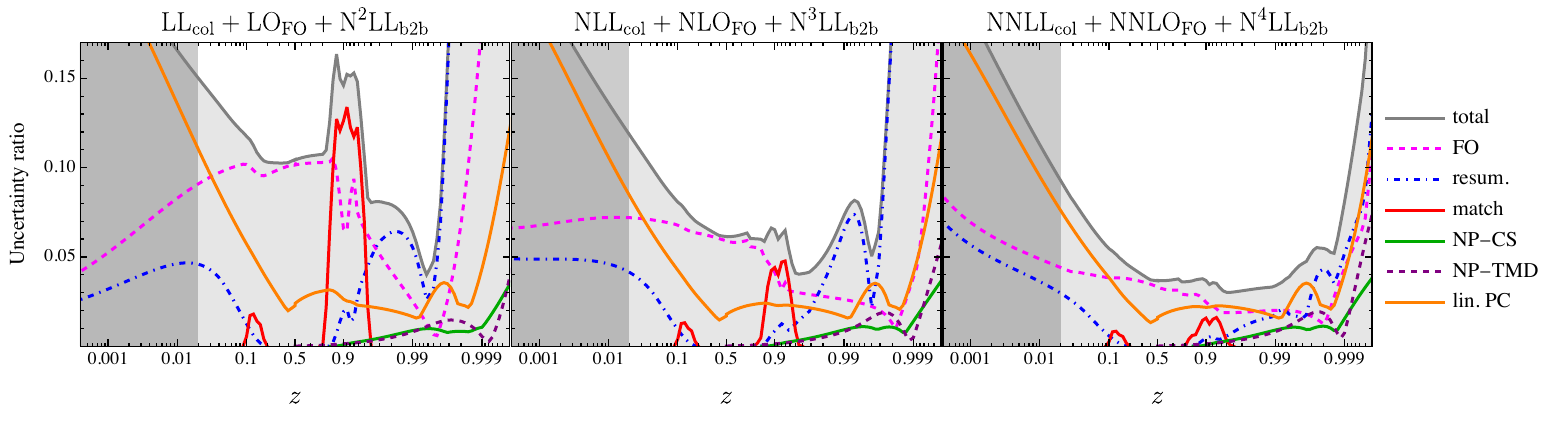}
    \caption{Contributions of different sources of uncertainty to the total theory uncertainty at $\text{LL}_\text{col}+\text{NNLL}_\text{b2b}$ (left), $\text{NLL}_\text{col}+\text{NNNLL}_\text{b2b}$ (middle) and $\text{NNLL}_\text{col}+\text{NNNNLL}_\text{b2b}$ (right). The total uncertainty is obtained by taking the quadrature sum of the uncertainties for the different sources.}
    \label{fig:uncertainties_convergence}
\end{figure}

\begin{figure}
\centering
\includegraphics[width=\textwidth]{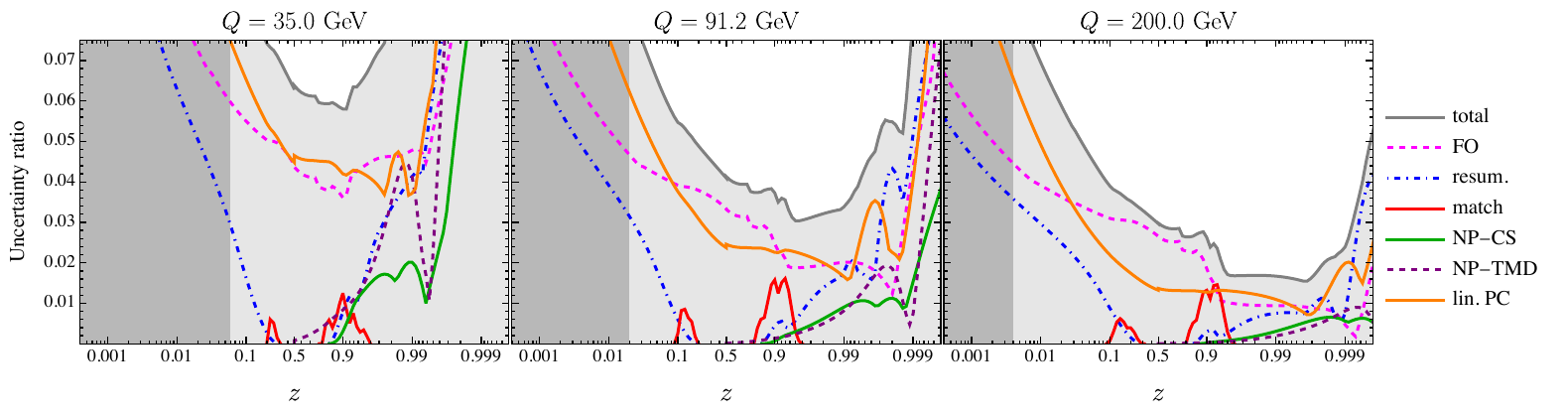}
\caption{Contributions of different sources of uncertainty to the total theory uncertainty to our highest order prediction, $\text{NNLL}_\text{col}+\text{NNNNLL}_\text{b2b}$, as a function of energy. Here we set $\Omega_{1g}=\Omega_{1q}$. As expected, as we increase the energy, the leading source of uncertainty transitions from being dominated by non-perturbative contributions to perturbative contributions.}
\label{fig:uncertainties_Qdependence}
\end{figure}

We now perform a detailed study of the composition of the uncertainties in our final result, and their dependence on perturbative order, and collision energy.

In \Fig{fig:uncertainties_sources}, we show a breakdown of the theoretical uncertainties for our highest order prediction, $\text{NNLO}_{\text{FO}}+\text{NNLL}_\text{col}+\text{NNNNLL}_\text{b2b}$, at $Q=91.2$ GeV. In the region on the left of the plot, we have shaded out the region where we lose control and transition to the collinear plateau. The uncertainties in this region are of a different nature. That region will also not be used for precision fits of $\alpha_s$, and therefore we do not consider it in detail in this section.

The different sources of uncertainty are as follows
\begin{itemize}
\item FO: Denotes the uncertainty from the variation of $\mu_{\text{FO}}$ contributions. As expected, this is one of the primary uncertainties in the bulk of the distribution, but significantly decreases as we go from LO to NLO to NNLO.
\item res: Denotes the resummation uncertainty, as estimated from scale variations in the collinear and back-to-back factorization theorems. We see that this is under good control by the time we reach $\text{NNLL}_\text{col}+\text{NNNNLL}_\text{b2b}$.
\item trans: Denotes the matching/profiling uncertainties for the transitions between collinear /back-to-back resummation regions, and the non-singular regions. This uncertainty is one of the smallest errors at $\text{NNLL}_\text{col}+\text{NNNNLL}_\text{b2b}$, as expected
\item lin.~PC: Denotes the uncertainties from the linear power corrections, $\Omega_{1q}$, and $\Omega_{1g}$. Importantly, this uncertainty is not just associated with the variation of the values of these parameters, but also with the variation in the scales of the factorization theorems in the collinear and back-to-back limit where they appear. 
\item NP-CS: Denotes the uncertainty from the non-perturbative parameterization of the Collins-Soper kernel.
\item NP-TMD: Denotes the uncertainty from the non-perturbative parameterization of TMD fragmentation in the back-to-back limit.
\end{itemize}

Overall, we see that for our highest order prediction, $\text{NNLO}+\text{NNLL}_\text{col}+\text{NNNNLL}_\text{b2b}$ we achieve quite good precision, namely a few percent, throughout the bulk of the distribution. 

There are a number of important conclusions that we can draw from this analysis that guide how we can improve our description of the EEC. From \Fig{fig:uncertainties_convergence}, we see that at $Q=91.2$ GeV, our theoretical uncertainties are minimized around $z\sim 0.95$, where we achieve an uncertainty of about three percent. This value of $z$ corresponds to the perturbative regime of the back-to-back resummation. This is consistent with the order of the theoretical ingredients in our calculation: the back-to-back limit of the energy correlator probes the $J\to \infty$ limit of the twist-2 anomalous dimensions, which are known to higher order than the values at low $J$, as probed in the collinear limit. For $\alpha_s$ fits, it will be important to expand the region in $z$ where we achieve this level of precision, both to smaller values of $z$ (i.e. towards the collinear limit), and to larger values of $z$. To achieve this, we can see that the driving sources of uncertainty that we need to improve are the linear power correction and fixed order perturbative contributions in the collinear limit, and the linear power corrections in the back-to-back limit. We discuss these each in turn.

The leading uncertainty in the back-to-back limit arises from the linear power correction. We re-emphasize that this is not dominated by the value of the $\Omega$ parameter, but rather by scale variations of the Sudakov that dresses it.  As emphasized earlier, due to its strong power law growth, $1/(1-z)^{3/2}$, the contributions from the linear power correction are extremely sensitive to the form of the Sudakov, necessitating its calculation to higher perturbative orders. It will be essential to improve the understanding of the resummation associated with this linear power correction to improve the uncertainty in this limit.

In the bulk ($z\sim 1/2$) region of the distribution, and moving into the collinear limit, we see from \Fig{fig:uncertainties_convergence}, that the fixed order perturbative contribution provide a large uncertainty. This is expected in the bulk of the distribution, but also in the collinear limit, due to the weak single logarithmic behavior that arises there. In the absence of N$^3$LO perturbative corrections for the EEC, one way of improving the perturbative calculation in the collinear limit, will be to extend the perturbative order of the resummation in the collinear limit. The three-loop single-inclusive hard function \cite{He:2025hin} has recently been calculated, and there has been significant progress in the calculation of the four-loop splitting functions \cite{Ruijl:2016pkm,Moch:2017uml,Herzog:2018kwj,Moch:2018wjh,Falcioni:2023luc,Falcioni:2023vqq,Falcioni:2023tzp,Moch:2023tdj,Falcioni:2024xyt,Gehrmann:2023cqm,Gehrmann:2023iah,Gehrmann:2023ksf}. While this provides only the singular contributions, the results of \cite{Dixon:2019uzg} suggest that these can provide a good approximation to the distribution to relatively large values of $z$, and could help to reduce this dominant uncertainty.

In the collinear limit, the other major uncertainty arises from  the relatively unconstrained value of $\Omega_{1g}$. We have included in our predictions a conservative error for this parameter. We hope that the excellent data from the LEP reanalyses can be used to further constrain this parameter. Additionally, in the collinear limit we only consider the resummation associated with the non-perturbative power corrections at LL, leading to a non-trivial scale variation. It will be important to extend the incorporation of the leading power corrections in the collinear limit beyond the leading logarithmic order.

The theoretical uncertainties from other ingredients of our calculation, such as those associated with the Collins-Soper kernel and matching are much smaller. To further understand these uncertainties, in \Fig{fig:uncertainties_convergence} we show a breakdown of the theoretical uncertainties for our calculations at the three different orders: $\text{LL}_\text{col}+\text{NNLL}_\text{b2b}$ (left), $\text{NLL}_\text{col}+\text{NNNLL}_\text{b2b}$ (middle) and $\text{NNLL}_\text{col}+\text{NNNNLL}_\text{b2b}$ (right). This nicely illustrates the large reduction in uncertainty achieved using the high order perturbative ingredients in this paper. At lower perturbative oders, one is largely dominated by perturbative uncertainties. Due to the efforts in this paper, we have pushed these down to the level of the non-perturbative uncertainties. 

In \Fig{fig:uncertainties_Qdependence} we further show the dependence of the uncertainties on the collision energy. These plots largely illustrate the expected features: as we go to higher $Q$ the linear power corrections are suppressed, allowing us to achieve a 1 percent uncertainty at 200 GeV. These higher energies significantly improve the behavior of the energy correlator in the back-to-back limit, further emphasizing that the uncertainties in the back-to-back limit are dominated by non-perturbative effects. Additionally, they improve the region over which the perturbative uncertainties dominate in the collinear limit. The energy dependence of the back-to-back limit of the energy correlator was studied by DELPHI in \cite{DELPHI:2003yqh}, up to a maximum energy of 202 GeV. Based on our results, and the exciting re-analysis of archival DELPHI data \cite{DELPHI:2003yqh}, it would be interesting to perform precision measurements above the Z-pole.

In this paper we have assessed our theoretical uncertainties using scale variations. It would be interesting to perform a complementary analysis using the approach of theory nuisance parameters \cite{Tackmann:2024kci,Cridge:2025wwo}. We believe that the energy correlator is ideal for such an analysis due to the fact that both the collinear and back-to-back limits are controlled by well understood anomalous dimensions. This approach would also be particularly fruitful in incorporating correlations. We look forward to considering this approach in future studies.

\subsection{$\alpha_s$ and $\Omega_1$ Variations}\label{sec:compare_as}

\begin{figure}
    \centering
    \includegraphics[width=\textwidth]{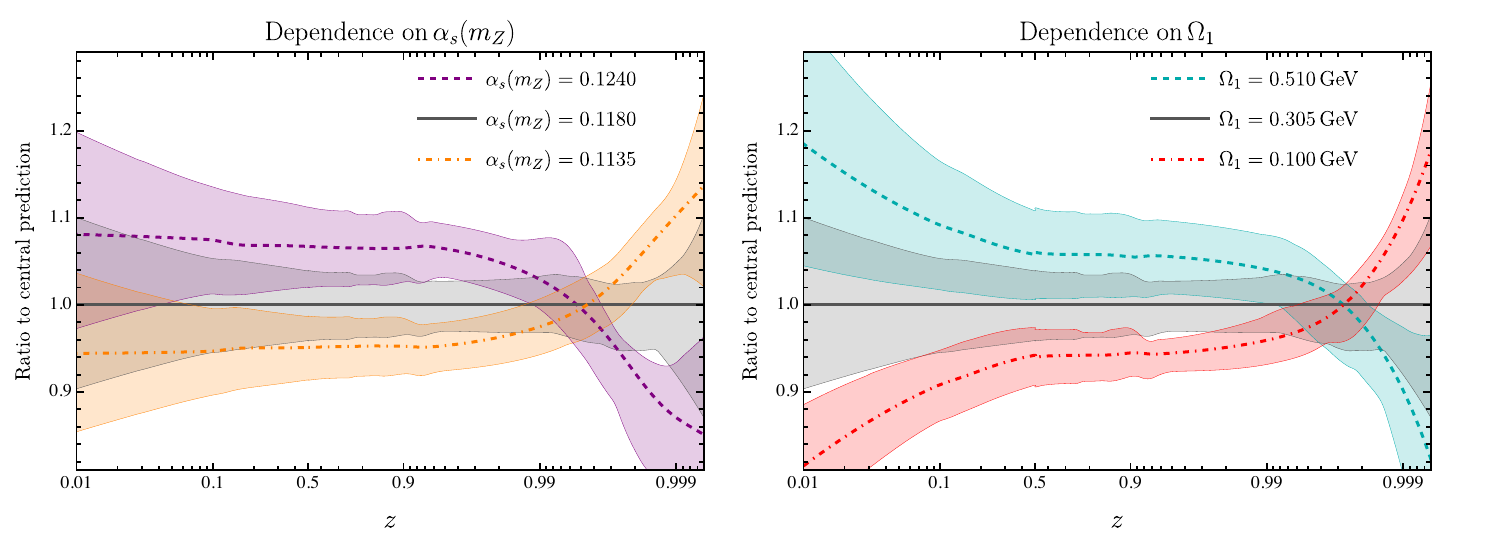}
    \caption{Three curves at $\text{NNLL}_\text{col}+\text{NNNNLL}_\text{b2b}$ precision with $Q=m_Z$ showing the dependence of the track-EEC on $\alpha_s(m_Z)$ (left) and $\Omega_{1q}$ (right), where $\Omega_{1g}$ is set to $\Omega_{1q}$. All curves are normalized to the central $\alpha_s(m_Z)=0.118$ curve. The different shape dependence of the $\Omega_{1q}$ and $\alpha_s$ variations in the back-to-back and collinear limit should aid in breaking their degeneracy.}
    \label{fig:as_dependence}
\end{figure}

Having achieved a precision calculation of the EEC on tracks, and given that it has now been measured using archical data from both ALEPH \cite{Bossi:2025xsi} and DELPHI \cite{Zhang:2025nlf}, it is interesting to evaluate the possibility of performing a precision extraction of the strong coupling constant from the EEC. To do so, we study the dependence of our results on $\alpha_s$ and $\Omega_1$. For simplicity, in this section we set $\Omega_{1q}=\Omega_{1g}$.  These studies reveal what we believe to be an appealing feature of the energy correlator observable for disentangling degeneracies in $\alpha_s$ and $\Omega_1$, motivating a program to perform a precision extraction of the strong coupling constant using the track-based EEC. Additionally, the fact that we are sensitive to both $\Omega_{1q}$ and $\Omega_{1g}$ makes the energy correlator a promising observable for the extraction of $\Omega_{1g}$.

 In the left panel of fig.~\ref{fig:as_dependence} we show the dependence of our highest precision results for the track EEC on the value of $\alpha_s$, by varying it up and down by $5\%$. In the right panel we similarly show the dependence on the non-perturbative parameter, $\Omega_1$. In precision fits of $\alpha_s$ from event shapes, these are typically fit simultaneously \cite{Abbate:2010xh,Hoang:2015hka}. In the Sudakov regime of thrust and C-parameter, where precision fits have been achieved, these two parameters are highly degenerate. This degeneracy is broken by comparisons to data at different values of $Q$.

 If we focus on the $z\to 1$ limit of the EEC, we see that this is also the case for the EEC, as expected. For $z\gtrsim 0.5$ there is a strong degeneracy between $\alpha_s$ and $\Omega_1$. However, a beautiful feature of the energy correlator is that one also has the collinear limit, which has a completely different resummation structure, but is still controlled by the same two-parameters. Due to the single logarithmic resummation in the collinear limit, the effect of $\alpha_s$ variations is much milder, amounting to an approximate $z$-independent shift, similar to the behavior in the central region of the energy correlator. However, the dependence on $\Omega_1$ in the collinear limit is more drastic. This is due to the fact that the non-perturbative power correction scales like $1/z^{3/2}$, as compared to the perturbative prediction, which scales like $1/z$. Therefore we see that in the collinear limit the variation in $\alpha_s$ and $\Omega_1$ are highly non-degenerate. We find this particularly appealing for performing fits for $\alpha_s$.

 Another interesting distinction between the EEC and thrust/C-parameter arises in the nature of the Sudakov region, which is different for the two-observables. This suggests that it should be possible to use the peak region of the EEC to fit for $\alpha_s$. Thrust and C-parameters are scalar sums, and therefore the peak region of these observables is genuinely non-perturbative. Fits for $\alpha_s$ from these observables are performed in the tail region, which is predominantly fixed order, but stabilized by high order resummation. On the other hand, the EEC has a structure similar to the $p_T$ distribution of color singlet bosons, making the peak region more perturbative. Precision fits of $\alpha_s$ from the $p_T$ spectrum often use the shape dependence in the peak region (see e.g. \cite{ATLAS:2023lhg}). One can see this relationship more quantitatively, since in the back-to-back limit, one can perform an approximate conversion between angle and $p_T$
\begin{align}
p_T \simeq Q \sqrt{1-z}\,.
\end{align}
The strong shape dependence occurs for $p_T\sim 5$ GeV, which agrees well with the case of the Z-boson $p_T$ distribution. Indeed, the dependence of our results on $\alpha_s$ is remarkably similar to that for the Z-boson $p_T$ distribution (see e.g. \cite{Cridge:2025wwo}). 

It is particularly appealing to attempt to fit $\alpha_s$ from the shape dependence of the EEC, since this largely mitigates uncertainties in the normalization arising from the track functions. In this region of the distribution, the use of tracks to achieve a precise angular resolution is particularly important. We observe that we achieve quite good theoretical control into the peak region, where there is a strong shape dependence. To further improve this, it will be important to improve the resummation order associated with the $\Omega_1$ power correction. Note that the presence of a linear power correction is the primary difference in the structure of the back-to-back limit of the EEC, as compared to $p_T$.

\begin{figure}
    \centering
    \includegraphics[width=0.65\textwidth]{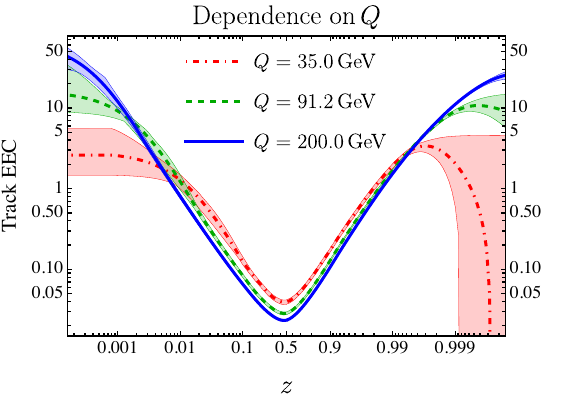}
    \caption{Our highest order prediction for the track EEC at three different COM energies. Here we have chosen $\Omega_{1g}=\Omega_{1q}$. Note that the region where the EEC is under perturbative control increases with increasing energies. }
    \label{fig:energy_dep}
\end{figure}

Beyond the case of fits for $\alpha_s$ using just the EEC, we find the relationship of the back-to-back limit of the EEC in $e^+e^-$ and the $Z$ boson $p_T$ spectrum particularly appealing for joint fits. Indeed, the factorization theorems for the two observables can be thought of as a form of ``crossing" of each other, as was emphasized in \cite{Moult:2018jzp}. The $Z$ $p_T$ spectrum is one of the most precisely measured observables at the LHC \cite{ATLAS:2014alx, ATLAS:2015iiu,ATLAS:2019zci, CMS:2011wyd, CMS:2016mwa, CMS:2019raw, LHCb:2015mad, LHCb:2016fbk}, and has attracted significant theoretical attention \cite{Billis:2021ecs, Ju:2021lah, Re:2021con, Chen:2022cgv, Neumann:2022lft, Camarda:2023dqn, Moos:2023yfa, Billis:2024dqq}, in particular enabling a precision extraction of the strong coupling constant \cite{ATLAS:2023lhg}\footnote{See \cite{Cridge:2025wwo} for a detailed study using theory nuisance parameters \cite{Tackmann:2024kci}.}. The fact that we now have extremely precise data for the Sudakov region of the energy correlator in $e^+e^-$ opens up the opportunity for a joint study of these two observables. An improved understanding of this physics also improves our theoretical understanding of the Higgs $q_T$ spectrum, whose experimental precision is rapidly improving  
\cite{CMS:2025ihj}, and which provides interesting constraints on Higgs interactions, such as light quark Yukawas \cite{Bishara:2016jga,Soreq:2016rae}.

We also want to comment on our conclusions, compared to a recent study \cite{Cuerpo:2025zde}, which concluded ``contrary to previous claims, we demonstrate that the \emph{current data} do not provide meaningful constraints on either the Collins-Soper kernel or $\alpha_s$" (the emphasis is ours). We emphasize two major developments since the publication of \cite{Cuerpo:2025zde}. First, is the availability of new high quality data \cite{Bossi:2025xsi,Zhang:2025nlf}, and second is the availability of calculations over the entire range of the energy correlator, to break degeneracies in $\alpha_s$ and $\Omega_1$. Ref.~\cite{Cuerpo:2025zde} only performed calculations in the back-to-back limit. With these developments, we believe that the energy correlator is now the best measured, and best theoretically understood QCD event shape, and is a promising direction for extractions of $\alpha_s$.

\subsection{Energy Dependence}\label{sec:resultsQs}

While our primary focus is on $Q=m_Z$, there is also LEP2 data at $Q=189-209$ GeV, as well as lower energies. To motivate renewed attention to these datasets, we study our calculation at three values, namely $Q = 35, 200$ GeV and $m_Z$. In \Fig{fig:energy_dep} we show the results of our highest precision predictions at $Q = 35, 200$ GeV and $m_Z$, as well as the ratio of the results at $Q = 35$ and $Q = 200$ GeV to the $m_Z$ result. In the bulk of the distribution, the impact of the $Q$ variation modifies the value of $\alpha_s$, and has a relatively $z$ independent effect on the distribution. The higher values of $Q$ extend the region of validity of perturbation theory, and therefore have a large impact on the behavior of the distribution in the collinear and back-to-back limits.  We also see that the uncertainty is significantly reduced at $Q=200$ GeV. The highest energy measurement of the back-to-back limit of the EEC that we are aware of is  at 202 GeV \cite{DELPHI:2003yqh}. The beautiful convergence of our results at 200 GeV strongly motivate measurements of the track-based EEC at this energy.

\subsection{Comparing Hadronic and Track-Based Calculations}\label{sec:compare}

\begin{figure}
    \centering
    \includegraphics[width=0.5\textwidth]{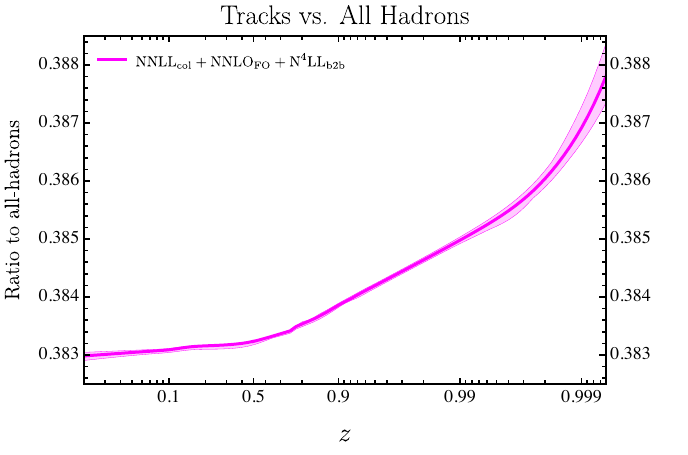}
    \caption{The ratio of the track-based two-point energy correlator with the all hadron energy correlator. The dependence on the use of tracks is extremely minimal particularly in the bulk of the distributions. Therefore, while the use of tracks is essential experimentally to measure the energy correlator in the asymptotic limits, it does not modify the underlying physics.}
    \label{fig:charged_all_ratio}
\end{figure}

We started this paper with a discussion of the physics of the energy correlator, and emphasizing that the change to performing measurements on tracks does not alter this physics. Measurements of energy correlators on tracks should therefore provide an ideal meeting between theoretical elegance and experimental realizability. Having performed a complete calculation of the EEC distribution on tracks at high perturbative orders, we are now in a position to justify this claim.

Since we performed the calculation using generic track functions, we can easily obtain the all-hadron result by setting the track functions to $T(x)=\delta(1-x)$.  In \Fig{fig:charged_all_ratio} we show the ratio between the track-based and all-hadron calculation for our highest order perturbative prediction. We see that the effect of measuring the observable on tracks is extremely small, at a fraction of a percent. This is positive, since it shows that we can take advantage of the exceptional resolution of track-based detectors, but that the measurement of the track-based EEC still isolates the same interesting physical effects as the all-hadron EEC. Going forward, it will also be interesting to perform a new measurement of the energy correlator on all-hadrons using archival ALEPH data.

\subsection{Partial Waves}\label{sec:partialwaves}

In phenomenological studies of the energy correlator, the primary focus has been on the distribution, $\text{EEC}(z)$. Much like scattering amplitudes, the energy correlators can also be decomposed in partial waves. For the case of the two-point energy correlator in a scalar source, in $d=4$, we have
\begin{align}
\text{EEC}(z)=\sum_J  \text{EEC}(J) P_J(1-2z)\,.
\end{align}
Here $P_J$ is a Legendre polynomial, and $\text{EEC}(J)$ is given explicitly by
\begin{align}
\text{EEC}(J)=\int\limits_0^1 \df z\, \text{EEC}(z)P_J(1-2z)\,. 
\end{align}
The partial wave coefficients exhibit nice positivity and boundedness conditions, namely in a unitary theory they satisfy
\begin{align}
0\leq \text{EEC}(J) \leq 1\,.
\end{align}
Such bounds were originally proven in \cite{Fox:1978vw}. The sum rules discussed in \Sec{sum} are examples of this with $J=0$ and $J=1$, where the bound is saturated. In a free theory, the bounds are saturated at $1$ for the even spin partial waves, and $0$ for the odd spin partial waves.

As compared to the full distribution in $z$, specific partial waves, particularly for low values of $J$ can be easier to compute using the numerical conformal bootstrap. Since part of our motivation for studying the energy correlators is to have an observable that can be accessed both in real experiments, as well as in simplified theories, we are motivated to also use our results to compute the partial waves of the energy correlator.\footnote{We thank Silviu Pufu, Ross Dempsey, Zahra Zahraee and Sasha Zhiboedov for discussions motivating us to also present the partial waves of the EEC.} To our knowledge, the spectrum of partial waves has not previously been experimentally measured, nor computed using modern techniques. For higher partial waves, it requires high angular resolution data, not available before \cite{Electron-PositronAlliance:2025fhk}. Additionally, on the theory side, it requires an understanding of both the collinear and back-to-back limits of the energy correlator, a complete calculation of which has been presented for QCD for the first time in this paper. 

\begin{figure}
    \centering
    \includegraphics[width=0.5\textwidth]{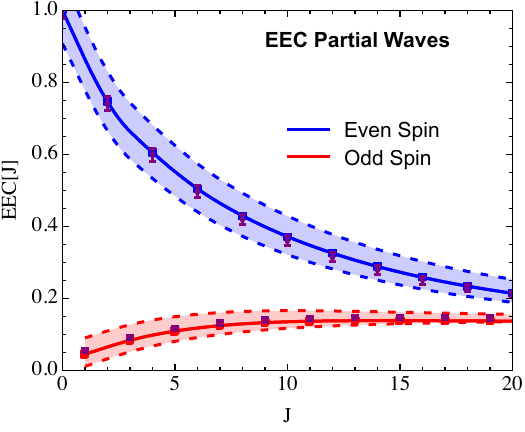}
    \caption{The even and odd spin partial waves for the energy correlator computed using our $\text{NNLO}+\text{NNLL}_\text{col}+\text{NNNNLL}_\text{b2b}$ predictions. Theoretical predictions are shown in blue and red. Partial waves computed from the data in \cite{Electron-PositronAlliance:2025fhk} are shown in purple, using the same conservative approach to computing the statistical+ systematic uncertainties. }
    \label{fig:partial_waves}
\end{figure}

Since our calculation, as well as the measurement of \cite{Electron-PositronAlliance:2025fhk} was performed on tracks, we lose the simple normalization of the partial waves. However, as shown in \Sec{compare} the use of tracks has an extremely small effect on the shape of the energy correlator distribution. We are therefore motivated to simply rescale the zeroth moment to unity, and extract the spectrum of partial waves. This is an approximation, but we expect that it is an extremely good one, and it enables us to use the precision data of \cite{Electron-PositronAlliance:2025fhk} to explore the partial waves.

In \Fig{fig:partial_waves} we show the first twenty partial waves for both the even and odd spin branches. Here we have made a crude and overly conservative estimate of the uncertainty, by evaluating the partial waves on the upper and lower boundaries of our theoretical predictions. Due to the complicated functional form of the Legendre polynomials, particularly for high spin, this should not be taken too seriously, but provides an exploratory look at the partial waves. The partial waves as computed using the data from \cite{Electron-PositronAlliance:2025fhk} are shown in purple. Good agreement between theory and data is observed. We hope to see more exploration of the EEC partial waves in QCD phenomenology in the future, as they may prove to be a nice target for the intersection of theoretical studies in simplified theories, and real world phenomenology.

\section{Opportunities for Improvement}\label{sec:improve}

The calculation presented in this paper is the first high precision calculation of the energy correlator over the entire angular region.  Although it incorporates a wealth of ingredients in QCD, there are still many directions in which it can be improved, which will be important to further improve the precision of the calculation. While some of these require new calculations, some simply require the incorporation of known ingredients. Here we summarize a number of these directions to motivate further work in these directions.

\medskip

\noindent {\bf{Perturbative Accuracy of Collinear Resummation:}} From our results we find that the convergence of the resummed results in collinear limit is not great, and requires higher-order resummation. This requires the calculation of the three-loop collinear jet function, the three-loop inclusive hard function, and four-loop timelike DGLAP. Unfortunately, currently none of these are known. However, there has been important recent progress:  \cite{Xu:2024rbt} achieved high order threshold resummation for the single inclusive hard function. There has also been a sustained effort to compute the DGLAP anomalous dimensions at four loops \cite{Ruijl:2016pkm,Moch:2017uml,Herzog:2018kwj,Moch:2018wjh}. This is being pursued both from the calculation of the partonic cross section \cite{Falcioni:2023luc,Falcioni:2023vqq,Falcioni:2023tzp,Moch:2023tdj,Falcioni:2024xyt}, as well as the direct calculation of the twist-2 matrix elements \cite{Gehrmann:2023cqm,Gehrmann:2023iah,Gehrmann:2023ksf}, and results for many moments, and phenomenological approximations are known. These results will then need to be crossed to the timelike anomalous dimensions for the energy correlator

\medskip

\noindent {\bf{Extraction of $\Omega_{1g}$:}} An interesting feature of the single logarithmic nature of the collinear resummation, is that the parameter $\Omega_{1g}$ appears already at leading logarithmic order. This parameter is poorly constrained, and we have pointed out that at the level of precision required to match the recent analysis of LEP data, it is important to take the mixing between $\Omega_{1q}$ and $\Omega_{1g}$ into account. However, this represents an opportunity to directly extract the parameter $\Omega_{1g}$ from data. It will be interesting to attempt this using LEP data, or to extract it from precision measurements of the energy correlators at the LHC.

\medskip

\noindent {\bf{Quark Mass Effects:}} In this paper we have performed our calculation using strictly massless quarks. This is not a good approximation for the $b$-quark, and it becomes important in the collinear and back-to-back regions when the scales $Q^2(1-z)$ or $Q^2 z$ become comparible to $m_b^2$. Resummed calculations for energy correlators including heavy quark effects exist both in the collinear \cite{Craft:2022kdo} and back-to-back limit \cite{Aglietti:2024zhg}, and it will be important to include these effects in future calculations. Many of the required ingredients, and the formalism exists, and has been studied in the context of the $Z$-boson $p_T$ distribution 
\cite{Pietrulewicz:2017gxc,Cal:2023mib,vonKuk:2024uxe}. 
\medskip

\noindent {\bf{Perturbative Power Corrections:}} To improve the uncertainties due to the matching between the fixed-order and resummation regions, it would be useful to have more information about the structure of perturbative power corrections, and the resummation of power-suppressed terms in both the back-to-back and collinear limits. In particular, in the back-to-back limit, since the resummation is strong enough to change the naive leading power scaling from $1/(1-z)$ to a flat plateau, one may worry that subleading power corrections could play an important role.  Due to the elegant theoretical properties of the energy correlators, we expect that this should be comparatively simple. Indeed some studies of power corrections in the back-to-back limit have been performed using SCET \cite{Moult:2019vou} or high-spin perturbation theory 
\cite{Chen:2023wah}.

\medskip

\noindent {\bf{Renormalon Subtractions:}} Although we have incorporated the leading non-perturbative correction, it is well known to suffer from a renormalon ambiguity. This can be remedied by using a short distance scheme, such as the R-scheme~\cite{Hoang:2007vb,Hoang:2008fs,Hoang:2009yr,Hoang:2017suc,Bachu:2020nqn}. This has been explored in the bulk and collinear limit of the EEC in refs.~\cite{Schindler:2023cww,Lee:2024esz}, and all the required ingredients were summarized in this paper. We will add this in a future calculation.

\medskip

\noindent {\bf{Improved Extraction of Collins-Soper Kernel:}} One exciting feature of our calculation of the energy correlators is that our back-to-back factorization theorem allows us to incorporate lattice data on the Collins-Soper Kernel. Further constraints on this kernel either from improved lattice calculations, or measurements, will therefore help to improve the description in the back-to-back limit. Alternatively, one can use the EEC to extract the Collins-Soper kernel, as was recently explored in ref.~\cite{Kang:2024dja}. In this paper we have not properly taken into account the effect of massive quarks on the Collins-Soper kernel. It will be interesting to consider this in more detail, particularly in light of the extremely precise data in the back-to-back region.

\medskip

\noindent {\bf{Improved Treatment of $\Omega_1$ in the Back-to-Back Limit:}} From our detailed study of the theoretical uncertainties in \Sec{sources}, the leading uncertainty in the back-to-back region, where there is a strong shape dependence on $\alpha_s$, arises from the interplay of the resummation with the linear power correction. The presence of this linear power correction is a primary difference between the energy correlator in the back-to-back limit, and the $Z$ $p_T$ spectrum. In this paper we have considered the resummation associated with these linear power corrections only at NLL. It will be important to extended this to NNLL by computing the perturbative matching coefficients. We are optimistic that incredibly precise data in the back-to-back limit will be helpful in this respect.

\medskip

\noindent {\bf{Improved Treatment of $\Omega_1$ in the Collinear Limit:}} In the collinear limit we have only considered the resummation of logarithms associated with the non-perturbative parameters $\Omega_{1q}$ and $\Omega_{1g}$ at LL order. It will be important to extend this beyond the LL order by computing the perturbative matching coefficients appearing in the factorization theorem.

\medskip

\noindent {\bf{Hadron Mass Effects:}} In our treatment of non-perturbative corrections, we have neglected hadron mass effects. These introduce additional non-perturbative parameters \cite{Mateu:2012nk}, which break universality. In measurements of a single observable, these should be absorbable into $\Omega_1$, but it would be interesting to treat these more carefully. Detailed studies in the case of thrust and C-parameter have been performed in 
\cite{Hoang:2014wka,Hoang:2015hka,Mateu:2012nk,Benitez:2024nav}.

\medskip

\noindent {\bf{QED Effects:}} In this paper we have focused only on QCD. However, for precision calculations one should also incorporate QED corrections in the final state. Such 
corrections play a non-trivial role in extractions of $\alpha_s$ \cite{Abbate:2010xh}, and are sometimes corrected for using old Monte Carlo generators. One potentially interesting and complementary feature of measurements of the EEC on tracks is the different treatment of photons in the measurement. It would be interesting to properly incorporate QED effects into our calculations using a track function for the photon $T_\gamma(x) $, which is perturbatively calculable and satisfies the initial condition $T_\gamma(x)=\delta(x)$. This would allow us to  account for QED effects throughout the entire distribution, including resummation regions, which would be interesting to consider in detail.

\medskip

\noindent {\bf{Sum Rules:}} As discussed in \Sec{overview}, the energy correlator satisfies a non-perturbative sum rule relating its integral to the total cross section. This is intriguing since the leading nonperturbative correction to the total cross section starts at $(\Lambda/Q)^4$. While the role of the sum rules have been explored in perturbation theory, and allow one to relate information about the bulk region with the endpoints, they have not been explored non-perturbatively. It would be particularly interesting to investigate how they can be used to constrain the full EEC. 

\medskip

\noindent {\bf{Independent Measurement of Contact Terms and Track Functions:}} The additional use of tracks is extremely advantageous for the angular resolution of the energy correlator measurement. However it introduces additional non-perturbative parameters namely the first two moments of the track functions. While these have been measured in ATLAS \cite{ATLAS:2024jrp}, it would be nice to measure them directly in $e^+e^-$. Alternatively, instead of the track function moments one could directly compute the one-point functions
\begin{align}
\text{E}^N\text{C}_{\text{tr}}(z)=\int \df ^4x\, e^{\img q\cdot x} \langle0| J(x) \cE_{\text{tr}}^N(\hat n_1) J(0)|0 \rangle \,,
\end{align}
which appear in the collinear limit of the EEC, as well as the non-perturbative parameter, $\langle  \left(  \sum\limits E_i \right)^2  \rangle$, which appears in the sum rule for the track-based energy correlator in eq.~\eqref{eq:sum_tracks}. These can both be computed in terms of the track functions, but having an independent measurement would be extremely useful, and enable the use of the sum rules.

\medskip

\noindent {\bf{Improved Treatment of the Collinear and Back-to-Back Transitions:}} It will be important to improve constraints on the transition regions into the free hadron scaling for both the back-to-back and collinear regions. While these are less important for precision $\alpha_s$ fits they provide insight into interesting physics. There has been recent progress in understanding the collinear transition \cite{Chang:2025kgq,Herrmann:2025fqy,Kang:2025zto,Lee:2025okn}. However, in both limits, non-perturbative functions (as opposed to parameters) are still needed. It would be appealing to be able to bootstrap these transitions using sum-rule constraints to provide model independent bounds.

\section{Conclusions}\label{sec:conclusion}

The study of energy flux in $e^+e^-$ collisions has a remarkable history, predating QCD itself. Due to tremendous efforts, it is now possible to re-analyze archival LEP data, with a modern perspective. Recently, the two-point energy correlator was measured with extremely high angular resolution on tracks using archival LEP data \cite{talk_ALEPH,Bossi:2025xsi,Bossi:2024qeu}. Inspired by these developments, we were motivated to achieve theoretical predictions with uncertainties matching those achieved experimentally.

In this paper we presented state of the art predictions for the energy correlator computed on tracks throughout the entire kinematic region. We achieve a record precision of $\text{NNLO}+\text{NNLL}_\text{col}+\text{NNNNLL}_\text{b2b}$, combined with the incorporation of leading non-perturbative corrections and their evolution. This is the state of the art for any event shape observable, but now we have extended this calculation to tracks. To achieve this precision, we have combined factorization theorems describing different kinematic limits, with non-perturbative inputs from the lattice, and high-loop perturbative ingredients. The precision of our results, at the level of a few percent, highlight the remarkable progress in perturbative QFT, and effective field theory techniques in the last decade.  We also provided a detailed analysis of the uncertainties in our calculation, and highlighted a number of ways in which they can be improved.

A key motivation for the precision calculation of the EEC is a precision extraction of the strong coupling constant $\alpha_s$. As mentioned in the introduction, there is currently a discrepancy between extractions of $\alpha_s$ from precision event shapes, and from the lattice. We are optimistic that our precision calculations, combined with archival measurements of the energy correlator, might prove useful in resolving this tension. While we did not perform a fit for the value of $\alpha_s$ in this paper, we studied the structure of the variation in $\alpha_s$ of our predictions. An interesting feature of the energy correlator is the different dependence of the collinear and back-to-back limit on the parameters $\alpha_s$ and $\Omega_1$. We believe that this will be particularly useful for breaking the degeneracy between these parameters. Additionally, we highlighted a close similarity between the energy correlators in the back-to-back limit and the transverse momentum of the $Z$ boson. The $Z$ $p_T$  distribution and the EEC distribution on tracks are now two of the most precise measurements of kinematic distribution sensitive to QCD. We believe that this relation presents an interesting opportunity to perform a simultaneous fit, and combine archival data, with modern LHC measurements. 

Beyond the case of precision extractions of QCD parameters, we believe that the EEC will be particularly interesting for improving the understanding of non-perturbative phenomena in QCD. We highlighted the breadth of physics probed by the energy correlator, including confining transitions and the physics of flux tubes. We will soon be in a position where we have for the first time both precision measurements of the correlator in QCD, combined with non-perturbative calculations of the EEC in closely related theories, such as planar $\mathcal{N}=4$ SYM \cite{N4_bootstrap}. This is a unique opportunity to study the same observable in these different theories, which we anticipate will lead to significant insight, for example, in the physics of confining vs. conformal flux tubes.

While we have focused on the two-point energy correlator, there are numerous  variants of the energy correlators that can also be measured on tracks at ALEPH. Examples include the three-point correlator, which has recently been computed for the first time \cite{Yan:2022cye,Yang:2022tgm}, or correlators incorporating angular dependencies \cite{Kang:2023big}. Thanks to the power of factorization, many of the techniques and perturbative ingredients introduced in this paper will be useful in this broader context.

The techniques developed in this paper to achieve high precision calculations of energy correlators are also important for precision QCD studies at future lepton colliders. At higher energies, the resummation (both Sudakov and collinear) become concentrated in increasingly small angular regions. Precision measurements of $\alpha_s$ at these higher energy colliders therefore necessarily require higher angular resolution, which is naturally provided by tracks. For a study of energy correlators at future $e^+e^-$ colliders, see \cite{Lin:2024lsj}.

The tremendous theoretical progress in the past decade allows us to look at QCD in a new light, and the re-analysis of archival ALEPH and DELPHI data provides a playground to confront theory with data to improve our understanding of QCD.

\acknowledgments

We thank Hannah Bossi, Yu-Chen Chen, Yi Chen, Yenjie Lee, Jingyu Zhang, for extensive discussions of the energy correlator re-analysis on ALEPH data \cite{Bossi:2025xsi,Bossi:2024qeu}. We additionally thank Jingyu Zhang for help with several figures used in this paper.  We thank Hao Chen, Lance Dixon,  Andre Hoang, Matt LeBlanc, Michael Peskin, Jennifer Roloff, Frank Tackmann, Gherardo Vita, Matt Walters, Zhen Xu,  Zahra Zahraee, Xiaoyuan Zhang, Alexander Zhiboedov for useful discussions. I.M. is supported by the DOE Early Career Award DE-SC0025581, and the Sloan Foundation. H.X.Z. is supported by the National Science Foundation of China under contract No.~12425505 and the Asian Young Scientist Fellowship.  M.J. and Y.L. are supported by the funding from the European Research Council (ERC) under the European Union’s Horizon 2022 Research and Innovation Program (ERC Advanced Grant agreement No.101097780, EFT4jets).  This work was performed in part at the Kavli Institute for Theoretical Physics, which is supported by the National Science Foundation grant PHY-2309135.

\appendix

\section{Perturbative Ingredients}\label{sec:ingredients}

In this appendix we summarize all the perturbative ingredients used in our calculations.

\subsection{QCD $\beta$ Function}

The QCD $\beta$ function is defined as
\begin{align}
    \frac{\df a_s}{\df\ln\mu^2}
    &=-\sum_{k=0}^\infty\beta_k a_s^{k+2}\,,
\end{align}
where $a_s=\alpha_s/(4\pi)$. The coefficients are given by \cite{Gross:1973id,Politzer:1973fx,Jones:1974mm,Caswell:1974gg,Egorian:1978zx,Tarasov:1980au,Larin:1993tp,vanRitbergen:1997va,Czakon:2004bu,Baikov:2016tgj}
\begin{align}
    \beta_0&=\frac{11}{3}\ca-\frac{4}{3}\nf\tf\,,\\
    \beta_1&=\frac{34}{3}\ca^2-\frac{20}{3}\nf\tf\ca-4\nf\tf\cf\,,
    \nonumber\\
    \beta_2&=\frac{2857}{54}\ca^3-\frac{1415}{27}\nf\tf\ca^2-\frac{205}{9}\nf\tf\ca\cf+2\nf\tf\cf^2+\frac{158}{27}\nf^2\tf^2\ca+\frac{44}{9}\nf^2\tf^2\cf\,,
    \nonumber\\
    \beta_3&=
    \Bigl(-\frac{44}{9}\zeta_3+\frac{150653}{486}\Bigr)\ca^4
    +\Bigl(\frac{136}{3}\zeta_3-\frac{39143}{81}\Bigr)\nf\tf\ca^3
    +\Bigl(-\frac{656}{9}\zeta_3+\frac{7073}{243}\Bigr)\nf\tf\ca^2\cf
    \nonumber\\
    &\quad
    +\Bigl(\frac{352}{9}\zeta_3-\frac{4204}{27}\Bigr)\nf\tf\ca\cf^2
    +46\nf\tf\cf^3
    +\Bigl(\frac{224}{9}\zeta_3+\frac{7930}{81}\Bigr)\nf^2\tf^2\ca^2
    \nonumber\\
    &\quad
    +\Bigl(\frac{448}{9}\zeta_3+\frac{17152}{243}\Bigr)\nf^2\tf^2\ca\cf
    +\Bigl(-\frac{704}{9}\zeta_3+\frac{1352}{27}\Bigr)\nf^2\tf^2\cf^2
    +\frac{424}{243}\nf^3\tf^3\ca
    \nonumber\\
    &\quad
    +\frac{1232}{243}\nf^3\tf^3\cf
    +\Bigl(\frac{704}{3}\zeta_3-\frac{80}{9}\Bigr)\frac{\dabcd_A\dabcd_A}{N_A}
    +\Bigl(-\frac{1664}{3}\zeta_3+\frac{512}{9}\Bigr)\nf\frac{\dabcd_F\dabcd_A}{N_A}
    \nonumber\\
    &\quad
    +\Bigl(\frac{512}{3}\zeta_3-\frac{704}{9}\Bigr)\nf^2\frac{\dabcd_F\dabcd_F}{N_A}\,.
    \nonumber
\end{align}
The color factors appearing in $\beta_3$ can be written in terms of $N_c$ as
\begin{align}
    \frac{\dabcd_A\dabcd_A}{N_A}&=\frac{\nc^2(\nc^2+36)}{24}\,,\\
    \frac{\dabcd_A\dabcd_F}{N_A}&=\frac{\nc(\nc^2+6)}{48}\,,\\
    \frac{\dabcd_A\dabcd_A}{N_A}&=\frac{\nc^4-6\nc^2+18}{96\nc^2}\,.
\end{align}
The five-loop result for the $\beta$ function can be found in \cite{Baikov:2016tgj,Herzog:2017ohr}.

\subsection{Ingredients for the Collinear Limit}\label{sec:colingredients}

The ingredients to achieve the NNLL resummation for the two-point energy correlator were computed in \cite{Dixon:2019uzg}. In the track case, only the jet function is different, which was presented in the main text. The hard function and anomalous dimensions are the same. We reproduce them here for completeness following the notation and conventions of \cite{Dixon:2019uzg}.

We expand the timelike splitting functions as
\begin{align}
P_{ij}(x)=\sum\limits_{L=0}^\infty
\left( \frac{\alpha_s}{4\pi} \right)^{L+1} P^{(L)}_{ij}(x)\,, 
\end{align}
and the $N=3$ moment as
\begin{align}
\gamma_{T,ij}^{(L)} = - \int\limits_0^1 \df x \, x^2 \, P^{(L)}_{ij}(x) \,,
\end{align}
where ``$T$'' simply denotes ``timelike''. 
It can be obtained to three loops from refs.~\cite{Mitov:2006wy,Mitov:2006ic,Moch:2007tx,Almasy:2011eq}. (Note that the pure singlet term is included in the $qq$ element.)  At LO, we have
\begin{align}
\gamma_{T,qq}^{(0)} &= \frac{25}{6} \, C_F\,, \qquad
\gamma_{T,gq}^{(0)}= -\frac{7}{6} \, C_F \,, \qquad
\gamma_{T,qg}^{(0)} = -\frac{7}{15} \, n_f \,, \qquad
\gamma_{T,gg}^{(0)} = \frac{14}{5}\, C_A + \frac{2}{3} \, n_f \,.
\end{align}
At NLO, we have
\begin{align}
\gamma_{T,qq}^{(1)}&=
\left( - 16 \zeta_3+ 24 \zeta_2-\frac{1693}{48} \right) C_F^2 
  + \left( 8\zeta_3 - \frac{86}{3} \zeta_2  +\frac{459}{8} \right) C_A C_F
  - \frac{5453}{1800} \, C_F n_f \,, \nn \\
\gamma_{T,gq}^{(1)} &= \left(\frac{28}{3}\zeta_2-\frac{2977}{432}  \right) C_F^2
+ \left(- \frac{14}{3}\zeta_2 -\frac{39451}{5400} \right) C_A C_F  \,, \nn \\
\gamma_{T,qg}^{(1)} &= \left(\frac{28}{15}\zeta_2 +\frac{619}{2700} \right) C_A n_f
- \frac{833}{216} \, C_F n_f - \frac{4}{25} \, n_f^2 \,, \nn \\
\gamma_{T,gg}^{(1)} &=
\left( - 8\zeta_3+ \frac{52}{15}\zeta_2+\frac{2158}{675}\right) C_A^2
+ \left(- \frac{16}{3}\zeta_2 +\frac{3803}{1350} \right) C_A n_f 
+ \frac{12839}{5400} \, C_F n_f \,.
\end{align}
At NNLO, we have
\begin{align}
\gamma_{T,qq}^{(2)}&=
\left(112\zeta_5+48\zeta_2\zeta_3- \frac{2083}{3} \zeta_4+ \frac{16153}{18} \zeta_3- \frac{13105}{72}\zeta_2- \frac{3049531}{31104} \right) C_F C_A^2  \nn \\
&+ \left(-432 \zeta_5-208 \zeta_2 \zeta_3+ \frac{8252}{3}\zeta_4- \frac{19424}{9} \zeta_3- \frac{16709}{27}\zeta_2+\frac{20329835}{15552} \right) C_F^2 C_A \nn \\
&+\left(416\zeta_5+224\zeta_2\zeta_3- \frac{6172}{3}\zeta_4+\frac{10942}{9}\zeta_3+\frac{11797}{18}\zeta_2- \frac{17471825}{15552} \right) C_F^3 \nn \\
&+\left(\frac{68}{3}\zeta_4-\frac{5803}{45}\zeta_3 + \frac{146971}{2700}\zeta_2-\frac{25234031}{1944000} \right) C_A C_F n_f \nn\\
&+\left(-\frac{136}{3}\zeta_4+\frac{8176}{45}\zeta_3-\frac{9767}{225}\zeta_2-\frac{4100189}{64800} \right) C_F^2 n_f -\frac{105799}{162000} \, C_F n_f^2 \,, \nn \\
\gamma_{T,gq}^{(2)} &=
\left(\frac{196}{3}\zeta_4-\frac{2791}{90}\zeta_3-\frac{50593}{600}\zeta_2-\frac{17093053}{777600} \right) C_F C_A^2 \nn \\
&+ \left(\frac{511}{3}\zeta_4-\frac{3029}{9}\zeta_3+\frac{123773}{900}\zeta_2+\frac{63294389}{388800} \right) C_F^2 C_A \nn \\
&+ \left(-308\zeta_4+\frac{2533}{9}\zeta_3+\frac{3193}{54}\zeta_2-\frac{647639}{3888} \right) C_F^3 \nn \\
&+ \left(\frac{182}{9}\zeta_3-\frac{73}{27}\zeta_2+\frac{246767}{60750} \right)C_A C_F n_f + \left(-\frac{28}{9}\zeta_3+\frac{4}{9}\zeta_2-\frac{419593}{81000} \right) C_F^2 n_f\,, \nn \\
\gamma_{T,qg}^{(2)} &=
\left(-\frac{252}{5}\zeta_4+\frac{343}{45}\zeta_3+\frac{333019}{13500}\zeta_2-\frac{1795237}{1944000} \right) C_A^2 n_f \nn \\
&+\left(-\frac{42}{5}\zeta_4+\frac{6208}{75}\zeta_3+\frac{24821}{1350}\zeta_2-\frac{3607891}{38880} \right) C_A C_F n_f \nn \\
&+\left(\frac{448}{15}\zeta_4-\frac{26102}{225}\zeta_3-\frac{2042}{225}\zeta_2+\frac{9397651}{97200}\right) C_F^2 n_f
+\left(-\frac{28}{9}\zeta_3-\frac{3616}{675}\zeta_2+\frac{1215691}{121500}\right) C_A n_f^2 \nn \\
&+ \left(\frac{3584}{675}\zeta_2-\frac{10657}{4050}\right) C_F n_f^2
 - \frac{172}{1125} \, n_f^3\,, \nn \\
\gamma_{T,gg}^{(2)} &=
\left(96\zeta_5+64\zeta_2 \zeta_3-\frac{2566}{15}\zeta_4-\frac{23702}{225}\zeta_3+\frac{66358}{1125}\zeta_2-\frac{5819653}{486000} \right)C_A^3 \nn \\
&+\left(104 \zeta_4+\frac{239}{9}\zeta_3-\frac{51269}{540}\zeta_2-\frac{12230737}{1944000} \right) C_A^2 n_f \nn \\
&+\left(\frac{282}{5}\zeta_3-\frac{16291}{675}\zeta_2-\frac{1700563}{108000} \right) C_A C_F n_f
+\left(-\frac{28}{9}\zeta_3+\frac{2411}{675}\zeta_2+\frac{219077}{194400} \right)C_F^2 n_f \nn \\
&+\left(-\frac{64}{9}\zeta_3+\frac{160}{27}\zeta_2-\frac{18269}{10125} \right) C_A n_f^2+\left(-\frac{196}{135}\zeta_2-\frac{2611}{162000} \right) C_F n_f^2\,.
\end{align}
Note that for $\gamma^{(2)}_{T,qg}$ the weight-2 terms with $C_A^2n_f, C_AC_Fn_f, C_An_f^2$ and $C_Fn_f^2$ are different from those listed in ref.~\cite{Dixon:2019uzg}, because we apply the updated $P_{qg}^{(2)}(x)$ calculated in ref.~\cite{Chen:2020uvt}.

We also require logarithmic moments of the timelike anomalous dimension, 
\begin{align}
\partial^n_N \gamma_{T,ij}^{(L)}\ =\  - \int\limits_0^1 \df x \, x^2 \, \ln^n \!x \, P^{(L)}_{ij}(x) \,,
\end{align}
which we denote with the shorthand $\dot\gamma \equiv \partial_N \gamma$
and $\ddot\gamma \equiv \partial^2_N \gamma$.
The required logarithmic moments to achieve NNLL accuracy are the first two moments of the LO splitting functions
\begin{align}
\dot{\gamma}_{T,qq}^{(0)}&= \left( 4\zeta_2 - \frac{385}{72} \right) C_F \,, \quad
\dot{\gamma}_{T,gq}^{(0)}= \frac{49}{72} \, C_F \,, \quad
\dot{\gamma}_{T,qg}^{(0)} = \frac{119}{900} \, n_f \,,
\dot{\gamma}_{T,gg}^{(0)} = \left( 4\zeta_2 - \frac{4319}{900} \right) C_A \,, \nn   \\
\ddot{\gamma}_{T,qq}^{(0)}&= \left( - 8\zeta_3 + \frac{3979}{432} \right) C_F \,, \quad
\ddot{\gamma}_{T,gq}^{(0)}= - \frac{331}{432} \, C_F \,, \quad
\ddot{\gamma}_{T,qg}^{(0)}= - \frac{2353}{27000} \, n_f  \,, \nn \\
\ddot{\gamma}_{T,gg}^{(0)}&= \left( - 8\zeta_3 + \frac{230353}{27000} \right) C_A \,, 
\end{align}
and the first moment of the NLO splitting functions
\begin{align}
\dot{\gamma}_{T,qq}^{(1)}&= \left(-56\zeta_4- \frac{158}{3}\zeta_3+\frac{385}{18}\zeta_2+\frac{152863}{1728}\right) C_F^2
+ \left(-12\zeta_4+\frac{41}{3}\zeta_3+\frac{307}{6}\zeta_2-\frac{35785}{432} \right) C_F C_A  \nn \\
&+ \left(\frac{16}{3}\zeta_3-\frac{40}{9}\zeta_2-\frac{101923}{108000} \right)  C_F n_f \,, \nn \\
\dot{\gamma}_{T,gq}^{(1)}&= \left(-\frac{49}{3}\zeta_3+\frac{59}{6}\zeta_2+\frac{956963}{108000} \right) C_F C_A
+ \left(14\zeta_3-\frac{275}{18}\zeta_2+\frac{8053}{1728} \right) C_F^2 \,, \nn \\
\dot{\gamma}_{T,qg}^{(1)} &= \left(\frac{42}{5}\zeta_3-\frac{92}{75}\zeta_2-\frac{1460321}{162000} \right) C_A n_f
+ \left(-\frac{28}{3}\zeta_3+\frac{178}{225}\zeta_2+\frac{46663}{4320} \right) C_F n_f  \nn \\
&+ \left(-\frac{28}{45}\zeta_2+\frac{18451}{20250} \right) n_f^2 \,, \nn \\
\dot{\gamma}_{T,gg}^{(1)} &= \left(-68\zeta_4-\frac{686}{15}\zeta_3+\frac{15338}{225}\zeta_2+\frac{3642257}{162000} \right) C_A^2 
+ \left(\frac{32}{3}\zeta_3-\frac{40}{9}\zeta_2-\frac{137323}{20250} \right) C_A n_f \nn \\
&- \frac{58247}{108000} \, C_F n_f .
\end{align}

We denote the moments of the hard function
for $e^+e^-$ annihilation  as
\begin{align}
\int_0^1 \df x \, x^2 \, H_{q,g}(x,\mu=Q)\ &=\ \sum\limits_{L=0}^\infty
\left( \frac{\alpha_s}{4\pi} \right)^{L} h_L^{q,g} \,, \nn\\
\int_0^1 \df x \, x^2 \, \ln x \, H_{q,g}(x,\mu=Q)\ &=\ \sum\limits_{L=1}^\infty
\left( \frac{\alpha_s}{4\pi} \right)^{L} \dot{h}_L^{q,g} \,.
\end{align}
These can be obtained from refs.~\cite{Mitov:2006ic,Moch:2007tx,Almasy:2011eq}.  To achieve NNLL accuracy, we require
\begin{align}
	h_0^q &= 2 \,, \qquad
	h_0^g = 0 \,, \qquad\qquad
	h_1^q = \frac{131}{4} \, C_F \,, \qquad
	h_1^g = - \frac{71}{12} \, C_F \,, \\
	h_2^q &= \left( 64 \zeta_4 - \frac{1172}{3} \zeta_3 - 166 \zeta_2
	+ \frac{2386397}{2592} \right) C_A C_F\nn\\
	&+ \left( - 128 \zeta_4 + \frac{1016}{3} \zeta_3 + \frac{1751}{18} \zeta_2
	- \frac{1105289}{5184} \right) C_F^2 
	+ \left( 32 \zeta_3 + \frac{118}{15} \zeta_2 - \frac{8530817}{54000} \right)
	C_F T_F n_f \,, \nn\\
	h_2^g &= \left( - \frac{76}{3} \zeta_3 + \frac{188}{45} \zeta_2
	- \frac{29802739}{324000} \right) C_A C_F
	+ \left( \frac{124}{3} \zeta_3 + \frac{523}{18} \zeta_2
	- \frac{674045}{5184} \right) C_F^2 \,, \nn\\
	\dot{h}_0^q&=0\,,\qquad \dot{h}_1^q = \left( 40 \zeta_3 + \frac{61}{3}  \zeta_2
	- \frac{5303}{72} \right) C_F  \,,\qquad	\dot{h}_0^g=0\,, \qquad
	\dot{h}_1^g = \left( - \frac{7}{3} \zeta_2 + \frac{31}{4} \right) C_F \,.\nn
\end{align}
Note that the normalization condition for the hard function in this paper is different from that in ref.~\cite{Dixon:2019uzg}, because here we use the energy weighting $E_i/Q$ in the jet function definition in contrast with $E_i/(Q/2)$ in ref.~\cite{Dixon:2019uzg}.

\subsection{Ingredients for the Back-to-Back Limit}\label{sec:b2bingredients}

In this appendix we provide the perturbative ingredients relevant for the back-to-back limit of the EEC. A detailed overview of all known perturbative data relevant for the description of the EEC in the back-to-back limit can be found in \cite{Moult:2022xzt}. For the perturbative expansion of the fixed-order ingredients $X$ and the corresponding anomalous dimensions $\gamma_X$ we use the following convention
\begin{align*}
    &X=\sum_{k=0}^\infty\biggl(\frac{\alpha_s}{4\pi}\biggr)^{k} X^{(k)}\,,
    &\gamma_X=\sum_{k=0}^\infty\biggl(\frac{\alpha_s}{4\pi}\biggr)^{k+1} \gamma_k^X\,.&
\end{align*}

The (quark) cusp anomalous dimension has been calculated up to four loops~\cite{Korchemsky:1987wg,Moch:2004pa,Moch:2018wjh,Moch:2017uml,Davies:2016jie,Henn:2019swt} and approximated at 5-loops~\cite{Herzog:2018kwj,Bruser:2019auj}. The results read
\begin{align}
    \Gamma_0^q&=4C_F\,,
    \\
    \Gamma_1^q&=\Bigl(-8\zeta_2+\frac{268}{9}\Bigr)\ca\cf
    +\Bigl(-\frac{80}{9}\Bigr)\nf\tf\cf
    \nonumber\\
    \Gamma_2^q&=\Bigl(88\zeta_4+\frac{88}{3}\zeta_3-\frac{1072}{9}\zeta_2+\frac{490}{3}\Bigr)\ca^2\cf
    +\Bigl(64\zeta_3-\frac{220}{3}\Bigr)\nf\tf\cf^2
    \nonumber\\
    &\quad+\Bigl(-\frac{224}{3}\zeta_3+\frac{320}{9}\zeta_2-\frac{1672}{27}\Bigr)\nf\tf\ca\cf
    +\Bigl(-\frac{64}{27}\Bigr)\nf^2\tf^2\cf
    \nonumber\\
    \Gamma_3^q&=
    \Bigl(-\frac{2504}{3}\zeta_6-16\zeta_3^2-\frac{3608}{9}\zeta_5-\frac{352}{3}\zeta_3\zeta_2+1804\zeta_4+\frac{20944}{27}\zeta_3-\frac{88400}{81}\zeta_2+\frac{84278}{81}\Bigr)\ca^3\cf
    \nonumber\\
    &\quad+\Bigl(\frac{4192}{9}\zeta_5+\frac{896}{3}\zeta_3\zeta_2-\frac{352}{3}\zeta_4-\frac{46208}{27}\zeta_3+\frac{40640}{81}\zeta_2-\frac{48274}{81}\Bigr)\nf\tf\ca^2\cf
    \nonumber\\
    &\quad
    +\Bigl(320\zeta_5-256\zeta_3\zeta_2-352\zeta_4+\frac{7424}{9}\zeta_3+\frac{880}{3}\zeta_2-\frac{68132}{81}\Bigr)\nf\tf\ca\cf^2
    \nonumber\\
    &\quad+\Bigl(-640\zeta_5+\frac{1184}{3}\zeta_3+\frac{1144}{9}\Bigr)\nf\tf\cf^3
    +\Bigl(128\zeta_4-\frac{2560}{9}\zeta_3+\frac{9568}{81}\Bigr)\nf^2\tf^2\cf^2
    \nonumber\\
    &\quad+\Bigl(-\frac{448}{3}\zeta_4+\frac{8960}{27}\zeta_3-\frac{2432}{81}\zeta_2+\frac{3692}{81}\Bigr)\nf^2\tf^2\ca\cf
    +\Bigl(\frac{512}{27}\zeta_3-\frac{256}{81}\Bigr)\nf^3\tf^3\cf
    \nonumber\\
    &\quad+\Bigl(-992\zeta_6-384\zeta_3^2+\frac{3520}{3}\zeta_5+\frac{128}{3}\zeta_3-128\zeta_2\Bigr)\frac{\dabcd_F\dabcd_A}{\nc}
    \nonumber\\
    &\quad+\Bigl(-\frac{1280}{3}\zeta_5-\frac{256}{3}\zeta_3+256\zeta_2\Bigr)\nf\frac{\dabcd_F\dabcd_F}{\nc}\,,
    \nonumber\\
    \Gamma_4^q&\approx 104.93\cf\,.
\nn\end{align}
The anomalous dimensions for the hard, jet and soft functions are given to 4-loop order. The relevant perturbative ingredients, and their assembly are detailed in \cite{Moult:2022xzt}. The relevant coefficients are given by
\begin{align}
    \gamma^H_0&=-6\cf\,,\\\nn
    \gamma^H_1&=\Bigl(52\zeta_3-22\zeta_2-\frac{961}{27}\Bigr)\ca\cf+\Bigl(-48\zeta_3+24\zeta_2-3\Bigr)\cf^2+\Bigl(8\zeta_2+\frac{260}{27}\Bigr)\nf\tf\cf\,,\\\nn
    \gamma^H_2&=\Bigl(-272\zeta_5-\frac{176}{3}\zeta_3\zeta_2-166\zeta_4+\frac{7052}{9}\zeta_3-\frac{14326}{81}\zeta_2-\frac{139345}{1458}\Bigr)\ca^2\cf\\\nn
    &\quad+\Bigl(-240\zeta_5-32\zeta_3\zeta_2+\frac{988}{3}\zeta_4-\frac{1688}{3}\zeta_3+\frac{820}{3}\zeta_2-\frac{151}{2}\Bigr)\ca\cf^2\\\nn
    &\quad+\Bigl(480\zeta_5+64\zeta_3\zeta_2-288\zeta_4-136\zeta_3-36\zeta_2-29\Bigr)\cf^3\\\nn
    &\quad+\Bigl(88\zeta_4-\frac{3856}{27}\zeta_3+\frac{10376}{81}\zeta_2-\frac{34636}{729}\Bigr)\nf\tf\ca\cf\\\nn
    &\quad+\Bigl(-\frac{560}{3}\zeta_4+\frac{1024}{9}\zeta_3-\frac{104}{3}\zeta_2+\frac{5906}{27}\Bigr)\nf\tf\cf^2\\\nn
    &\quad+\Bigl(-\frac{64}{27}\zeta_3-\frac{160}{9}\zeta_2+\frac{19336}{729}\Bigr)\nf^2\tf^2\cf\,,\\\nn
    \gamma^H_3&=3616.73\,,
    \\[3ex]
    \gamma^J_0&=3C_F\,,\\\nn
    \gamma^J_1&=\Bigl(24\zeta_3-12\zeta_2+\frac{3}{2}\Bigr)\cf^2+\Bigl(-12\zeta_3+\frac{44}{3}\zeta_2+\frac{17}{6}\Bigr)\ca\cf+\Bigl(-\frac{16}{3}\zeta_2-\frac{2}{3}\Bigr)\cf\nf\tf\,,\\\nn
    \gamma^J_2&=\Bigl(-240\zeta_5-32\zeta_3\zeta_2+144\zeta_4+68\zeta_3+18\zeta_2+\frac{29}{2}\Bigr)\cf^3\\\nn
    &\quad+\Bigl(120\zeta_5+16\zeta_3\zeta_2-\frac{494}{3}\zeta_4+\frac{844}{3}\zeta_3-\frac{410}{3}\zeta_2+\frac{151}{4}\Bigr)\ca\cf^2\\\nn
    &\quad+\Bigl(40\zeta_5-5\zeta_4-\frac{1552}{9}\zeta_3+\frac{4496}{27}\zeta_2-\frac{1657}{36}\Bigr)\ca^2\cf\\\nn
    &\quad+\Bigl(4\zeta_4+\frac{400}{9}\zeta_3-\frac{2672}{27}\zeta_2+40\Bigr)\ca\cf\nf\tf\\\nn
    &\quad+\Bigl(\frac{232}{3}\zeta_4-\frac{272}{3}\zeta_3+\frac{40}{3}\zeta_2-46\Bigr)\cf^2\nf\tf\\\nn
    &\quad+\Bigl(-\frac{64}{9}\zeta_3+\frac{320}{27}\zeta_2-\frac{68}{9}\Bigr)\cf\nf^2\tf^2\,,\\\nn
    \gamma^J_3&=707.276\,,
    \\[3ex]
    \gamma^S_0&=0\\\nn
    \gamma^S_1&=\Bigl(28\zeta_3+\frac{22}{3}\zeta_2-\frac{808}{27}\Bigr)\ca\cf
    +\Bigl(-\frac{8}{3}\zeta_2+\frac{224}{27}\Bigr)\cf\nf\tf\\\nn
    \gamma^S_2&=\Bigl(192\zeta_5-\frac{176}{3}\zeta_3\zeta_2-176\zeta_4+\frac{1316}{3}\zeta_3+\frac{12650}{81}-\frac{136781}{729}\Bigr)\ca^2\cf\\\nn
    &\quad+\Bigl(96\zeta_4-\frac{1456}{27}\zeta_3-\frac{5656}{81}\zeta_2+\frac{23684}{729}\Bigr)\ca\cf\nf\tf\\\nn
    &\quad+\Bigl(-32\zeta_4-\frac{608}{9}\zeta_3-8\zeta_2+\frac{3422}{27}\Bigr)\cf^2\nf\tf\\\nn
    &\quad+\Bigl(-\frac{448}{27}\zeta_3+\frac{160}{27}\zeta_2+\frac{8320}{729}\Bigr)\cf\nf^2\tf^2\,,\\\nn
    \gamma^S_3&=5031.28\,.
\end{align}
The perturbative boundary condition of the Collins-Soper kernel (see eq.~\eqref{eq:CS_boundary}) is described by the rapidity anomalous dimension. It has been calculated up to 4-loop order~\cite{Li:2016ctv,Moult:2022xzt,Duhr:2022yyp} with the coefficients given by
\begin{align}
    \gamma^r_0&=0\,,\\\nn
    \gamma^r_1&=\Bigl(28\zeta_3-\frac{808}{27}\Bigr)\cf\ca+\frac{112}{27}\nf\cf\\\nn
    \gamma^r_2&=\Bigl(192\zeta_5-\frac{176}{3}\zeta_3\zeta_2+\frac{154}{3}\zeta_4+\frac{12328}{27}\zeta_3+\frac{6392}{81}\zeta_2-\frac{297029}{729}\Bigr)\ca^2\cf\\\nn
    &\quad+\Bigl(\frac{20}{3}\zeta_4-\frac{904}{27}\zeta_3-\frac{824}{81}\zeta_2+\frac{62626}{729}\Bigr)\nf\ca\cf+\Bigl(-16\zeta_4-\frac{304}{9}\zeta_3+\frac{1711}{27}\Bigr)\nf\cf^2\\\nn
    &\quad+\Bigl(-\frac{32}{9}\zeta_3-\frac{1856}{729}\Bigr)\nf^2\cf\,,\\\nn
    \gamma^r_3&=6681.4\,.
\end{align}

The moments of the TMD matching coefficients can be written as a polynomial in the coupling, logarithms involving $b_\perp$ and $\mu$ and logarithms involving $Q$ and $\nu$. We write
\begin{align}
    \tilde{\cC}_{ji}(n,b_\perp,Q,\mu,\nu)
    &=
    \sum_{k=0}^\infty \sum_{l,m=0}^k
    a_s^k(\mu)\,
    \ln^l\Bigl(\frac{b_\perp^2 \mu^2}{\mu_0^2}\Bigr)
    \ln^m\Bigl(\frac{Q^2}{\nu^2}\Bigr)
    \tilde{\cC}_{ji}^{(k,l,m)}(n)\,.
\end{align}
The numerical values for the $n=1$ coefficients for SU(3) and with $n_f=5$ are provided in the ancillary file ``TMDC\_moments.m" up to three loops. 

The jet function that appears in the back-to-back factorization formula, or more specifically the TMD matching coefficients, can be written as a polynomial of logarithms involving the different scales. At any given order, all logarithmic terms can be constructed from the $\gamma_J$, $\gammacusp$ and the beta function. The constant terms can only be determined via explicit computation. They are obtained from moments of the TMD fragmentation functions, weighted by first moments of the quark or gluon track functions. The TMD fragmentation functions were computed to NNLO in \cite{Echevarria:2016scs,Luo:2019hmp,Luo:2019bmw}, and NNNLO in \cite{Ebert:2020qef,Luo:2020epw}

Since we believe that they will be of more general utility,  we present the analytic results for the constant terms of jet functions in the back-to-back limit on tracks to two loops. The jet function constants $c_k^{J_i}$ are defined by 
\begin{align}\label{eq:b2b_J_constants}
    J_i\bigl(b_\perp,Q,\tfrac{b_0}{b_\perp},Q\bigr)
    &=\sum_j T_j\bigl(1,\tfrac{b_0}{b_\perp}\bigr)\,
    \tilde{\cC}_{ji}(1,b_\perp,Q,\tfrac{b_0}{b_\perp},Q\bigr)
    =\sum_n c_k^{J_i} a_s^k\bigl(\tfrac{b_0}{b_\perp}\bigr)\,.
\end{align}
For the quark jet function to 2-loop order we have,
\begin{align}
    c_0^{J_q}&=T_q(1)\,, \nn \\
    c_1^{J_q}&=T_q(1) C_F\left(\frac{88}{9}-\frac{4\pi^2}{3}\right)-T_g^{(1)}C_F \frac{52}{9} \,, \nn\\
    c_2^{J_q}&=
    T_q(1) \left\{C_AC_F \left(-\frac{22 \zeta_3}{3}+\frac{55649}{648}-\frac{73 \pi ^2}{9}-\frac{\pi ^4}{18}\right)
    +C_F^2\left(-106\zeta_3+\frac{75293}{648}\right.\right.\nonumber\\
    &\quad\left.\left.-\frac{146 \pi^2}{9}+\frac{14\pi^4}{9}\right)
    +C_Fn_fT_F\left(\frac{8 \zeta_3}{3}-\frac{4837}{162}+\frac{20 \pi^2}{9}\right)+C_FT_F\left(\frac{1748}{81}+\frac{8\pi^2}{9}\right)\right\}\nonumber\\
    &\quad
    +T_g(1) \left\{C_AC_F\left(32\zeta_3-\frac{10427}{162}-\frac{16\pi^2}{9}\right)
    +C_F^2\left(32\zeta_3-\frac{17885}{162}+\frac{104\pi^2}{9}\right)\right\}\nonumber\\
    &\quad+\sum_{\substack{Q\\ (Q\neq q)}}T_{Q}(1) \ C_FT_F\left(\frac{1748}{81}+\frac{8\pi^2}{9}\right)
    \,.
\end{align}
For gluon jets we must consider both the unpolarized and linearly-polarized contribution, for which we denote the corresponding jet function constants by  $c_i^{J,g}$ and $c_i^{J^\prime,g}$, respectively.  The 2-loop results for the gluon jet functions read,
\begin{align}
    c_0^{J,g}=&T_g(1)\,,\\
    c_1^{J,g}=& T_g(1)\ C_A\left(\frac{65}{18}-\frac{4\pi^2}{3}\right)+\sum_q\left[T_q(1)+T_{\bar{q}}(1)\right]\ T_F\left(-\frac{5}{18}\right)\, , \\
    c_1^{J^\prime,g}=& T_g(1)\ C_A\frac{1}{3}+\sum_q\left[T_q(1)+T_{\bar{q}}(1)\right]\ T_F\left(-\frac{1}{3}\right)\,, \\
    c_2^{J,g}=& T_g(1)\left\{C_A^2 \left(-\frac{176 \zeta_3}{3}+\frac{2269}{81}-\frac{727 \pi ^2}{54}+\frac{3\pi ^4}{2}\right)\right.\\
    &\left.+N_fT_F\left[C_A\left(-\frac{32 \zeta_3}{3}+\frac{122}{27}+\frac{40 \pi ^2}{27}\right)+\frac{674}{81}C_F\right]\right\}\nonumber\\
    &+\sum_q\left[T_q(1)\!+\!T_{\bar{q}}(1)\right]\left\{T_F\left[C_A\left(8 \zeta_3\!-\!\frac{511}{162}\!+\!\frac{5 \pi^2}{6}\right)\!+\!C_F\left(8\zeta_3-\frac{1997}{162}\right)\right]\!-\!N_fT_F^2\frac{28}{81} \right\}\, ,\nonumber\\
    c_2^{J^\prime,g}=& T_g(1)\left\{\left(\frac{107}{27}-\frac{4 \pi ^2}{9}\right)C_A^2+N_fT_F\left(\frac{64}{9}C_F-\frac{70}{27}C_A\right)\right\}\nonumber\\
    &+\sum_q\left[T_q(1)+T_{\bar{q}}(1)\right]\left\{T_F\left[\left(\frac{4 \pi ^2}{9}-\frac{128}{27}\right)C_A-\frac{14}{9}C_F\right]
    +\frac{4}{27}N_fT_F^2\right\}
    \,.
\end{align}
 The results for gluon jets are not needed for the $e^+e^-$ collisions considered in this paper, but enter in a description of the TEEC at the LHC. Another application is to the EIC, for which we note that the track function formalism applies equally well to flavored particles (e.g.~strange) and can thus be used to probe the flavor structure of PDFs, as emphasized in ref.~\cite{Li:2021txc}.

The soft function for the EEC has been calculated up to three loops. We expand the soft function as
\begin{align}
S(b_\perp,\mu,\nu)&=1+\sum_{n=1}^{\infty}\biggl(\frac{\as(\mu)}{4\pi}\biggr)^n S^{(n)}(b_\perp,\mu,\nu)\,.
\end{align}
These coefficients can in turn be written in terms of the perturbative ingredients provided above, and the soft function constants $c_k^S$ 
\begin{align}
    S^{(1)}&=
    \frac{1}{2} \Gamma_0 L_\mu^2
    +\Bigl(-\Gamma_0 L_\nu-\gamma^S_0\Bigr)L_\mu
    +\gamma^r_0 L_\nu
    +c^S_1\,,\nn
    \\
    S^{(2)}&=
    \frac{1}{8} \Gamma_0^2 L_\mu^4
    +\biggl[-\frac{1}{2}\Gamma_0^2 L_\nu+\frac{1}{6}\beta_0 \Gamma_0-\frac{1}{2}\Gamma_0 \gamma^S_0\biggr]L_\mu^3
    \nonumber\\
    &\quad
    +\biggl[\frac{1}{2} \Gamma_0^2L_\nu^2
    +\Bigl(\Gamma_0 \gamma^S_0-\frac{1}{2} \beta_0\Gamma_0+\frac{1}{2} \Gamma_0\gamma^r_0\Bigr)L_\nu
    -\frac{1}{2} \beta_0 \gamma^S_0+\frac{1}{2}\Gamma_1+\frac{1}{2} \Gamma_0 c^S_1+\frac{1}{2}(\gamma^S_0)^2\biggr]L_\mu^2 
    \nonumber\\
    &\quad
    +\biggl[-\Gamma_0 \gamma^r_0L_\nu^2+\beta_0c^S_1-c^S_1 \gamma^S_0
    +\Bigl(\beta_0 \gamma^r_0-\Gamma_1-\gamma^r_0 \gamma^S_0-\Gamma_0 c^S_1
    \Bigr)L_\nu-\gamma^S_1\biggr]L_\mu
    \nonumber\\
    &\quad
    +\frac{1}{2} (\gamma^r_0)^2 L_\nu^2
    +\Bigl(\gamma^r_1+\gamma^r_0 c^S_1\Bigr)L_\nu
    +\frac{1}{2}(c^S_1)^2+c^S_2\,,\nn
    \\
    S^{(3)}&=
    \frac{1}{48} \Gamma_0^3 L_\mu^6
    +\biggl[-\frac{1}{8} \Gamma_0^3 L_\nu +\frac{1}{12} \beta_0 \Gamma_0^2-\frac{1}{8} \gamma^S_0 \Gamma_0^2\biggr]L_\mu^5
    \nonumber\\
    &\quad
    +\biggl[\frac{1}{4} \Gamma_0^3 L_\nu^2
    +\Bigl(-\frac{5}{12}\beta_0 \Gamma_0^2+\frac{1}{8}\gamma^r_0 \Gamma_0^2+\frac{1}{2}\gamma^S_0 \Gamma_0^2\Bigr)L_\nu 
    \nonumber\\
    &\quad\qquad
    +\frac{1}{8} c^S_1 \Gamma_0^2+\frac{1}{12} \beta_0^2 \Gamma_0+\frac{1}{4} (\gamma^S_0)^2 \Gamma_0+\frac{1}{4} \Gamma_1 \Gamma_0-\frac{5}{12} \beta_0 \gamma^S_0 \Gamma_0\biggr]L_\mu^4
    \nonumber\\
    &\quad
    +\biggl[
    -\frac{1}{6} L_\nu^3 \Gamma_0^3
    +\Bigl(\frac{1}{2} \beta_0\Gamma_0^2-\frac{1}{2} \gamma^r_0\Gamma_0^2-\frac{1}{2} \gamma^S_0\Gamma_0^2\Bigr)L_\nu^2 
    \nonumber\\
    &\quad\qquad
    +\Bigl(-\frac{1}{3}\Gamma_0 \beta_0^2+\frac{2}{3}\Gamma_0 \gamma^r_0 \beta_0+\Gamma_0 \gamma^S_0\beta_0-\frac{1}{2} c^S_1 \Gamma_0^2-\frac{1}{2} \Gamma_0 (\gamma^S_0)^2-\Gamma_0 \Gamma_1-\frac{1}{2} \Gamma_0 \gamma^r_0 \gamma^S_0\Bigr)L_\nu 
    \nonumber\\
    &\quad\qquad
    +\frac{2}{3} c^S_1 \beta_0\Gamma_0+\frac{1}{6} \beta_1\Gamma_0-\frac{1}{2} c^S_1 \gamma^S_0 \Gamma_0-\frac{1}{2} \gamma^S_1\Gamma_0-\frac{1}{6}(\gamma^S_0)^3+\frac{1}{2} \beta_0 (\gamma^S_0)^2 \nn\\
    &\hspace{8cm}+\frac{1}{3} \beta_0 \Gamma_1-\frac{1}{3}\beta_0^2 \gamma^S_0-\frac{1}{2} \Gamma_1 \gamma^S_0\biggr]L_\mu^3
    \nonumber\\
    &\quad
    +\biggl[\frac{1}{2} \Gamma_0^2 \gamma^r_0 L_\nu^3
    +\Bigl(\frac{1}{2} c^S_1\Gamma_0^2+\frac{1}{4} (\gamma^r_0)^2\Gamma_0+\Gamma_1 \Gamma_0-\frac{3}{2} \beta_0 \gamma^r_0 \Gamma_0+\gamma^r_0\gamma^S_0 \Gamma_0\Bigr)L_\nu^2
    \nonumber\\
    &\quad\qquad
    +\Bigl(\gamma^r_0 \beta_0^2-\frac{3}{2} c^S_1 \Gamma_0\beta_0-\Gamma_1 \beta_0-\frac{3}{2} \gamma^r_0 \gamma^S_0 \beta_0+\frac{1}{2} \gamma^r_0 (\gamma^S_0)^2\nn \\
    &\hspace{2cm}-\frac{1}{2} \beta_1 \Gamma_0+\frac{1}{2} c^S_1\Gamma_0 \gamma^r_0+\frac{1}{2}\Gamma_1 \gamma^r_0+\frac{1}{2}\Gamma_0 \gamma^r_1+c^S_1\Gamma_0 \gamma^S_0+\Gamma_1 \gamma^S_0+\Gamma_0\gamma^S_1 \Bigr) L_\nu
    \nonumber\\
    &\quad\qquad
    +c^S_1 \beta_0^2+\frac{1}{2} c^S_1 (\gamma^S_0)^2+\frac{1}{4} (c^S_1)^2 \Gamma_0+\frac{1}{2} c^S_2 \Gamma_0+\frac{1}{2} c^S_1 \Gamma_1+\frac{\Gamma_2}{2}-\frac{3}{2}c^S_1 \beta_0 \gamma^S_0\nn \\
    &\hspace{4cm}-\frac{1}{2} \beta_1 \gamma^S_0-\beta_0 \gamma^S_1+\gamma^S_0 \gamma^S_1\biggr]L_\mu^2
    \nonumber\\
    &\quad
    +\biggl[-\frac{1}{2} \Gamma_0 (\gamma^r_0)^2 L_\nu^3
    +\Bigl(\beta_0(\gamma^r_0)^2-\frac{1}{2} \gamma^S_0(\gamma^r_0)^2-c^S_1 \Gamma_0\gamma^r_0-\Gamma_1 \gamma^r_0-\Gamma_0 \gamma^r_1\Bigr) L_\nu^2
    \nonumber\\
    &\quad\qquad
    +\Bigl(-\frac{1}{2} \Gamma_0 (c^S_1)^2-\Gamma_1 c^S_1+2\beta_0 \gamma^r_0c^S_1-\gamma^r_0 \gamma^S_0c^S_1-c^S_2 \Gamma_0-\Gamma_2+\beta_1 \gamma^r_0 \nn \\
    &\hspace{6cm}+2\beta_0 \gamma^r_1-\gamma^r_1 \gamma^S_0-\gamma^r_0\gamma^S_1\Bigr) L_\nu
    \nonumber\\
    &\quad\qquad
    +(c^S_1)^2\beta_0+2 c^S_2 \beta_0+c^S_1 \beta_1-\frac{1}{2}(c^S_1)^2 \gamma^S_0-c^S_2\gamma^S_0-c^S_1 \gamma^S_1-\gamma^S_2\biggr]L_\mu
    \nonumber\\
    &\quad
    +\frac{(c^S_1)^3}{6}+\frac{1}{6} L_\nu^3(\gamma^r_0)^3+c^S_1c^S_2+c^S_3+L_\nu^2 \left(\frac{1}{2}c^S_1 (\gamma^r_0)^2+\gamma^r_1 \gamma^r_0\right)\nn \\
    &\quad+L_\nu\left(\frac{1}{2} \gamma^r_0(c^S_1)^2+\gamma^r_1c^S_1+c^S_2 \gamma^r_0+\gamma^r_2\right)
\end{align}
where the soft function constants are given by, \
\begin{align}
    c_1^S&=-2\zeta_2\cf\,, \nn\\
    c_2^S&=\Bigl(10\zeta_4-\frac{154}{9}\zeta_3-\frac{67}{3}\zeta_2+\frac{2428}{81}\Bigr)\ca\cf
    +\Bigl(\frac{56}{9}\zeta_3+\frac{20}{3}\zeta_2-\frac{656}{81}\Bigr)\nf\tf\cf\,,\nn \\
    c_3^S&=\Bigl(-\frac{3086}{27}\zeta_6+\frac{928}{9}\zeta_3^2+\frac{1804}{9}\zeta_5+\frac{1100}{9}\zeta_3\zeta_2+\frac{3649}{27}\zeta_4 \nn \\
    &\hspace{4cm} -\frac{151132}{243}\zeta_3-\frac{297481}{729}\zeta_2+\frac{5211949}{13122}\Bigr)\cf\ca^2 \nn \\
    &\quad+\Bigl(-\frac{184}{3}\zeta_5+\frac{40}{9}\zeta_3\zeta_2-\frac{416}{27}\zeta_4+\frac{8152}{81}\zeta_3+\frac{74530}{729}\zeta_2-\frac{412765}{6561}\Bigr)\nf\cf\ca \nn \\
    &\quad+\Bigl(\frac{224}{9}\zeta_5-\frac{80}{3}\zeta_3\zeta_2+\frac{152}{9}\zeta_4+\frac{3488}{81}\zeta_3+\frac{275}{9}\zeta_2-\frac{42727}{486}\bigr)\nf\cf^2 \nn \\
    &\quad+\Bigl(-\frac{44}{27}\zeta_4-\frac{560}{243}\zeta_3-\frac{136}{27}\zeta_2-\frac{256}{6561}\Bigr)\nf^2\cf\,,
\end{align}
and
\begin{align}
    &L_\mu=\ln\Bigl(\frac{b_\perp^2\mu^2}{b_0^2}\Bigr)\,,
    &L_\nu=\ln\Bigl(\frac{b_\perp^2\nu^2}{b_0^2}\Bigr)\,.&
\end{align}

The hard function for the back-to-back limit of the energy correlator has been calculated up to three loops \cite{Baikov:2009bg,Lee:2010cga,Gehrmann:2010ue}. The results read
\begin{align}
    H^{(1)}
    &=
    -\frac{1}{2}\Gamma_0 L_h^2 +\gamma^H_1 L_h +c^H_1\,, \nn
    \\
    H^{(2)}
    &=
    \frac{1}{8}\Gamma_0^2 L_h^4
    +\Bigl(-\frac{1}{6}\beta_0\Gamma_0-\frac{1}{2}\gamma^H_0\Gamma_0\Bigr)L_h^3
    +\Bigl(-\frac{1}{2}\Gamma_1+\frac{1}{2}\beta_0\gamma^H_0+\frac{1}{2}(\gamma^H_0)^2-\frac{1}{2}c^H_1\Gamma_0\Bigr)L_h^2
    \nonumber\\
    &\quad
    +\Bigl(\gamma_1^H + c^H_1 \beta_0 + c^H_1 \gamma^H_0\Bigr) L_h
    +c^H_2\,,\nn
    \\
    H^{(3)}
    &=
    -\frac{1}{48}\Gamma_0^3 L_h^6
    +\Bigl(\frac{1}{12}\beta_0 \Gamma_0^2 + \frac{1}{8}\Gamma_0^2 \gamma^H_0\Bigr)L_h^5
    \nonumber\\
    &\quad
    +\Bigl(-\frac{1}{12}\beta_0^2 \Gamma_0 + \frac{1}{8}c^H_1 \Gamma_0^2+\frac{1}{4}\Gamma_0\Gamma_1-\frac{5}{12}\beta_0\Gamma_0\gamma^H_0-\frac{1}{4}\Gamma_0(\gamma^H_0)^2\Bigr)L_h^4
    \nonumber\\
    &\quad
    +\Bigl(-\frac{2}{3}c^H_1\beta_0\Gamma_0-\frac{1}{6}\beta_1\Gamma_0-\frac{1}{3}\beta_0\Gamma_1+\frac{1}{3}\beta_0^2 \gamma^H_0-\frac{1}{2}c^H_1\Gamma_0\gamma^H_0\nn \\
    &\hspace{4cm} -\frac{1}{2}\Gamma_1\gamma^H_0+\frac{1}{2}\beta_0(\gamma^H_0)^2+\frac{1}{6}(\gamma^H_0)^3-\frac{1}{2}\Gamma_0\gamma^H_1\Bigr)L_h^3
    \nonumber\\
    &\quad
    +\Bigl(c^H_1 \beta_0^2 -\frac{1}{2}c^H_2\Gamma_0-\frac{1}{2}c^H_1\Gamma_1-\frac{1}{2}\Gamma_2+\frac{3}{2}c^H_1\beta_0\gamma^H_0+\frac{1}{2}\beta_1\gamma^H_0\nn\\ 
    & \hspace{4cm}+\frac{1}{2}c^H_1 (\gamma^H_0)^2 + \beta_0\gamma^H_1+\gamma^H_0\gamma^H_1\Bigr)L_h^2
    \nonumber\\
    &\quad
    +\Bigl(2c^H_2\beta_0+c^H_1\beta_1+c^H_2\gamma^H_0+c^1H\gamma^H_1
    +\gamma^H_2\Bigr)L_h
    +c^H_3\,,
\end{align}
with
\begin{align}
    c_1^H
    &=
    (14\zeta_2-16)\cf\,,\nn
    \\
    c_2^H
    &=
    \Bigl(-16\zeta_4+\frac{626}{9}\zeta_3+\frac{1061}{9}\zeta_2-\frac{51157}{324}\Bigr)\cf^2
    +\Bigl(201\zeta_4-60\zeta_3-166\zeta_2+\frac{511}{4}\Bigr)\ca\cf
    \nonumber\\
    &\quad
    +\Bigl(\frac{8}{9}\zeta_3-\frac{364}{9}\zeta_2+\frac{4085}{81}\Bigr)\nf\tf\cf\,,\nn
    \\
    c_3^H&=8998.080\,,
\end{align}
and
\begin{align}
    L_h=\ln\Bigl(\frac{\mu^2}{Q^2}\Bigr)\,.
\end{align}

\subsection{Total Cross Section}\label{sec:total_xsec}

For comparison with experimental results, it is convenient to normalize the energy correlator to the total cross section, $\sigma$.  An excellent review of knowledge of the total cross section is \cite{Chetyrkin:1996ela}. Since we have only performed the calculation on massless quarks in QCD, here we restrict ourselves to providing the R-ratio for massless quarks. This could be improved in more sophisticated future treatments.

We expand the non-singlet R-ratio as
\begin{align}
r(Q)=\frac{\sigma}{\sigma_0}=1+\sum_n \biggl( \frac{\alpha_s(Q)}{4\pi} \biggr)^n r_n\,.
\end{align}
The coefficients to three-loops are given by \cite{Herzog:2017dtz}
\begin{align}
r_1&=3 C_F\,, \\
r_2&=-\frac{3}{2}C_F^2 +C_A C_F \left(\frac{123}{2}-44 \zeta_3 \right) -C_F n_f (11-8 \zeta_3)\,,\nn \\
r_3 &=
       - \frac{69}{2}\, C_F^3 
       \,- \,\ca C_F^2 \* 
        \, \*  \Big[
            127
          + 572\, \* \zeta_3
          - 880\, \* \zeta_5
          \Big]\nn
 \\ &  \quad
       + C_A^2 \cf \, \*  \Bigg[
            \frac{90445}{54}
          - \frac{242}{3}\, \* \zeta_2
          - \frac{10948}{9}\, \* \zeta_3
          - \frac{440}{3}\, \* \zeta_5
          \Bigg]
       \,-\,C_F^2
       \, \* \nf \, \*  \Bigg[
            \frac{29}{2}
          - 152\, \* \zeta_3
          + 160\, \* \zeta_5
          \Bigg]
\nn \\ & \quad
       - \,\ca \* \cf\, \* \nf \, \*  \Bigg[
            \frac{15520}{27}
          - \frac{88}{3}\, \* \zeta_2
          - \frac{3584}{9}\, \* \zeta_3
          - \frac{80}{3}\, \* \zeta_5
          \Bigg]
       \,+\,\cf\, \* n_f^2
        \, \*  \Bigg[
            \frac{1208}{27}
          - \frac{8}{3}\, \* \zeta_2
          - \frac{304}{9}\, \* \zeta_3
          \Bigg]\,. \nn
\end{align}
The singlet contributions for both the vector and axial current are also known. Since we do not include them in our NNLO calculation of the bulk distribution of the energy correlator, we also do not include them in the calculation of the total cross section, so as to preserve the sum rule. This can also easily be improved in a future analysis.

\bibliography{EEC_ref.bib}
\bibliographystyle{JHEP}

\end{document}